\documentclass[aps,prb,twocolumn,amsmath,amssymb,superscriptaddress,floatfix,nofootinbib,]{revtex4-2}
\usepackage{amsthm}
\usepackage{amsmath}
\usepackage{graphicx}
\usepackage{amssymb}
\usepackage{bm} 
\usepackage{flafter}
\usepackage{slashed} 
\usepackage{color} 
\usepackage{bbm}
\usepackage{amsfonts}
\usepackage{cases}
\usepackage{float}
\usepackage{graphicx}
\usepackage{physics}
\usepackage{subfigure}
\usepackage{xurl}
\usepackage{framed}
\usepackage{bbold}
\usepackage{cancel,soul,ulem}
\usepackage{appendix}
\usepackage{mathtools} 
\usepackage[breaklinks=true,colorlinks,citecolor=blue,linkcolor=blue,urlcolor=blue]{hyperref}

\newcommand{\blue}[1]{{\textcolor{blue}{#1}}}

\newcommand{\red}[1]{{\textcolor{red}{#1}}} 

\def\=={\raisebox{0.35pt}{$\mathrm{:}$}\!\!=}

\newcommand{\bea}{\begin{eqnarray}}
	\newcommand{\eea}{\end{eqnarray}}

\newcommand{\bracket}[1]{\langle #1 \rangle}

\theoremstyle{definition}

\theoremstyle{remark}

\begin{document}

\title{
Incommensurate gapless ferromagnetism connecting competing\linebreak
symmetry-enriched deconfined quantum phase transitions
      }

\author{Anthony Rey}
\affiliation{
Condensed Matter Theory Group,
PSI Center for Scientific Computing,
Theory and Data,
5232 Villigen PSI, Switzerland
            }

\author{\"{O}mer M. Aksoy}
\affiliation{
Department of Physics, Massachusetts Institute of Technology, Cambridge,
Massachusetts 02139, USA
            }

\author{Daniel P. Arovas}
\affiliation{
Department of Physics,
University of California at San Diego, La Jolla,
California 92093, USA
            }

\author{Claudio Chamon}
\affiliation{
Department of Physics, Boston University, Boston,
Massachusetts 02215, USA
            }
\affiliation{
Department of Physics and Astronomy, Purdue University,
West Lafayette, Indiana 47907, USA
            }

\author{Christopher Mudry}
\affiliation{
Condensed Matter Theory Group,
PSI Center for Scientific Computing,
Theory and Data,
5232 Villigen PSI, Switzerland
            }
\affiliation{
Institut de Physique, EPF Lausanne, CH-1015 Lausanne, Switzerland
            }  

\date{\today}

\begin{abstract}
We present a scenario, in which
a gapless extended phase serves as a ``hub'' connecting multiple
symmetry-enriched deconfined quantum critical points. As a concrete
example, we construct a lattice model with
$\mathbb{Z}^{\,}_{2}\times
\mathbb{Z}^{\,}_{2}\times
\mathbb{Z}^{\,}_{2}$
symmetry for quantum spin-1/2 degrees of freedom that realizes four
distinct gapful phases supporting antiferromagnetic long-range order
and one extended incommensurate gapless ferromagnetic phase. The
quantum phase transition between any two of the four gapped and
antiferromagnetic phases goes through either a (deconfined) quantum
critical point, a quantum tricritical point, or the incommensurate
gapless ferromagnetic phase. In this phase diagram, it is
possible to interpolate between four deconfined quantum critical
points by passing through the extended gapless ferromagnetic phase.
We identify the phases in the model and the nature of the transitions
between them through a combination of analytical arguments and density
matrix renormalization group studies.
\end{abstract}

\maketitle


\textit{Introduction.}
In one-dimensional space,
when periodic boundary conditions (PBCs) hold, 
there is a single quantum phase of bosonic matter
that is described by a nondegenerate and gapped ground state%
~\cite{Chen10083745}. For example,
any local Hamiltonian
hosting quantum spin degrees of freedom, 
obeying PBCs, and supporting 
a nondegenerate gapped ground state
can be adiabatically connected without gap closing,
up to adding or removing ancilla degrees of freedom,
to a trivial quantum paramagnet whose ground state
is a product state of the quantum spins.
This picture changes in the presence of symmetries, namely,
when the adiabatic evolution is restricted to be invariant under a
symmetry transformation from a group $G$.
There are then degenerate and gapped ground states
that cannot be connected to a $G$-symmetric product state
without gap closing or spontaneous breaking of $G$.
This possibility leads to the concept of
symmetry-protected topological (SPT) 
phases~\cite{Gu09031069,Pollmann09094059,Pollmann09101811,
Chen10083745,Chen11033323,Fidkowski10084138}.

\begin{figure}
\includegraphics[width=1\columnwidth]{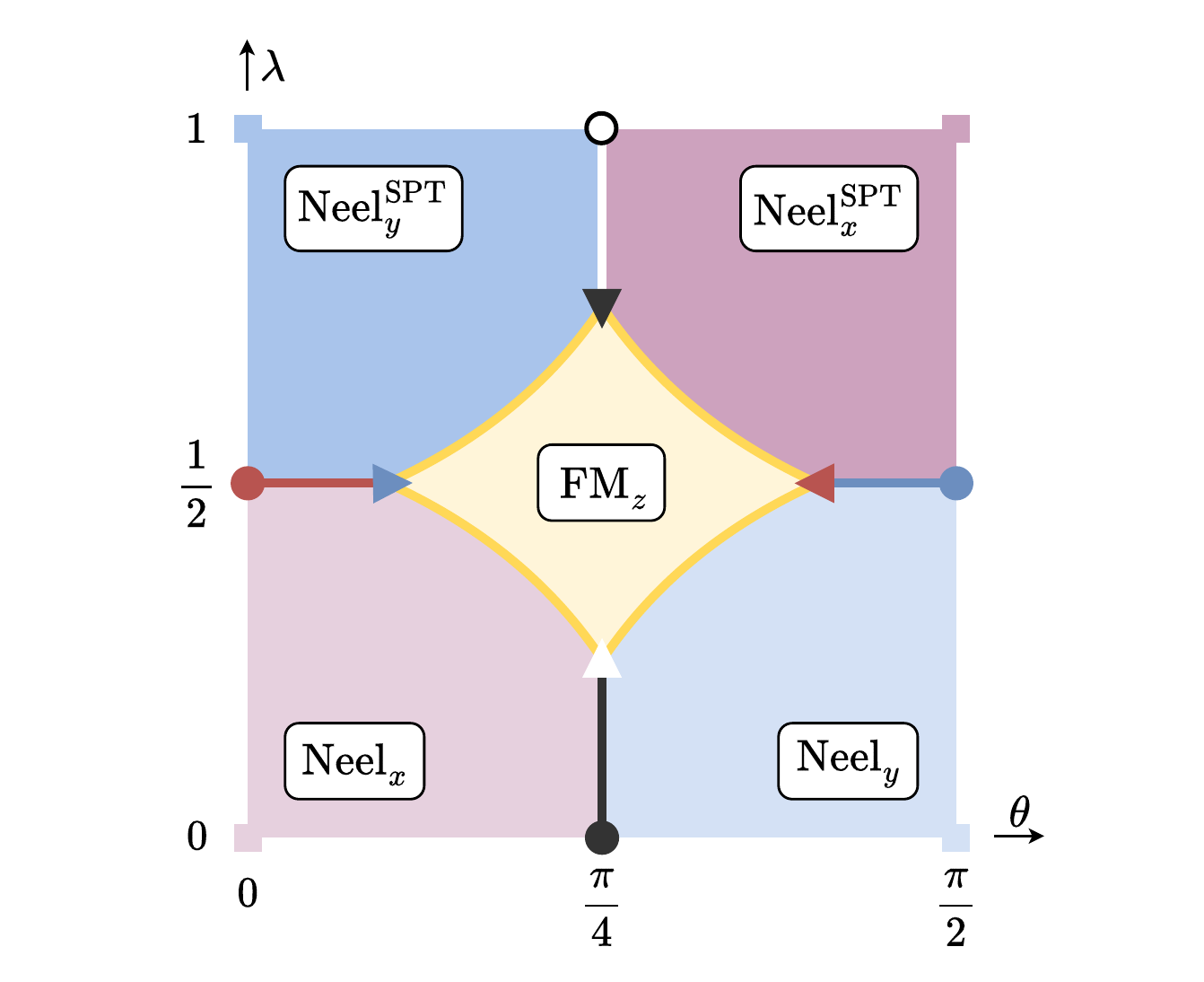}
\vspace{-20pt}
\caption{(Color online)
Phase diagram of Hamiltonian~\eqref{eq:def H}.
Squares (circles) on the boundary denote stable (unstable) fixed points
(DQCPs). 
}
\label{fig:phase diagram}
\vspace{-10pt}
\end{figure}

Just like gapped phases, quantum critical 
points or gapless phases can also be enriched by
symmetries~\cite{Kestner11,Sau11,Fidkowski11,Keselman150202037,
Ruhman17,Scaffidi170501557,Jiang18,Verresen18,Parker18,
Keselman18,Jones19,Ben-Zion20,Wang20,verresen2020topology,Duque21,Thorngren21,
Borla21,Verresen21,Hidaka22,Yu22,chang2022absence,prembabu2022boundary,Yang23,
Wen23,Mondal23,Yang23b,wang2023stability,li2023intrinsicallypurely,
huang2023topological,wen2023classification,Su24,yu2024universal,
ando2024gauging,zhong2024topological,yu2024quantum,li2024decorated,
zhang2024quantum,yang2024gifts}.
For example, when enriched by a global
$\mathbb{Z}^{\,}_{2}\times\mathbb{Z}^{\,}_{2}$ symmetry,
there are two kinds of Ising conformal field theories 
(CFT)~\cite{Scaffidi170501557}.
It is believed that any smooth path in the coupling space of a microscopic,
local, and $\mathbb{Z}^{\,}_{2}\times\mathbb{Z}^{\,}_{2}$-symmetric Hamiltonian
that interpolates between these two Ising CFT must go through either
a multicritical point or a gapped phase.

In this work we offer a scenario where a gapless
extended phase serves as a ``hub'' connecting multiple symmetry-enriched
deconfined quantum critical points (DQCPs). We study the interplay of four
symmetry-enriched DQCPs, with each one
describing a continuous quantum phase transition
between two gapped phases of matter supporting distinct long-range orders.
While two of the four DQCPs separate two gapped phases realizing
the same SPT phase,
the other two DQCPs separate those phases
that realize distinct SPT gapped phases of matter.
Hence, the latter two DQCPs can also be interpreted as examples of
quantum topological phase transitions (QTPTs).
We construct a local,
$\mathbb{Z}^{\,}_{2}\times\mathbb{Z}^{\,}_{2}\times\mathbb{Z}^{\,}_{2}$-symmetric
lattice Hamiltonian [Eq.~\eqref{eq:def H}]
for quantum spin-1/2 degrees of freedom
that depends on two dimensionless couplings and
that interpolates between all four DQCPs.
We show that any two of the four DQCPs
cannot be smoothly connected in the parameter space
of Fig.\ \ref{fig:phase diagram}
without encountering a multicritical point or a gap opening.
Surprisingly, we find that the competition
between the four symmetry-enriched DQCPs leads to an extended gapless 
phase in the center of the phase diagram
(see Fig.~\ref{fig:phase diagram}), 
which is connected to all four symmetry-enriched
DQCPs through four tricritical points. This phase
(i) supports a pair of degenerate ground states that are
ferromagnetically ordered along the quantization $Z$ axis in spin space;
(ii) is incommensurate in that 
the longitudinal spin-spin correlation function
is long-ranged with, for both periodic and
open boundary conditions (OBCs),
an oscillatory factor
$e^{\mathrm{i}q^{z}s}$,
whereby the wave vector $q^{z}$ is not commensurate 
with the lattice and $s$ is the separation between two spins; and
(iii) is gapless in that the transverse spin-spin
correlation functions decay algebraically with the wave vector $\pi$.
We call this phase the incommensurate gapless ferromagnetic phase.
Our work thus provides an explicit example of
an exotic gapless phase resulting from
the competition between different symmetry-enriched DQCPs.

\textit{Model and its symmetries.}
We define the family of Hamiltonians parametrized by two dimensionless 
couplings $\theta\in[0,\pi/2]$ and $\lambda\in [0,1]$ by 
\begin{subequations}
\label{eq:def H}
\begin{align}
\label{eq:def Ha}
&
\widehat{H}^{\,}_{b}(\theta,\lambda)
\coloneqq
(1-\lambda)
\sum_{j=1}^{2N-b}
\widehat{h}^{\,}_{j}(\theta)
+
\lambda
\sum_{j=1}^{2N-3b}
\widehat{Z}^{\,}_{j}\,
\widehat{h}^{\,}_{j+1}(\theta)\,
\widehat{Z}^{\,}_{j+3},
\end{align}
with
\begin{align}
\label{eq:def Hb}
&
\widehat{h}^{\,}_{j}(\theta)\coloneqq\,
\cos\theta\,
\widehat{X}^{\,}_{j}\,
\widehat{X}^{\,}_{j+1}
+
\sin\theta\,
\widehat{Y}^{\,}_{j}\,
\widehat{Y}^{\,}_{j+1}.
\end{align}
\end{subequations}
Here,
$\widehat{X}^{\,}_{j}$,
$\widehat{Y}^{\,}_{j}$,
and
$\widehat{Z}^{\,}_{j}$
are on-site Pauli operators that
commute on unequal sites and
satisfy the Pauli algebra on the same site, e.g.,
$\widehat{X}^{\,}_{j}\,\widehat{Y}^{\,}_{j}=\mathrm{i}\widehat{Z}^{\,}_{j}$.
Periodic boundary conditions follow from choosing
$b=0$ under the assumption that
$\widehat{X}^{\,}_{j+2N}\equiv \widehat{X}^{\,}_{j}$
and $\widehat{Y}^{\,}_{j+2N}\equiv\widehat{Y}^{\,}_{j}$. 
Open boundary conditions follow from choosing $b=1$.
The number of sites $2N$, with
$N$ being a positive integer, is explicitly even.

When $\theta\in[0,\pi/2]$ and $\lambda=0$,
Hamiltonian~\eqref{eq:def H}
reduces to that of
the anisotropic, nearest-neighbor, and antiferromagnetic
quantum spin-1/2 $XY$ model.
As the coupling $\lambda$ is tuned from $0$
to $1$, the two-body $XY$ interactions in Hamiltonian~\eqref{eq:def H} 
compete with the four-body interactions obtained by dressing each
two-body $XY$ interaction with a two-body third-neighbor $ZZ$ interaction.
We call the limit $\lambda=1$
of Hamiltonian~\eqref{eq:def H}
the ``dressed'' quantum spin-1/2 $XY$ model.

For any pair $\theta\in[0,\pi/2]$ and $\lambda\in [0,1]$,
Hamiltonian~\eqref{eq:def H} 
is invariant under a
$\mathbb{Z}^{X}_{2}\times\mathbb{Z}^{Y}_{2}\times\mathbb{Z}^{T}_{2}$
symmetry generated by the operators 
\begin{align}
\widehat{X}\coloneqq
\prod_{j=1}^{2N}
\widehat{X}^{\,}_{j},
\quad
\widehat{Y}\coloneqq
\prod_{j=1}^{2N}
\widehat{Y}^{\,}_{j},
\quad
\widehat{T}\coloneqq
\widehat{Y}\,
\mathsf{K},
\label{eq:def UX UY UT}
\end{align}
respectively, where $\mathsf{K}$ denotes complex conjugation.
The subgroup
$\mathbb{Z}^{X}_{2}\times\mathbb{Z}^{Y}_{2}$
implements global $\pi$ rotations around the $X$, $Y$, and $Z$
axes in spin space, while the subgroup $\mathbb{Z}^{T}_{2}$ implements
the reversal of time. Because the number of sites $2N$ is even,
the symmetry generators in Eq.~\eqref{eq:def UX UY UT}
commute pairwise.

\textit{Gapped phases.}
In the coupling space $(\theta,\lambda)$
from Fig.~\ref{fig:phase diagram},
Hamiltonian~\eqref{eq:def H}
realizes four gapped phases, which we label as
$\mathrm{\hbox{N\'eel}}^{\,}_{x}$,
$\mathrm{\hbox{N\'eel}}^{\,}_{y}$,
$\mathrm{\hbox{N\'eel}}^{\mathrm{SPT}}_{y}$,
and $\mathrm{\hbox{N\'eel}}^{\mathrm{SPT}}_{x}$.
The properties of each gapped phase are captured by the exactly
solvable point in the corresponding corner of the
phase diagram, marked by a square in Fig.~\ref{fig:phase diagram}.

The $\mathrm{\hbox{N\'eel}}^{\,}_{x(y)}$ phase is controlled by 
the exactly solvable point at ${\theta=0\,(\pi/2)}$ and $\lambda=0$, 
i.e., the lower left (right) corner in Fig.~\ref{fig:phase diagram}.
At this point, Hamiltonian~\eqref{eq:def H} reduces to the classical
Ising antiferromagnet with $XX$ ($YY$) couplings. 
Consequently, the $\mathrm{\hbox{N\'eel}}^{\,}_{x(y)}$ phase is long-range ordered 
with spontaneously broken 
$\widehat{Y}$ ($\widehat{X}$)
and
$\widehat{T}$
symmetries. In the thermodynamic limit, the ground states 
are twofold degenerate for both phases independently of the choice
between PBCs or OBCs.

The $\mathrm{\hbox{N\'eel}}^{\mathrm{SPT}}_{y(x)}$ phase is controlled by 
the exactly solvable stable fixed point at
${\theta=0\,(\pi/2)}$ and $\lambda=1$, 
i.e., the upper left (right) corner in Fig.~\ref{fig:phase diagram}.
At this point, Hamiltonian~\eqref{eq:def H} reduces to that
of a cluster model~\cite{Suzuki1971}
with $ZXXZ$ ($ZYYZ$) couplings. When PBCs are chosen,
the $\mathrm{\hbox{N\'eel}}^{\mathrm{SPT}}_{y(x)}$
phase describes the same long-range order
as the $\mathrm{\hbox{N\'eel}}^{\,}_{y(x)}$ phase does
in that the same internal symmetries, 
$\widehat{X}$ ($\widehat{Y}$) and $\widehat{T}$,
are spontaneously broken by the selection of one out of
two degenerate ground states.
However, each one of the two degenerate ground states
in the $\mathrm{\hbox{N\'eel}}^{\mathrm{SPT}}_{y(x)}$ phase realizes a nontrivial SPT 
state protected by the unbroken
$\widehat{Y}$ and $\widehat{X}\,\widehat{T}$ 
($\widehat{X}$ and $\widehat{Y}\,\widehat{T}$) 
symmetries~\cite{Verresen170705787,Chen11064752,Aksoy220410333,Levin12023120,Else14095436,SM}. 
Hence, when OBCs are chosen, the $\mathrm{\hbox{N\'eel}}^{\mathrm{SPT}}_{y(x)}$
phase supports two decoupled quantum spin-1/2 degrees of freedom localized
on the left and right boundaries, respectively; i.e.,
the ground states are now eightfold (instead of twofold)
degenerate in the thermodynamic limit.
While the $\mathrm{\hbox{N\'eel}}^{\mathrm{SPT}}_{y}$ and $\mathrm{\hbox{N\'eel}}^{\mathrm{SPT}}_{x}$
phases support distinct long-range orders,
their symmetry-breaking ground states realize the same SPT phases. 

\begin{figure}[t!]
\includegraphics[width=1\columnwidth]{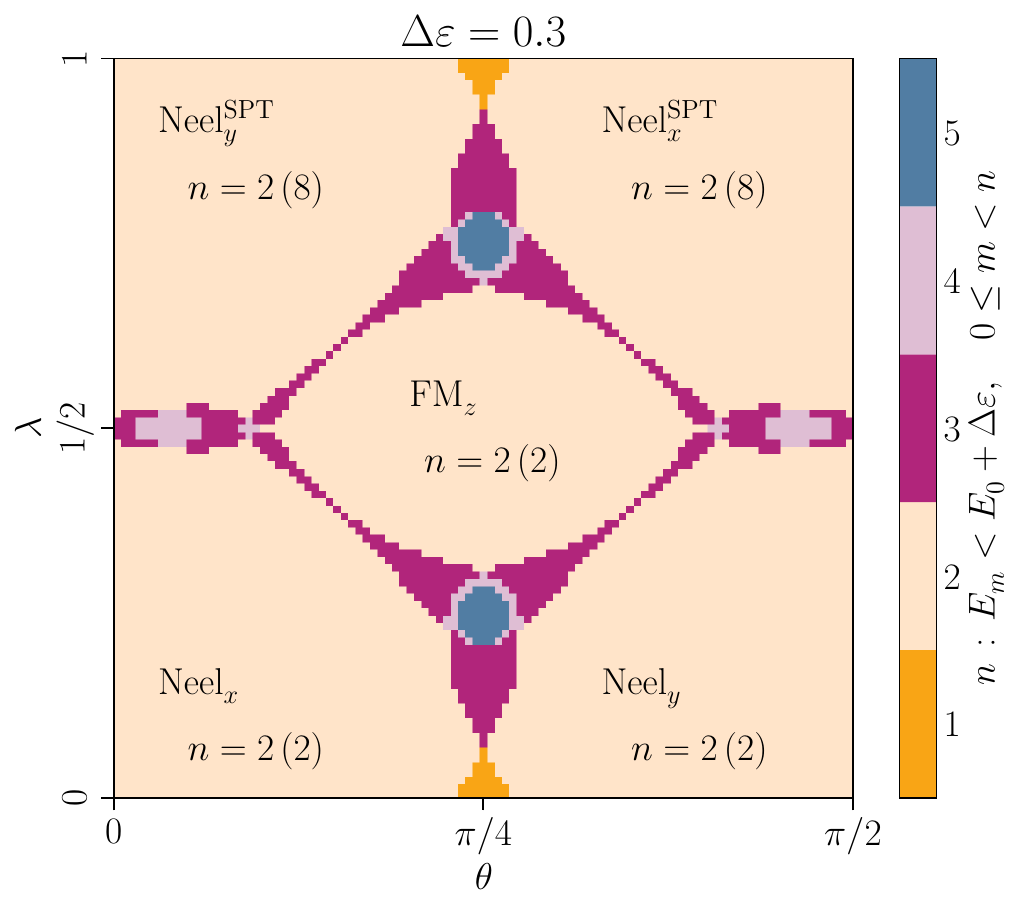}
\vspace{-20pt}
\caption{(Color online)
Number $n$ of eigenenergies $E^{\,}_{m}$ 
with an energy smaller than
$E^{\,}_{0}+\Delta\varepsilon$, where $E^{\,}_{0}$ is the ground-state energy calculated using exact diagonalization with PBCs for $2N=10$ sites,
$\Delta\varepsilon=0.3$,
and the integer $0\leq m < n$.
We also quote the value of $n$
for the five distinct phases computed with OBCs in parentheses.
}
\label{fig:ed_10}
\vspace{-10pt}
\end{figure}

\textit{Symmetry-enriched DQCPs.} 
The four corners describing the gapped phases in the coupling space
in Fig.~\ref{fig:phase diagram} are pairwise connected by four edges.
When restricted to any of these edges, 
Hamiltonian~\eqref{eq:def H} describes a linear interpolation
between two exactly solvable models that support distinct long-range orders. 
The midpoint of each edge, marked by a circle in
Fig.~\ref{fig:phase diagram}, 
is then a DQCP that separates two gapped phases with competing
long-range orders, as we now explain.

At the midpoint $(\pi/4,0)$ of the lower horizontal edge
in Fig.~\ref{fig:phase diagram},
Hamiltonian~\eqref{eq:def H}
reduces to the well-known isotropic spin-1/2 quantum $XY$ model,
which is the DQCP between
the $\mathrm{\hbox{N\'eel}}^{\,}_{x}$ and $\mathrm{\hbox{N\'eel}}^{\,}_{y}$
phases as shown in Refs.\ \cite{Jiang19,Mudry19,Roberts19};
a $\mathsf{c}=1$ CFT with a nondegenerate gapless ground state~\cite{Lieb61}.
The two exactly solvable corners controlling the
$\mathrm{\hbox{N\'eel}}^{\,}_{x}$ and $\mathrm{\hbox{N\'eel}}^{\,}_{y}$
phases are related to each other by the unitary transformation
$\widehat{U}^{\,}_{\mathrm{R}}=\widehat{U}^{\,}_{\pi/2}\,\widehat{X}$,
where $\widehat{U}^{\,}_{\pi/2}$ and $\widehat{X}$
implement rotations in spin space 
by $\pi/2$ and $\pi$ around the
$Z$ and $X$ axes, respectively; i.e., 
the only nontrivial action of 
$\widehat{U}^{\,}_{\mathrm{R}}$ on the Pauli operators is given by
$
(\widehat{X}^{\,}_{j},\widehat{Y}^{\,}_j)
\mapsto
(\widehat{Y}^{\,}_{j},\widehat{X}^{\,}_{j})
$
for any site $j$.

The two midpoints
$(0,1/2)$
and
$(\pi/2,1/2)$
of the left and right vertical edges in Fig.~\ref{fig:phase diagram}
are the two DQCPs between the $\mathrm{\hbox{N\'eel}}^{\,}_{x(y)}$ and
$\mathrm{\hbox{N\'eel}}^{\mathrm{SPT}}_{y(x)}$ phases, respectively.
These DQCPs are also QTPTs between distinct topological phases~\cite{SM}.
Indeed, when PBCs are chosen, the two corners
in the phase diagram of Fig.~\ref{fig:phase diagram} that control the
$\mathrm{\hbox{N\'eel}}^{\,}_{x(y)}$ and $\mathrm{\hbox{N\'eel}}^{\mathrm{SPT}}_{y(x)}$ phases are
related to each other by the unitary transformation
$\widehat{U}^{\,}_{\mathrm{E}}=
\widehat{U}^{\,}_{\mathrm{CZ}}\,
\widehat{U}^{\,}_{\pi/2}\,
\widehat{X}$.
Here, $\widehat{U}^{\,}_{\mathrm{E}}$ 
is the composition of $\widehat{U}^{\,}_{\mathrm{R}}$
with the global controlled-$Z$ unitary,
whose only nontrivial action on the Pauli operators
is defined by
$
\widehat{A}^{\,}_{j}
\mapsto
\widehat{Z}^{\,}_{j-1}\,\widehat{A}^{\,}_{j}\,\widehat{Z}^{\,}_{j+1}
$,
with $A=X$ and $Y$ for any site $j$.
This unitary equivalence breaks down for OBCs.
The unitary $\widehat{U}^{\,}_{\mathrm{E}}$
is an example of an entangler, that exchanges distinct SPT phases 
when PBCs are chosen~\cite{Santos15,Tantivasadakarn2023}.
Here, it becomes an anomalous $\mathbb{Z}^{\,}_{2}$
symmetry of the two DQCPs at the midpoints of the vertical edges~\cite{SM}.

The midpoint $(\pi/4,1)$ of the top horizontal edge in 
Fig.~\ref{fig:phase diagram} is the DQCP between the
$\mathrm{\hbox{N\'eel}}^{\mathrm{SPT}}_{y}$
and
$\mathrm{\hbox{N\'eel}}^{\mathrm{SPT}}_{x}$ phases. When PBCs are chosen, this DQCP is
unitarily equivalent to that at $(\pi/4,0)$
under the action of the
entangler $\widehat{U}^{\,}_{\mathrm{E}}$. Hence, it is also
described by the $\mathsf{c}=1$ CFT and has the same bulk properties.
However, when OBCs are chosen, this unitary equivalence no longer holds. As
opposed to the spin-1/2 quantum $XY$ model at $(\pi/4,0)$,
the dressed spin-1/2 quantum $XY$ model at $(\pi/4,1)$
has four degenerate ground states associated with zero modes localized
on the boundaries~\cite{SM}.

\begin{figure}[t!]
\includegraphics[width=1\columnwidth]{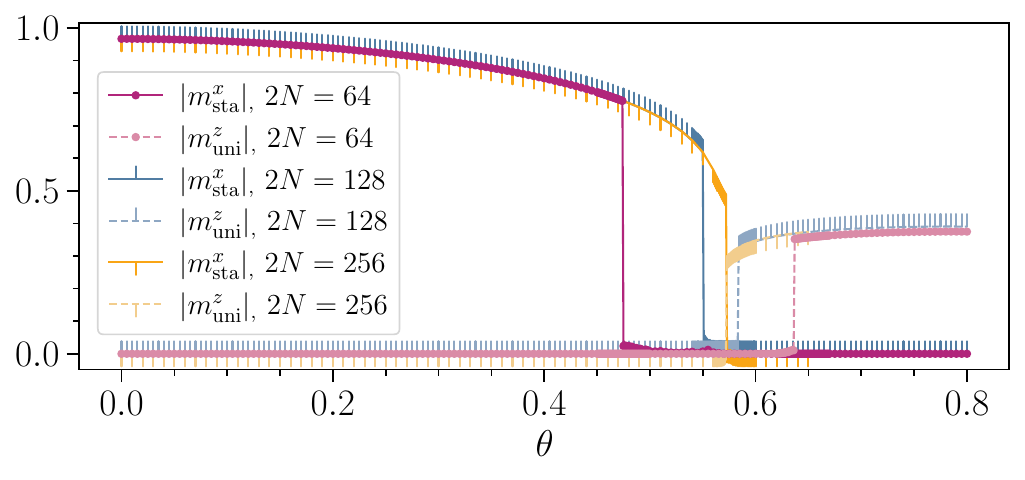}
\vspace{-20pt}
\caption{(Color online)
Magnitudes of the staggered magnetization per site along the
$X$ axis~\eqref{eq:mx_sta} and the uniform magnetization per site along
the $Z$ axis~\eqref{eq:mz_uni}, obtained along the line $\lambda=0.4$
using DMRG with OBCs for $2N=64,128$, and $256$ sites. The sharp suppression
of the order parameters on both sides of the quantum phase transition
around $\theta=0.57$ indicates that the correlation length
is larger than the length of the chain. 
        }
\label{fig:dmrg_op}
\vspace{-10pt}
\end{figure}

Finally, the point $(\pi/4,1/2)$ is invariant under the action of
$\widehat{U}^{\,}_{\mathrm{R}}$
for both periodic and open boundary conditions,
while it is only invariant under the action of
$\widehat{U}^{\,}_{\mathrm{E}}$
for periodic boundary conditions.
Remarkably this point
displays the phenomenon of Hilbert-space
fragmentation~\cite{Yang20},
is Bethe Ansatz integrable,
realizes a CFT with central charge $1$,
and supports ferromagnetic long-range order
as shown in
Refs.~\cite{Zadnik21a,Zadnik21b,Pozsgay21,Jones22}.

\textit{Phase diagram.}
We now turn our attention to the fate of the four gapped
phases labeled by squares and the associated four DQCPs
labeled by circles in Fig.~\ref{fig:phase diagram}.
Each corner of the phase diagram in Fig.~\ref{fig:phase diagram}
realizes a gapped phase of matter supporting long-range order.
Associated with the long-range order is a
degeneracy of $2$ that originates from time-reversal symmetry and
that is independent of the choice between PBCs and
OBCs. Additionally, the degeneracy of the upper corners increases by a
factor of $4$ upon choosing OBCs.  The locus of points in coupling
space along which such a degeneracy changes defines the phase boundary
of each one of these four phases.  An estimate of the location of
these phase boundaries is achieved by diagonalizing Hamiltonian
\eqref{eq:def H} for $2N=10$ sites and counting the number of
eigenenergies within a distance of $\Delta\varepsilon=0.3>1/(2N)$
from the ground-state energy, as is illustrated in Fig.\
\ref{fig:ed_10}.
In doing so, we find a fifth phase centered around the point
$(\pi/4,1/2)$
that is shaped like a diamond with a degeneracy of two for both PBCs and OBCs.
We show using
density matrix renormalization group (DMRG) studies
that, remarkably, this fifth phase is gapless,
while it supports long-range ferromagnetic order
along the quantization $Z$ axis in spin space.

\textit{Incommensurate gapless ferromagnetic phase.}
The DMRG~\cite{schollwock2011} numerical study was performed
using our own code as well as the open-source Julia library
ITensor~\cite{ITensors, ITensorsCode}.  While confirming the
existence of the nonvanishing value taken by the antiferromagnetic
order parameter (the staggered magnetization)~\cite{3footnote}
\begin{equation}
m^{x}_{\mathrm{sta}} \coloneqq
\lim_{\bm{B}^{\,}_{\mathrm{sta}}\to0}
\lim_{\beta\to\infty}
\lim_{2N\to\infty}
\frac{1}{2N}
\sum_{j=1}^{2N}\,
(-1)^{j}
\langle\widehat{X}^{\,}_{j}\rangle^{\,}_{\beta,\bm{B}^{\,}_{\mathrm{sta}}}
\label{eq:mx_sta}
\end{equation}
in the $\mathrm{\hbox{N\'eel}}^{\,}_{x}$ phase from Fig.~\ref{fig:phase diagram},
we establish with Fig.~\ref{fig:dmrg_op}
the nonvanishing value taken by the ferromagnetic (FM) order parameter
(the uniform magnetization)~\cite{3footnote}
\begin{equation}
m^{z}_{\mathrm{uni}} \coloneqq
\lim_{\bm{B}^{\,}_{\mathrm{uni}}\to0}
\lim_{\beta\to\infty}
\lim_{2N\to\infty}
\frac{1}{2N}
\sum_{j=1}^{2N}\,
\langle\widehat{Z}^{\,}_{j}\rangle^{\,}_{\beta,\bm{B}^{\,}_{\mathrm{uni}}}
\label{eq:mz_uni}
\end{equation}
inside the $\mathrm{FM}^{\,}_{z}$ phase from Fig.~\ref{fig:phase diagram}.
The maximum value taken by $m^{z}_{\mathrm{uni}}$
inside the $\mathrm{FM}^{\,}_{z}$ phase from Fig.~\ref{fig:phase diagram}
is $4/10$,
i.e., 40\% of the classical uniform magnetization per site
and greater than the lower bound
$1/3\leq |m^{z}_{\mathrm{uni}}|$
found in Ref.~\cite{Zadnik21a},
independently of $2N>32$
and the choice between PBCs and OBCs.

\begin{figure}[t!]
\subfigure{\includegraphics[width=1\columnwidth]{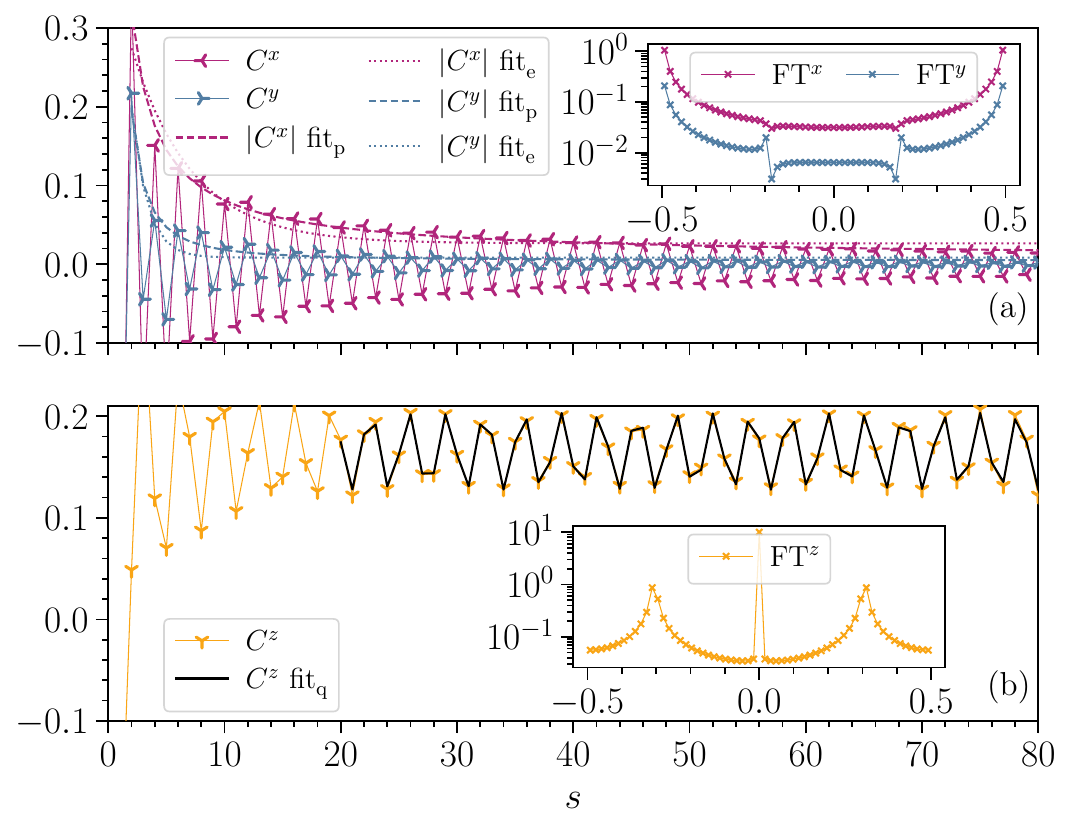} \label{fig:dmrg_corr a}} \\
\vspace{-25pt}
\subfigure{\relax \label{fig:dmrg_corr b}} 
\subfigure{\relax \label{fig:dmrg_corr c}} 
\subfigure{\includegraphics[width=1\columnwidth]{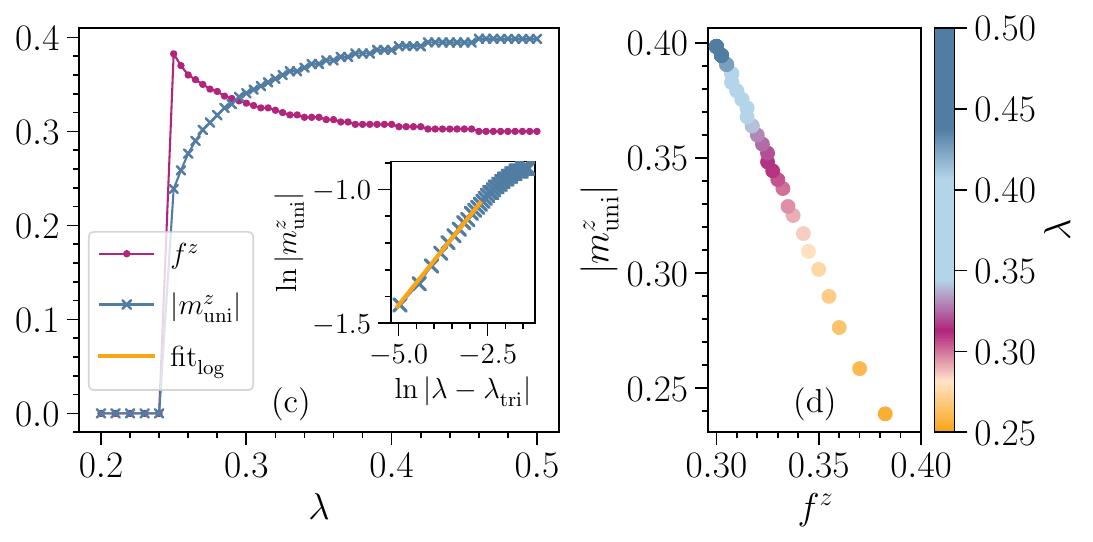} \label{fig:dmrg_corr_ft}\label{fig:dmrg_corr d}}
\vspace{-20pt}
\caption{(Color online)
Spin-spin correlation functions $C^{\alpha}$ defined in Eq.\
\eqref{eq:correlations} computed using DMRG with OBCs,
for $2N=128$, $\lambda=0.4$, and $\theta=\pi/4-0.1$ with
(a) $C^{x}$ and $C^{y}$ and (b)
$C^{z}$. Fits used are labeled ``p'' for power law,
``e'' for exponential, and ``q'' for cosine form. We find
$q^{z}\simeq2\pi\times3/10$.
The goodness of fit for p is always greater than that for
e. The insets show the Fourier transforms of the respective
correlations, where the abscissa is the frequency
$f^{\alpha}\equiv q^{\alpha}/(2\pi)$
and the ordinate is the expansion coefficient $A(f^{\alpha})$,
$\alpha=x,y,$ and $z$.
(c) Dependence on $\lambda$ of the uniform magnetization
$m^{z}_{\mathrm{uni}}$ defined in Eq.~\eqref{eq:mz_uni} and the frequency
$f^{z}$ of the Fourier transform 
of the spin-spin correlations $C^{z}$ in the bulk
of a chain with $2N=512$ sites obtained using DMRG with OBCs
at $\theta=\pi/4$. 
The inset shows the fit$^{\,}_{\mathrm{log}}$ ansatz 
$\ln|m^{z}_{\mathrm{uni}}(\lambda)|\sim 1/6\,\ln|\lambda-\lambda^{\,}_{\mathrm{tri}}|$,
where $\lambda^{\,}_{\mathrm{tri}}\approx0.25$.
(d) Proportionality between $f^{z}(\lambda)$ and
$m^{z}_{\mathrm{uni}}(\lambda)$
on a chain of $2N=512$ sites obtained 
using DMRG with OBCs at $\theta=\pi/4$.
The different values of $\lambda$ are specified by different colors.
        }
\label{fig:dmrg_corr}
\vspace{-10pt}
\end{figure}

We argue that the $\mathrm{FM}^{\,}_{z}$ phase in
Fig.~\ref{fig:phase diagram}
is critical as follows~\cite{SM}.
First, the spin-spin correlation functions
\begin{equation}
C^{\alpha}_{j,j+{s}}\coloneqq 
\lim_{\beta\to\infty}\lim_{2N\to\infty}
\langle
\widehat{A}^{\,}_{j}\,
\widehat{A}^{\,}_{j+{s}}
\rangle^{\,}_{\beta}
\label{eq:correlations}
\end{equation}
with $A=X,Y$, and $Z$,
decay algebraically as
$(-1)^{{s}}\,{s}^{-\eta^{\alpha}(\theta,\lambda)}$
for $\alpha=x,y$ with both PBCs and OBCs,
if ${s}\gg 1$~\cite{1footnote}.
The exponents $\eta^{x}$ and
$\eta^{y}$ are identical on the $\mathrm{O}(2)$ symmetric line
$\theta=\pi/4$, while they differ when this symmetry is explicitly
broken for $\theta \neq \pi/4$, as is illustrated in
Fig.~\ref{fig:dmrg_corr}.
We quote the values $\eta^{x}=\eta^{y}=1$ at $(\pi/4,1/2)$.
Second, any point from the $\mathrm{FM}^{\,}_{z}$
phase in Fig.~\ref{fig:phase diagram} realizes a $\mathsf{c}=1$ CFT,
where the central charge $\mathsf{c}$
is obtained from the bipartite entanglement entropy~\cite{calabrese2004}.
Finally,
any energy spacing between two consecutive nondegenerate
eigenenergies scales as $1/(2N)$ in the thermodynamic limit.

The incommensurability of the $\mathrm{FM}^{\,}_{z}$ phase
is corroborated by the form
$C^{z}_{j,j+{s}}\sim(m^{z}_{\mathrm{uni}})^{2}+B\,\cos(q^{z}\,{s}+\phi)$,
with ${s}\gg1$~\cite{1footnote},
of the correlation function,
as is illustrated in Figs.~\ref{fig:dmrg_corr a} and~\ref{fig:dmrg_corr b}.
We find that the dependence on $\lambda$ of $q^{z}$
is symmetric with respect to the $\lambda=1/2$ line,
while it is independent of $\theta$~\cite{SM}.
Moreover, $q^{z}$ is a monotonically decreasing function of
increasing $\lambda$ in the interval from the tricritical point up to
$\lambda= 1/2$. The wave vector $q^{z}$
saturates to the approximate value $q^{z}\simeq2\pi\times3/10$
for $\lambda\in [0.45, 0.5]$ at $\theta=\pi/4$. We could
not establish if this dependence on $\lambda$ is continuous or
displays a nonanalytic behavior, reminiscent of disorder points~\cite{schollwock1996}. 
Hereto, the amplitude $B$ is independent of
$\theta$, but is sensitive to the choice between OBCs and PBCs
\cite{Ye150205049}
and is a decreasing function of $\lambda$ upon approaching
the phase boundary of the $\mathrm{FM}^{\,}_{z}$ phase in
Fig.~\ref{fig:phase diagram}.
The uniform magnetization $m^{z}_{\mathrm{uni}}$
along the vertical cut at $\theta=\pi/4$
is a function of $\lambda$ that behaves in the vicinity of the
tricritical value $\lambda^{\,}_{\mathrm{tri}}\approx0.25$ as
$(\lambda-\lambda^{\,}_{\mathrm{tri}})^{1/6}$~\cite{SM}
[see Fig.~\ref{fig:dmrg_corr c}].
Remarkably, Fig.~\ref{fig:dmrg_corr d} shows
that $q^{z}(\lambda)$ is proportional
to $m^{z}_{\mathrm{uni}}(\lambda)$ along
the vertical cut at $\theta=\pi/4$. 
This proportionality extends to the scaling exponent
$\eta^{x}(\lambda)$ and to the scaling exponent
$\eta^{z,\mathrm{c}}(\lambda)$
of the connected two-point function for $\widehat{Z}^{\,}_{j}$, 
as \cite{SM}
\begin{equation}
q^{z}=\pi(1-|m^{z}_{\mathrm{uni}}|)
\propto\eta^{x}\propto\eta^{z,\mathrm{c}}.
\end{equation}
As such this proportionality is reminiscent of the critical phase called the
Tomonaga-Luttinger-Liquid phase with central charge one (TLL1) 
in the study done in Ref.~\cite{Hikihara10}
of the quantum spin-1/2 chain
with nearest- and next-nearest-neighbor Heisenberg exchange couplings
in a uniform magnetic field.

\textit{Phase boundaries.}
The phase boundary of any one of the four gapped phases in
Fig.~\ref{fig:phase diagram} terminates at two of the four DQCPs
located at $(\pi/4,0)$, $(\pi/2,1/2)$, $(\pi/4,1)$, and $(0,1/2)$
through straight segments emanating from the four tricritical points.
The DQCPs $(\pi/4,0)$ and $(\pi/4,1)$
are known analytically to realize a $\mathsf{c}=1$ CFT. Whereas it
was possible to confirm numerically the value of the central
charge $\mathsf{c}=1$ at the two DCQP $(\pi/4,0)$ and $(\pi/4,1)$
with a systematic uncertainty of less than 3\%, the same estimate of
$\mathsf{c}=1$ at the two DCQP $(0,1/2)$ and $(\pi/2,1/2)$
suffers a systematic uncertainty of order 30\% due to strong
finite-size effects~\cite{SM}.
The same systematic uncertainty of order 30\%
also plagues the numerical estimate of $\mathsf{c}=1$
upon exiting the $\mathrm{FM}^{\,}_{z}$ phase~\cite{SM}.

Spin-spin correlations $C^{x}_{j,j+{s}}$ and 
$C^{y}_{j,j+{s}}$ at $(\pi/4,0)$ and $(\pi/4,1)$
are observed numerically to decay as $(-1)^{{s}}\,{s}^{-1/2}$,
as expected from the Jordan-Wigner transformation
to a free fermion theory \cite{SM}. 
Strong finite-size effects prevented us from
estimating the rate of decay of correlation functions
at either $(0,1/2)$ or upon exiting the
$\mathrm{FM}^{\,}_{z}$ phase \cite{SM}.

The resolution of the phase boundaries in 
Fig.~\ref{fig:ed_10} can be improved by measuring the
competing order parameters along a cut in the phase diagram, as is
done in Fig.~\ref{fig:dmrg_op}.
For example, we locate in this way the two
tricritical points attached to the $\mathrm{\hbox{N\'eel}}^{\,}_{x}$ phase at
$(0.3\pm0.04,1/2)$ and $(\pi/4,0.25\pm0.005)$.  
Exiting the $\mathrm{FM}^{\,}_{z}$ to anyone
of the four gapped phases is done through
a smooth concave curve~\cite{2footnote}
that connects a pair of
tricritical points as is qualitatively captured by
Fig.~\ref{fig:phase diagram}.

The location of the
latter tricritical point is consistent with the following
exact diagonalization study.
The dependence on $2N$ of the spacing $\delta E^{\,}_{1,0}$
between the first excited
energy $E^{\,}_{1}$ above the ground-state energy $E^{\,}_{0}$ at the
point $(\pi/4,0)$ is computed for increasing values of $\lambda$. The
position of the tricritical point is the value of $\lambda$ at which
$\delta E^{\,}_{1,0}$ changes from a power to an exponential suppression as a
function of $2N$~\cite{SM}.

\textit{Degeneracies at the phase boundaries with OBCs.}
With PBCs, the degeneracy of the ground-state energy is onefold at the
four DQCPs. This fact has been verified numerically with DMRG~\cite{SM}.
The SPT nature of the gapped phases at the upper-left and right corners
of the phase diagram in Fig.~\ref{fig:phase diagram} has a
measurable signature at the DQCP $(\pi/4,1)$
as its ground-state degeneracy with OBCs is fourfold~\cite{SM}.
The vertical boundary separating the
$\mathrm{\hbox{N\'eel}}^{\mathrm{SPT}}_{y}$ and $\mathrm{\hbox{N\'eel}}^{\mathrm{SPT}}_{x}$
phases with
$\lambda<1$
exhibits a twofold ground-state degeneracy
in the thermodynamic limit. Indeed,
the energy difference between the first excited
state and the ground state scales like $(2N)^{-\kappa}$ with
$\kappa\approx6.6$ when computed using the Lanczos algorithm with
$2N=14,16,18,20,$ and $22$~\cite{SM}.
We have also verified using DMRG
that the finite-size spacing between
the second and the first energies above the ground-state energy
scales as $\alpha^{\,}_{2}/(2N)$
with $\alpha^{\,}_{2}\sim(1-\lambda)^{2}$ in the vicinity of
$\lambda=1$ along $\theta=\pi/4$~\cite{SM}.
The same analysis shows that the spacing between the fourth
and third excited energies scales as $\alpha^{\,}_{4}/(2N)$ with
$\alpha^{\,}_{4}-4.66\sim(\lambda-1)^{1}$, which is consistent
with a mean-field analysis.
Strong finite-size effects did not allow
a reliable measurement of the ground-state degeneracies
under OBCs along the phase boundaries
$(\theta,1/2)$
and
$(\pi/2-\theta,1/2)$,
where
$0\leq\theta\leq\theta^{\,}_{\mathrm{tri}}$
with the tricritical value $\theta^{\,}_{\mathrm{tri}}\approx0.3$~\cite{SM}.

\textit{Dynamical exponent $z$.}
For any point in Fig.~\ref{fig:phase diagram} realizing a
gapless phase of matter, the dynamical critical exponent can be
computed by measuring the first nonvanishing finite-size spacing
between low-lying excitation energies above the ground state's energy. 
Using DMRG, we observe the scaling 
$(2N)^{-z}$~\cite{sandvik2010} with $z=1$ in the $\mathrm{FM}^{\,}_{z}$
phase as well as at the points
$(\pi/4,0)$,
$(\pi/2,1/2)$,
$(\pi/4,1)$, and
$(0,1/2)$
\cite{SM}. 

\textit{Discussion.} We introduced a Hamiltonian that interpolates
between four DQCPs enriched by
$\mathbb{Z}^{X}_{2}\times\mathbb{Z}^{Y}_{2}\times\mathbb{Z}^{T}_{2}$-
symmetry. We numerically obtained the phase diagram inside the coupling
space, showing
that the four symmetry-enriched DQCPs cannot be
connected to one another without gap-opening
transitions or passing
through multicritical points.  Importantly, we found that the four
DQCPs are all connected to an extended gapless phase that is
ferromagnetically ordered with incommensurate oscillatory
longitudinal correlations.
In closing, we conjecture
that the topology of the phase diagram we found for the
one-dimensional model~\eqref{eq:def H} would be similar for its
natural generalization in higher-dimensions; a combination of an
$XY$ model and its version dressed with products of $\widehat{Z}$ on
all neighbors to sites sharing a bond.  We note that such
generalization of the model, in any dimension, does not have a sign
problem for $\lambda \le 1/2$ (and its dual for $\lambda \ge 1/2$) and
could therefore be studied via quantum Monte Carlo simulations
(see also Dupont \textit{et al}.\ in Refs.\ \cite{Dupont21a,Dupont21b},
who found using Monte Carlo simulations
an extended gapless phase separating two
topologically ordered phases on the triangular lattice).

\textit{Acknowledgments.}
We thank Akira Furusaki, Toshiya Hikihara,
Saran Prembabu, Subir Sachdev, and Rahul Sahay for discussion.
This work is supported
by SNSF Grant No. 219339 (A.R.),
by NSF Grant No. DMR-2022428 (\"O.M.A.),
and
by DOE Grant No. DE-FG02-06ER46316 (C.C.).
We thank one of the referees for pointing out that
the center $(\pi/4,1/2)$ of the phase diagram was
treated in Refs.\
\cite{Yang20,Zadnik21a,Zadnik21b,Pozsgay21,Jones22}.

\textit{Data availability.}
The data that support the findings of this article are
openly available \cite{dataset_Rey25}.

\addtocontents{toc}{\string\tocdepth@munge}
\bibliography{references}

\begin{thebibliography}{77}
\bibitem [{3fo()}]{3footnote}%
  \BibitemOpen
  \href@noop {} {}\bibinfo {note} {The angular bracket denotes averaging in the
  canonical ensemble at the inverse temperature $\beta$ and in the presence of
  the source term that couples linearly with the order parameter.}\BibitemShut
  {Stop}%
\bibitem [{1fo()}]{1footnote}%
  \BibitemOpen
  \href@noop {} {}\bibinfo {note} {For OBCs, $j$ and $j+s$ are chosen far away
  from the boundaries}\BibitemShut {NoStop}%
\end{thebibliography}
\addtocontents{toc}{\string\tocdepth@restore}



\newpage
\onecolumngrid


\vspace{1cm}
\begin{center}
\textbf{
\begin{large}    
SUPPLEMENTAL MATERIAL
\end{large}
       }
\end{center}
\vspace{1cm}

\tableofcontents

\newpage

\section{Definitions}

\begin{figure}[t]
\includegraphics[width=0.5\columnwidth]{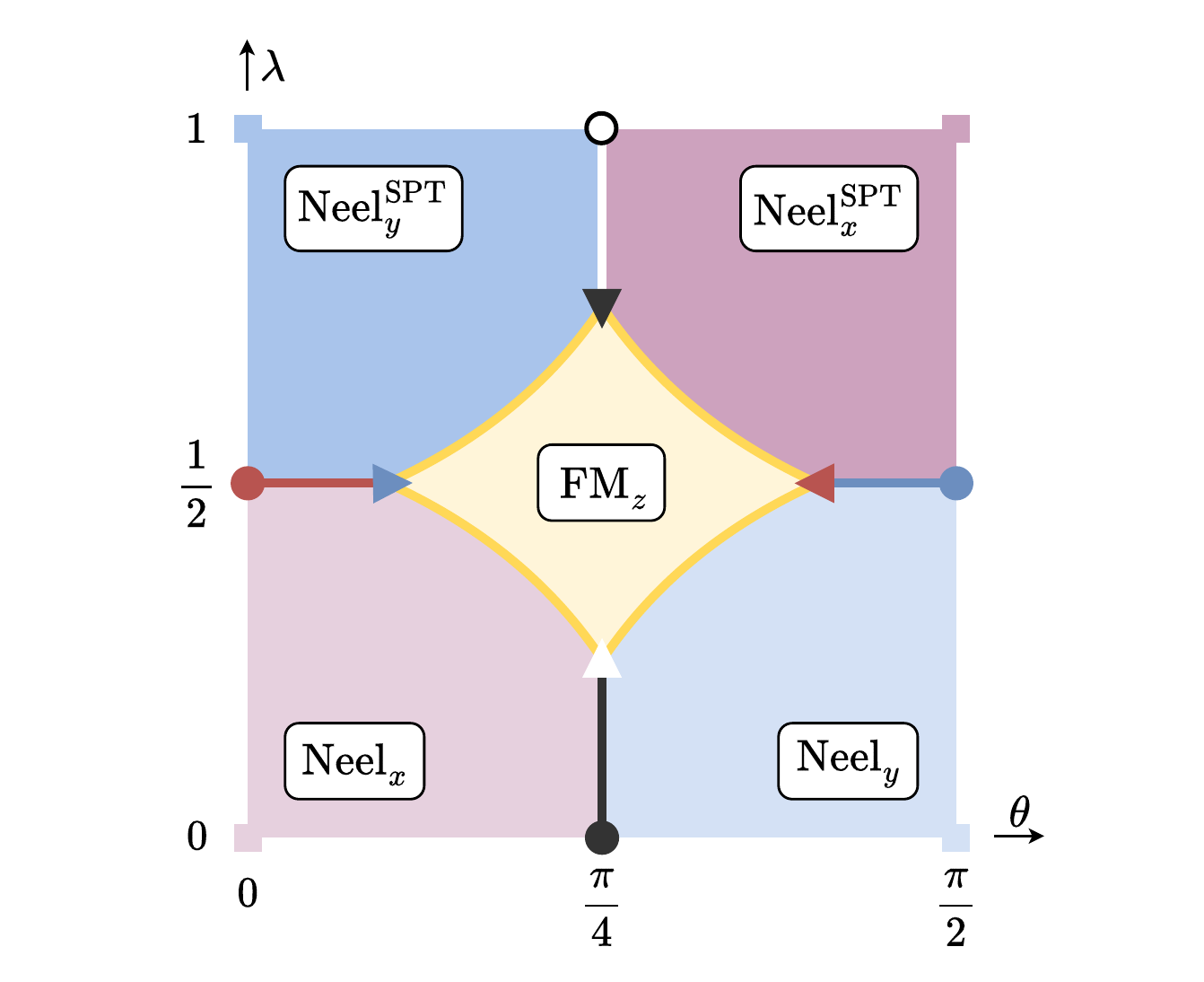}
\caption{(Color online)
Phase diagram of Hamiltonian \eqref{suppeq:def H}.
Squares (circles) on the boundary denote stable (unstable) fixed points
(DQCP). 
}

\label{suppfig:phase diagram}
\end{figure}

The goal of the Supplementary Material is twofold.
First, a symmetry analysis of Hamiltonian (\ref{suppeq:def H})
is provided. Second, numerical results are presented
in support of the conjectured phase diagram
from Fig.~\ref{suppfig:phase diagram}.

We consider the Hamiltonian
\begin{subequations}
\label{suppeq:def H}
\begin{align}
\widehat{H}^{\,}_{b}(\theta,\lambda)
\coloneqq&\,
(1-\lambda)
\sum_{j=1}^{2N-b}
\left(
\cos\theta\,
\widehat{X}^{\,}_{j}\,
\widehat{X}^{\,}_{j+1}
+
\sin\theta\,
\widehat{Y}^{\,}_{j}\,
\widehat{Y}^{\,}_{j+1}
\right)
\nonumber\\
&\,
\hspace{11pt}
+
\lambda
\sum_{j=1}^{2N-3b}
\widehat{Z}^{\,}_{j}\,
\left(
\cos\theta\,
\widehat{X}^{\,}_{j+1}\,
\widehat{X}^{\,}_{j+2}
+
\sin\theta\,
\widehat{Y}^{\,}_{j+1}\,
\widehat{Y}^{\,}_{j+2}
\right)
\widehat{Z}^{\,}_{j+3},
\label{suppeq:def H a}
\end{align}
where coupling space is defined by
\begin{equation}
0\leq\theta\leq\pi/2,
\qquad
0\leq\lambda\leq1,
\label{suppeq:def H b}
\end{equation}
${b=0}$ for periodic boundary conditions (PBC),
${b=1}$ for open boundary conditions (OBC),
and
\begin{equation}
\widehat{X}^{\,}_{j}\,\widehat{Y}^{\,}_{k}=
(-1)^{\delta^{\,}_{j,k}}\,
\widehat{Y}^{\,}_{k}\,
\widehat{X}^{\,}_{j},
\qquad
\widehat{X}^{\,}_{j}\,\widehat{Y}^{\,}_{j}
=
\mathrm{i}\widehat{Z}^{\,}_{j},
\qquad
\widehat{X}^{2}_{j}=
\widehat{Y}^{2}_{j}=
\widehat{Z}^{2}_{j}=
\mathbbm{1},
\label{suppeq:def H c}
\end{equation}
\end{subequations}
for any $j,k=1,\cdots,2N$.
We choose a basis for the $2^{2N}$-dimensional global
Hilbert space for which any $\widehat{Z}^{\,}_{j}$
operators is diagonal. To this end, we denote by $\bm{a}$
the string ${(a^{\,}_1,a^{\,}_2,\cdots,a^{\,}_{2N})}$,
where $a^{\,}_j=0,1$ at every site.
Then, the global Hilbert space $\mathcal{H}\cong\mathbb{C}^{2^{2N}}$
is spanned by the states
$\ket*{\bm{a}}$ that satisfy
\begin{align}
\label{suppeq:basis Z qubits}
\widehat{Z}^{\,}_{j}\, 
\ket*{\bm{a}}=
(-1)^{a^{\,}_{j}}\,
\ket*{\bm{a}},
\qquad
\widehat{X}^{\,}_{j}\, 
\ket*{\bm{a}}=
\ket*{\bm{a}+\bm{\delta}^{(j)}},
\qquad
\widehat{Y}^{\,}_{j}\,
\ket*{\bm{a}}=
\mathrm{i}
(-1)^{a^{\,}_{j}}\,
\ket*{\bm{a}+\bm{\delta}^{(j)}},
\end{align}
where $\bm{\delta}^{(j)}$ is a string with
all its entries vanishing except the one at $j$, which is the number one,
and the addition of the components in
$\bm{a}+\bm{\delta}^{(j)}$ is defined modulo 2.

\subsection{Symmetries}
\label{suppsec:Symmetries}

Hamiltonian
\eqref{suppeq:def H}
commutes with the pair of unitary operators
\begin{equation}
\widehat{X}\coloneqq
\prod_{j=1}^{2N}
\widehat{X}^{\,}_{j},
\qquad
\widehat{Y}\coloneqq
\prod_{j=1}^{2N}
\widehat{Y}^{\,}_{j}.
\end{equation}
Hamiltonian
\eqref{suppeq:def H}
also commutes with the antinunitary operator
\begin{subequations}
\begin{equation}
\widehat{T}^{\prime}\coloneqq
\mathsf{K},
\end{equation}
where $\mathsf{K}$ denotes complex conjugation under which
\begin{equation}
\mathsf{K}\,
\widehat{X}^{\,}_{j}\,
\mathsf{K}=
\widehat{X}^{\,}_{j},
\qquad
\mathsf{K}\,
\widehat{Y}^{\,}_{j}\,
\mathsf{K}=
-
\widehat{Y}^{\,}_{j},
\qquad
\mathsf{K}\,
\widehat{Z}^{\,}_{j}\,
\mathsf{K}=
\widehat{Z}^{\,}_{j},
\qquad
j=1,\cdots,2N.
\end{equation}
\end{subequations}
Operators $\widehat{X}$, $\widehat{Y}$, and $\widehat{T}^{\prime}$
generate the symmetry group
\begin{subequations}
\begin{equation}
\mathbb{Z}^{X}_{2}\times 
\mathbb{Z}^{Y}_{2}\times
\mathbb{Z}^{T}_{2},
\label{suppeq:symmetry group}
\end{equation}
whereby
$\mathbb{Z}^{X}_{2}$ is generated by $\widehat{X}$,
$\mathbb{Z}^{Y}_{2}$ is generated by $\widehat{Y}$,
and
$\mathbb{Z}^{T}_{2}$ is generated by reversal of time
\begin{equation}
\widehat{T}\coloneqq
\widehat{Y}\,\widehat{T}^{\prime}.
\end{equation}
\end{subequations}

\section{Gapped and gapless fixed-points}

There are are four stable fixed points denoted by squares
in Fig.~\ref{suppfig:phase diagram}
and four unstable fixed points denoted by circles
in Fig.~\ref{suppfig:phase diagram}.
We claim that all stable and unstable fixed points are pairwise distinct,
respectively.
This claim is to be proved by analyzing the symmetries preserved
by each fixed-point and the pattern by which the corresponding set of symmetries
are fractionalized.

Central to our discussion will be the following two unitary transformations
\begin{subequations}
\label{suppeq:two unitaries}
\begin{align}
&
\widehat{U}^{\,}_{\mathrm{R}}\coloneqq
\widehat{U}^{\,}_{\pi/2}\,\widehat{X},
\qquad
\widehat{U}^{\,}_{\pi/2}\coloneqq
\prod_{j=1}^{2N}
e^{-\mathrm{i}\frac{\pi}{4}\widehat{Z}^{\,}_{j}},
\qquad
\widehat{X}\coloneqq
\prod_{j=1}^{2N}
\widehat{X}^{\,}_{j},
\label{suppeq:two unitaries a}
\\
&
\widehat{U}^{\,}_{\mathrm{E}}\coloneqq
\widehat{U}^{\,}_{\mathrm{CZ}}\,
\widehat{U}^{\,}_{\mathrm{R}},
\qquad
\widehat{U}^{\,}_{\mathrm{CZ}}\coloneqq
\prod_{j=1}^{2N}
\frac{1}{2}
\left(
1
+
\widehat{Z}^{\,}_{j}
+
\widehat{Z}^{\,}_{j+1}
-
\widehat{Z}^{\,}_{j}\,
\widehat{Z}^{\,}_{j+1}
\right),
\label{suppeq:two unitaries b}
\end{align}
which act on the Pauli operators by conjugation as
\footnote{
Unitarity of $\protect{\widehat{U}^{\,}_{\mathrm{CZ}}}$
is a consequence of the identity
$$
\protect{
\hphantom{\,}
\left(
1
+
\widehat{Z}^{\,}_{j}
+
\widehat{Z}^{\,}_{j+1}
-
\widehat{Z}^{\,}_{j}\,
\widehat{Z}^{\,}_{j+1}
\right)
\left(
1
+
\widehat{Z}^{\,}_{j}
+
\widehat{Z}^{\,}_{j+1}
-
\widehat{Z}^{\,}_{j}\,
\widehat{Z}^{\,}_{j+1}
\right)=
4\,
\widehat{\mathbb{1}}.
\hphantom{{}^{\,}_{j+1}}
          }
$$
The effect of conjugation of either
$\protect{\widehat{X}^{\,}_{j}}$
or
$\protect{\widehat{Y}^{\,}_{j}}$
by $\protect{\widehat{U}^{\,}_{\mathrm{CZ}}}$
requires the sequential use of the pair of identities
$$
\protect{
\left(
1
+
\widehat{Z}^{\,}_{j}
+
\widehat{Z}^{\,}_{j+1}
-
\widehat{Z}^{\,}_{j}\,
\widehat{Z}^{\,}_{j+1}
\right)
\left(
1
-
\widehat{Z}^{\,}_{j}
+
\widehat{Z}^{\,}_{j+1}
+
\widehat{Z}^{\,}_{j}\,
\widehat{Z}^{\,}_{j+1}
\right)=
4\,
\widehat{Z}^{\,}_{j+1},
          }
$$
and
$$
\protect{
\left(
1
+
\widehat{Z}^{\,}_{j-1}
+
\widehat{Z}^{\,}_{j}
-
\widehat{Z}^{\,}_{j-1}\,
\widehat{Z}^{\,}_{j}
\right)
\left(
1
+
\widehat{Z}^{\,}_{j-1}
-
\widehat{Z}^{\,}_{j}
+
\widehat{Z}^{\,}_{j-1}\,
\widehat{Z}^{\,}_{j}
\right)=
4\,
\widehat{Z}^{\,}_{j-1}.
        }
$$
It follows from the Pauli algebra \eqref{suppeq:def H c} that
$$
\protect{
\widehat{U}^{\,}_{\mathrm{CZ}}\,
\widehat{A}^{\,}_{j}\,
\widehat{U}^{\dag}_{\mathrm{CZ}}=
\widehat{Z}^{\,}_{j-1}\,
\widehat{A}^{\,}_{j}\,
\widehat{Z}^{\,}_{j+1},
\qquad
\widehat{A}^{\,}_{j}=
\widehat{X}^{\,}_{j},
\widehat{Y}^{\,}_{j},
}
$$
for $j=1+b,\cdots,2N-b$.
}
\begin{align}
&
\widehat{U}^{\,}_{\mathrm{R}}
\colon
\left(
\widehat{X}^{\,}_{j},\,
\widehat{Y}^{\,}_{j},\,
\widehat{Z}^{\,}_{j}
\right)
\mapsto
\widehat{U}^{\,}_{\mathrm{R}}
\left(
\widehat{X}^{\,}_{j},\,
\widehat{Y}^{\,}_{j},\,
\widehat{Z}^{\,}_{j}
\right)
\widehat{U}^{\dag}_{\mathrm{R}}=
\left(
\widehat{Y}^{\,}_{j},\,
\widehat{X}^{\,}_{j},\,
-
\widehat{Z}^{\,}_{j}
\right),
\\
&
\widehat{U}^{\,}_{\mathrm{E}}
\colon
\left(
\widehat{X}^{\,}_{j},\,
\widehat{Y}^{\,}_{j},\,
\widehat{Z}^{\,}_{j}
\right)
\mapsto
\widehat{U}^{\,}_{\mathrm{E}}
\left(
\widehat{X}^{\,}_{j},\,
\widehat{Y}^{\,}_{j},\,
\widehat{Z}^{\,}_{j}
\right)
\widehat{U}^{\dag}_{\mathrm{E}}=
\left(
\widehat{Z}^{\,}_{j-1}\,
\widehat{Y}^{\,}_{j}\,
\widehat{Z}^{\,}_{j+1},\,
\widehat{Z}^{\,}_{j-1}\,
\widehat{X}^{\,}_{j}\,
\widehat{Z}^{\,}_{j+1},\,
-
\widehat{Z}^{\,}_{j}
\right),
\end{align}
\end{subequations}
for any $j=1,\cdots,2N$
provided PBC are chosen ${(b=0)}$.
When PBC are chosen ${(b=0)}$,
conjugation of Hamiltonian \eqref{suppeq:def H} by
these unitary transformations gives
\begin{align}
\label{suppeq:two symmetries of phase diagram}
\widehat{U}^{\,}_{\mathrm{R}}\,
\widehat{H}^{\,}_{0}(\theta,\lambda)\,
\widehat{U}^{\dag}_{\mathrm{R}}=
\widehat{H}^{\,}_{0}(\pi/2-\theta,\lambda),
\qquad
\widehat{U}^{\,}_{\mathrm{E}}\,
\widehat{H}^{\,}_{0}(\theta,\lambda)\,
\widehat{U}^{\dag}_{\mathrm{E}}=
\widehat{H}^{\,}_{0}(\theta,1-\lambda).
\end{align}
Hence, $\widehat{U}^{\,}_{\mathrm{R}}$ and $\widehat{U}^{\,}_{\mathrm{E}}$
generate $\mathbb{Z}^{\,}_{2}$
reflections of the phase diagram \ref{suppfig:phase diagram}
about the lines ${\theta=\pi/4}$ and ${\lambda=1/2}$, respectively.
When the parameter space \eqref{suppeq:def H b}
is restricted to any one of these two lines, the
corresponding unitary transformation in Eq.~\eqref{suppeq:two unitaries} 
becomes a symmetry of the Hamiltonian \eqref{suppeq:def H}.
Along any of these two lines, the Abelian symmetry group
(\ref{suppeq:symmetry group})
is enlarged to a discrete non-Abelian symmetry group owing to the algebra
\begin{equation}
\widehat{U}^{\,}_{\mathrm{R}}\,
\widehat{X}\,
\widehat{U}^{\dag}_{\mathrm{R}}=
\widehat{Y},
\qquad
\widehat{U}^{\,}_{\mathrm{E}}\,
\widehat{X}\,
\widehat{U}^{\dag}_{\mathrm{E}}=
\widehat{Y},
\end{equation}
when PBC are chosen ${(b=0)}$.

When OBC are chosen $(b=1)$,
$\widehat{U}^{\,}_{\mathrm{R}}$ still implements a reflection
about the $\theta=\pi/4$ line,
i.e.,
\begin{equation}
\widehat{U}^{\,}_{\mathrm{R}}\,
\widehat{H}^{\,}_{1}(\theta,\lambda)\,
\widehat{U}^{\dag}_{\mathrm{R}}=
\widehat{H}^{\,}_{1}(\pi/2-\theta,\lambda),
\end{equation}
while the unitary equivalence under the action of 
$\widehat{U}^{\,}_{\mathrm{E}}$ defined by
\begin{align}
\widehat{U}^{\,}_{\mathrm{E}}
\colon
\left(
\widehat{X}^{\,}_{j},\,
\widehat{Y}^{\,}_{j},\,
\widehat{Z}^{\,}_{j}
\right)
\mapsto
\begin{cases}
\left(
\hphantom{\widehat{Z}^{\,}_{j-1}\,}
\widehat{Y}^{\,}_{j}\hphantom{,}
\widehat{Z}^{\,}_{j+1},\,
\hphantom{\widehat{Z}^{\,}_{j-1}\,}
\widehat{X}^{\,}_{j}\hphantom{,}
\widehat{Z}^{\,}_{j+1},\,
{-\widehat{Z}^{\,}_{j}}
\right),
& j=1,\\
\left(
\widehat{Z}^{\,}_{j-1}\,
\widehat{Y}^{\,}_{j}\hphantom{,}
\widehat{Z}^{\,}_{j+1},\,
\widehat{Z}^{\,}_{j-1}\,
\widehat{X}^{\,}_{j}\hphantom{,}
\widehat{Z}^{\,}_{j+1},\,
{-\widehat{Z}^{\,}_{j}}
\right),
& j=2,\cdots,2N-1,\\
\left(
\widehat{Z}^{\,}_{j-1}\,
\widehat{Y}^{\,}_{j},
\hphantom{\widehat{Z}^{\,}_{j+1},\,}
\widehat{Z}^{\,}_{j-1}\,
\widehat{X}^{\,}_{j},
\hphantom{\widehat{Z}^{\,}_{j+1},\,}
{-\widehat{Z}^{\,}_{j}}
\right),
& j=2N,
\end{cases}
\end{align}
is lost.

\subsection{Gapped stable fixed-points}
\label{suppsubsec: Gapped stable fixed-points}

When $\lambda=0$, the two gapped stable fixed points $(0,0)$ 
and $(\pi/2,0)$ correspond to the $\mathrm{\hbox{N\'eel}}^{\,}_{x}$
and $\mathrm{\hbox{N\'eel}}^{\,}_{y}$ phases, respectively. At these points,
Hamiltonian \eqref{suppeq:def H}
reduces to the pair of Ising antiferromagnets
\begin{subequations}
\begin{align}
&
\widehat{H}^{\,}_{b}(0,0)=
\sum_{j=1}^{2N-b}
\widehat{X}^{\,}_{j}\,
\widehat{X}^{\,}_{j+1},
\label{suppeq:Hamiltonians at lambda=0 a}
\\
&
\widehat{H}^{\,}_{b}(\pi/2,0)=
\sum_{j=1}^{2N-b}
\widehat{Y}^{\,}_{j}\,
\widehat{Y}^{\,}_{j+1},
\label{suppeq:Hamiltonians at lambda=0 b}
\end{align}
with the pair of orthonormal ground states 
\begin{align}
&
\ket*{\mathrm{\hbox{N\'eel}}^{\,}_{x},\pm}\coloneqq
\frac{1}{2^{N}}
\sum_{\{a^{\,}_{j}=0,1\}}
(-1)^{\sum\limits_{j=1}^{2N} j\,a^{\,}_{j}}\,
(\pm1)^{\sum\limits_{j=1}^{2N} a^{\,}_{j}}\,
\ket*{\bm{a}},
\label{suppeq:GSs at lambda=0 a} 
\\
&
\ket*{\mathrm{\hbox{N\'eel}}^{\,}_{y},\pm}\coloneqq
\frac{1}{2^{N}}
\sum_{\{a^{\,}_{j}=0,1\}}
(-1)^{\sum\limits_{j=1}^{2N} j\,a^{\,}_{j}}\,
(\pm\mathrm{i})^{\sum\limits_{j=1}^{2N} a^{\,}_{j}}\,
\ket*{\bm{a}},
\label{suppeq:GSs at lambda=0 b}
\end{align}
and the pair of order parameters
\begin{align}
&
m^{x}_{\mathrm{sta}}[\mathrm{\hbox{N\'eel}}^{\,}_{x},\pm]\coloneqq
\frac{1}{2N}
\sum_{j=1}^{2N}
(-1)^{j}\,
\bra*{\mathrm{\hbox{N\'eel}}^{\,}_{x},\pm}
\widehat{X}^{\,}_{j}
\ket*{\mathrm{\hbox{N\'eel}}^{\,}_{x},\pm}=
\pm1,
\label{suppeq:OPs at lambda=0 a}
\\
&
m^{y}_{\mathrm{sta}}[\mathrm{\hbox{N\'eel}}^{\,}_{y},\pm]\coloneqq
\frac{1}{2N}
\sum_{j=1}^{2N}
(-1)^{j}\,
\bra*{\mathrm{\hbox{N\'eel}}^{\,}_{y},\pm}
\widehat{Y}^{\,}_{j}
\ket*{\mathrm{\hbox{N\'eel}}^{\,}_{y},\pm}=
\pm1,
\label{suppeq:OPs at lambda=0 b}
\end{align}
\end{subequations}
respectively, and for any choice of boundary conditions
$b\in\{0,1\}$, respectively.

The ground states of both Hamiltonians
(\ref{suppeq:Hamiltonians at lambda=0 a})
and
(\ref{suppeq:Hamiltonians at lambda=0 b})
span a two-dimensional vector space.
One possible orthonormal and complete basis for
their respective ground states is the choice
(\ref{suppeq:GSs at lambda=0 a})
and
(\ref{suppeq:GSs at lambda=0 b}),
respectively.
In either case, these basis elements
are interchanged by the action of time-reversal $\widehat{T}$,
\begin{subequations}
\begin{align}
&
\ket*{\mathrm{\hbox{N\'eel}}^{\,}_{x},\pm}=
\widehat{T}\,
\ket*{\mathrm{\hbox{N\'eel}}^{\,}_{x},\mp},
\\
&
\ket*{\mathrm{\hbox{N\'eel}}^{\,}_{y},\pm}=
\widehat{T}\,
\ket*{\mathrm{\hbox{N\'eel}}^{\,}_{y},\mp}.
\end{align}
\end{subequations}
Distinct subgroups of the symmetry group
(\ref{suppeq:symmetry group})
survive its spontaneous symmetry breaking,
as measured by the order parameters
(\ref{suppeq:OPs at lambda=0 a})
and
(\ref{suppeq:OPs at lambda=0 b}).
To see this, we first observe that
either one of the pair of ground states 
$\ket*{\mathrm{\hbox{N\'eel}}^{\,}_{x},\pm}$
and
$\ket*{\mathrm{\hbox{N\'eel}}^{\,}_{y},\pm}$
in Eq.~(\ref{suppeq:GSs at lambda=0 a})
and
Eq.~(\ref{suppeq:GSs at lambda=0 b})
breaks the symmetry group (\ref{suppeq:symmetry group})
down to the unbroken symmetry subgroups
\begin{subequations}
\label{suppeq:pattern SSB}
\begin{align}
&
\mathbb{Z}^{X}_{2}\times\mathbb{Z}^{YT}_{2},
\label{suppeq:pattern SSB a}
\\
&
\mathbb{Z}^{Y}_{2}\times\mathbb{Z}^{XT}_{2},
\label{suppeq:pattern SSB b}
\end{align}
\end{subequations}
respectively,
where
$\mathbb{Z}^{YT}_{2}$ ($\mathbb{Z}^{XT}_{2}$)
is the diagonal subgroup of
$\mathbb{Z}^{Y}_{2}\times \mathbb{Z}^{T}_{2}$ 
$(\mathbb{Z}^{X}_{2}\times \mathbb{Z}^{T}_{2})$.
Second, we observe that
$\widehat{U}^{\,}_{\mathrm{R}}$ does not commute with the global symmetry
subgroup generated by $\widehat{X}$ and $\widehat{Y}$
since
\begin{align}
\label{suppeq:UR exchanges X and Y ops}
\widehat{U}^{\,}_{\mathrm{R}}\,
\widehat{X}=
\widehat{Y}\,
\widehat{U}^{\,}_{\mathrm{R}},
\end{align}
i.e., it acts non-trivially on the full symmetry group  
(\ref{suppeq:symmetry group}).
Combining both observations, we conclude that
the two gapped stable fixed-point Hamiltonians in Eqs.~\eqref{suppeq:Hamiltonians at lambda=0 a}
and~\eqref{suppeq:Hamiltonians at lambda=0 b}
are exchanged under the action of $\widehat{U}^{\,}_{\mathrm{R}}$. 
Correspondingly,
the unitary operator $\widehat{U}^{\,}_{\mathrm{R}}$
interchanges the symmetry-breaking
$\mathrm{\hbox{N\'eel}}^{\,}_{x}$ and $\mathrm{\hbox{N\'eel}}^{\,}_{y}$ orders
or, equivalently, the unbroken symmetry subgroups
$\mathbb{Z}^{X}_{2}\times\mathbb{Z}^{YT}_{2}$
and
$\mathbb{Z}^{Y}_{2}\times\mathbb{Z}^{XT}_{2}$.

The other two gapped stable fixed points when $\lambda=1$ are located
at $(0,1)$ and $(\pi/2,1)$ and given by
the quantum spin-1/2 $r=3$ cluster chains%
~\cite{Suzuki1971,Verresen170705787,Aksoy220410333}
\begin{subequations}
\begin{align}
&
\widehat{H}^{\,}_{b}(0,1)=
\sum_{j=1}^{2N-3b}
\widehat{Z}^{\,}_{j}\,
\widehat{X}^{\,}_{j+1}\,
\widehat{X}^{\,}_{j+2}\,
\widehat{Z}^{\,}_{j+3},
\label{suppeq:Hamiltonians at lambda=1 a}
\\
&
\widehat{H}^{\,}_{b}(\pi/2,1)=
\sum_{j=1}^{2N-3b}
\widehat{Z}^{\,}_{j}\,
\widehat{Y}^{\,}_{j+1}\,
\widehat{Y}^{\,}_{j+2}\,
\widehat{Z}^{\,}_{j+3},
\label{suppeq:Hamiltonians at lambda=1 b}
\end{align}
\end{subequations}
respectively.
When PBC are chosen ($b=0$), the
fixed-point Hamiltonians
(\ref{suppeq:Hamiltonians at lambda=1 a})
and
(\ref{suppeq:Hamiltonians at lambda=1 b})
are unitarily equivalent to Hamiltonians
\eqref{suppeq:Hamiltonians at lambda=0 a}
and
\eqref{suppeq:Hamiltonians at lambda=0 b},
respectively,
owing to the relation
\eqref{suppeq:two symmetries of phase diagram}.
Thus, with PBC ($b=0$), 
they have the pair of orthonormal ground states
\begin{align}
\begin{split}
&
\ket*{\mathrm{\hbox{N\'eel}}^{\mathrm{SPT}}_{y},\pm}=
(-1)^{N}\,
\widehat{U}^{\,}_{\mathrm{E}}\,
\ket*{\mathrm{\hbox{N\'eel}}^{\,}_{x},\pm}=
\frac{1}{2^{N}}
\sum_{\{a^{\,}_{j}=0,1\}}
(-1)^{\sum\limits_{j=1}^{2N} a^{\,}_{j} (j+a^{\,}_{j+1}) }\,
(\pm\mathrm{i})^{\sum\limits_{j=1}^{2N} a^{\,}_{j}}
\ket*{\bm{a}},
\\
&
\ket*{\mathrm{\hbox{N\'eel}}^{\mathrm{SPT}}_{x},\pm}=
(-1)^{N}\,
\widehat{U}^{\,}_{\mathrm{E}}\,
\ket*{\mathrm{\hbox{N\'eel}}^{\,}_{y},\pm}=
\frac{1}{2^{N}}
\sum_{\{a^{\,}_{j}=0,1\}}
(-1)^{\sum\limits_{j=1}^{2N} a^{\,}_{j} (j+a^{\,}_{j+1}) }\,
(\pm1)^{\sum\limits_{j=1}^{2N} a^{\,}_{j}}
\ket*{\bm{a}},
\end{split}
\label{suppeq:NeelSPT gds}
\end{align}
which describe the $\mathrm{\hbox{N\'eel}}^{\mathrm{SPT}}_{y}$ and 
$\mathrm{\hbox{N\'eel}}^{\mathrm{SPT}}_{x}$ phases, respectively. 
Notice that just like $\widehat{U}^{\,}_{\mathrm{R}}$,
the unitary transformation $\widehat{U}^{\,}_{\mathrm{E}}$
interchanges the underlying long-range order since 
\begin{align}
\widehat{U}^{\,}_{\mathrm{E}}\,
\widehat{X}=
\widehat{Y}\,
\widehat{U}^{\,}_{\mathrm{E}}
\end{align}
holds for PBC ($b=0$).

In addition to exchanging the subgroups
$\mathbb{Z}^{Y}_{2}\times\mathbb{Z}^{XT}_{2}$
and
$\mathbb{Z}^{X}_{2}\times\mathbb{Z}^{YT}_{2}$
of the symmetry group (\ref{suppeq:symmetry group}),
that are left unbroken by the symmetry-breaking ground states
(\ref{suppeq:NeelSPT gds}), respectively,
the unitary transformation $\widehat{U}^{\,}_{\mathrm{E}}$
also introduces non-zero (short-range) entanglement to the ground state.
Therefore, the ground states of the resulting $\mathrm{\hbox{N\'eel}}^{\mathrm{SPT}}_{y}$ and 
$\mathrm{\hbox{N\'eel}}^{\mathrm{SPT}}_{x}$ phases describe non-trivial SPT phases. To see that, 
we follow the strategy adopted in Refs.\
\cite{Verresen170705787,Aksoy220410333}
and write the fixed point Hamiltonians
\eqref{suppeq:Hamiltonians at lambda=1 a}
and
\eqref{suppeq:Hamiltonians at lambda=1 b}
as
\begin{subequations}
\begin{align}
&
\widehat{H}^{\,}_{b}(0,1)=
\sum_{j=1}^{2N-3b}
\left(
\widehat{Z}^{\,}_{j}\,
\widehat{Y}^{\,}_{j+1}\,
\widehat{Z}^{\,}_{j+2}
\right)
\left(
\widehat{Z}^{\,}_{j+1}\,
\widehat{Y}^{\,}_{j+2}\,
\widehat{Z}^{\,}_{j+3}
\right),
\label{suppeq:Hamiltonians at lambda=1 a bis}
\\
&
\widehat{H}^{\,}_{b}(\pi/2,1)=
\sum_{j=1}^{2N-3b}
\left(
\widehat{Z}^{\,}_{j}\,
\widehat{X}^{\,}_{j+1}\,
\widehat{Z}^{\,}_{j+2}
\right)
\left(
\widehat{Z}^{\,}_{j+1}\,
\widehat{X}^{\,}_{j+2}\,
\widehat{Z}^{\,}_{j+3}
\right),
\label{suppeq:Hamiltonians at lambda=1 b bis}
\end{align}
\end{subequations}
respectively.
In both Hamiltonians
(\ref{suppeq:Hamiltonians at lambda=1 a bis})
and
(\ref{suppeq:Hamiltonians at lambda=1 b bis}),
all three-body terms in the parentheses commute
pairwise.  Hence, the ground-state properties can be captured by the
mean-field Hamiltonians
\begin{subequations}
\begin{align}
&
\widehat{H}^{\pm}_{b}(0,1)=
\sum_{j=1}^{2N-2b}
(-1)^{j}
(\pm 1)\,
\widehat{Z}^{\,}_{j}\,
\widehat{Y}^{\,}_{j+1}\,
\widehat{Z}^{\,}_{j+2},
\label{suppeq:NeelSPT MF Hams a}
\\
&
\widehat{H}^{\pm}_{b}(\pi/2,1)=
\sum_{j=1}^{2N-2b}
(-1)^{j}
(\pm 1)\,
\widehat{Z}^{\,}_{j}\,
\widehat{X}^{\,}_{j+1}\,
\widehat{Z}^{\,}_{j+2},
\label{suppeq:NeelSPT MF Hams b}
\end{align}
\end{subequations}
which admit the $\ket*{\mathrm{\hbox{N\'eel}}^{\mathrm{SPT}}_{y},\pm}$
and $\ket*{\mathrm{\hbox{N\'eel}}^{\mathrm{SPT}}_{x},\pm}$ states as their non-degenerate
gapped ground-state, respectively. 
Both pairs of mean-field Hamiltonians
\eqref{suppeq:NeelSPT MF Hams a}
and
\eqref{suppeq:NeelSPT MF Hams b}
are that of the quantum spin-1/2 $r=2$ cluster chain%
~\cite{Suzuki1971,Verresen170705787,Aksoy220410333}
with fourfold degenerate ground states when OBC are chosen (${b=1}$).
Importantly, this fourfold ground-state degeneracy is protected 
by the symmetries preserved
by the pairs of mean-field Hamiltonians
\eqref{suppeq:NeelSPT MF Hams a}
and
\eqref{suppeq:NeelSPT MF Hams b},
which
are the
$\mathbb{Z}^{Y}_{2}\times \mathbb{Z}^{XT}_{2}$
and 
$\mathbb{Z}^{X}_{2}\times \mathbb{Z}^{YT}_{2}$
symmetry groups, respectively.

Next, we claim that the pattern encoding the symmetry fractionalization 
is the same for all four mean-field Hamiltonians 
\eqref{suppeq:NeelSPT MF Hams a}
and
\eqref{suppeq:NeelSPT MF Hams b}.

We are going to verify
this claim first for the pair of Hamiltonians $\widehat{H}^{\pm}_{b}(0,1)$.
When OBC ($b=1$) are imposed, the terms 
$\widehat{Z}^{\,}_{2N-1}\,\widehat{Y}^{\,}_{2N}\,\widehat{Z}^{\,}_{1}$
and $\widehat{Z}^{\,}_{2N}\,\widehat{Y}^{\,}_{1}\,\widehat{Z}^{\,}_{2}$
do not appear
in either one of the pair of Hamiltonians $\widehat{H}^{\pm}_{b}(0,1)$.
Therefore, choosing the left boundary
around site $j=1$ without loss of generality,
we may define the boundary spin-1/2 degrees of freedom
\begin{align}
\widehat{S}^{X}_{\mathrm{L}}\coloneqq 
\widehat{X}^{\,}_{1}\,\widehat{Z}^{\,}_{2},
\qquad
\widehat{S}^{Y}_{\mathrm{L}}\coloneqq
\widehat{Y}^{\,}_{1}\,\widehat{Z}^{\,}_{2},
\qquad
\widehat{S}^{Z}_{\mathrm{L}}\coloneqq
\widehat{Z}^{\,}_{1},
\label{suppeq:def left triplet spin operators}
\end{align}
for either one of the pair of Hamiltonians $\widehat{H}^{\pm}_{b}(0,1)$.
One verifies that this triplet of boundary operators
satisfies the Pauli algebra \eqref{suppeq:def H c}.
Moreover, one verifies that,
under the remaining $\mathbb{Z}^{Y}_{2}\times\mathbb{Z}^{XT}_{2}$
symmetry, this triplet of operators transforms as
\begin{align}
\begin{split}
\widehat{Y}
&\colon
\left(
\widehat{S}^{X}_{\mathrm{L}},\,
\widehat{S}^{Y}_{\mathrm{L}},\,
\widehat{S}^{Z}_{\mathrm{L}}
\right)
\mapsto
\left(
+\widehat{S}^{X}_{\mathrm{L}},\,
-\widehat{S}^{Y}_{\mathrm{L}},\,
-\widehat{S}^{Z}_{\mathrm{L}}
\right),
\\
\widehat{X}\,
\widehat{T}
&\colon
\left(
\widehat{S}^{X}_{\mathrm{L}},\,
\widehat{S}^{Y}_{\mathrm{L}},\,
\widehat{S}^{Z}_{\mathrm{L}}
\right)
\mapsto
\left(
-\widehat{S}^{X}_{\mathrm{L}},\,
+\widehat{S}^{Y}_{\mathrm{L}},\,
+\widehat{S}^{Z}_{\mathrm{L}}
\right).
\end{split}
\end{align}
These transformation rules can be satisfied by defining the boundary
symmetry operators
\begin{align}
\widehat{Y}^{\,}_{\mathrm{L}}\coloneqq
\widehat{S}^{X}_{\mathrm{L}},
\qquad
\left(
\widehat{X}\,
\widehat{T}
\right)^{\,}_{\mathrm{L}}\coloneqq
\widehat{S}^{Z}_{\mathrm{L}}\,
\mathsf{K}^{\,}_{\mathrm{L}}.
\label{suppeq:pair of operators realizing projectove algebra}
\end{align}
Here, the symbol $\mathsf{K}^{\,}_{\mathrm{L}}$
acts as complex conjugation does on $\mathbb{C}$ numbers,
it treats the pair of operators $\widehat{S}^{X}_{\mathrm{L}}$ and
$\widehat{S}^{Z}_{\mathrm{L}}$ as complex conjugation does with real numbers,
and it treats operator 
$\widehat{S}^{Y}_{\mathrm{L}}$ as complex conjugation does with imaginary numbers.
One then verifies that the boundary symmetries realize the projective
algebra
\begin{align}
\widehat{Y}^{\,}_{\mathrm{L}}\,
\left(
\widehat{X}\,
\widehat{T}
\right)^{\,}_{\mathrm{L}}
=
-
\left(
\widehat{X}\,
\widehat{T}
\right)^{\,}_{\mathrm{L}}\,
\widehat{Y}^{\,}_{\mathrm{L}},
\qquad
\left(
\widehat{X}\,
\widehat{T}
\right)^{\,}_{\mathrm{L}}
\left(
\widehat{X}\,
\widehat{T}
\right)^{\,}_{\mathrm{L}}=
\mathbbm{1},
\label{suppeq:projective algebra on L bd}
\end{align}
for each of the Hamiltonians $\widehat{H}^{\pm}_{b}(0,1)$.
Hence, the global $\mathbb{Z}^{Y}_{2}\times\mathbb{Z}^{XT}_{2}$
symmetry with a non-projective representation fractionalizes at the 
left boundary
for either one of the pair of Hamiltonians $\widehat{H}^{\pm}_{b}(0,1)$.
A similar relation can be realized at the
right boundary as well~\cite{Aksoy220410333}
for either one of the pair of Hamiltonians $\widehat{H}^{\pm}_{b}(0,1)$.

Second, we are going to show that
the same symmetry fractionalization pattern occurs for the pair of
Hamiltonians $\widehat{H}^{\pm}_{b}(\pi/2,1)$. We notice that 
the boundary degrees of freedom are exactly the same as for the pair of
Hamiltonians  $\widehat{H}^{\pm}_{b}(0,1)$.
However, now the bulk symmetries act as
\begin{align}
\begin{split}
\widehat{X}
&\colon
\left(
\widehat{S}^{X}_{\mathrm{L}},\,
\widehat{S}^{Y}_{\mathrm{L}},\,
\widehat{S}^{Z}_{\mathrm{L}}
\right)
\mapsto
\left(
-\widehat{S}^{X}_{\mathrm{L}},\,
+\widehat{S}^{Y}_{\mathrm{L}},\,
-\widehat{S}^{Z}_{\mathrm{L}}
\right),
\\
\widehat{Y}\,\widehat{T}
&\colon
\left(
\widehat{S}^{X}_{\mathrm{L}},\,
\widehat{S}^{Y}_{\mathrm{L}},\,
\widehat{S}^{Z}_{\mathrm{L}}
\right)
\mapsto
\left(
+\widehat{S}^{X}_{\mathrm{L}},\,
-\widehat{S}^{Y}_{\mathrm{L}},\,
+\widehat{S}^{Z}_{\mathrm{L}}
\right).
\end{split}
\end{align}
These transformation laws suggest that we define,
with the help of Eq.~(\ref{suppeq:def left triplet spin operators}),
the boundary symmetry operators through
\begin{align}
\widehat{X}^{\,}_{\mathrm{L}}\coloneqq
\widehat{S}^{Y}_{\mathrm{L}},
\qquad
\left(
\widehat{Y}\,
\widehat{T}
\right)^{\,}_{\mathrm{L}}\coloneqq
\mathsf{K}^{\,}_{\mathrm{L}}.
\end{align}
While these operators are distinct from the ones defined in
Eq.~(\ref{suppeq:pair of operators realizing projectove algebra}),
one verifies that they satisfy
the same projective algebra found in
Eq.~(\ref{suppeq:projective algebra on L bd}),
namely,
\begin{align}
\widehat{X}^{\,}_{\mathrm{L}}\,
\left(
\widehat{Y}
\widehat{T}
\right)^{\,}_{\mathrm{L}}
=
-
\left(
\widehat{Y}\,
\widehat{T}
\right)^{\,}_{\mathrm{L}}\,
\widehat{X}^{\,}_{\mathrm{L}},
\qquad
\left(
\widehat{Y}\,
\widehat{T}
\right)^{\,}_{\mathrm{L}}\,
\left(
\widehat{Y}\,
\widehat{T}
\right)^{\,}_{\mathrm{L}}
=
\mathbbm{1}.
\end{align}

We conclude that all four mean-field Hamiltonians
\eqref{suppeq:NeelSPT MF Hams a}
and
\eqref{suppeq:NeelSPT MF Hams b},
describe the same SPT phase of matter.
The stable fixed-point Hamiltonians
\eqref{suppeq:Hamiltonians at lambda=1 a}
and
\eqref{suppeq:Hamiltonians at lambda=1 b}
each have eightfold degenerate gapped ground states when OBC are chosen. 

The unitary transformation
$\widehat{U}^{\,}_{\mathrm{E}}$
defined in Eq.\
\eqref{suppeq:two symmetries of phase diagram}
is the composition of the pair of operators
$\widehat{U}^{\,}_{\mathrm{R}}$ and $\widehat{U}^{\,}_{\mathrm{CZ}}$. 
The operator $\widehat{U}^{\,}_{\mathrm{CZ}}$
is also known as the SPT ``entangler,'' which 
is a globally symmetric (finite-depth) unitary transformation 
that maps the trivial SPT phase to the non-trivial
cluster SPT state
\cite{Santos15,Tantivasadakarn2023}.
Hence, the effect of the
unitary transformation $\widehat{U}^{\,}_{\mathrm{E}}$
is to interchange the  $\mathrm{\hbox{N\'eel}}^{\,}_{x}$ and $\mathrm{\hbox{N\'eel}}^{\,}_{y}$
long-range orders and then ``paste'' the cluster state by introducing
non-zero entanglement on each symmetry breaking ground state. 
We therefore label the images of the
$\mathrm{\hbox{N\'eel}}^{\,}_{x}$ and $\mathrm{\hbox{N\'eel}}^{\,}_{y}$ phases 
under this transformation by $\mathrm{\hbox{N\'eel}}^{\mathrm{SPT}}_{y}$
and $\mathrm{\hbox{N\'eel}}^{\mathrm{SPT}}_{x}$, respectively.

We close this discussion
by presenting in closed form the full (not only the ground states) spectral
decomposition of all Hamiltonians denoted by squares
in Fig.~\ref{suppfig:phase diagram}.
To this end, we will use a different
basis than the basis (\ref{suppeq:basis Z qubits}).

Each corner of the phase diagram in Fig.~\ref{suppfig:phase diagram}
can be written as the Hamiltonian
\begin{subequations}
\label{suppeq:cluster rep four corneres}
\begin{equation}
\widehat{H}^{(b^{\,}_{r})}_{r}\coloneqq
\sum_{j=1}^{2N-b^{\,}_{r}}
\widehat{C}^{\,}_{j},
\label{suppeq:cluster rep four corneres a}
\end{equation}
where the spin-1/2 cluster $r$ operator $\widehat{C}^{\,}_{j}$ is given by
\begin{equation}
\widehat{C}^{\,}_{j}=
\widehat{X}^{\,}_{j}\,
\widehat{X}^{\,}_{j+1}\equiv
\widehat{M}^{(1)}_{j}\,
\widehat{M}^{(1)}_{j+1},
\qquad
\widehat{M}^{(1)}_{j}\equiv
\widehat{X}^{\,}_{j},
\label{suppeq:cluster rep four corneres b}
\end{equation}
at the corner $(0,0)$ with $r=1$ and $b^{\,}_{r}=0,1$ for PBC and OBC,
respectively,
\begin{equation}
\widehat{C}^{\,}_{j}=
\widehat{Y}^{\,}_{j}\,
\widehat{Y}^{\,}_{j+1}\equiv
\widehat{M}^{(1)}_{j}\,
\widehat{M}^{(1)}_{j+1},
\qquad
\widehat{M}^{(1)}_{j}\equiv
\widehat{Y}^{\,}_{j},
\label{suppeq:cluster rep four corneres c}
\end{equation}
at the corner $(\pi/2,0)$ with $r=1$ and with $b^{\,}_{r}=0,1$ for PBC and OBC,
respectively,
\begin{equation}
\widehat{C}^{\,}_{j}=
\widehat{Z}^{\,}_{j}\,
\widehat{Y}^{\,}_{j+1}\,
\widehat{Y}^{\,}_{j+2}\,
\widehat{Z}^{\,}_{j+3}\equiv
\widehat{M}^{(3)}_{j}\,
\widehat{M}^{(3)}_{j+1},
\qquad
\widehat{M}^{(3)}_{j}\equiv
\widehat{Z}^{\,}_{j}\,
\widehat{X}^{\,}_{j+1}\,
\widehat{Z}^{\,}_{j+2},
\label{suppeq:cluster rep four corneres d}
\end{equation}
at the corner $(\pi/1,1)$ with $r=3$ and $b^{\,}_{r}=0,3$ for PBC and OBC,
respectively,
and
\begin{equation}
\widehat{C}^{\,}_{j}=
\widehat{Z}^{\,}_{j}\,
\widehat{X}^{\,}_{j+1}\,
\widehat{X}^{\,}_{j+2}\,
\widehat{Z}^{\,}_{j+3}\equiv
\widehat{M}^{(3)}_{j}\,
\widehat{M}^{(3)}_{j+1},
\qquad
\widehat{M}^{(3)}_{j}\equiv
\widehat{Z}^{\,}_{j}\,
\widehat{Y}^{\,}_{j+1}\,
\widehat{Z}^{\,}_{j+2},
\label{suppeq:cluster rep four corneres e}
\end{equation}
\end{subequations}
at the corner $(0,1)$ with $r=3$ and $b^{\,}_{r}=0,3$ for PBC and OBC,
respectively.
As $\widehat{C}^{\,}_{j}$ is Hermitean and squares to the identity,
its eigenvalues are $\pm1$.
As any pair $\widehat{C}^{\,}_{j},\widehat{C}^{\,}_{j'}$ commutes for any
pair of sites $j,j'$, all $\widehat{C}^{\,}_{j}$ labeled by $j$ can
be simultaneously diagonalized. As the product of
all $\widehat{C}^{\,}_{j}$ labeled by $j=1,\cdots,2N$ is the identity,
only $2N-1$ of the $2N$ operators $\widehat{C}^{\,}_{j}$
with $j=1,\cdots,2N$
are independent.
To indentify the last quantum number that is needed to label all
the energy eigenstates,
we observe that
$\widehat{M}^{(3)}_{j}$ is Hermitean,
squares to the identity, and commutes with $\widehat{M}^{(3)}_{j'}$
for any $j,j'=1,\cdots,2N$.
Hence, all $2^{2N}$ energy eigenvalues and energy eigenfunctions
of $\widehat{H}^{(b^{\,}_{r})}_{r}$
are given by
\begin{subequations}
\label{suppeq:case spectral decomposition cluster models}
\begin{equation}
E^{(b^{\,}_{r})}_{\mathrm{m}^{\,}_{1},\mathrm{c}^{\,}_{1},\cdots,\mathrm{c}^{\,}_{2N-1}}=
\sum_{j=1}^{2N-b^{\,}_{r}}\mathrm{c}^{\,}_{j},
\qquad
\mathrm{m}^{\,}_{1}=\pm1,
\qquad
\mathrm{c}^{\,}_{j}=\pm1,
\qquad
\mathrm{c}^{\,}_{2N}=\prod_{j=1}^{2N-1}\mathrm{c}^{\,}_{j},
\label{suppeq:case spectral decomposition cluster models a}
\end{equation}
and
\begin{equation}
|E^{(b^{\,}_{r})}_{\mathrm{m}^{\,}_{1},\mathrm{c}^{\,}_{1},\cdots,\mathrm{c}^{\,}_{2N-1}}\rangle
\equiv  
|\mathrm{m}^{\,}_{1},\mathrm{c}^{\,}_{1},\cdots,\mathrm{c}^{\,}_{2N-1}\rangle,
\label{suppeq:case spectral decomposition cluster models b}
\end{equation}
\end{subequations}
respectively,
where
$\mathrm{m}^{\,}_{1}=\pm1$
is the eigenvalue of $\widehat{M}^{(c)}_{1}$
and
$\mathrm{c}^{\,}_{1},\cdots,\mathrm{c}^{\,}_{2N-1}=\pm1$
are the $2N-1$ independent eigenvalues
of the independent operators
$\widehat{C}^{\,}_{1},\cdots,\widehat{C}^{\,}_{2N-1}$.
As the energy eigenvalues are independent of the quantum number
$\mathrm{m}^{\,}_{1}=\pm1$,
it follows that all energy eigenstates are at least two-fold degenerate
for any number of sites $2N$ and for both PBC and OPB.
For PBC, one finds the equally spaced energy eigenvalues
\footnote{
Notice the following subtlety. 
In the case $n$ odd, 
the constraint 
$c^{\,}_{2N} = \prod_{j=1}^{2N-1} c^{\,}_{j}$ 
always enforces $c^{\,}_{2N}=+1$ so that 
an additional degeneracy arises from 
the equality of the energies 
$E^{\,}_{2m-1} = E^{\,}_{2m}$ for any
$m=1,2,\cdots$
}
\begin{subequations}
\label{suppeq:case PBC spectral decomposition cluster models}
\begin{equation}
E^{(0)}_{n}=
-\left(2N-2n-2\,\delta^{\,}_{\mathrm{c}^{\,}_{2N},+1}\right),
\label{suppeq:case PBC spectral decomposition cluster models a}
\end{equation}
whereby the label $n=0,1,\cdots,2N-1$ is
the number of $\mathrm{c}^{\,}_{j}=+1$ for
$j=1,\cdots,2N-1$
and with the degeneracies
\begin{equation}
d^{(0)}_{n}=2\times\frac{(2N-1)!}{n!\,(2N-1-n)!},
\qquad
n=0,1,\cdots,2N-1.
\label{suppeq:case PBC spectral decomposition cluster models b}
\end{equation}
\end{subequations}
For OBC, one finds instead the equally spaced energy eigenvalues
\begin{subequations}
\label{suppeq:case OBC spectral decomposition cluster models}
\begin{equation}
E^{(c)}_{n}=
-\left(2N-c-2n\right),
\label{suppeq:case OBC spectral decomposition cluster models a}
\end{equation}
with $n=0,\cdots,2N-c$ the number of $\mathrm{c}^{\,}_{j}=+1$
for $j=1,\cdots,2N-c$.
The degeneracy $d^{(c)}_{n}$ of the eigenvalue $E^{(c)}_{n}$ is 
the factor $2^{c}$
(that accounts for the fact that $E^{(c)}_{n}$ does not depend on
$\mathrm{m}^{\,}_{1}$,
$\mathrm{c}^{\,}_{2N-c+1}$,
$\cdots$,
$\mathrm{c}^{\,}_{2N-1}$)
times the number of ways to choose $n$ out of $2N-c$, namely,
\begin{equation}
d^{(c)}_{n}=
2\times2^{c-1}\times
\frac{(2N-c)!}{n!\,(2N-c-n)!},
\quad
n=0,\cdots,2N-c.
\label{suppeq:case OBC spectral decomposition cluster models b}
\end{equation}
\end{subequations}

With PBC, the degeneracy $d^{(0)}_{n}$
for any finite size $2N$ is an artifact
of the integrability of Hamiltonian
(\ref{suppeq:cluster rep four corneres a}).
As soon as one moves away from the
corners of the phase diagram in Fig.~\ref{suppfig:phase diagram},
the degeneracy $d^{(0)}_{n}$ is completed lifted
due to the fact that the Hamiltonian
(\ref{suppeq:def H})
is no longer the sum of pairwise commuting operators.
All eigenstates are non-degenerate for any finite size $2N$ and
it is only in the thermodynamic limit $2N\to\infty$ that a
minimum degeneracy of two is recovered sufficiently close to the
corners of the phase diagram in Fig.~\ref{suppfig:phase diagram}.

\subsection{Gapless unstable fixed-points}

The locations of the four unstable gapless fixed points in the phase diagram
\ref{suppfig:phase diagram} can be understood as the 
fixed points under the two unitary transformations 
\eqref{suppeq:two symmetries of phase diagram}.
We claim that any pair of these four unstable gapless fixed points are distinct.

To show this, we focus first on the $\widehat{U}^{\,}_{\mathrm{R}}$
operator, that interchanges the pair of
$\mathrm{\hbox{N\'eel}}^{\,}_{x}$ ($\mathrm{\hbox{N\'eel}}^{\mathrm{SPT}}_{x}$)
and 
$\mathrm{\hbox{N\'eel}}^{\,}_{y}$ ($\mathrm{\hbox{N\'eel}}^{\mathrm{SPT}}_{y}$)
orders.
This operator also exchanges the pair of symmetry operators
$\widehat{X}$ and $\widehat{Y}$
according to \eqref{suppeq:UR exchanges X and Y ops}.
Consider any Hamiltonian that commutes with
$\widehat{X}$,
$\widehat{Y}$,
and
$\widehat{U}^{\,}_{\mathrm{R}}$,
as is the case along the line $\theta=\pi/4$.
Consequently, there cannot be a symmetry breaking pattern by which
a $\widehat{U}^{\,}_{\mathrm{R}}$-symmetric ground state
of a $\widehat{U}^{\,}_{\mathrm{R}}$-symmetric Hamiltonian
breaks spontaneously the symmetry generated by $\widehat{X}$,
while preserving the symmetry generated by $\widehat{Y}$, or vice versa.
Any $\widehat{U}^{\,}_{\mathrm{R}}$-symmetric
ground state of a $\widehat{U}^{\,}_{\mathrm{R}}$-symmetric
Hamiltonian must either break spontaneously or preserve simultaneously
both the $\widehat{X}$ and $\widehat{Y}$
symmetries. For any boundary conditions ($b=0,1$),
the $\widehat{U}^{\,}_{\mathrm{R}}$-symmetric fixed-point
Hamiltonian
\begin{subequations}
\begin{align}
\label{suppeq:DQCP at 0 pi/4}
\widehat{H}^{\,}_{b}(\pi/4,0)=
\frac{1}{\sqrt{2}}
\sum\limits_{j=1}^{2N-b}
\left(
\widehat{X}^{\,}_{j}\,
\widehat{X}^{\,}_{j+1}
+
\widehat{Y}^{\,}_{j}\,
\widehat{Y}^{\,}_{j+1}
\right),
\end{align}
corresponds to the isotropic point of the
spin-1/2 quantum $XY$ model and thus
realizes the scenario by which the ground state 
is symmetric under the group generated by
both $\widehat{X}$ and $\widehat{Y}$
and non-degenerate for any finite chain of length $2N$
for any boundary conditions ($b=0,1$),
while gapless in the thermodynamic limit.
The low-energy theory is described by a
conformal field theory
(CFT) with central charge $\mathsf{c}=1$.
Being unitarily equivalent by $\widehat{U}^{\,}_{\mathrm{E}}$
to this fixed-point when PBC are chosen,
the $\widehat{U}^{\,}_{\mathrm{R}}$-symmetric fixed-point Hamiltonian 
\begin{align}
\label{suppeq:DQCP at 1 pi/4}
\widehat{H}^{\,}_{b}(\pi/4,1)
=
\frac{1}{\sqrt{2}}
\sum\limits_{j=1}^{2N-3b}
\widehat{Z}^{\,}_{j}\,
\left(
\widehat{X}^{\,}_{j+1}\,
\widehat{X}^{\,}_{j+2}
+
\widehat{Y}^{\,}_{j+1}\,
\widehat{Y}^{\,}_{j+2}
\right)
\widehat{Z}^{\,}_{j+3},
\end{align}
\end{subequations}
also realizes the scenario by which the ground state 
is symmetric under the group generated by both
$\widehat{X}$ and $\widehat{Y}$,
non-degenerate and gapped for any finite chain of length $2N$
for PBC ($b=0$), while gapless in the thermodynamic limit.
Hereto, the low-energy theory is described by a
CFT with central charge $\mathsf{c}=1$
when PBC are selected.
Since both Hamiltonians
\eqref{suppeq:DQCP at 0 pi/4}
and
\eqref{suppeq:DQCP at 1 pi/4}
describe transitions  between gapped phases that are identical
from the SPT classification point of view,
but support distinct long-range orders,
the two gapless fixed points are examples of deconfined quantum critical
points (DQCP). We direct the reader to Refs.\
\cite{Jiang19,Mudry19,Roberts19}
if the reader is not familiar with DQCP in one-dimensional spin-1/2 chains.

The distinction between the DQCP at
$(\pi/4,0)$ and $(\pi/4,1)$
becomes apparent when OBC are chosen. In this case,
for any finite value of the chain length $2N$,
the Hamiltonian \eqref{suppeq:DQCP at 0 pi/4}
supports a non-degenerate ground state,
whereas Hamiltonian \eqref{suppeq:DQCP at 1 pi/4} 
supports fourfold-degenerate ground states,
owing to the presence of boundary zero modes. Hence, the gapless fixed-point 
Hamiltonian \eqref{suppeq:DQCP at 1 pi/4}
can be thought of as realizing a DQCP with
non-trivial symmetry fractionalization compared to the 
gapless fixed-point
Hamiltonian \eqref{suppeq:DQCP at 0 pi/4}.
We will argue on the basis of numerics in the next section that,
as we tune $\lambda$ away from $\lambda=1$,
the effect of the small perturbation \eqref{suppeq:DQCP at 0 pi/4}
on the unperturbed Hamiltonian \eqref{suppeq:DQCP at 1 pi/4} 
is to lower the ground-state degeneracy of four of
Hamiltonian \eqref{suppeq:DQCP at 1 pi/4}
to a ground-state degeneracy of two when OBC are selected
for any finite value of the chain length $2N$. Furthermore,
when $\lambda$ is lowered from the value $\lambda=1$
below a value at which three distinct phases meet,
there is a quantum phase transition to an extended 
gapless phase that breaks spontaneously down to the identity
the non-Abelian symmetry group present along the $\theta=\pi/4$
line that is generated by
$\widehat{X}$, $\widehat{Y}$, $\widehat{T}$, 
and $\widehat{U}^{\,}_{\mathrm{R}}$.

The remaining two unstable fixed-points are located at the 
$\widehat{U}^{\,}_{\mathrm{E}}$-symmetric points
$(0,1/2)$
and
$(\pi/2,1/2)$, the end points of the line
$(\lambda=1/2$, $0\leq\theta\leq\pi/2)$
along which Hamiltonian
\eqref{suppeq:def H}
has an enlarged non-Abelian symmetry group that is generated by
$\widehat{X}$, $\widehat{Y}$, $\widehat{T}$, 
and $\widehat{U}^{\,}_{\mathrm{E}}$ when PBC ($b=0$) are selected.
At these fixed points
[which are $\widehat{U}^{\,}_{\mathrm{E}}$-symmetric for PBC ($b=0$)],
Hamiltonian 
\eqref{suppeq:def H}
becomes
\begin{subequations}
\label{suppeq:DQCPs at lambda 1/2}
\begin{align}
&
\widehat{H}^{\,}_{b}(0,1/2)=
\sum\limits_{j=1}^{2N-b}
\widehat{X}^{\,}_{j}\,
\widehat{X}^{\,}_{j+1}
+
\sum\limits_{j=1}^{2N-3b}
\widehat{Z}^{\,}_{j}\,
\widehat{X}^{\,}_{j+1}
\widehat{X}^{\,}_{j+2}\,
\widehat{Z}^{\,}_{j+3},
\label{suppeq:DQCP at 1/2 0}
\\
&
\widehat{H}^{\,}_{b}(\pi/2,1/2)=
\sum\limits_{j=1}^{2N-b}
\widehat{Y}^{\,}_{j}\,
\widehat{Y}^{\,}_{j+1}
+
\sum\limits_{j=1}^{2N-3b}
\widehat{Z}^{\,}_{j}\,
\widehat{Y}^{\,}_{j+1}
\widehat{Y}^{\,}_{j+2}\,
\widehat{Z}^{\,}_{j+3},
\label{suppeq:DQCP at 1/2 pi/2}
\end{align}
\end{subequations}
respectively.
Hamiltonians
(\ref{suppeq:DQCP at 1/2 0})
and
(\ref{suppeq:DQCP at 1/2 pi/2})
are unitarily equivalent under 
the action of $\widehat{U}^{\,}_{\mathrm{R}}$
for any boundary conditions ($b=0,1$).
Application of the Jordan-Wigner transformation
to $\widehat{H}^{\,}_{b}(0,1/2)$
[and hence $\widehat{H}^{\,}_{b}(\pi/2,1/2)$]
yields a strongly interacting Majorana chain.
We could not show by analytical means that this
strongly interacting Majorana chain
realizes a CFT at low energies.

Instead,
we shall present numerical evidences that support the conjecture that,
in the thermodynamic limit,
$\widehat{H}^{\,}_{b}(0,1/2)$
[and hence $\widehat{H}^{\,}_{b}(\pi/2,1/2)$]
realizes a continuous phase transition with two attributes of a
CFT, namely a dynamical exponent $z=1$ and a
central charge $\mathsf{c}=1$.
Finite-size corrections and incommensuration effects for both OBC and PBC
were too strong to estimate scaling exponents
entering two-point functions at criticality
or the onset of long-range order away from criticality.

The phase boundaries
$(0,1/2)$ and $(\pi/2,1/2)$
each realize simultaneously a DQCP and a
quantum topological phase transition (QTPT). This can be understood 
from the presence of the $\widehat{U}^{\,}_{\mathrm{E}}$ symmetry
when PBC ($b=0$) are selected, which 
not only exchanges the
$\mathrm{\hbox{N\'eel}}^{\,}_{x}$ and $\mathrm{\hbox{N\'eel}}^{\,}_{y}$ orders,
but also exchanges the topological attribute 
realized by the quantum spin-1/2, $r=1$ cluster ring
with that realized by the quantum spin-1/2, $r=3$ cluster ring.
In fact, the presence of the $\widehat{U}^{\,}_{\mathrm{E}}$
symmetry obeyed by both Hamiltonians 
(\ref{suppeq:DQCP at 1/2 0})
and
(\ref{suppeq:DQCP at 1/2 pi/2})
when PBC ($b=0$) are selected
puts a stronger constraint than the presence of the
$\widehat{U}^{\,}_{\mathrm{R}}$ symmetry
obeyed by both Hamiltonians
(\ref{suppeq:DQCP at 0 pi/4})
and
(\ref{suppeq:DQCP at 1 pi/4}). 
According to Eq.~\eqref{suppeq:two symmetries of phase diagram}, 
the unitary symmetry $\widehat{U}^{\,}_{\mathrm{E}}$ is nothing 
but the composition of $\widehat{U}^{\,}_{\pi/2}$ with the generator
$\widehat{U}^{\,}_{\mathrm{CZ}}\,\widehat{X}$  
of the so-called CZX symmetry, which was first
introduced in Ref.~\cite{Chen11064752}. 
The latter is known to be an anomalous symmetry,
i.e., any Hamiltonian that commutes
with the CZX operator $\widehat{U}^{\,}_{\mathrm{CZ}}$
cannot have a non-degenerate and gapped 
ground state~\cite{Chen11064752,Levin12023120,Else14095436}, 
unlike with a $\widehat{U}^{\,}_{\mathrm{R}}$-symmetric Hamiltonian. 
The additional composition with the generator
$\widehat{U}^{\,}_{\pi/2}$ of the $\widehat{U}^{\,}_{\mathrm{R}}$ symmetry
does not change the anomaly associated to the CZX symmetry. 
Hence, the gaplessness of the two presumed DQCP
at $(0,1/2)$ and $(\pi/2,1/2)$
is consistent with the presence of the anomalous $\mathbb{Z}^{\,}_{2}$ symmetry 
generated by $\widehat{U}^{\,}_{\mathrm{E}}$
when PBC ($b=0$) are selected.
The two DQCP \eqref{suppeq:DQCP at 1/2 0} and \eqref{suppeq:DQCP at 1/2 pi/2}
are not to be distinguished by their boundary signatures.
Rather, they are to be distinguished
by the fact that they are related by the unitary
$\widehat{U}^{\,}_{\mathrm{R}}$,
which does not commute with the global  
$\mathbb{Z}^{X}_{2}\times\mathbb{Z}^{Y}_{2}$ symmetry. 

\section{Numerics}

We have studied Hamiltonian \eqref{suppeq:def H} through exact
diagonalization (ED) and density matrix renormalization group (DMRG)
with either open boundary conditions (OBC) or periodic boundary
conditions (PBC).

The DMRG study was mostly done with the Julia library ITensor.
The input to DMRG is an initial variational state
$|\Psi^{\,}_{\mathrm{ini}}\rangle$
in the form of a matrix product state.
The output after $n$ DMRG sweeps is the state
$|\Psi^{[n]}_{\mathrm{ini}}\rangle$.
The desired state $|\Psi\rangle$ is the state
$|\Psi^{[n]}_{\mathrm{ini}}\rangle$
such that the three conditions
\begin{subequations}
\begin{align}
&
\left\langle\Psi^{[n]}_{\mathrm{ini}}\left|
\left[
\widehat{H}^{\,}_{b}(\theta,\lambda)
-
\left\langle\Psi^{[n]}_{\mathrm{ini}}\left|
\widehat{H}(\theta,\lambda)
\right|\Psi^{[n]}_{\mathrm{ini}}\right\rangle
\right]^{2}
\right|\Psi^{[n]}_{\mathrm{ini}}\right\rangle<\epsilon^{\,}_{\mathrm{variance}},
\label{suppeq:three conditions for DMRG to have converged a}
\\
&
\left|
\left\langle\Psi^{[n]}_{\mathrm{ini}}\left|
\widehat{H}^{\,}_{b}(\theta,\lambda)
\right|\Psi^{[n]}_{\mathrm{ini}}\right\rangle
-
\left\langle\Psi^{[n-1]}_{\mathrm{ini}}\left|
\widehat{H}^{\,}_{b}(\theta,\lambda)
\right|\Psi^{[n-1]}_{\mathrm{ini}}\right\rangle
\right|<\epsilon^{\,}_{\mathrm{energy}},
\label{suppeq:three conditions for DMRG to have converged b}
\\ 
&
\left|
\max_{{l}}\,S^{\,}_{\Psi^{[n]}_{\mathrm{ini}}}({l})
-
\max_{{l}}\,S^{\,}_{\Psi^{[n-1]}_{\mathrm{ini}}}({l})
\right|
<\epsilon^{\,}_{\mathrm{entropy}},
\label{suppeq:three conditions for DMRG to have converged c}
\end{align}
hold.
Here, the entanglement entropy
$S^{\,}_{\Psi^{[n]}_{\mathrm{ini}}}({l})$ 
for the state
$\ket*{\Psi^{[n]}_{\mathrm{ini}}}$
is defined in Eq.~(\ref{suppeq:def S(ell) for MPS}).
Unless stated otherwise, we choose
\begin{equation}
\epsilon^{\,}_{\mathrm{variance}}=10^{-6},
\qquad
\epsilon^{\,}_{\mathrm{energy}}=10^{-9},
\qquad
\epsilon^{\,}_{\mathrm{entropy}}=10^{-6}.
\end{equation}
\end{subequations}

\begin{figure}[t!]
\centering
\subfigure[]{\includegraphics[width=0.5\columnwidth]{FIGURES/ED/GRID_101/pbc=1_arpack=0_N=10/DEGENERACIES/eps=0.3.pdf} \label{suppfig:ed_10 a}}%
\subfigure[]{\includegraphics[width=0.5\columnwidth]{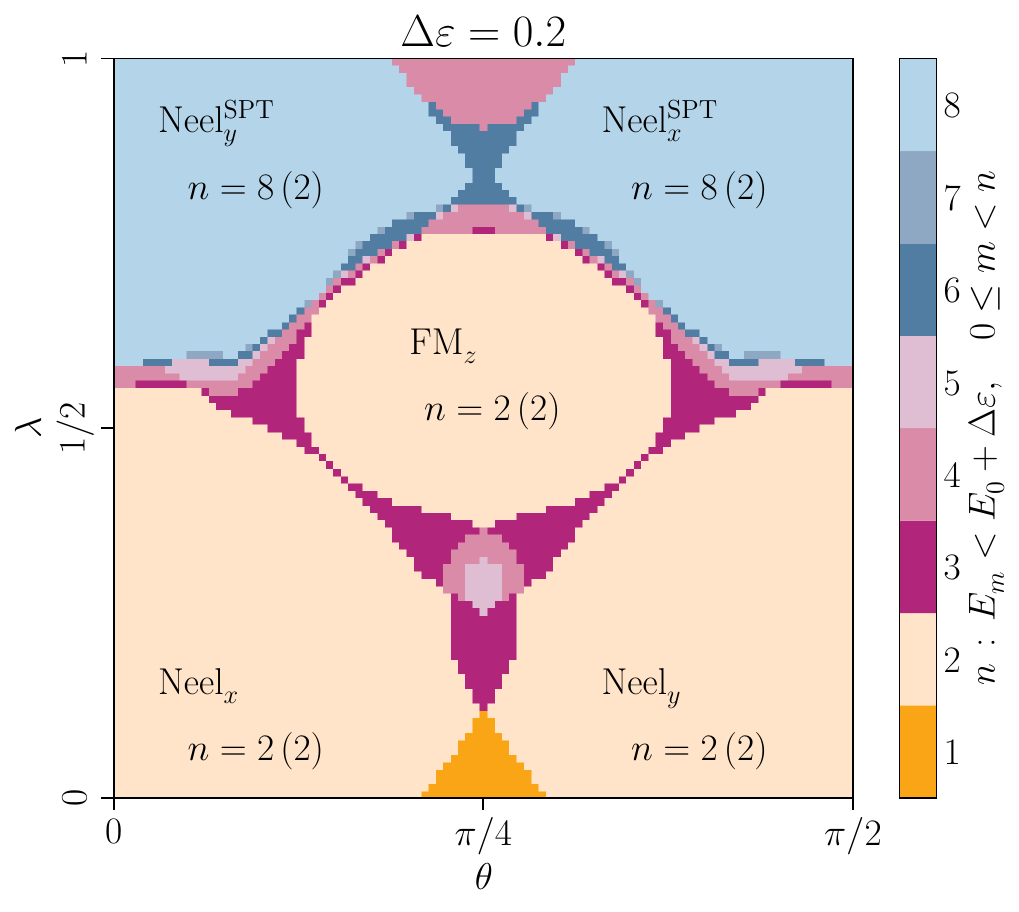} \label{suppfig:ed_10 b}}
\vspace{-5pt}
\caption{(Color online)
(a) Number $n$ of eigenenergies $E^{\,}_{m}$ with an energy smaller
than $E^{\,}_{0}+\Delta\varepsilon$, where $E^{\,}_{0}$ is the
ground state energy calculated using ED with PBC for $2N=10$ sites,
$\Delta\varepsilon=0.3$, and the integer $0\leq m < n$. We also
quote the value of $n$ for the five distinct phases computed with
OBC in parentheses.
(b) Number $n$ of eigenenergies $E^{\,}_{m}$
with an energy smaller than $E^{\,}_{0}+\Delta\varepsilon$, where
$E^{\,}_{0}$ is the ground state energy calculated using ED with OBC
for $2N=10$ sites, $\Delta\varepsilon=0.2$, and the integer
$0\leq m< n$. We also quote the value of $n$ for the five distinct phases
computed with PBC in parentheses. Under OBC, we observe
very strong finite-size
effects that not only broadens the width 
but also moves the position of the putative phase transition.
}
\label{suppfig:ed_10}
\end{figure}

\subsection{Five phases}

We begin the study of Hamiltonian \eqref{suppeq:def H} by examining
the fate of the four gapped phases labeled by squares and the four
fixed points labeled by circles in
Fig.~\ref{suppfig:phase diagram}.

The phase boundaries of the four phases adiabatically connected to
the corners of the phase diagram in Fig.~\ref{suppfig:phase diagram}
are identified as the loci
in coupling space where the degeneracies
of the ground states in the thermodynamic limit change.
To estimate the
location of these phase boundaries, we have diagonalized
Hamiltonian
\eqref{suppeq:def H} with PBC for $2N=10$ sites and counted the
eigenenergies within a window of energy $\Delta\varepsilon=0.3>1/(2N)$
above the ground-state energy,
as shown in Fig.~\ref{suppfig:ed_10 a}.
A less precise estimate can also be obtained by diagonalizing the
Hamiltonian with OBC for $2N=10$ sites and by counting the eigenenergies
within a smaller window of energy $\Delta \varepsilon = 0.2 > 1/(2N)$
above the ground-state energy,
as shown in Fig.~\ref{suppfig:ed_10 b}.
Through this analysis, we have identified
a fifth phase centered around
$(\pi/4,1/2)$, shaped like a diamond, with a degeneracy of two under
both PBC and OBC.

According to the transformation laws
(\ref{suppeq:two symmetries of phase diagram})
of Hamiltonian (\ref{suppeq:def H a})
under the unitary transformations
$\widehat{U}^{\,}_{\mathrm{R}}$ and $\widehat{U}^{\,}_{\mathrm{E}}$
defined in Eq.\ \eqref{suppeq:two unitaries}
that apply for PBC, any point in the phase diagram in
Fig.
\ref{suppfig:phase diagram}
has the same ground-state degeneracy
as its image under
(i) a $\pi$-rotation about the center $(\pi/4, 1/2)$
of the phase diagram and
(ii) a reflection about the line
$(\pi/4,\lambda)$ of the phase diagram.
In contrast, under OBC,
any point in the phase diagram in
Fig.\
\ref{suppfig:phase diagram}
has the same ground-state degeneracy
as its image solely under a reflection about the line
$(\pi/4,\lambda)$
in Fig.\
\ref{suppfig:phase diagram}.
We have confirmed numerically these two predictions using ED in 
Figs.\
\ref{suppfig:ed_10 a}
and
\ref{suppfig:ed_10 b},
respectively.
This $\pi$-rotational relation is
reduced to a mirror symmetry along the U(1)-symmetric line
$(\pi/4,\lambda)$
under OBC, as
$\widehat{U}^{\,}_{\mathrm{E}}$
is no longer a symmetry of Hamiltonian \eqref{suppeq:def H}
[see Fig.~\ref{suppfig:ed_10 b}].
The loss of the $\widehat{U}^{\,}_{\mathrm{E}}$ symmetry under OBC
allows for the area labeled
$\mathrm{\hbox{N\'eel}}^{\,}_{x}$ 
or
$\mathrm{\hbox{N\'eel}}^{\,}_{y}$
to be larger than the area labeled
$\mathrm{\hbox{N\'eel}}^{\mathrm{SPT}}_{x}$ 
or
$\mathrm{\hbox{N\'eel}}^{\mathrm{SPT}}_{y}$
in Fig.~\ref{suppfig:ed_10 b}
for any given finite number of sites $2N$.

We have confirmed numerically using DMRG
the expectation that there are four antiferromagnetic Ising-like
gaped phases,
each smoothly connected to one of the four corners of
Fig.~\ref{suppfig:phase diagram}.
These phases are identified as $\mathrm{\hbox{N\'eel}}^{\,}_{x}$, $\mathrm{\hbox{N\'eel}}^{\,}_{y}$,
$\mathrm{\hbox{N\'eel}}^{\mathrm{SPT}}_{x}$, and $\mathrm{\hbox{N\'eel}}^{\mathrm{SPT}}_{y}$, labeled
counterclockwise starting from the point $(0,0)$
in Fig.~\ref{suppfig:phase diagram}.
Additionally, we have determined
the nature of the phase labeled
$\mathrm{FM}^{\,}_{z}$ in Fig.\ \ref{suppfig:phase diagram}.
To achieve these conclusions,
we define a set of observables
that will serve as probes throughout our
exploration of the phase diagram in Fig.\ \ref{suppfig:phase diagram}.

\subsubsection{Order parameters and correlation functions}

Label with $\alpha=x,y,z$
the $X$-, $Y$-, $Z$-axis in spin space, respectively.
For each $\alpha=x,y,z$,
define on the site $j$ the Pauli operator
\begin{equation}
\widehat{\sigma}^{\alpha}_{j}\coloneqq
\begin{cases}
\widehat{X}^{\,}_{j},&\alpha=x,\\
\widehat{Y}^{\,}_{j},&\alpha=y,\\
\widehat{Z}^{\,}_{j},&\alpha=z,\\
\end{cases}
\end{equation}
in the $X$-, $Y$-, $Z$-axis in spin space, respectively.

For any state $\ket*{\Psi}\in\mathbb{C}^{2^{2N}}$,
with $\mathbb{C}^{2^{2N}}$
the domain of definition of Hamiltonian
\eqref{suppeq:def H},
the following correlation functions are defined. 

For any $\alpha=x,y,z$,
the two-point correlation function
$C^{\alpha}$,
the connected component of the two-point
correlation function
$C^{\alpha,\mathrm{c}}$,
and the dressed two-point correlation function
$C^{\mathrm{SPT},\alpha}$
in the state $\ket*{\Psi}$
are defined 
by the expectation values
\begin{subequations}
\label{suppeq:Cs}
\begin{align}
&\label{suppeq:Calpha}
C^{\alpha}_{j^{\,}_{0},j^{\,}_{0}+s}[\Psi]\coloneqq
\ev*{\, \widehat{\sigma}^{\alpha}_{j^{\,}_{0}}\,
\widehat{\sigma}^{\alpha}_{j^{\,}_{0}+s}}{\Psi},
\\
&\label{suppeq:CalphaConn}
C^{\alpha,\mathrm{c}}_{j^{\,}_{0},j^{\,}_{0}+s}[\Psi]\coloneqq
\ev*{\,\widehat{\sigma}^{\alpha}_{j^{\,}_{0}}\,
\widehat{\sigma}^{\alpha}_{j^{\,}_{0}+s}}{\Psi}
-
\ev*{\,\widehat{\sigma}^{\alpha}_{j^{\,}_{0}}}{\Psi}
\ev*{\,\widehat{\sigma}^{\alpha}_{j^{\,}_{0}+s}}{\Psi},
\\
&\label{suppeq:CalphaSPT}
C^{\mathrm{SPT},\alpha}_{j^{\,}_{0},j^{\,}_{0}+s}[\Psi]\coloneqq
\ev*{\,
(
\widehat{\sigma}^{z}_{j^{\,}_{0}-1}\,
\widehat{\sigma}^{\alpha}_{j^{\,}_{0}}\,
\widehat{\sigma}^{z}_{j^{\,}_{0}+1}
)\,
(
\widehat{\sigma}^{z}_{j^{\,}_{0}-1+s}\,
\widehat{\sigma}^{\alpha}_{j^{\,}_{0}+s}\,
\widehat{\sigma}^{z}_{j^{\,}_{0}+1+s}
)\,
}{\Psi},
\end{align}
\end{subequations}
respectively. When OBC are selected, these two-point correlation
functions depend on both $j^{\,}_{0}$ and $j^{\,}_{0}+s$.
Notice that $C^{\mathrm{SPT},\alpha}$ is the
correlation between two monomials of order three. 
For $\alpha=x,y$, the correlation $C^{\mathrm{SPT},\alpha}$ is
the relevant order parameter for one of the SPT phases governed by the
$r=3$ cluster spin-$1/2$ chain
(\ref{suppeq:Hamiltonians at lambda=1 a})
and
(\ref{suppeq:Hamiltonians at lambda=1 b})
in the phase diagram shown in
Fig.~\ref{suppfig:phase diagram}.
The correlation function $C^{\mathrm{SPT},\alpha}$
in Eq.\ (\ref{suppeq:CalphaSPT})
for $\alpha=x,y$
is obtained by conjugation of the spin operators
by $\widehat{U}^{\,}_{\mathrm{CZ}}$
on the right-hand side of Eq.~(\ref{suppeq:Calpha}),
\begin{align}
C^{\mathrm{SPT},\alpha}_{j^{\,}_{0},j^{\,}_{0}+s}[\Psi]=&\,
\langle\Psi|
\,\big(
\widehat{\sigma}^{z}_{j^{\,}_{0}-1}\,
\widehat{\sigma}^{\alpha}_{j^{\,}_{0}}\,
\widehat{\sigma}^{z}_{j^{\,}_{0}+1}
\big)\,
\big(
\widehat{\sigma}^{z}_{j^{\,}_{0}-1+s}\,
\widehat{\sigma}^{\alpha}_{j^{\,}_{0}+s}\,
\widehat{\sigma}^{z}_{j^{\,}_{0}+1+s}
\big)\,
|\Psi\rangle
\nonumber\\
=&\,
\langle\Psi|
\,\big(
\widehat{U}^{\,}_{\mathrm{CZ}}\,
\widehat{\sigma}^{\alpha}_{j^{\,}_{0}}\,
\widehat{U}^{\dag}_{\mathrm{CZ}}
\big)\,
\big(
\widehat{U}^{\,}_{\mathrm{CZ}}\,
\widehat{\sigma}^{\alpha}_{j^{\,}_{0}+s}\,
\widehat{U}^{\dag}_{\mathrm{CZ}}
\big)\,
|\Psi\rangle
\nonumber\\
=&\,
\big(
\langle\Psi|\,
\widehat{U}^{\,}_{\mathrm{CZ}}
\big)\,
\widehat{\sigma}^{\alpha}_{j^{\,}_{0}}\,
\widehat{\sigma}^{\alpha}_{j^{\,}_{0}+s}\,
\big(
\widehat{U}^{\dag}_{\mathrm{CZ}}\,
|\Psi\rangle
\big)
\nonumber\\
\equiv&\,
\langle\Psi^{\,}_{\mathrm{CZ}}|\,
\widehat{\sigma}^{\alpha}_{j^{\,}_{0}}\,
\widehat{\sigma}^{\alpha}_{j^{\,}_{0}+s}\,
|\Psi^{\,}_{\mathrm{CZ}}\rangle
\nonumber\\
=&\,
C^{\alpha}_{j^{\,}_{0},j^{\,}_{0}+s}[\Psi^{\,}_{\mathrm{CZ}}].
\end{align}
For any $r=1,2,\cdots$,
we also define the expectation value
$G^{r,\alpha}$ 
of $r$-string operators in the state $\ket*{\Psi}$ by
the expectation value
\begin{align}
\label{suppeq:Gr}
G^{r,\alpha}_{j^{\,}_{0},\cdots,j^{\,}_{0}+r-1}[\Psi]\coloneqq
\begin{cases}
\ev*{\,
\widehat{\sigma}^{\alpha}_{j^{\,}_{0}}\,
\widehat{\sigma}^{\alpha}_{j^{\,}_{0}+1}
}{\Psi},
&
r=2,
\\&\\
\ev*{\,
\widehat{\sigma}^{z}_{j^{\,}_{0}}\,
\widehat{\sigma}^{\alpha}_{j^{\,}_{0}+1}\,
\cdots\,
\widehat{\sigma}^{\alpha}_{j^{\,}_{0}+r-2}\,
\widehat{\sigma}^{z}_{j^{\,}_{0}+r-1}}{\Psi},
&
r>2.\\
\end{cases}
\end{align}
These string-expectation values are
used to identify the topological character 
of the state $\ket*{\Psi}$.

Finally, for any $\alpha=x,y,z$,
the uniform (FM) magnetization
$m^{\alpha}_{\mathrm{uni}}$,
the staggered (AFM) magnetization
$m^{\alpha}_{\mathrm{sta}}$,
the uniform (FM) SPT magnetization
$m^{\alpha}_{\mathrm{uni,SPT}}$,
and
the staggered (AFM) SPT magnetization
$m^{\alpha}_{\mathrm{stag,SPT}}$
in the state $\ket*{\Psi}$
are defined by the expectation values
\footnote{
We do not index $m^{\alpha}_{\mathrm{uni}}$ and $m^{\alpha}_{\mathrm{sta}}$
with the label $b=0,1$ defining the boundary conditions,
as
$m^{\alpha}_{\mathrm{uni}}$ and $m^{\alpha}_{\mathrm{sta}}$ 
were independent of $b=0,1$ when non-vanishing.
}
\begin{subequations}
\label{suppeq:def order parameters}
\begin{align}
&
m^{\alpha}_{\mathrm{uni}}[\Psi]\coloneqq
\frac{1}{2N}\,
\sum_{j=1}^{2N} \ev*{\,\widehat{\sigma}^{\alpha}_{j}}{\Psi},
\label{suppeq:muni}\\
&
m^{\alpha}_{\mathrm{sta}}[\Psi]\coloneqq
\frac{1}{2N}\,
\sum_{j=1}^{2N}
(-1)^{j}\,
\ev*{\,\widehat{\sigma}^{\alpha}_{j}}{\Psi},
\label{suppeq:msta}\\
&
m^{\alpha}_{\mathrm{uni,SPT}}[\Psi]\coloneqq
\frac{1}{2N-2b}\,
\sum_{j=1+b}^{2N-b}
\ev*{\,
\widehat{\sigma}^{z}_{j-1}\,
\widehat{\sigma}^{\alpha}_{j}\,
\widehat{\sigma}^{z}_{j+1}
}{\Psi},
\label{suppeq:muniSPT}\\
&
m^{\alpha}_{\mathrm{sta,SPT}}[\Psi]\coloneqq
\frac{1}{2N-2b}\,
\sum_{j=1+b}^{2N-b}
(-1)^{j}\,
\ev*{\,
\widehat{\sigma}^{z}_{j-1}\,
\widehat{\sigma}^{\alpha}_{j}\,
\widehat{\sigma}^{z}_{j+1}
}{\Psi},
\label{suppeq:mstaSPT}
\end{align}
\end{subequations}
respectively. The parameter $b=0\ (1)$ dictates
the choice of PBC (OBC). One verifies that
\begin{equation}
m^{\alpha}_{\mathrm{uni,SPT}}[\Psi]=
m^{\alpha}_{\mathrm{uni}}[\Psi^{\,}_{\mathrm{CZ}}],
\qquad
m^{\alpha}_{\mathrm{sta,SPT}}[\Psi]=
m^{\alpha}_{\mathrm{sta}}[\Psi^{\,}_{\mathrm{CZ}}],
\qquad
|\Psi^{\,}_{\mathrm{CZ}}\rangle=
\widehat{U}^{\dag}_{\mathrm{CZ}}\,
|\Psi\rangle,
\qquad
\alpha=x,y,
\end{equation}
hold for PBC.
We shall use the short-hand notation
\begin{equation}
\ev*{\, \cdot \,}^{\,}_{\Psi} \equiv \ev*{\, \cdot \,}{\Psi}
\label{suppeq:bracket psi}
\end{equation}
for any expectation value in the state $|\Psi\rangle\in\mathbb{C}^{2^{2N}}$.

\begin{figure}[t!]
\centering
\subfigure[]{\includegraphics[width=0.33\columnwidth]{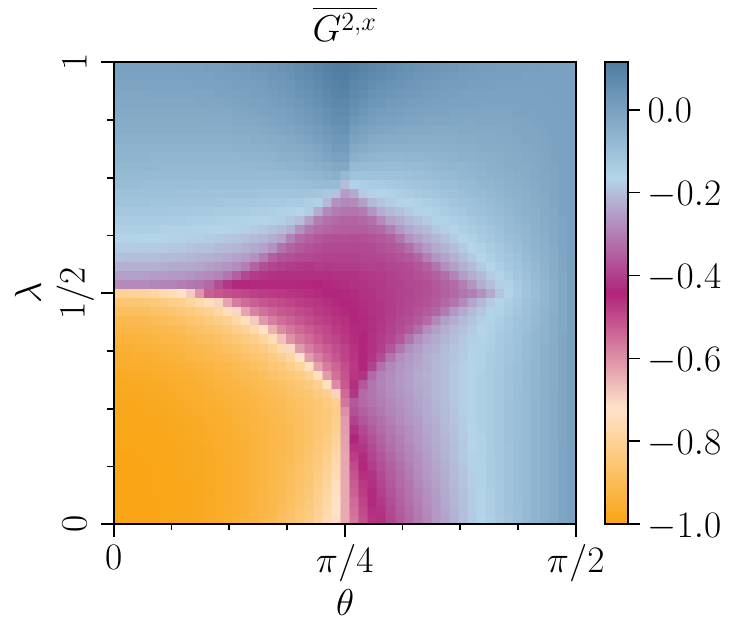} \label{suppfig:DMRG OBC Gs a}}%
\subfigure[]{\includegraphics[width=0.33\columnwidth]{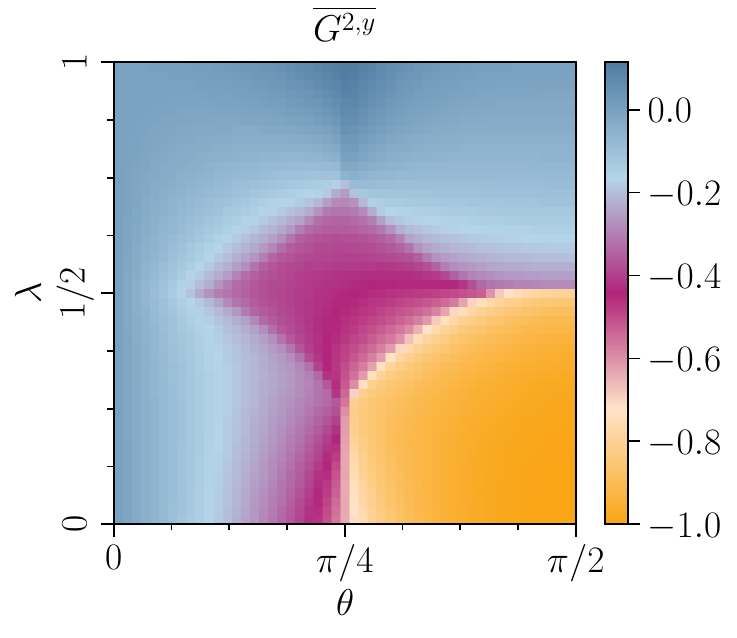} \label{suppfig:DMRG OBC Gs b}}%
\subfigure[]{\includegraphics[width=0.33\columnwidth]{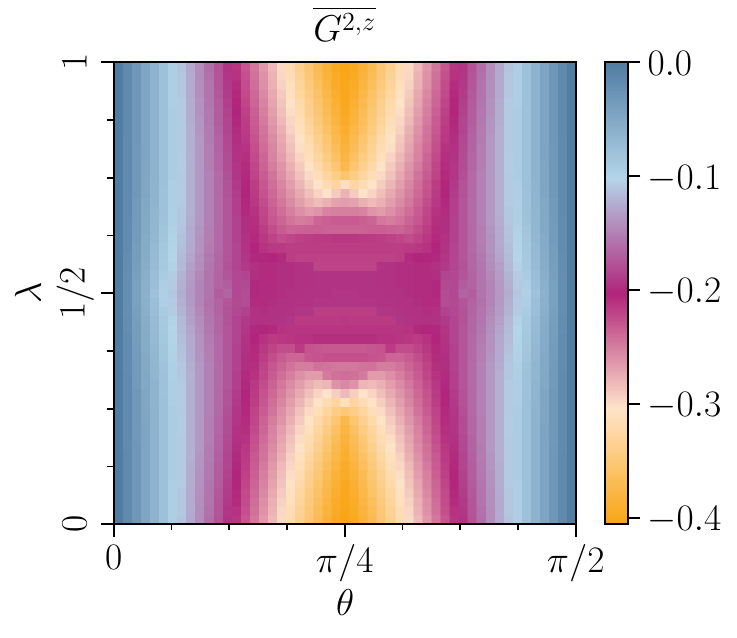} \label{suppfig:DMRG OBC Gs c}}\\
\subfigure[]{\includegraphics[width=0.33\columnwidth]{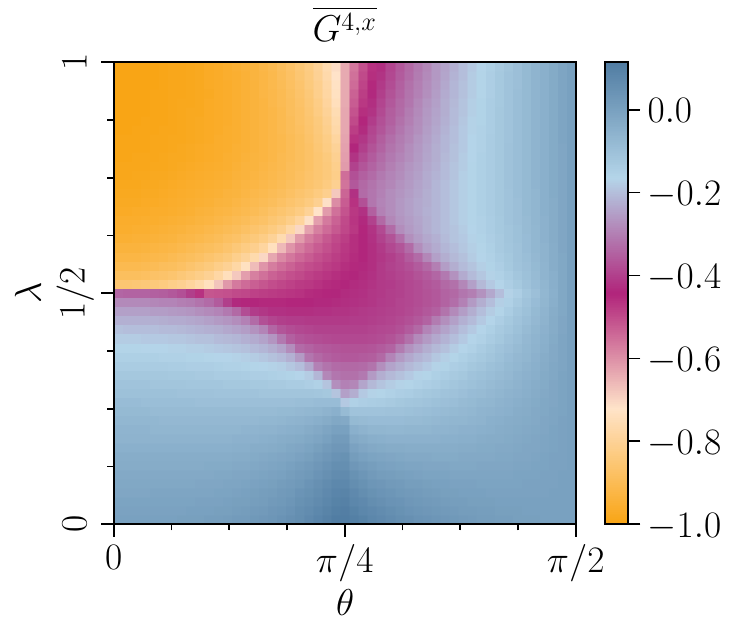} \label{suppfig:DMRG OBC Gs d}}%
\subfigure[]{\includegraphics[width=0.33\columnwidth]{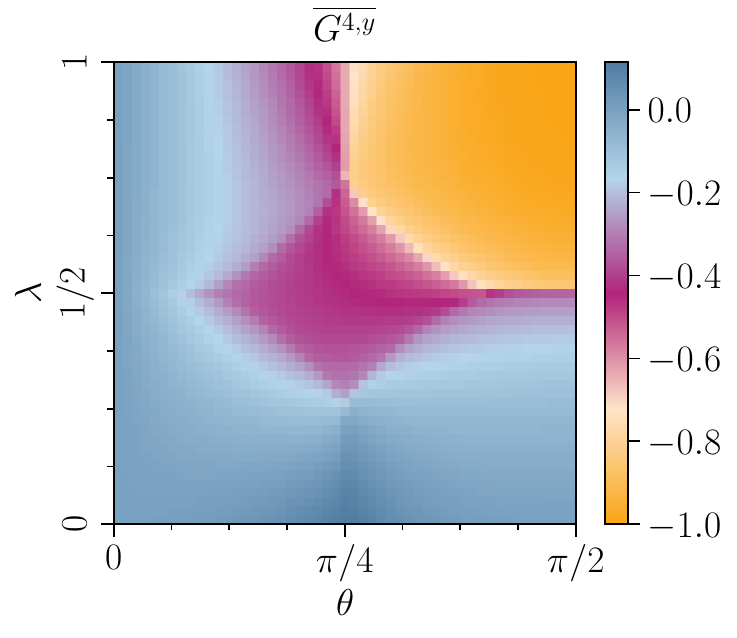} \label{suppfig:DMRG OBC Gs e}}%
\subfigure[]{\includegraphics[width=0.33\columnwidth]{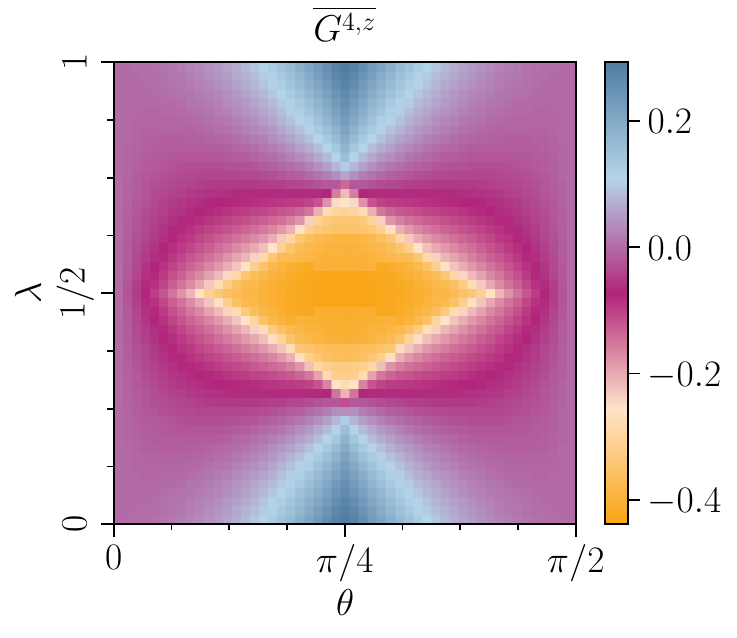} \label{suppfig:DMRG OBC Gs f}}\\
\subfigure[]{\includegraphics[width=0.33\columnwidth]{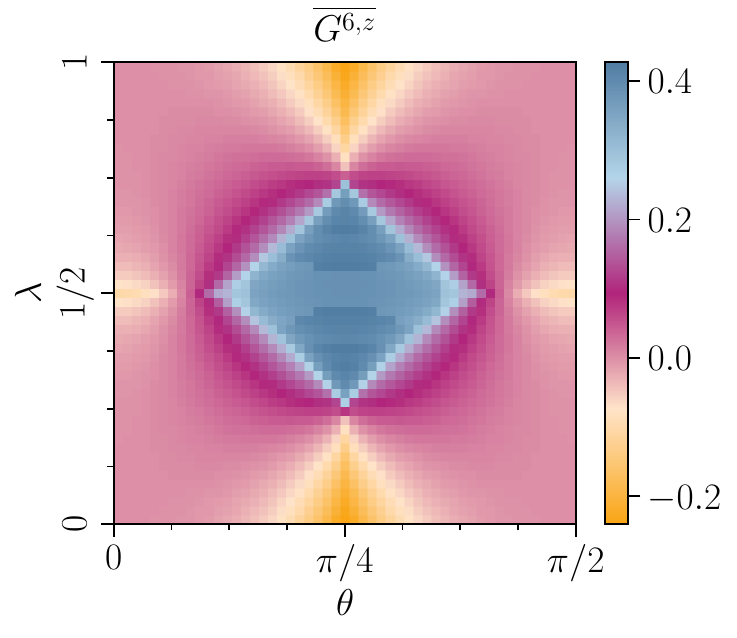} \label{suppfig:DMRG OBC Gs g}}%
\subfigure[]{\includegraphics[width=0.33\columnwidth]{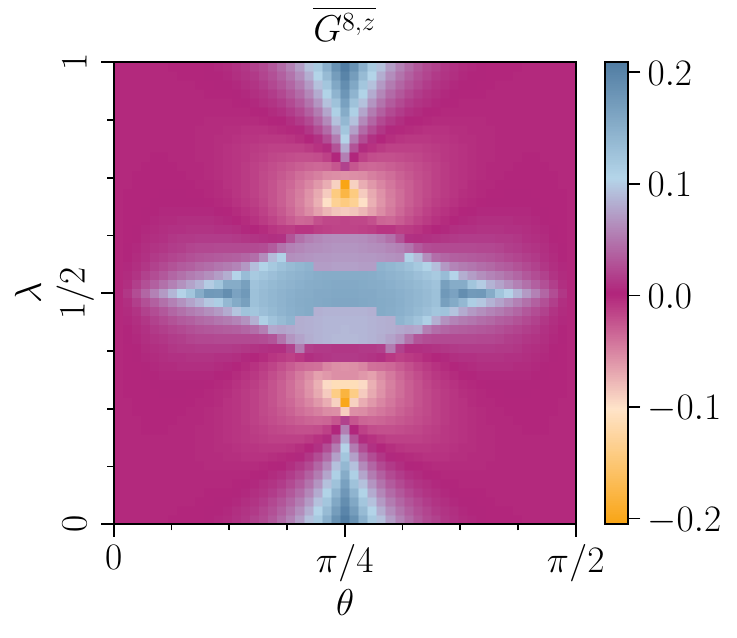} \label{suppfig:DMRG OBC Gs h}}%
\subfigure[]{\includegraphics[width=0.33\columnwidth]{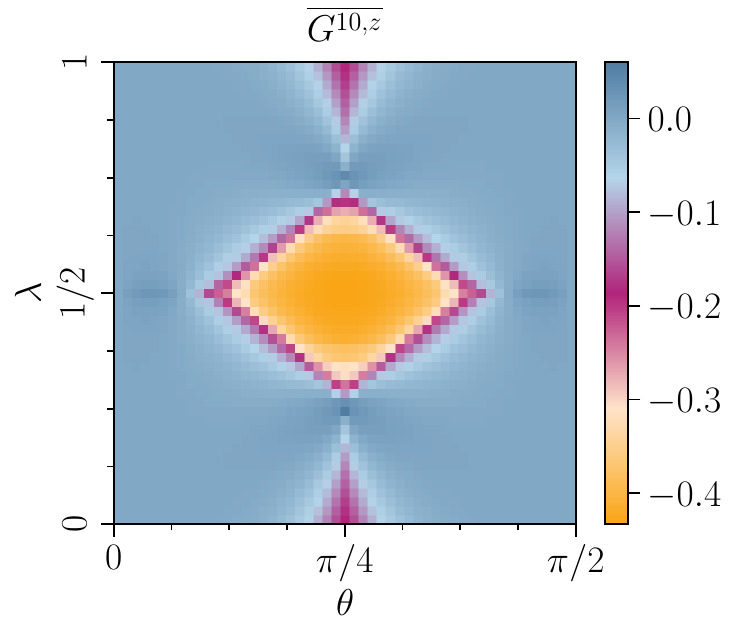} \label{suppfig:DMRG OBC Gs i}}
\caption{(Color online)
The dependence on $\theta$ and $\lambda$ of
(a) $\overline{G^{2,x}}[\Psi]$,
(b) $\overline{G^{2,y}}[\Psi]$,
(c) $\overline{G^{2,z}}[\Psi]$,
(d) $\overline{G^{4,x}}[\Psi]$,
(e) $\overline{G^{4,y}}[\Psi]$,
(f) $\overline{G^{4,z}}[\Psi]$,
(g) $\overline{G^{6,z}}[\Psi]$,
(h) $\overline{G^{8,z}}[\Psi]$,
and
(i) $\overline{G^{10,z}}[\Psi]$,
defined in Eqs.~(\ref{suppeq:Gr})
and~(\ref{suppeq:def overline  Gr,alpha}), is shown through intensity plots.
Here, the state $|\Psi\rangle$ that approximates the ground state is
computed using DMRG
for a chain of length $2N=128$ with OBC.
The bond dimension $\chi=128$ is used in the DMRG.
}
\label{suppfig:DMRG OBC Gs}
\end{figure}

Figure~\ref{suppfig:DMRG OBC Gs} serves as a guide in the
exploration of the five phases of Hamiltonian \eqref{suppeq:def H}
with the help of numerical
estimates for the thermodynamic limit of the order parameters
(\ref{suppeq:def order parameters})
and of the correlation functions
(\ref{suppeq:Cs}) and (\ref{suppeq:Gr}).
We hereby define the space average
\begin{equation}
\overline{G^{r,
\alpha}}[\Psi]\coloneqq
\frac{1}{N-r+1}
\sum_{j^{\,}_{0}=1}^{N-r+1}
G^{r,\alpha}_{j^{\,}_{0},\cdots,j^{\,}_{0}+r-1}[\Psi]
\label{suppeq:def overline  Gr,alpha}
\end{equation}
for any state $|\Psi\rangle\in\mathbb{C}^{2^{2N}}$. This correlation function
is useful when choosing OBC.

\begin{figure}[t!]
\centering
\includegraphics[width=0.7\columnwidth]{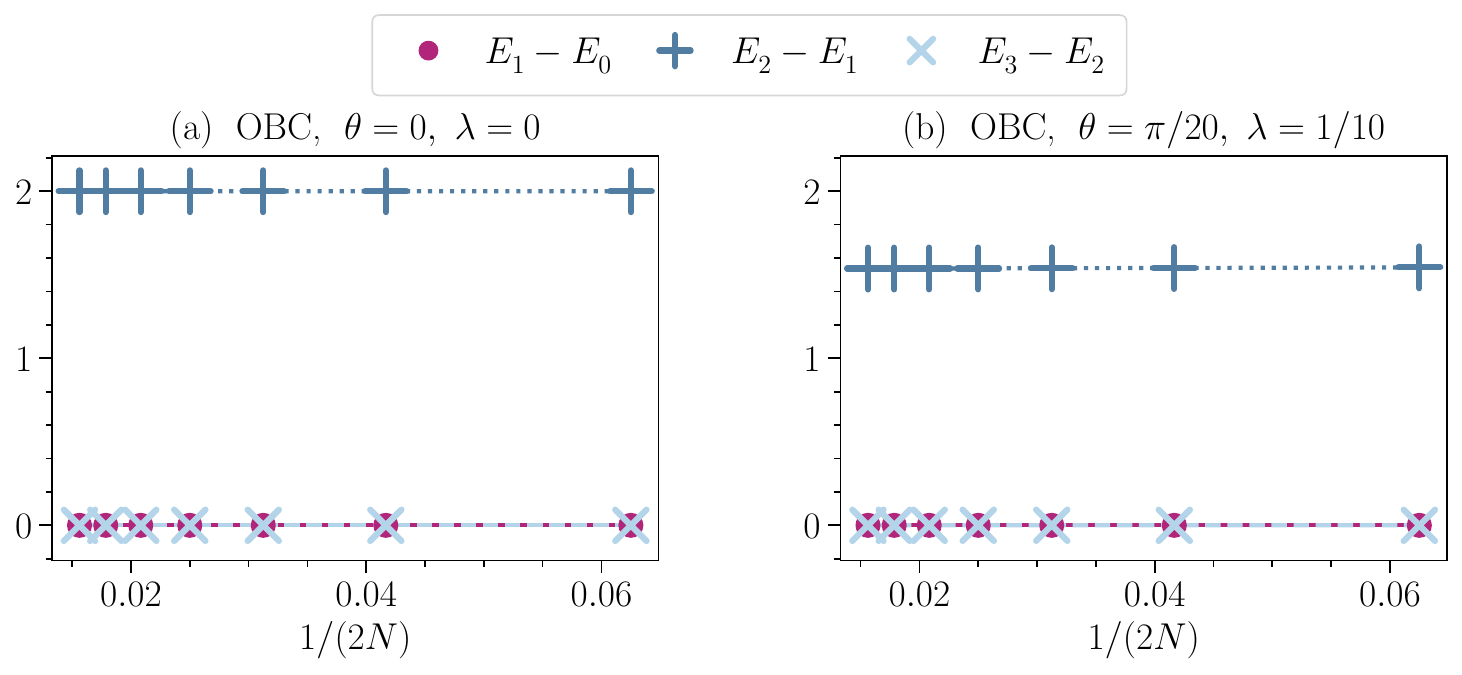}
\subfigure{\label{suppfig:DMRG OBC NeelX gaps a}}
\subfigure{\label{suppfig:DMRG OBC NeelX gaps b}}
\vspace{-0.5cm}
\caption{(Color online)
The case of OBC. Dependence on $1/(2N)$ of the
energy spacings $E^{\,}_{n}-E^{\,}_{n-1}$, $n=1,2,3$,
above the ground-state energy $E^{\,}_{0}$ obtained with DMRG
($\chi=128$) for
(a) the point $(0,0)$ and
(b) the point $(\pi/20,1/10)$
in the $\mathrm{\hbox{N\'eel}}^{\,}_{x}$ phase from Fig.~\ref{suppfig:phase diagram}.
}
\label{suppfig:DMRG OBC NeelX gaps}
\end{figure}

\begin{figure}[t!]
\centering
\includegraphics[width=0.7\columnwidth]{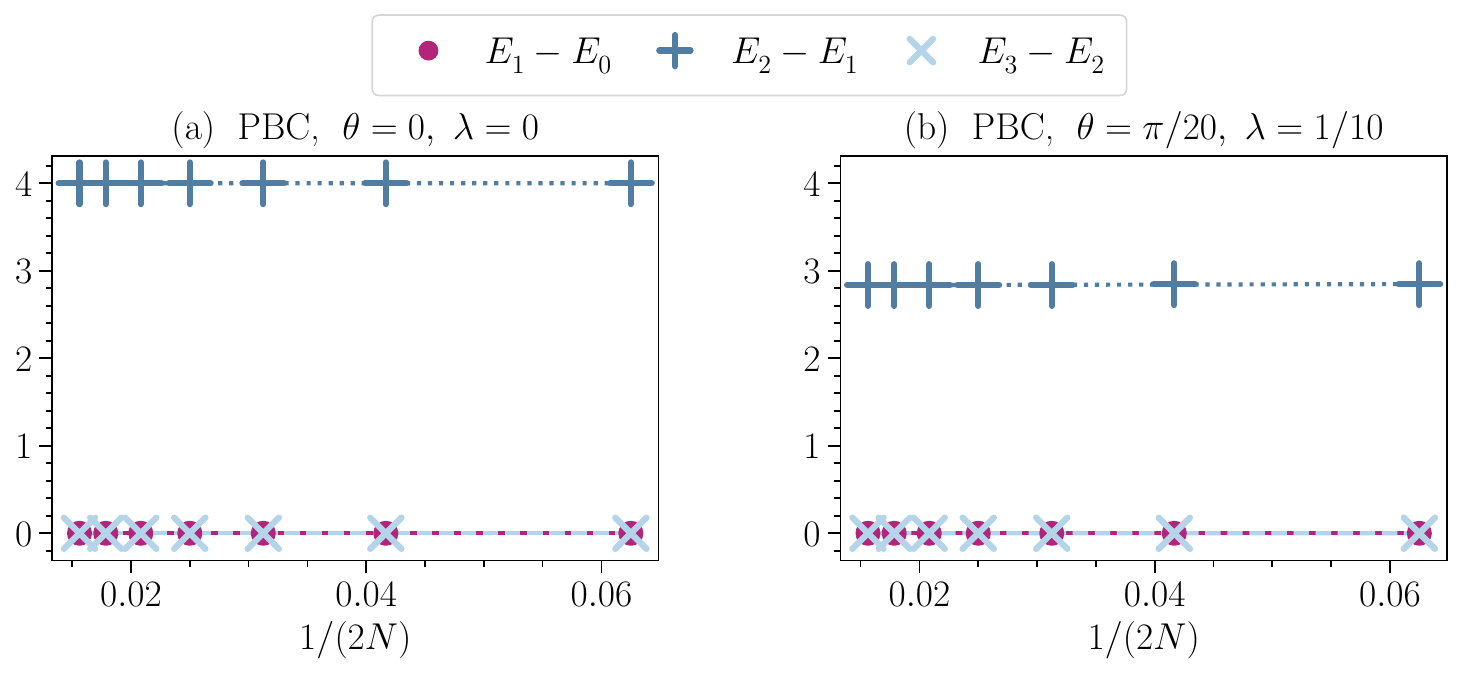}
\subfigure{\label{suppfig:DMRG PBC NeelX gaps a} \relax}
\subfigure{\label{suppfig:DMRG PBC NeelX gaps b} \relax}
\vspace{-0.5cm}
\caption{(Color online)
The case of PBC.
Dependence on $1/(2N)$ of the
energy spacings $E^{\,}_{n}-E^{\,}_{n-1}$, $n=1,2,3$,
above the ground-state energy $E^{\,}_{0}$ obtained with DMRG
($\chi=128$) for
(a) the point $(0,0)$ and
(b) the point $(\pi/20,1/10)$
in the $\mathrm{\hbox{N\'eel}}^{\,}_{x}$ phase from Fig.~\ref{suppfig:phase diagram}.
}
\label{suppfig:DMRG PBC NeelX gaps}
\end{figure}

\subsubsection{The $\mathrm{\hbox{N\'eel}}^{\,}_{x}$ phase}
\label{suppsec:NeelX}

The $\mathrm{\hbox{N\'eel}}^{\,}_{x}$ phase is the only one of the four gapped phases
for which it is possible to find a ground state $|\Psi\rangle$ such that,
in the thermodynamic limit,
\begin{equation}
\overline{G^{2,x}}[\Psi]\neq0,
\qquad
\overline{G^{2,y}}[\Psi]=0,
\qquad
\overline{G^{4,x}}[\Psi]=0,
\qquad
\overline{G^{4,y}}[\Psi]=0,
\end{equation}
in Fig.~\ref{suppfig:DMRG OBC Gs}.

With PBC, any one of the four corners in Fig.~\ref{suppfig:phase diagram}
is unitarily equivalent to a classical antiferromagnetic
nearest-neighbor Ising chain with the Hamiltonian
$\widehat{H}^{\,}_{0}(0,0)$ defined in Eq.~(\ref{suppeq:Hamiltonians at lambda=0 a}).

For the classical antiferromagnetic nearest-neighbor Ising chain
$\widehat{H}^{\,}_{0}(0,0)$, all ordered energy eigenvalues
\begin{equation}
E^{\,}_{0}\leq
E^{\,}_{1}\leq
\cdots\leq
E^{\,}_{n}\leq
E^{\,}_{n+1}\leq
\cdots            
\end{equation}
with the orthonormal eigenfunctions
\begin{equation}
|\Psi^{\,}_{0}\rangle,\
|\Psi^{\,}_{1}\rangle,\
\cdots,\
|\Psi^{\,}_{n}\rangle,\
|\Psi^{\,}_{n+1}\rangle,\
\cdots             
\end{equation}
are at least two-fold degenerate, for any spin configuration must share the
same energy as the one obtained by reversing the direction of all
spins through the operation of time reversal $\widehat{T}$.
Moreover, any two distinct but consecutive energy eigenvalues
are equally spaced.
This spacing is the gap $\Delta^{\,}_{0}(0,0)$ that can be
computed from subtracting the energy of any one of the two
$\mathrm{\hbox{N\'eel}}^{\,}_{x}$ states
\footnote{
The state $\ket*{\Psi^{\,}_{0}}$
is the representation
in the basis that diagonalizes
$\protect{\widehat{\sigma}^{x}_{j}}$
of the state $\ket*{\mathrm{\hbox{N\'eel}}^{\,}_{x},+}$
defined by Eq.~(\ref{suppeq:GSs at lambda=0 a})
in the basis that diagonalizes
$\protect{\widehat{\sigma}^{z}_{j}}$
}
\begin{subequations}
\label{suppeq:NeelX states}
\begin{align}
&
\ket*{\Psi^{\,}_{0}}\coloneqq
|
\leftarrow\rightarrow\cdots
\leftarrow\rightarrow\
\leftarrow\rightarrow\
\leftarrow\rightarrow\cdots
\leftarrow\rightarrow
\rangle,
\label{suppeq:NeelX states 0}
\\
&
\ket*{\Psi^{\,}_{1}}\coloneqq
\widehat{T}\,\ket*{\Psi^{\,}_{0}},
\qquad 
\widehat{T}\,|\rightarrow\rangle
=
\mathrm{i}|\leftarrow\rangle,
\quad
\widehat{T}\,|\leftarrow\rangle
=
-\mathrm{i}|\rightarrow\rangle,
\label{suppeq:NeelX states 1}
\end{align}
in the basis for which
\begin{align}
&
\widehat{\sigma}^{x}_{j}\,|\rightarrow\rangle^{\,}_{j}=
+|\rightarrow\rangle^{\,}_{j},
\label{suppeq:NeelX states 00}
\\
&
\widehat{\sigma}^{x}_{j}\,|\leftarrow\rangle^{\,}_{j}=
-|\leftarrow\rangle^{\,}_{j},
\label{suppeq:NeelX states 11}
\end{align}
from the energy of one of the excited states, say
\begin{align}
&
\ket*{\Psi^{\,}_{2}}\coloneqq
\begin{cases}
|
\leftarrow\rightarrow\cdots
\leftarrow\rightarrow\
\leftarrow\red{\!\!\rightarrow}\
\red{\rightarrow\!\!}\leftarrow\cdots
\rightarrow\leftarrow
\rangle,
&\hbox{ for OBC,}
\\ &\\
|
\leftarrow\rightarrow\cdots
\leftarrow\!\red{\!\rightarrow}\
\red{\rightarrow}\blue{\leftarrow}\
\blue{\leftarrow\!\!}\rightarrow\cdots
\leftarrow\rightarrow
\rangle,
&\hbox{ for PBC,}
\label{suppeq:NeelX states 2}
\end{cases}
\\
&
\ket*{\Psi^{\,}_{3}}\coloneqq\widehat{T}\,\ket*{\Psi^{\,}_{2}},
\label{suppeq:NeelX states 3}
\end{align}
\end{subequations}
The existence of the energy gap $E^{\,}_{2}-E^{\,}_{1}$
guarantees the stability of
the $\mathrm{\hbox{N\'eel}}^{\,}_{x}$ phase in the phase diagram in Fig.~\ref{suppfig:phase diagram}.

The energy with respect of Hamiltonian $\widehat{H}^{\,}_{1}(0,0)$
of the $\mathrm{\hbox{N\'eel}}^{\,}_{x}$ state
(\ref{suppeq:NeelX states 0})
and its time-reversed counterpart
(\ref{suppeq:NeelX states 1})
is that of the twofold-degenerate
ground-state energy
$E^{\,}_{0}=E^{\,}_{1}$
of $\widehat{H}^{\,}_{1}(0,0)$.
Under OBC, this energy is
[see Eq.~(\ref{suppeq:case OBC spectral decomposition cluster models})]
\begin{subequations}
\begin{equation}
E^{\,}_{0}=-2N+1
\label{suppeq:E0=-2N+1} 
\end{equation}
for $\widehat{H}^{\,}_{1}(0,0)$
and it is the reference energy in
Fig.~\ref{suppfig:DMRG OBC NeelX gaps}.
Under PBC, this energy is
[see Eq.~(\ref{suppeq:case PBC spectral decomposition cluster models})]
\begin{equation}
E^{\,}_{0}=-2N 
\end{equation}
\end{subequations}
for $\widehat{H}^{\,}_{0}(0,0)$
and it is the reference energy in
Fig.~\ref{suppfig:DMRG PBC NeelX gaps}.
The first excited state defined in Eq.~(\ref{suppeq:NeelX states 2})
of Hamiltonian $\widehat{H}^{\,}_{1}(0,0)$
is
[see Eq.~(\ref{suppeq:case OBC spectral decomposition cluster models})]
\begin{subequations}
\begin{equation}
E^{\,}_{2}=-2N+3  
\end{equation}
when OBC are chosen.
It follows that the differences in the energies
$E^{\,}_{1}-E^{\,}_{0}$ and $E^{\,}_{2}-E^{\,}_{1}$
of Fig.~\ref{suppfig:DMRG OBC NeelX gaps a}
are independent of $2N$.
The first excited state defined in Eq.~(\ref{suppeq:NeelX states 2})
of Hamiltonian $\widehat{H}^{\,}_{0}(0,0)$
is
[see Eq.~(\ref{suppeq:case PBC spectral decomposition cluster models})]
\begin{equation}
E^{\,}_{2}=-2N+4
\end{equation}
\end{subequations}
when PBC are chosen.
Hereto, it follows that the differences in the energies
$E^{\,}_{1}-E^{\,}_{0}$ and $E^{\,}_{2}-E^{\,}_{1}$
of Fig.~\ref{suppfig:DMRG PBC NeelX gaps a}
are independent of $2N$.
The dependence on the choice between OBC and PBC
for the value of $E^{\,}_{2}-E^{\,}_{1}$
can be understood as follows.
Under OBC, excitations can support an odd number of domain walls, e.g.,
Eq.~(\ref{suppeq:NeelX states 2}),
with each domain wall contributing the
energy $+2$, e.g., Fig.~\ref{suppfig:DMRG OBC NeelX gaps a}.  
In contrast, 
under PBC, excitations must support an even number of domain walls,
e.g.,
Eq.~(\ref{suppeq:NeelX states 2}),
with each domain wall
contributing an energy of $+2$, e.g.,
Fig.~\ref{suppfig:DMRG PBC NeelX gaps a}.
This difference in the structure of the excitations
affects the degeneracy of the excitation energy $E^{\,}_{2}$.
For OBC, the degeneracy for the lowest lying excited states is
\footnote{
With OBC, select either one of the two ground states
$|\Psi^{\,}_{0}\rangle$
or
$|\Psi^{\,}_{1}\rangle$
by fixing the eigenvalue of $\protect{\widehat{X}^{\,}_{1}}$
on the first site. There are
``choose one sites out of $2N-1$ sites'' ways
of creating one domain wall labeled by $j=2,\cdots,2N$
by reversing the eigenvalues of
$\protect{\widehat{X}^{\,}_{k}}$
for all $k=j,j+1,\cdots,2N$.
This is the degeneracy $d^{(1)}_{1}$
defined in
Eq.~(\ref{suppeq:case OBC spectral decomposition cluster models b}) 
}
\begin{subequations}
\begin{equation}
2(2N-1).
\label{suppeq:degeneracy lowest excited states Ising OBC}  
\end{equation}
For PBC, the degeneracy for the lowest lying excited states is
\footnote{
With PBC, starting from either the classical antiferromagnetic states
$|\Psi^{\,}_{0}\rangle$
or
$|\Psi^{\,}_{1}\rangle$,
there are
``choose two sites out of $2N$ sites'' ways
of creating two domain walls
labeled by $j=1,\cdots,2N$ and $j'=j+1,\cdots,2N$
by reversing the eigenvalues of
$\protect{\widehat{X}^{\,}_{k}}$ with $j\leq k\leq j'$.
The same degeneracy $2N(2N-1)$ is obtained from observing that
$E^{(0)}_{n}=E^{(0)}_{n+1}$ for $n=1,3,5,\cdots,2N-5,2N-3$ in 
Eq.~(\ref{suppeq:case PBC spectral decomposition cluster models a})
to derive the number of ways there are to insert two domain walls by
adding $d^{(0)}_{2}$ to $d^{(0)}_{1}$ defined in
Eq.~(\ref{suppeq:case PBC spectral decomposition cluster models b})
}
\begin{equation}
2N(2N-1).
\label{suppeq:degeneracy lowest excited states Ising PBC}  
\end{equation}
\end{subequations}
These distinctions highlight the significant impact
of boundary conditions on the energy eigenvalue spectrum and the corresponding
physical properties of the $\mathrm{\hbox{N\'eel}}^{\,}_{x}$ phase.

When moving slightly away from the corner $(0,0)$ in Fig.~\ref{suppfig:phase diagram},
the $\mathrm{\hbox{N\'eel}}^{\,}_{x}$ state
(\ref{suppeq:NeelX states 0})
and its time-reversed counterpart
(\ref{suppeq:NeelX states 1})
are no longer exact ground states.
Moreover, for any finite number $2N$ of sites, the energy $E^{\,}_{0}$
is strictly lower than the energy $E^{\,}_{1}$ of the first excited state.
However, the energy spacing $E^{\,}_{1}-E^{\,}_{0}$
between the ground state energy $E^{\,}_{0}$ and the first
excited state energy $E^{\,}_{1}$
is observed numerically to vanish exponentially fast with $2N$
so that a two-fold degeneracy of the ground state is recovered
in the thermodynamic limit $2N\to\infty$.
For example, for a chain made of $2N=32$ sites at $(\pi/20,1/10)$,
$E^{\,}_{1}-E^{\,}_{0}\approx0$
within numerical precision under both PBC and OBC.

We have tracked the excitation energy $E^{\,}_{2}$,
which remains non-degenerate with $E^{\,}_{0}$ in the thermodynamic limit
$2N\to\infty$, upon approaching the phase boundary to the
$\mathrm{\hbox{N\'eel}}^{\,}_{x}$ phase. The energy gap $E^{\,}_{2}-E^{\,}_{1}$ at $(\pi/20,1/10)$
is smaller than the energy gap $E^{\,}_{2}-E^{\,}_{1}$ at $(0,0)$
for any system size $2N$.
This difference is evident when comparing Figs.~\ref{suppfig:DMRG OBC NeelX gaps a} 
and~\ref{suppfig:DMRG OBC NeelX gaps b} for OBC and in
Figs.~\ref{suppfig:DMRG PBC NeelX gaps a}
and~\ref{suppfig:DMRG PBC NeelX gaps b} for PBC.
This difference aligns with the
expectation of a gap closing upon exiting the $\mathrm{\hbox{N\'eel}}^{\,}_{x}$ phase.

\begin{figure}[t!]
\centering
\includegraphics[width=0.71\columnwidth]{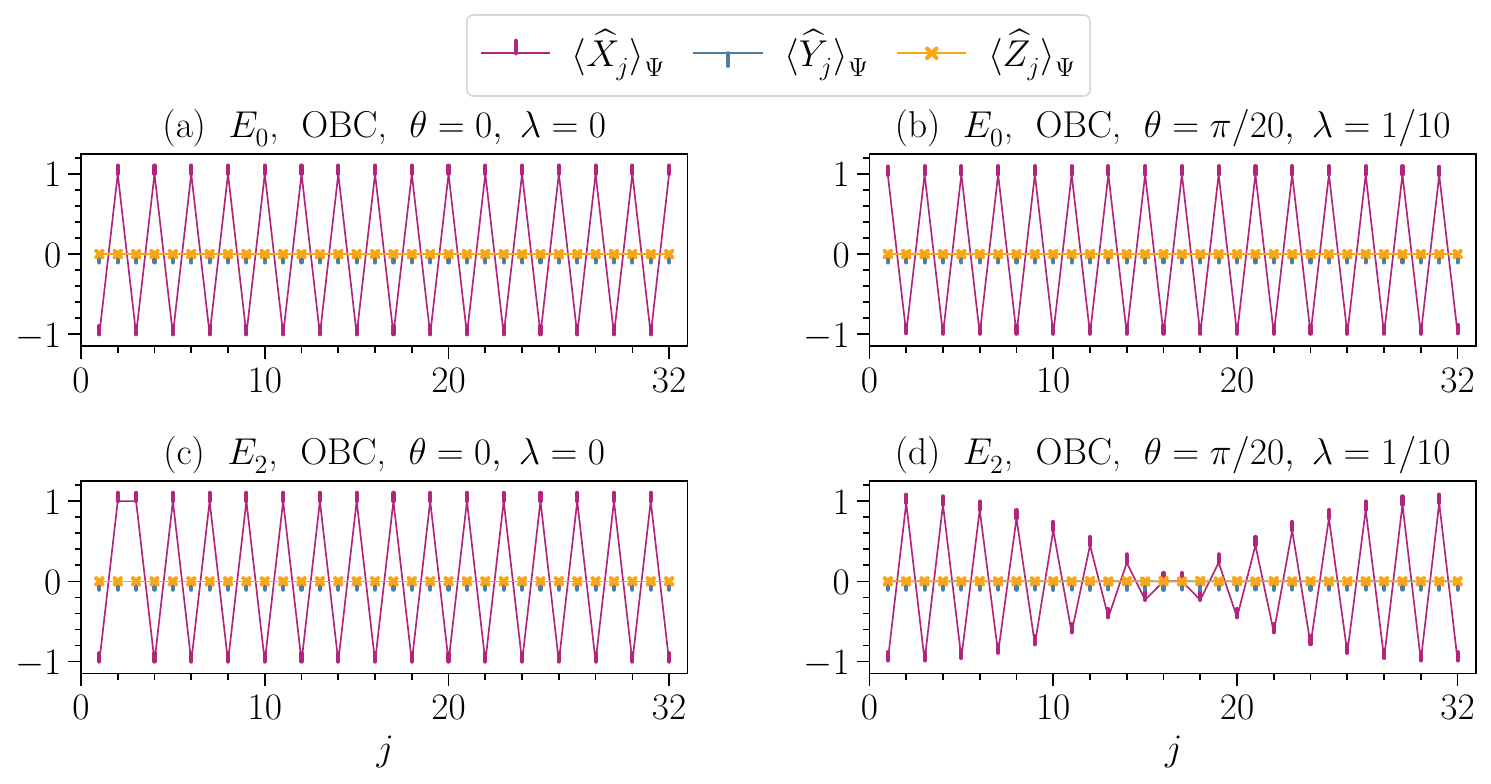}
\subfigure{\label{suppfig:DMRG OBC NeelX spins a}}
\subfigure{\label{suppfig:DMRG OBC NeelX spins b}}
\subfigure{\label{suppfig:DMRG OBC NeelX spins c}}
\subfigure{\label{suppfig:DMRG OBC NeelX spins d}}
\vspace{-0.5cm}
\caption{(Color online)
The case of OBC.
Spin expectations values in the states
obtained with DMRG
($\chi=128$ and $2N=32$)
as approximations to the exact eigenstates with energies
(a) $E^{\,}_{0}$ at $(0,0)$,
(d) $E^{\,}_{0}$ at $(\pi/20,1/10)$,
(c) $E^{\,}_{2}$ at $(0,0)$, and
(d) $E^{\,}_{2}$ at $(\pi/20,1/10)$.
}
\label{suppfig:DMRG OBC NeelX spins}
\end{figure}

\begin{figure}[t!]
\centering
\includegraphics[width=0.71\columnwidth]{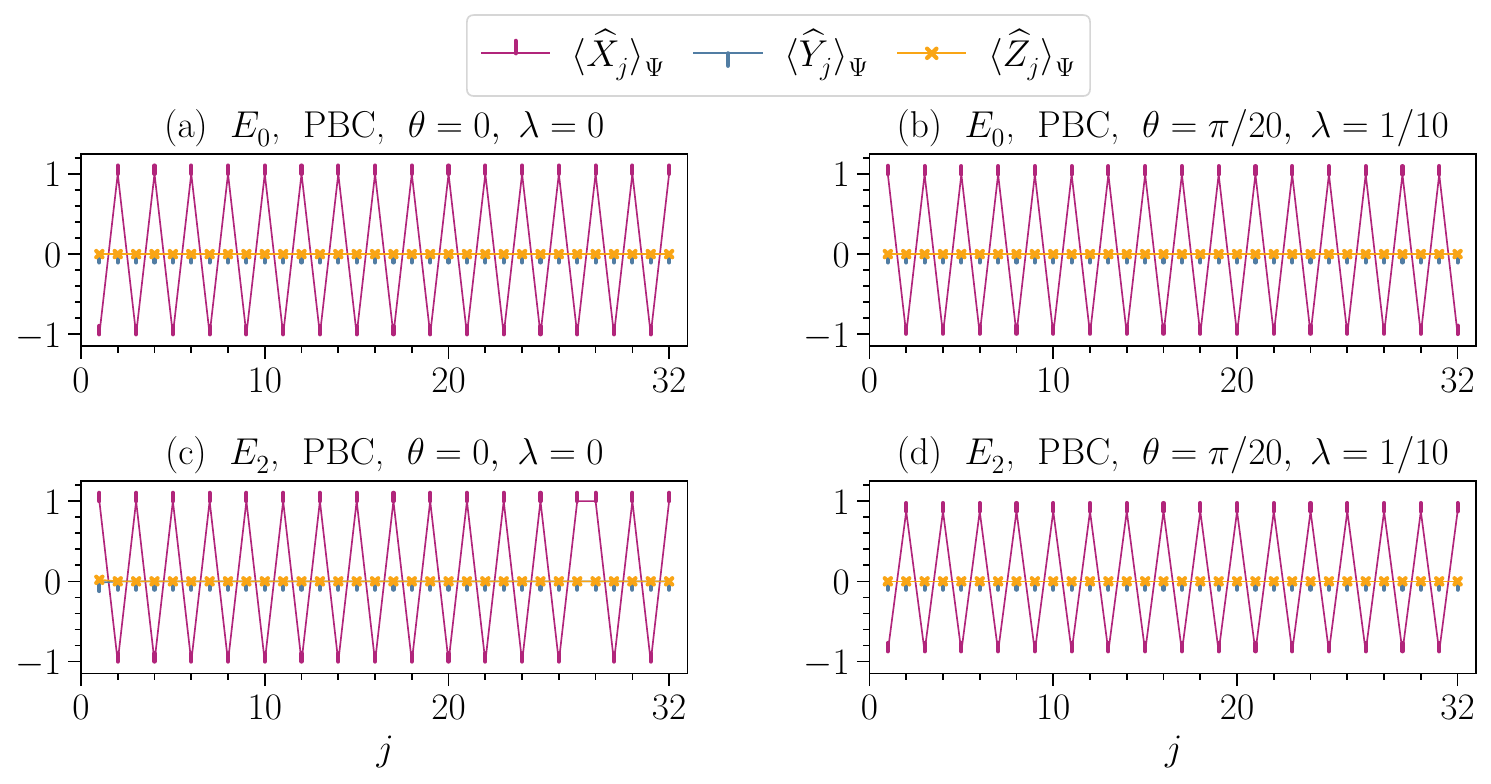}
\subfigure{\label{suppfig:DMRG PBC NeelX spins a} \relax}
\subfigure{\label{suppfig:DMRG PBC NeelX spins b} \relax}
\subfigure{\label{suppfig:DMRG PBC NeelX spins c} \relax}
\subfigure{\label{suppfig:DMRG PBC NeelX spins d} \relax}
\vspace{-0.5cm}
\caption{(Color online)
The case of PBC.
Spin expectations values in the states
obtained with DMRG
($\chi=128$ and $2N=32$)
as approximations to the exact eigenstates with energies
(a) $E^{\,}_{0}$ at $(0,0)$,
(b) $E^{\,}_{0}$ at $(\pi/20,1/10)$,
(c) $E^{\,}_{2}$ at $(0,0)$, and
(d) $E^{\,}_{2}$ at $(\pi/20,1/10)$.
}
\label{suppfig:DMRG PBC NeelX spins}
\end{figure}

At the corner $(0,0)$,
the two orthogonal ground states
$|\Psi^{\,}_{0}\rangle$
and
$|\Psi^{\,}_{1}\rangle$
of Hamiltonian
\eqref{suppeq:def H}
exhibit the staggered magnetization
\begin{subequations}
\label{suppeq:classical value sta mag}
\begin{equation}
|m^{x}_{\mathrm{sta}}[\Psi^{\,}_{0}]|=
|m^{x}_{\mathrm{sta}}[\Psi^{\,}_{1}]|=1,
\label{suppeq:classical value sta mag a}
\end{equation}
indicating perfect classical ($\mathrm{\hbox{N\'eel}}^{\,}_{x}$) antiferromagnetic order.
This is shown in
Fig.~\ref{suppfig:DMRG OBC NeelX spins a}
for OBC and in
Fig.~\ref{suppfig:DMRG PBC NeelX spins a}
for PBC.
More specifically, the ground
state $\ket*{\Psi^{\,}_{0}}$ in Eq.~(\ref{suppeq:NeelX states 0})
has the staggered magnetization
\begin{equation}
m^{x}_{\mathrm{sta}}[\Psi^{\,}_{0}]=+1,
\label{suppeq:classical value sta mag b}
\end{equation}
while its time-reversed image
$\ket*{\Psi^{\,}_{1}}$ in Eq.~(\ref{suppeq:NeelX states 1})
has the staggered magnetization
\begin{equation}
m^{x}_{\mathrm{sta}}[\Psi^{\,}_{1}]=-1.
\label{suppeq:classical value sta mag c}
\end{equation}
\end{subequations}

The states $\ket*{\Psi^{\,}_{0}}$
in Eq.~(\ref{suppeq:NeelX states 0})
and $\ket*{\Psi^{\,}_{1}}$
in Eq.~(\ref{suppeq:NeelX states 1})
define an orthonormal basis for the ground states
of the Hamiltonian $\widehat{H}^{\,}_{0}(0,0)$.
Another basis of this two-dimensional
subspace associated with the eigenvalues $E^{\,}_{0}=E^{\,}_{1}$ is
formed by the cat
states $\ket*{\Psi^{\,}_{+}}$ and $\ket*{\Psi^{\,}_{-}}$
defined by
\begin{equation}
\ket*{\Psi^{\,}_{\pm}}\coloneqq
\frac{1}{\sqrt{2}}
\left(
\ket*{\Psi^{\,}_{0}}
\pm
\ket*{\Psi^{\,}_{1}}
\right)=
\frac{1}{\sqrt{2}}
\left(
\widehat{\mathbb{1}}
\pm
\widehat{T}
\right)
\ket*{\Psi^{\,}_{0}}.
\end{equation}
It follows that both states $\ket*{\Psi^{\,}_{\pm}}$ 
obey (here we make use of the hypothesis that $2N$ is even to deduce that
$\widehat{T}^{2}=\widehat{\mathbb{1}}$)
\begin{equation}
\widehat{T}\,\ket*{\Psi^{\,}_{\pm}}\coloneqq
\frac{1}{\sqrt{2}}
\left(
\widehat{T}
\pm
\widehat{T}^{2}
\right)
\ket*{\Psi^{\,}_{0}}=
\frac{1}{\sqrt{2}}
\left(
\widehat{T}
\pm
\widehat{\mathbb{1}}
\right)
\ket*{\Psi^{\,}_{0}}=
\pm
\ket*{\Psi^{\,}_{\pm}}.
\end{equation}
As $\ket*{\Psi^{\,}_{\pm}}$ are symmetric and antisymmetric linear combinations
of orthonormal eigenstates of opposite staggered magnetization
\begin{equation}
m^{x}_{\mathrm{sta}}[\Psi^{\,}_{0}]=-m^{x}_{\mathrm{sta}}[\Psi^{\,}_{1}],
\end{equation}  
the staggered magnetization of $\ket*{\Psi^{\,}_{\pm}}$
vanishes for any finite system size $2N$, i.e.,
\begin{equation}
m^{x}_{\mathrm{sta}}[\Psi^{\,}_{\pm}]=0.
\end{equation}
For any finite number of sites $2N$,
it is only at the corner $(0,0)$ of the
$\mathrm{\hbox{N\'eel}}^{\,}_{x}$ phase
that it is possible to choose a ground state with
non-vanishing magnetization,
for it is only there in the $\mathrm{\hbox{N\'eel}}^{\,}_{x}$ phase
that the ground state is exactly two-fold degenerate.
For any finite number of sites $2N$ and at any point in the
$\mathrm{\hbox{N\'eel}}^{\,}_{x}$ phase distinct from the corner $(0,0)$,
the ground state $|\Psi^{\,}_{0}\rangle$ is 
a non-degenerate eigenstate of $\widehat{T}$,
and thus exhibits
$m^{x}_{\mathrm{sta}}[\Psi^{\,}_{0}]=0$.
This prevents a priori the use of
$m^{x}_{\mathrm{sta}}[\Psi^{\,}_{0}]$
as a probe to locate the phase transition upon exiting the
$\mathrm{\hbox{N\'eel}}^{\,}_{x}$
phase.

In DMRG, the initial state is chosen such that it is not an
eigenstate of the time-reversal operator $\widehat{T}$.
Consequently, the DMRG state $|\Psi^{\mathrm{DMRG}}_{0}\rangle$
that approximates the energy of the true ground state $|\Psi^{\,}_{0}\rangle$
to exponential accuracy can exhibit a non-vanishing value for
$m^{x}_{\mathrm{sta}}[\Psi^{\mathrm{DMRG}}_{0}]$
and can be used as a probe to survey the extend of the
$\mathrm{\hbox{N\'eel}}^{\,}_{x}$ phase. In effect, DMRG mimics the standard procedure
in statistical physics to detect spontaneous symmetry breaking
by which (i) the order parameter is coupled linearly to a source that is
then added to the Hamiltonian in order the break explicitly
the symmetry that is broken spontaneously by the order parameter acquiring
a non-vanishing expectation value,
(ii) the thermodynamic limit is taken, and
(iii) the source is switched off, i.e.,
\begin{subequations}
\begin{equation}
m^{x}_{\mathrm{sta}}\coloneqq
\lim_{B\to 0}
\lim_{\beta \to \infty}
\lim_{2N \to \infty}\,
\frac{1}{2N}
\sum_{j=1}^{2N}
(-1)^{j}
\ev*{\widehat{\sigma}^{x}_{j}}^{\,}_{\beta,B},
\end{equation}
where
\begin{equation}
\ev*{\widehat{\sigma}^{x}_{j}}^{\,}_{\beta,B}\coloneqq
\frac{
\mathrm{Tr}\,
\left\{
e^{
-
\beta
\left[
\widehat{H}^{\,}_{b}(\theta,\lambda)
-
B
\sum\limits_{j=1}^{2N}
(-1)^{j}\,\widehat{\sigma}^{x}_{j}
\right]
}\,
\widehat{\sigma}^{x}_{j}
\right\}
}
{
\mathrm{Tr}\,
\left\{       
e^{
-
\beta
\left[
\widehat{H}^{\,}_{b}(\theta,\lambda)
-
B
\sum\limits_{j=1}^{2N}
(-1)^{j}\,\widehat{\sigma}^{x}_{j}
\right]
}
\hphantom{\,\widehat{\sigma}^{x}_{j}}
\right\}
}.
\end{equation}
\end{subequations}
Henceforth, we will use the symbol $\Psi^{\,}_{0}$ as a reference to
the DMRG state that approximates the exact target eigenstate
in the thermodynamic limit that breaks spontaneously time-reversal symmetry.

As stated earlier,
when moving away from the corner $(0,0)$
in Fig.~\ref{suppfig:phase diagram},
the $\mathrm{\hbox{N\'eel}}^{\,}_{x}$ state
(\ref{suppeq:NeelX states 0})
and its time-reversed counterpart
(\ref{suppeq:NeelX states 1})
are no longer exact ground states.
However, sufficiently close to $(0,0)$, for instance at
$(\pi/20,1/10)$, the DMRG approximation to the
ground state of Hamiltonian
\eqref{suppeq:def H} remains qualitatively similar to the
$\mathrm{\hbox{N\'eel}}^{\,}_{x}$ states in the sense that it also supports a non-vanishing
magnetization. The DMRG approximation $|\Psi^{\,}_{0}\rangle$
to the ground state 
at $(\pi/20,1/10)$ from Fig.~\ref{suppfig:phase diagram}
displays a reduced staggered magnetization of
\begin{equation}
|m^{x}_{\mathrm{sta}}[\Psi^{\,}_{0}]|\simeq0.99<1,
\end{equation}
as shown for both OBC in Fig.~\ref{suppfig:DMRG OBC NeelX spins b}
and PBC in Fig.~\ref{suppfig:DMRG PBC NeelX spins b},
relative to the classical value of one
at $(0,0)$ from Fig.~\ref{suppfig:phase diagram}, shown in Fig.~\ref{suppfig:DMRG OBC NeelX spins a}
for OBC and Fig.~\ref{suppfig:DMRG PBC NeelX spins a}
for PBC.
This reduction from
the classical ($\mathrm{\hbox{N\'eel}}^{\,}_{x}$) value
(\ref{suppeq:classical value sta mag})
to
$|m^{x}_{\mathrm{sta}}[\Psi^{\,}_{0}]|\simeq0.99$
reflects the effects of quantum fluctuations that arises
as the first line on the right-hand side of
Eq.~(\ref{suppeq:def H a})
does not commute with the second line.

At the corner $(0,0)$,
the excited state $|\Psi^{\,}_{2}\rangle$ with the energy $E^{\,}_{2}$
in Eq.~(\ref{suppeq:NeelX states 2})
supports one (two) domain wall(s) for OBC (PBC).
The existence of the domain walls shows up in the
dependence on $j$ of the local expectation value
\begin{equation}
\langle\widehat{X}^{\,}_{j}\rangle^{\,}_{\Psi^{\,}_{2}}\equiv
\langle\widehat{\sigma}^{x}_{j}\rangle^{\,}_{\Psi^{\,}_{2}}\coloneqq
\langle\Psi^{\,}_{2}|\,
\widehat{\sigma}^{x}_{j}\,
|\Psi^{\,}_{2}\rangle.
\end{equation}
Whereas Fig.~\ref{suppfig:DMRG OBC NeelX spins a}
shows the dependence $(-1)^{j}$ on $j$ for
$\langle\widehat{X}^{\,}_{j}\rangle^{\,}_{\Psi^{\,}_{0}}$,
Fig.~\ref{suppfig:DMRG OBC NeelX spins c}
shows one domain wall between $j=2$ and $j=3$ when OBC are selected.
Similarly, whereas Fig.~\ref{suppfig:DMRG PBC NeelX spins a}
shows the dependence $(-1)^{j}$ on $j$ for
$\langle\widehat{X}^{\,}_{j}\rangle^{\,}_{\Psi^{\,}_{2}}$,
Fig.~\ref{suppfig:DMRG PBC NeelX spins c}
shows two domain walls; one located between $j=32$ and $j=1$
and one located between $j=27$ and $j=28$
when PBC are selected.

At $(\pi/20,1/10)$ from Fig.~\ref{suppfig:phase diagram},
the dependence on $j$ of the local spin expectation value
$\langle\widehat{X}^{\,}_{j}\rangle^{\,}_{\Psi^{\,}_{0}}$
is $0.99\times(-1)^{j+1}$ for OBC in Fig.~\ref{suppfig:DMRG OBC NeelX spins b}
and $0.99\times(-1)^{j+1}$ for PBC in Fig.~\ref{suppfig:DMRG PBC NeelX spins b}.
At $(\pi/20,1/10)$ from Fig.~\ref{suppfig:phase diagram},
the difference between OBC and PBC for 
the excited state $|\Psi^{\,}_{2}\rangle$ with the energy $E^{\,}_{2}$
is more subtle. For OBC, the single domain wall between $j=16$ and $j=17$
has a vanishing local magnetization
that corresponds to the periodic modulation
$\cos\big(\pi j/2N\big)$ present in
$\langle\widehat{X}^{\,}_{j}\rangle^{\,}_{\Psi^{\,}_{2}}$,
as can be seen in
Fig.~\ref{suppfig:DMRG OBC NeelX spins d}.
For PBC, $\langle\widehat{X}^{\,}_{j}\rangle^{\,}_{\Psi^{\,}_{2}}$
is domain-wall free, it is only affected by
the uniform suppression 
$|m^{x}_{\mathrm{sta}}[\Psi^{\,}_{2}]|\simeq0.87$
of its magnitude as is shown in
Fig.~\ref{suppfig:DMRG PBC NeelX spins d}.

\subsubsection{The $\mathrm{\hbox{N\'eel}}^{\,}_{y}$ phase}

The $\mathrm{\hbox{N\'eel}}^{\,}_{y}$ phase is the only one of the four gapped phases
for which it is possible to find a ground state $|\Psi\rangle$ such that,
in the thermodynamic limit,
\begin{equation}
\overline{G^{2,x}}[\Psi]=0,
\qquad
\overline{G^{2,y}}[\Psi]\neq0,
\qquad
\overline{G^{4,x}}[\Psi]=0,
\qquad
\overline{G^{4,y}}[\Psi]=0,
\end{equation}
in Fig.~\ref{suppfig:DMRG OBC Gs}.

The unitary transformation $\widehat{U}^{\,}_{\mathrm{R}}$
defined in Eq.~\eqref{suppeq:two unitaries a}
realizes the reflection about the line $\theta=\pi/4$
from the phase diagram in Fig.~\ref{suppfig:phase diagram}
for both PBC and OBC, since
\begin{equation}
\widehat{U}^{\,}_{\mathrm{R}}\,
\widehat{H}^{\,}_{b}(\theta,\lambda)\,
\widehat{U}^{\dagger}_{\mathrm{R}}
=\widehat{H}^{\,}_{b}(\pi/2-\theta,\lambda)
\end{equation}
holds for $b=0,1$.
Consequently, the numerical characterization
of the $\mathrm{\hbox{N\'eel}}^{\,}_{x}$ phase
made in Sec.~\ref{suppsec:NeelX} must also hold for
the $\mathrm{\hbox{N\'eel}}^{\,}_{y}$ phase if the replacements 
\begin{equation}
(\theta,\lambda) \to (\pi/2-\theta,\lambda),\qquad
\widehat{X}^{\,}_{j} \to \widehat{Y}^{\,}_{j}, \qquad
m^{x}_{\mathrm{sta}} \to m^{y}_{\mathrm{sta}},
\label{suppeq:changes NeelX to NeelY}
\end{equation}
are made. We have verified that both the ED and DMRG
simulations agree with this prediction.

\subsubsection{The $\mathrm{\hbox{N\'eel}}^{\mathrm{SPT}}_{x}$ phase}
\label{suppsubsubsec:NeelXSPT}

The $\mathrm{\hbox{N\'eel}}^{\mathrm{SPT}}_{x}$ phase is the only one of the four gapped phases
for which it is possible to find a ground state $|\Psi\rangle$ such that,
in the thermodynamic limit,
\begin{equation}
\overline{G^{2,x}}[\Psi]=0,
\qquad
\overline{G^{2,y}}[\Psi]=0,
\qquad
\overline{G^{4,x}}[\Psi]=0,
\qquad
\overline{G^{4,y}}[\Psi]\neq0,
\end{equation}
in Fig.~\ref{suppfig:DMRG OBC Gs}.

The unitary transformation $\widehat{U}^{\,}_{\mathrm{E}}$
defined in Eq.~\eqref{suppeq:two unitaries b}
realizes the reflection about the line $\lambda=1/2$
from the phase diagram in Fig.~\ref{suppfig:phase diagram}
for PBC, since
\begin{equation}
\widehat{U}^{\,}_{\mathrm{E}}\,
\widehat{H}^{\,}_{b=0}(\theta,\lambda)\,
\widehat{U}^{\dagger}_{\mathrm{E}}
=\widehat{H}^{\,}_{b=0}(\theta,1/2-\lambda)
\end{equation}
holds if and only if $b=0$.
The composition
$
\widehat{U}^{\,}_{\mathrm{CZ}}\equiv
\widehat{U}^{\,}_{\mathrm{E}}\,
\widehat{U}^{\,}_{\mathrm{R}}
$
realizes the rotation by $\pi$ about the center $(\pi/4,1/2)$
from the phase diagram in Fig.~\ref{suppfig:phase diagram}
for PBC, since
\begin{equation}
\widehat{U}^{\,}_{\mathrm{CZ}}\,
\widehat{H}^{\,}_{b=0}(\theta,\lambda)\,
\widehat{U}^{\dagger}_{\mathrm{CZ}}
=\widehat{H}^{\,}_{b=0}(\pi/2-\theta,1/2-\lambda)
\end{equation}
holds if and only if $b=0$.
Hence, when PBC are imposed, the properties of
the $\mathrm{\hbox{N\'eel}}^{\mathrm{SPT}}_{x}$ phase
are unitarily equivalent
under conjugation by
$\widehat{U}^{\,}_{\mathrm{CZ}}$
to the ones of the $\mathrm{\hbox{N\'eel}}^{\,}_{x}$ phase. 

After imposing PBC at the corner $(\pi/2,1)$
from Fig.~\ref{suppfig:phase diagram},
we know from Sec.~\ref{suppsubsec: Gapped stable fixed-points}
that the ordered energy eigenvalues
\begin{equation}
E^{\,}_{0}\leq
E^{\,}_{1}\leq
\cdots\leq
E^{\,}_{n}\leq
E^{\,}_{n+1}\leq
\cdots            
\end{equation}
with the orthonormal eigenfunctions
\begin{equation}
|\Psi^{\,}_{0}\rangle,\
|\Psi^{\,}_{1}\rangle,\
\cdots,\
|\Psi^{\,}_{n}\rangle,\
|\Psi^{\,}_{n+1}\rangle,\
\cdots             
\end{equation}
of Hamiltonian
$\widehat{H}^{\,}_{0}(\pi/2,1)$
defined in Eq.~\eqref{suppeq:Hamiltonians at lambda=1 b},
are at least two-fold degenerate
for any number $2N$ of sites.
Indeed, any ``spin'' configuration must share the
same energy as the one obtained by reversing the direction of all ``spins''
through the operation of reversal of time $\widehat{T}$. Moreover,
any two distinct but consecutive energy eigenvalues are equally spaced.
This spacing is the gap
$\Delta^{\,}_{0}(\pi/2,1)=\Delta^{\,}_{0}(0,0)$
that can be computed from subtracting the energy of any one of the two
$\mathrm{\hbox{N\'eel}}^{\mathrm{SPT}}_{x}$ states
\footnote{
The state $\ket*{\Psi^{\,}_{0}}$
is the image by $\protect{\widehat{U}^{\,}_{\mathrm{CZ}}}$
of the representation
in the basis that diagonalizes
$\protect{\widehat{\sigma}^{x}_{j}}$
of the state $\ket*{\mathrm{\hbox{N\'eel}}^{\,}_{x},+}$
defined by Eq.~(\ref{suppeq:GSs at lambda=0 a})
in the basis that diagonalizes
$\protect{\widehat{\sigma}^{z}_{j}}$
}
\begin{subequations}
\label{suppeq:NeelXSPT PBC states}
\begin{align}
&
\ket*{\Psi^{\,}_{0}}\coloneqq
|
\Leftarrow\Rightarrow\cdots
\Leftarrow\Rightarrow\
\Leftarrow\Rightarrow\
\Leftarrow\Rightarrow\cdots
\Leftarrow\Rightarrow
\rangle,
\label{suppeq:NeelXSPT states 0}
\\
&
\ket*{\Psi^{\,}_{1}}\coloneqq
\widehat{T}\,\ket*{\Psi^{\,}_{0}},
\qquad
\widehat{T}\,
|\Rightarrow\rangle^{\,}_{j}=
\mathrm{i}|\Leftarrow\rangle^{\,}_{j},
\quad
\widehat{T}\,
|\Rightarrow\rangle^{\,}_{j}=
-\mathrm{i}|\Leftarrow\rangle^{\,}_{j},
\label{suppeq:NeelXSPT states 1}
\end{align}
in the ``spin'' basis for which
\begin{align}
&
\left(  
\widehat{\sigma}^{z}_{j-1}\,  
\widehat{\sigma}^{x}_{j}\,
\widehat{\sigma}^{z}_{j+1}
\right)
|\Rightarrow\rangle^{\,}_{j}=
+|\Rightarrow\rangle^{\,}_{j},
\qquad
\widehat{\sigma}^{z}_{j-1}\,  
\widehat{\sigma}^{x}_{j}\,
\widehat{\sigma}^{z}_{j+1}=
\widehat{U}^{\,}_{\mathrm{CZ}}\,
\widehat{\sigma}^{x}_{j}\,
\widehat{U}^{\dag}_{\mathrm{CZ}},
\qquad
|\Rightarrow\rangle^{\,}_{j}=
\widehat{U}^{\,}_{\mathrm{CZ}}\,
|\rightarrow\rangle,
\label{suppeq:NeelXSPT states 00}
\\
&
\left(
\widehat{\sigma}^{z}_{j-1}\,
\widehat{\sigma}^{x}_{j}\,
\widehat{\sigma}^{z}_{j+1}
\right)
|\Leftarrow\rangle^{\,}_{j}=
-|\Leftarrow\rangle^{\,}_{j},
\qquad
\widehat{\sigma}^{z}_{j-1}\,  
\widehat{\sigma}^{x}_{j}\,
\widehat{\sigma}^{z}_{j+1}=
\widehat{U}^{\,}_{\mathrm{CZ}}\,
\widehat{\sigma}^{x}_{j}\,
\widehat{U}^{\dag}_{\mathrm{CZ}},
\qquad
|\Leftarrow\rangle^{\,}_{j}=
\widehat{U}^{\,}_{\mathrm{CZ}}\,
|\leftarrow\rangle,
\label{suppeq:NeelXSPT states 11}
\end{align}
from the energy of one of the excited states, say
\begin{align}
&
\ket*{\Psi^{\,}_{2}}\coloneqq
|
\Leftarrow\Rightarrow\cdots
\Leftarrow\!\red{\!\Rightarrow}\
\red{\Rightarrow}\blue{\Leftarrow}\
\blue{\Leftarrow\!\!}\Rightarrow\cdots
\Leftarrow\Rightarrow
\rangle,
\label{suppeq:NeelXSPT states 2}
\\
&
\ket*{\Psi^{\,}_{3}}\coloneqq
\widehat{T}\,\ket*{\Psi^{\,}_{2}},
\qquad
\widehat{T}\,
|\Rightarrow\rangle^{\,}_{j}=
\mathrm{i}|\Leftarrow\rangle^{\,}_{j},
\quad
\widehat{T}\,
|\Rightarrow\rangle^{\,}_{j}=
-\mathrm{i}|\Leftarrow\rangle^{\,}_{j},
\label{suppeq:NeelXSPT states 3}
\end{align}
\end{subequations}
The existence of this gap guarantees the stability of
the $\mathrm{\hbox{N\'eel}}^{\mathrm{SPT}}_{x}$ phase in the phase diagram in Fig.~\ref{suppfig:phase diagram}.

\begin{figure}[t!]
\centering
\includegraphics[width=0.7\columnwidth]{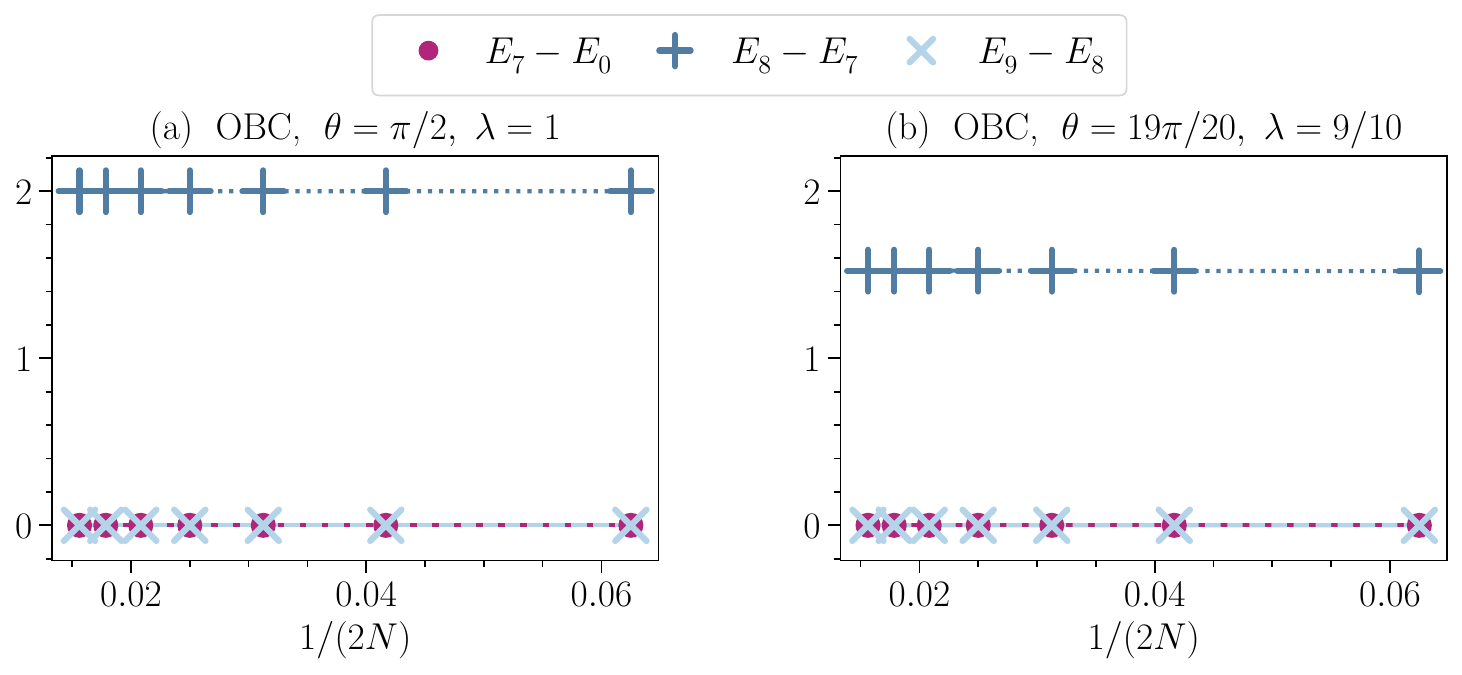}
\subfigure{\label{suppfig:DMRG OBC NeelXSPT gaps a} \relax}
\subfigure{\label{suppfig:DMRG OBC NeelXSPT gaps b} \relax}
\caption{(Color online)
The case of OBC. Dependence on $1/(2N)$ of the
energy spacings $E^{\,}_{n}-E^{\,}_{n-1}$,
$n=1,\cdots,9$,
obtained with DMRG and OBC
($\chi=128$) for
(a) the point $(\pi/2,1)$
and (b) the point
$(19\pi/20,9/10)$ in the $\mathrm{\hbox{N\'eel}}^{\mathrm{SPT}}_{x}$ phase
from Fig.~\ref{suppfig:phase diagram}.
}
\label{suppfig:DMRG OBC NeelXSPT gaps}
\end{figure}     

\begin{figure}[t!]
\centering
\includegraphics[width=0.7\columnwidth]{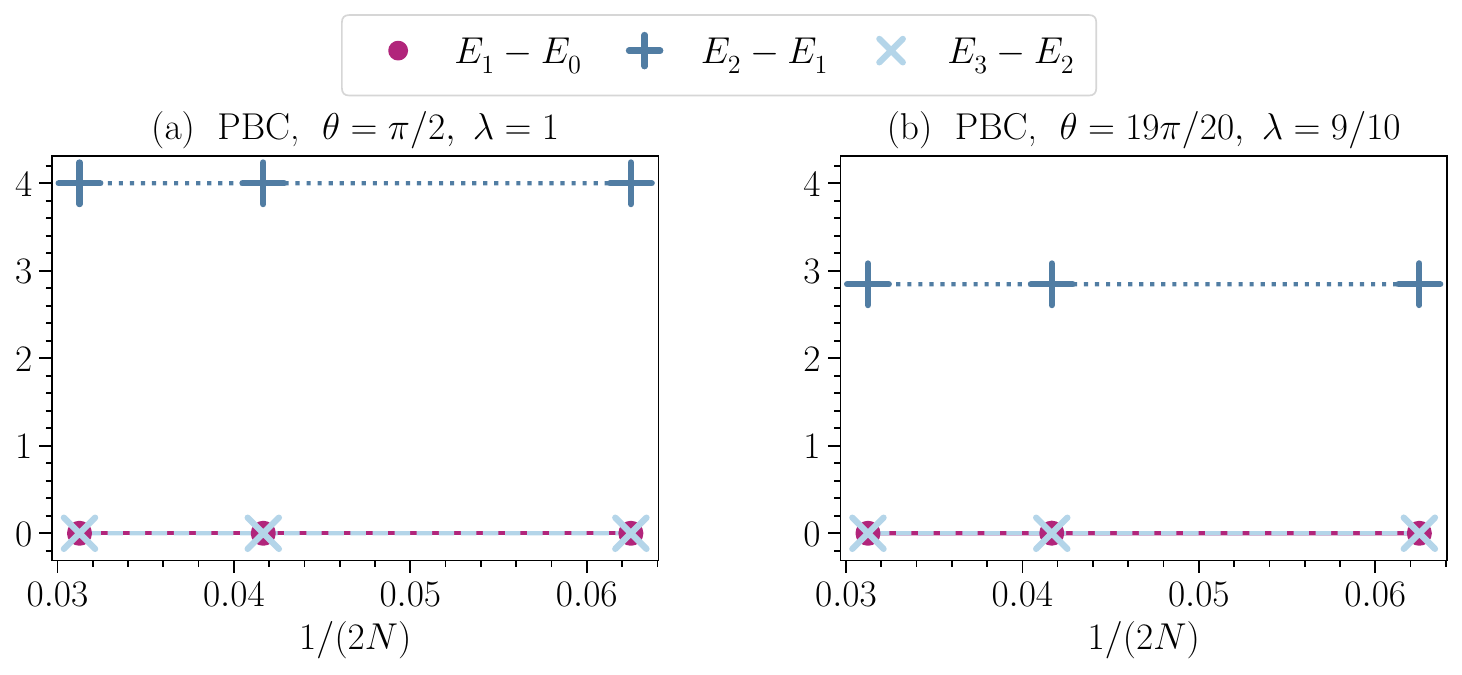}
\subfigure{\label{suppfig:DMRG PBC NeelXSPT gaps a} \relax}
\subfigure{\label{suppfig:DMRG PBC NeelXSPT gaps b} \relax}
\caption{(Color online)
The case of PBC. Dependence on $1/(2N)$ of the
energy spacings $E^{\,}_{n}-E^{\,}_{n-1}$,
$n=1,\cdots,3$,
obtained with DMRG and PBC
($\chi=128$) for
(a) the point $(\pi/2,1)$
and (b) the point
$(19\pi/20,9/10)$ in the $\mathrm{\hbox{N\'eel}}^{\mathrm{SPT}}_{x}$ phase
from Fig.~\ref{suppfig:phase diagram}.
}
\label{suppfig:DMRG PBC NeelXSPT gaps}
\end{figure} 

In contrast, after imposing OBC at the corner $(\pi/2,1)$
from Fig.~\ref{suppfig:phase diagram},
we know from Sec.~\ref{suppsubsec: Gapped stable fixed-points}
that the ground states of
Hamiltonian $\widehat{H}^{\,}_{1}(\pi/2,1)$
defined in Eq.~\eqref{suppeq:Hamiltonians at lambda=1 b}
are eightfold degenerate for any number $2N$ of sites.
An orthonormal basis for these ground states is
\begin{subequations}
\label{suppeq:NeelXSPT OBC state}
\begin{align}
&
\ket*{\Psi^{\,}_{0}}\coloneqq
|
\uparrow\,
\Leftarrow\Rightarrow
\cdots
\Leftarrow\Rightarrow
\cdots
\Leftarrow\Rightarrow\,
\uparrow
\rangle,
\\
&
\ket*{\Psi^{\,}_{1}}\coloneqq
\widehat{T}\,
\ket*{\Psi^{\,}_{0}},
\\
&
\ket*{\Psi^{\,}_{2}}\coloneqq
|
\uparrow\,
\Leftarrow\Rightarrow
\cdots
\Leftarrow\Rightarrow
\cdots
\Leftarrow\Rightarrow\,
\downarrow
\rangle,
\\
&
\ket*{\Psi^{\,}_{3}}\coloneqq
\widehat{T}\,
\ket*{\Psi^{\,}_{2}},
\\
&
\ket*{\Psi^{\,}_{4}}\coloneqq
|
\downarrow\,
\Leftarrow\Rightarrow
\cdots
\Leftarrow\Rightarrow
\cdots
\Leftarrow\Rightarrow\,
\uparrow
\rangle,
\\
&
\ket*{\Psi^{\,}_{5}}\coloneqq
\widehat{T}\,
\ket*{\Psi^{\,}_{4}},
\\
&
\ket*{\Psi^{\,}_{6}}\coloneqq
|
\downarrow\,
\Leftarrow\Rightarrow
\cdots
\Leftarrow\Rightarrow
\cdots
\Leftarrow\Rightarrow\,
\downarrow
\rangle,
\\
&
\ket*{\Psi^{\,}_{7}}\coloneqq
\widehat{T}\,
\ket*{\Psi^{\,}_{6}},
\end{align}
where the state $\ket*{\uparrow}^{\,}_{j}$ and $\ket*{\downarrow}^{\,}_{j}$ 
are defined by the action
\begin{align}
&
\widehat{\sigma}^{z}_{j}\,\ket*{\uparrow}^{\,}_{j}=
+\ket*{\uparrow}^{\,}_{j},
\qquad
\widehat{T}\ket*{\uparrow}^{\,}_{j} = \mathrm{i}\ket*{\downarrow}^{\,}_{j},
\\
&
\widehat{\sigma}^{z}_{j}\,\ket*{\downarrow}^{\,}_{j}=
-\ket*{\downarrow}^{\,}_{j},
\qquad
\widehat{T}\ket*{\downarrow}^{\,}_{j} = -\mathrm{i}\ket*{\uparrow}^{\,}_{j},
\end{align}
\end{subequations}
for $j=1,\cdots,2N$.

\begin{subequations}
When OBC are chosen,
the ground-state (first excited) energy of
$\widehat{H}^{\,}_{1}(\pi/2,1)$ is
[see Eq.~(\ref{suppeq:case OBC spectral decomposition cluster models})
and Fig.~\ref{suppfig:DMRG OBC NeelXSPT gaps a}]
\begin{equation}
E^{\,}_{0}=-2N+3
\qquad
(E^{\,}_{8}=-2N+5).
\label{suppeq:E0 and E2 for NeelSPTx with OBC}
\end{equation}
When PBC are chosen, 
the ground-state (first excited) energy of
$\widehat{H}^{\,}_{0}(\pi/2,1)$ is
[see Eq.~(\ref{suppeq:case PBC spectral decomposition cluster models})
and Fig.~\ref{suppfig:DMRG PBC NeelXSPT gaps a}]
\begin{equation}
E^{\,}_{0}=-2N
\qquad
(E^{\,}_{2}=-2N+4).
\label{suppeq:E0 and E2 for NeelSPTx with PBC}
\end{equation}
\label{suppeq:E0 and E2 for NeelSPTx}
\end{subequations}

When moving slightly away from the corner $(\pi/2,1)$ in Fig.~\ref{suppfig:phase diagram} under OBC,
the $\mathrm{\hbox{N\'eel}}^{\mathrm{SPT}}_{x}$ state (\ref{suppeq:NeelXSPT states 0})
and its time-reversed counterpart (\ref{suppeq:NeelXSPT states 1})
[the eight states (\ref{suppeq:NeelXSPT OBC state})]
are no longer exact ground states.
Moreover, for any finite number $2N$ of sites, the energy
$E^{\,}_{0}$
in Eq.~(\ref{suppeq:E0 and E2 for NeelSPTx})
is strictly lower than the energy $E^{\,}_{1}$
of the first excited state.
However, the energy spacing $E^{\,}_{1}-E^{\,}_{0}$
between the ground-state energy $E^{\,}_{0}$
in Eq.~(\ref{suppeq:E0 and E2 for NeelSPTx})
and the first excited state energy $E^{\,}_{1}$
is observed numerically to vanish exponentially fast with $2N$
so that a two-fold degeneracy of the ground state is recovered
in the thermodynamic limit $2N\to\infty$.
For example, for a chain made of $2N=32$ sites at $(19\pi/20,9/10)$,
$E^{\,}_{1}-E^{\,}_{0}\approx0$
within numerical precision under both PBC and OBC.

We have tracked the excitation energy
$E^{\,}_{8}$
($E^{\,}_{2}$)
in Eq.~(\ref{suppeq:E0 and E2 for NeelSPTx with OBC})
[Eq.~(\ref{suppeq:E0 and E2 for NeelSPTx with PBC})].
It remains non-degenerate with $E^{\,}_{0}$ in the thermodynamic limit
$2N\to\infty$ upon approaching the phase boundary of the
$\mathrm{\hbox{N\'eel}}^{\mathrm{SPT}}_{x}$ phase. The energy gap
$E^{\,}_{8}-E^{\,}_{7}$ 
($E^{\,}_{2}-E^{\,}_{1}$) at
$(19\pi/20,9/10)$
is smaller than the energy gap
$E^{\,}_{8}-E^{\,}_{7}$
($E^{\,}_{2}-E^{\,}_{1}$)
at $(\pi/2,1)$ for any system size $2N$.
This difference is evident when comparing
Figs.~\ref{suppfig:DMRG OBC NeelXSPT gaps a}
and~\ref{suppfig:DMRG OBC NeelXSPT gaps b}
for OBC 
[Figs.~\ref{suppfig:DMRG PBC NeelXSPT gaps a}
and~\ref{suppfig:DMRG PBC NeelXSPT gaps b}
for PBC].
This difference aligns with the
expectation of a gap closing upon exiting the $\mathrm{\hbox{N\'eel}}^{\mathrm{SPT}}_{x}$ phase.

\begin{figure}[t!]
\centering
\subfigure{\includegraphics[width=0.71\columnwidth]{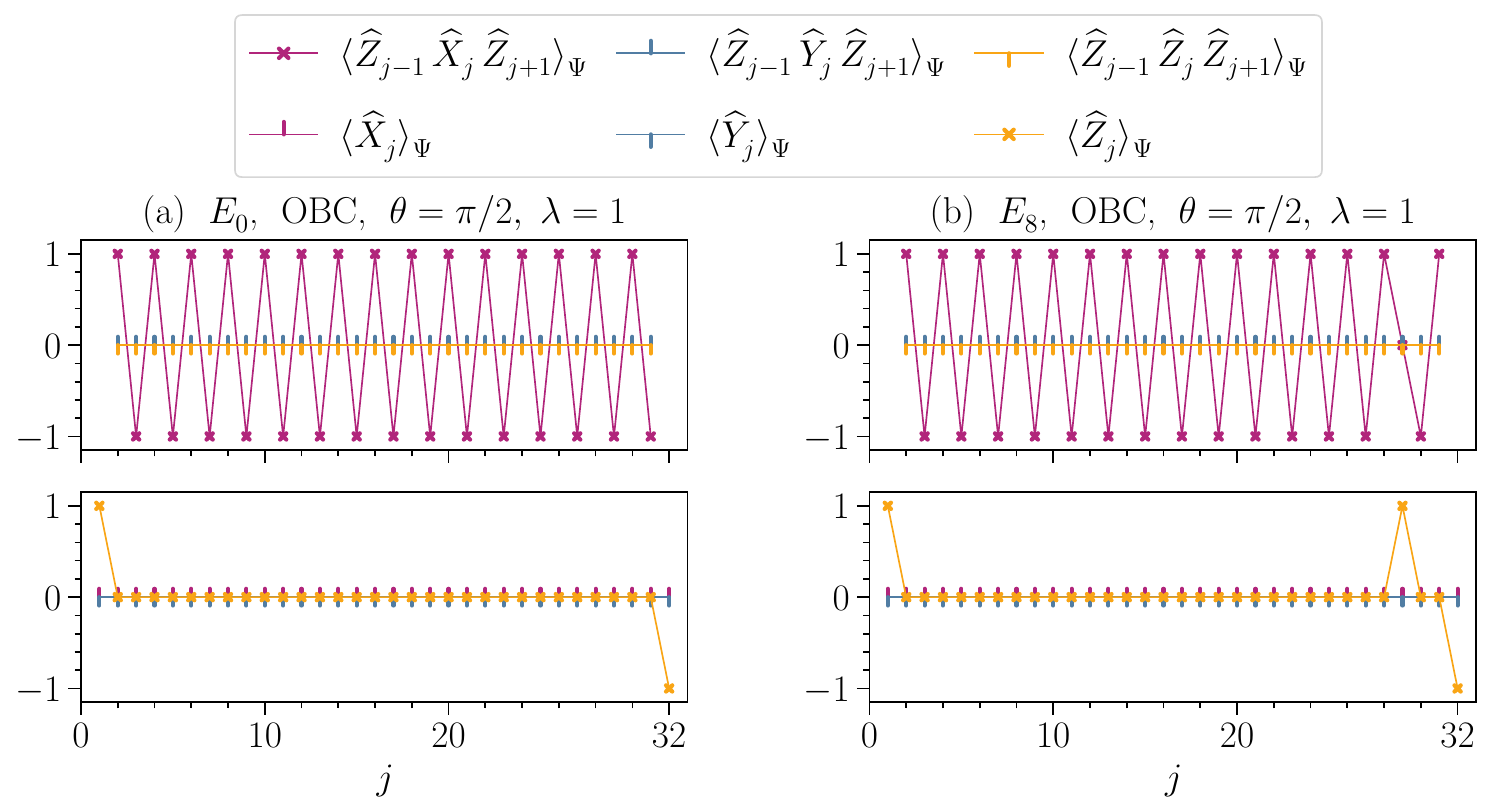} \label{suppfig:DMRG OBC NeelXSPT spins a}}
\subfigure{\includegraphics[width=0.71\columnwidth]{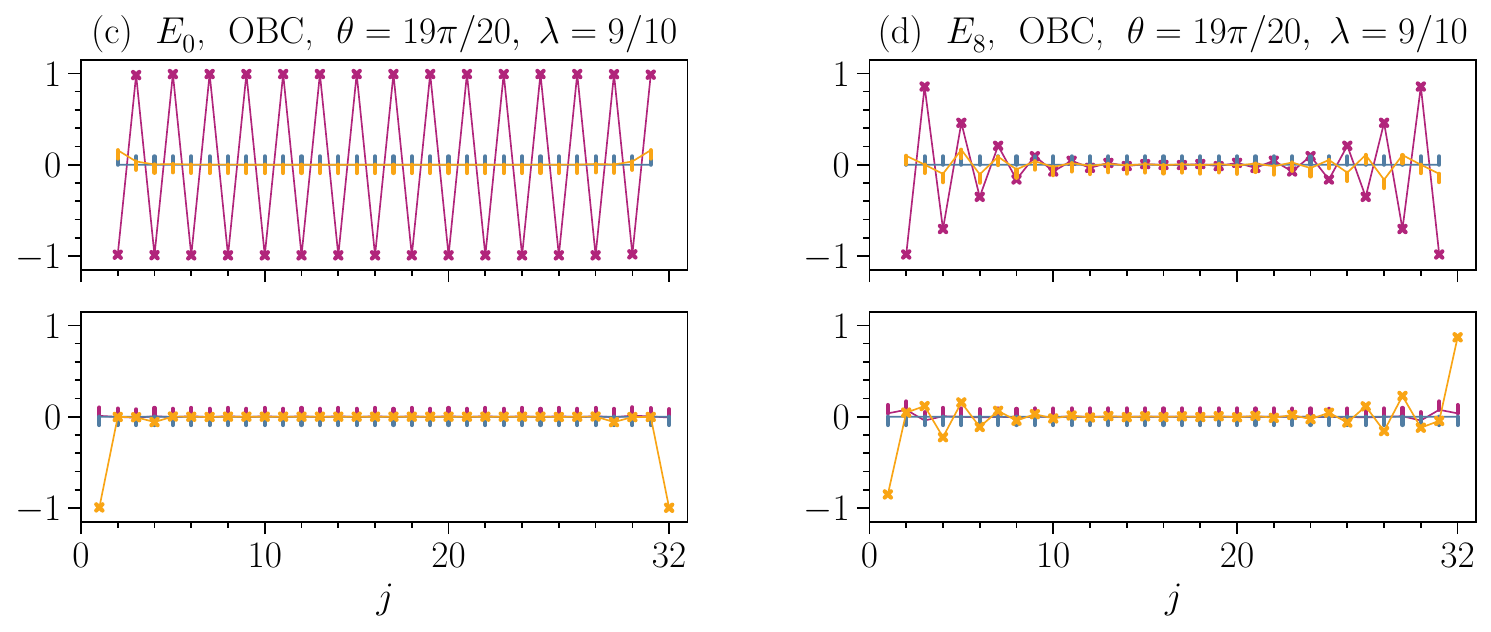} \label{suppfig:DMRG OBC NeelXSPT spins b}}
\\
\subfigure{\label{suppfig:DMRG OBC NeelXSPT spins c} \relax}
\subfigure{\label{suppfig:DMRG OBC NeelXSPT spins d} \relax}
\caption{(Color online)
The case of OBC.
Dependence on the site index $j=1,\cdots,2N$ of the local
cluster and spin expectations values
in the states obtained with DMRG ($\chi=128$ and $2N=32$)
as approximations to the exact eigenstates with energies
(a)
$E^{\,}_{0}$ at $(\pi/2,1)$,
(b)
$E^{\,}_{8}$ at $(\pi/2,1)$,
(c)
$E^{\,}_{0}$ at $(19\pi/20,9/10)$,
and
(d)
$E^{\,}_{8}$ at $(19\pi/20,9/10)$.
}
\label{suppfig:DMRG OBC NeelXSPT spins}
\end{figure}

\begin{figure}[t!]
\subfigure{\includegraphics[width=0.71\columnwidth]{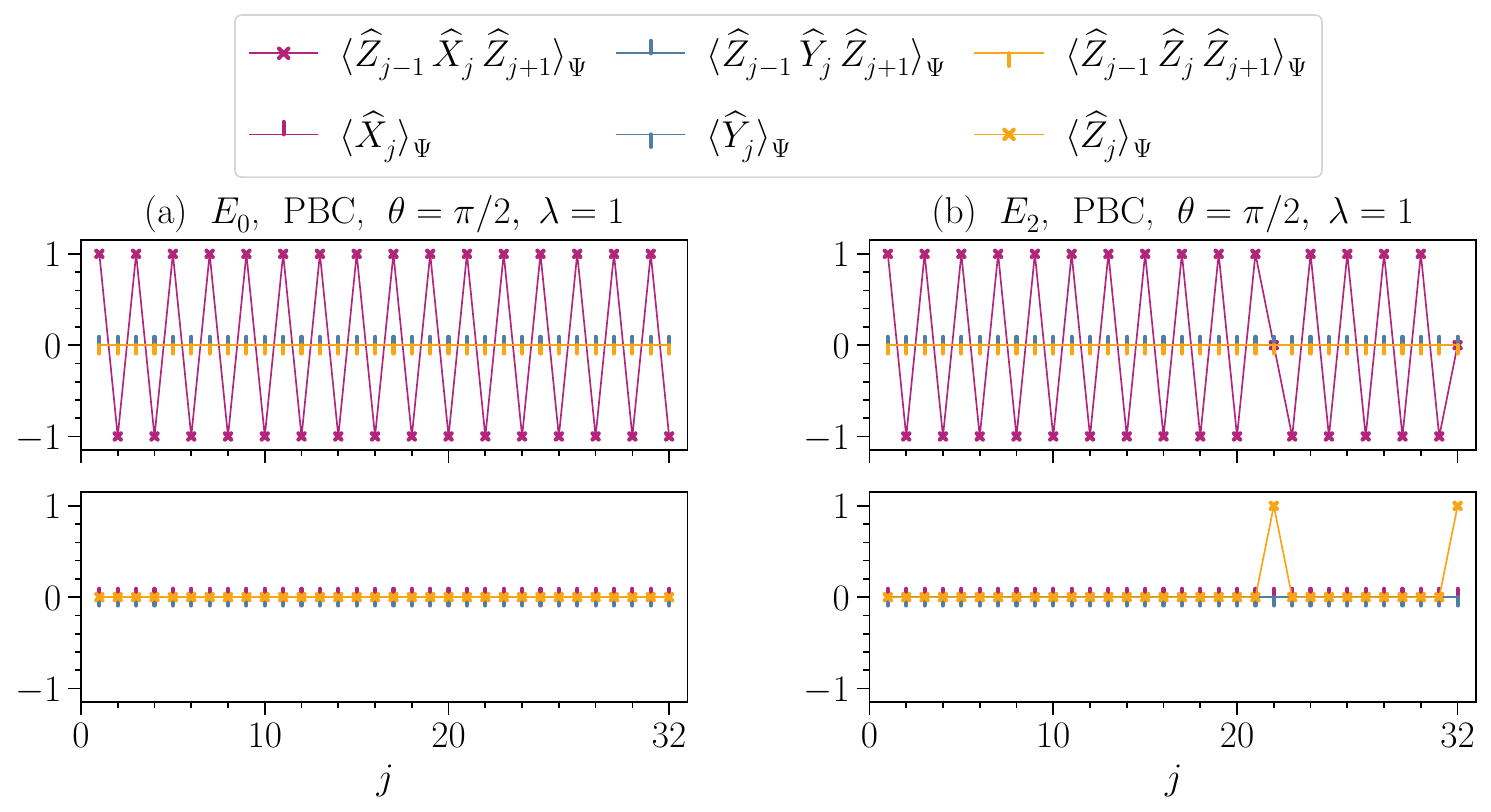} \label{suppfig:DMRG PBC NeelXSPT spins a}}
\subfigure{\includegraphics[width=0.71\columnwidth]{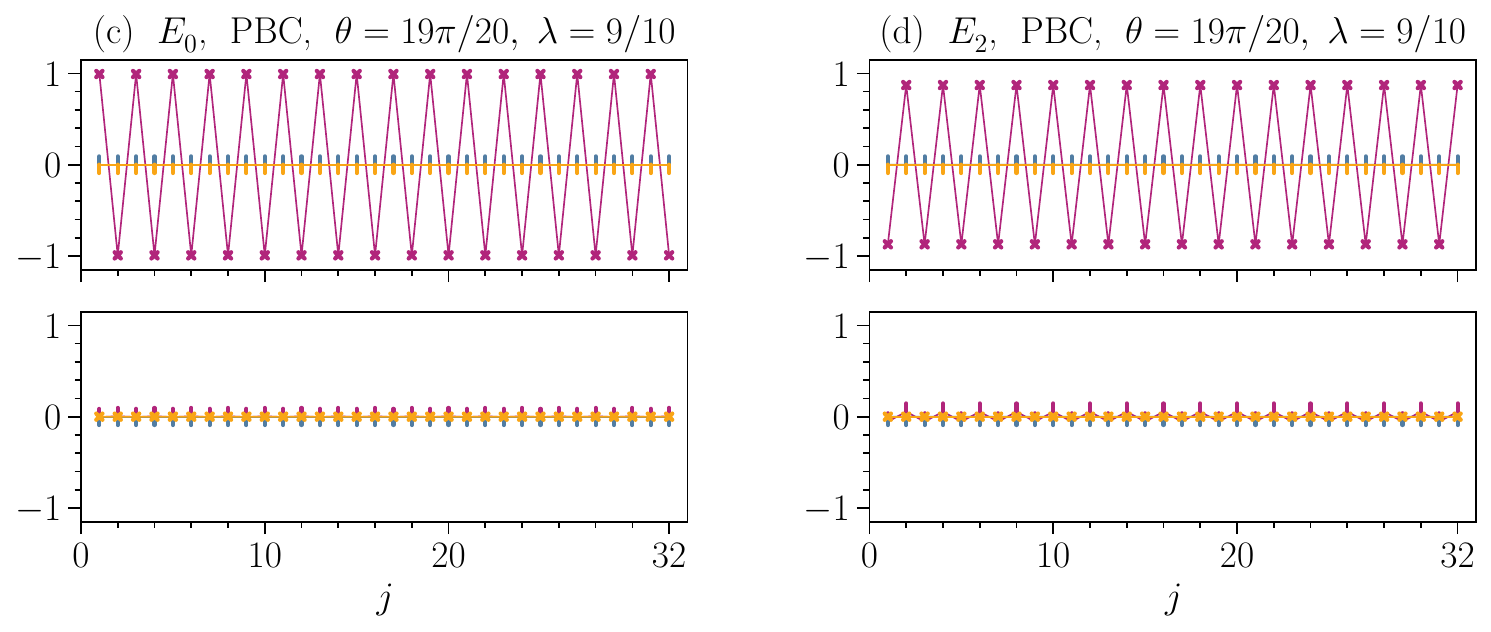} \label{suppfig:DMRG PBC NeelXSPT spins b}}
\\
\subfigure{\label{suppfig:DMRG PBC NeelXSPT spins c} \relax}
\subfigure{\label{suppfig:DMRG PBC NeelXSPT spins d} \relax}
\caption{(Color online)
The case of PBC.
Dependence on the site index $j=1,\cdots,2N$ of the local
cluster and spin expectations values
in the states obtained with DMRG ($\chi=128$ and $2N=32$)
as approximations to the exact eigenstates with energies
(a)
$E^{\,}_{0}$ at $(\pi/2,1)$,
(b)
$E^{\,}_{2}$ at $(\pi/2,1)$,
(c)
$E^{\,}_{0}$ at $(19\pi/20,9/10)$,
and
(d)
$E^{\,}_{2}$ at $(19\pi/20,9/10)$.
}
\label{suppfig:DMRG PBC NeelXSPT spins}
\end{figure}

At the corner $(\pi/2,1)$ from Fig.~\ref{suppfig:phase diagram},
the eight orthogonal ground states
$\ket*{\Psi^{\,}_{n}}$, $n=0,\cdots,7$, defined in
Eq.~\eqref{suppeq:NeelXSPT OBC state}
exhibit the staggered magnetization
[see Eq.~(\ref{suppeq:mstaSPT})]
\begin{equation}
|m^{x}_{\mathrm{sta,SPT}}[\Psi^{\,}_{n}]|=1,
\qquad
n=0,\cdots,7.
\end{equation}
This saturation of the staggered magnetization
indicates perfect ($\mathrm{\hbox{N\'eel}}^{\mathrm{SPT}}_{x}$) antiferromagnetic order
for the local order parameter
\begin{equation}
\widehat{U}^{\,}_{\mathrm{CZ}}\,
\widehat{\sigma}^{x}_{j}\,
\widehat{U}^{\dag}_{\mathrm{CZ}}=    
\widehat{\sigma}^{z}_{j-1}\,
\widehat{\sigma}^{x}_{j}\,
\widehat{\sigma}^{z}_{j+1},
\qquad
j=2,\cdots,2N-1.
\end{equation} 
This is shown in Fig.~\ref{suppfig:DMRG OBC NeelXSPT spins a}.
More specifically, four of
the ground states \eqref{suppeq:NeelXSPT OBC state} have the staggered
magnetization
\begin{equation}
m^{x}_{\mathrm{sta,SPT}}[\Psi^{\,}_{n}]=-1,
\qquad
n=0,2,4,6,
\end{equation}
while their time-reversed images have the staggered magnetization
\begin{equation}
m^{x}_{\mathrm{sta,SPT}}[\Psi^{\,}_{n}]=+1,
\qquad
n=1,3,5,7.
\end{equation} 
In fact, the four ground states
$|\Psi^{\,}_{n}\rangle$ with $n=0,1,2,4,6$ ($n=1,2,5,7$)
from Eq.~(\ref{suppeq:NeelXSPT OBC state})  
are eigenstates with eigenvalue
$-1$ ($+1$) of the staggered SPT ordered parameter
\begin{equation}
\widehat{m}^{x}_{\mathrm{sta,SPT}}\coloneqq
\frac{1}{2N-2}\,
\sum_{j=2}^{2N-1}
(-1)^{j}\,
\widehat{\sigma}^{z}_{j-1}\,
\widehat{\sigma}^{x}_{j}\,
\widehat{\sigma}^{z}_{j+1}.
\end{equation}
The fourfold degeneracy of those ground states 
that are eigenstates of $\widehat{m}^{x}_{\mathrm{sta,SPT}}$
arises from the free quantum spin-$1/2$
degrees of freedom at both ends the chain.
These free quantum spin-1/2
contribute to the expectation value of magnitude one for 
$\widehat{Z}^{\,}_{1}\equiv\widehat{\sigma}^{z}_{1}$
and
$\widehat{Z}^{\,}_{2N}\equiv\widehat{\sigma}^{z}_{2N}$
as is shown in Fig.~\ref{suppfig:DMRG OBC NeelXSPT spins a}.
In contrast, the two ground states
(\ref{suppeq:NeelXSPT PBC states})
selected by PBC are non-degenerate eigenstates
of the staggered SPT ordered parameter
\begin{equation}
\widehat{m}^{x}_{\mathrm{sta,SPT}}\coloneqq
\frac{1}{2N}\,
\sum_{j=1}^{2N}
(-1)^{j}\,
\widehat{\sigma}^{z}_{j-1}\,
\widehat{\sigma}^{x}_{j}\,
\widehat{\sigma}^{z}_{j+1}=
\widehat{U}^{\,}_{\mathrm{CZ}}\,
\widehat{m}^{x}_{\mathrm{sta}}\,
\widehat{U}^{\dag}_{\mathrm{CZ}},
\qquad
\widehat{m}^{x}_{\mathrm{sta}}\coloneqq
\frac{1}{2N}\,
\sum_{j=1}^{2N}
(-1)^{j}\,
\widehat{\sigma}^{x}_{j}=
-
\widehat{T}\,
\widehat{m}^{x}_{\mathrm{sta}}\,
\widehat{T}.
\end{equation}
This is shown in Fig.~\ref{suppfig:DMRG PBC NeelXSPT spins a}.
As there are no free quantum spin-1/2 at the ends of the chains,
the expectation value of
$\widehat{Z}^{\,}_{j}\equiv\widehat{\sigma}^{z}_{j}$
vanishes for all sites $j=1,\cdots,2N$,
as is shown in Fig.~\ref{suppfig:DMRG PBC NeelXSPT spins a}.

Figure~\ref{suppfig:DMRG OBC NeelXSPT spins b}
for OBC
and
Fig.~\ref{suppfig:DMRG PBC NeelXSPT spins b}
for PBC show that,
at the corner $(\pi/2,1)$ from Fig.~\ref{suppfig:phase diagram},
there are excited states
$|\Psi\rangle$
and sites $j$
such that the expectation value
\begin{equation}
\langle
\widehat{Z}^{\,}_{j-1}\,\widehat{X}^{\,}_{j}\,\widehat{Z}^{\,}_{j+1}
\rangle^{\,}_{\Psi}
\equiv
\langle
\widehat{\sigma}^{z}_{j-1}\,\widehat{\sigma}^{x}_{j}\,\widehat{\sigma}^{z}_{j+1}
\rangle^{\,}_{\Psi}=0
\end{equation}
vanishes, while the expectation value
\begin{equation}
\left|\left\langle
\widehat{Z}^{\,}_{j}
\right\rangle^{\,}_{\Psi}\right|
\equiv
\left|\left\langle
\widehat{\sigma}^{z}_{j}
\right\rangle^{\,}_{\Psi}\right|=1
\end{equation}
saturates to its classical value, respectively.
This property is a signature of the extensive degeneracies
(\ref{suppeq:case OBC spectral decomposition cluster models b})
for OBC
and
(\ref{suppeq:case PBC spectral decomposition cluster models b})
for PBC, respectively,
of the excited states of the quantum spin-1/2 cluster Hamiltonians,
as we now explain.

Without loss of generality, we consider
the corner $(0,0)$ from Fig.~\ref{suppfig:phase diagram}
under OBC. We choose the ground state 
\begin{subequations}
\begin{equation}
\ket*{\mathrm{\hbox{N\'eel}}^{\,}_{x}, +} \coloneqq |
\leftarrow\rightarrow\cdots
\leftarrow\rightarrow\
\leftarrow\rightarrow\
\leftarrow\rightarrow\cdots\
\leftarrow\rightarrow\
\rangle
\end{equation}
in the basis (\ref{suppeq:NeelX states 00}) and (\ref{suppeq:NeelX states 11}).
Its energy is $E^{\,}_{0}=-2N+1$ [see Eq.~(\ref{suppeq:E0=-2N+1})].
Equipped with this ground state,
we define the pair of orthonormal excited eigenstates
\begin{align}
&
\ket*{+^{\,}_{k}}\coloneqq
\widehat{Z}^{\,}_{k}\,
\widehat{Z}^{\,}_{k+1}\,
\cdots\,
\widehat{Z}^{\,}_{2N-1}\,
\widehat{Z}^{\,}_{2N}\,
\ket*{\mathrm{\hbox{N\'eel}}^{\,}_{x},+}
\nonumber\\
&
\hphantom{\ket*{+^{\,}_{k}}}=
|
\leftarrow\rightarrow\cdots
\leftarrow\red{\!\!\rightarrow}\
\red{\rightarrow\!\!}\leftarrow\
\rightarrow\leftarrow\
\cdots\
\rightarrow\leftarrow\
\rangle,
\\
&
\ket*{+^{\,}_{k+1}}\coloneqq
\widehat{\mathbb{1}}^{\,}_{k}\,
\widehat{Z}^{\,}_{k+1}\,
\cdots\,
\widehat{Z}^{\,}_{2N-1}\,
\widehat{Z}^{\,}_{2N}\,
\ket*{\mathrm{\hbox{N\'eel}}^{\,}_{x},+}
\nonumber\\
&
\hphantom{\ket*{+^{\,}_{k+1}}}=
|
\leftarrow\rightarrow\
\cdots\
\leftarrow\rightarrow
\red{\leftarrow}
\red{\leftarrow\!\!}\
\rightarrow\leftarrow\
\cdots\
\rightarrow\leftarrow\
\rangle,
\end{align}
\end{subequations}
for any $k=2,\cdots,2N$.
The excited energy eigenstate $\ket*{+^{\,}_{k}}$
is a domain wall between the sites $k-1$ and $k$.
The excited energy eigenstate $\ket*{+^{\,}_{k+1}}$
is a domain wall between the sites $k$ and $k+1$.
The pair of energy eigenstates $\ket*{+^{\,}_{k}}$ and $\ket*{+^{\,}_{k+1}}$
are degenerate with the energy eigenvalue $E^{\,}_{0}+2$.
Hence, the symmetric and antisymmetric linear combinations
\begin{subequations}
\begin{align}
&
\ket*{+^{(\mathrm{s})}_{k,k+1}}\coloneqq
\frac{1}{\sqrt{2}}
\left(
\ket*{+^{\,}_{k}}
+
\ket*{+^{\,}_{k+1}}
\right),
\\
&
\ket*{+^{(\mathrm{a})}_{k,k+1}}\coloneqq
\frac{1}{\sqrt{2}}
\left(
\ket*{+^{\,}_{k}}
-
\ket*{+^{\,}_{k+1}}
\right),
\end{align}
\end{subequations}
are an orthonormal pair of excited energy eigenstates with
the energy eigenvalue $E^{\,}_{0}+2$.
The pair of identities
\begin{subequations}
\begin{align}
&
\ket*{+^{\,}_{k}}=\widehat{Z}^{\,}_{k} \ket*{+^{\,}_{k+1}},
\\
&
\ket*{+^{\,}_{k+1}}=\widehat{Z}^{\,}_{k} \ket*{+^{\,}_{k}},
\end{align}
\end{subequations}
implies that
\begin{subequations}
\begin{align}
&
\widehat{Z}^{\,}_{k}\,\ket*{+^{(\mathrm{s})}_{k,k+1}}=
\frac{1}{\sqrt{2}}
\left(
\widehat{Z}^{2}_{k}\,\ket*{+^{\,}_{k+1}}
+
\widehat{Z}^{\,}_{k}\,\ket*{+^{\,}_{k+1}}
\right)=
\frac{1}{\sqrt{2}}
\left(
\ket*{+^{\,}_{k+1}}
+
\ket*{+^{\,}_{k}}
\right)=
+\ket*{+^{(\mathrm{s})}_{k,k+1}},
\\
&
\widehat{Z}^{\,}_{k}\,\ket*{+^{(\mathrm{a})}_{k,k+1}}=
\frac{1}{\sqrt{2}}
\left(
\widehat{Z}^{2}_{k}\,\ket*{+^{\,}_{k+1}}
-
\widehat{Z}^{\,}_{k}\,\ket*{+^{\,}_{k+1}}
\right)=
\frac{1}{\sqrt{2}}
\left(
\ket*{+^{\,}_{k+1}}
-
\ket*{+^{\,}_{k}}
\right)=
-
\ket*{+^{(\mathrm{a})}_{k,k+1}}.
\end{align}
\end{subequations}
In turn, the Pauli algebra obeyed by
$\widehat{X}^{\,}_{j}$,
$\widehat{Y}^{\,}_{j}$,
and
$\widehat{Z}^{\,}_{j}$
implies that
\begin{subequations}
\begin{align}
&
\widehat{X}^{\,}_{k}\,\ket*{+^{(\mathrm{s})}_{k,k+1}}=
\ket*{+^{(\mathrm{a})}_{k,k+1}},
\\
&
\widehat{X}^{\,}_{k}\,\ket*{+^{(\mathrm{a})}_{k,k+1}}=
\ket*{+^{(\mathrm{s})}_{k,k+1}}.
\end{align}
\end{subequations}
In particular, it follows that
\begin{subequations}
\begin{align}
&
\mel*{+^{(\mathrm{s})}_{k,k+1}}{\,\widehat{Z}^{\,}_{k}\,}{+^{(\mathrm{s})}_{k,k+1}}=+1,
\\
&
\mel*{+^{(\mathrm{a})}_{k,k+1}}{\,\widehat{Z}^{\,}_{k}\,}{+^{(\mathrm{a})}_{k,k+1}}=-1,
\end{align}
while
\begin{align}
\mel*{+^{(\mathrm{s})}_{k,k+1}}{\,\widehat{X}^{\,}_{k}\,}{+^{(\mathrm{s})}_{k,k+1}}=
\mel*{+^{(\mathrm{a})}_{k,k+1}}{\,\widehat{X}^{\,}_{k}\,}{+^{(\mathrm{a})}_{k,k+1}}=0.
\end{align}
\end{subequations}
On the other hand,
one verifies that, for any $l\neq k$,
\footnote{
The states
$\protect{\widehat{Z}}^{\,}_{l}\,|+^{\,}_{k}\rangle$
and
$\protect{\widehat{Z}}^{\,}_{l}\,\,|+^{\,}_{k+1}\rangle$
are orthogonal to both
$|+^{\,}_{k}\rangle$
and
$|+^{\,}_{k+1}\rangle$
if $l\neq k$
}
\begin{subequations}
\begin{align}
&
\mel*{+^{(\mathrm{s})}_{k,k+1}}{\,\widehat{Z}^{\,}_{l}\,}{+^{(\mathrm{s})}_{k,k+1}}=0,
\\
&
\mel*{+^{(\mathrm{a})}_{k,k+1}}{\,\widehat{Z}^{\,}_{l}\,}{+^{(\mathrm{a})}_{k,k+1}}=0,
\end{align}
while
\footnote{
After expressing the states
$|+^{\mathrm{(s)}}_{k,k+1}\rangle$
and
$|+^{\mathrm{(a)}}_{k,k+1}\rangle$
in terms of the state $\ket*{\mathrm{\hbox{N\'eel}}^{\,}_{x},+}$,
one may replace the operator
$\protect{\widehat{X}}^{\,}_{l}$
by the $\mathbb{C}$-number
$(-1)^{\delta^{\,}_{1,\mathrm{sgn}(l-k)}}\,(-1)^{l}$
if $l\neq k$
}
\begin{align}
&
\mel*{+^{(\mathrm{s})}_{k,k+1}}{\,\widehat{X}^{\,}_{l}\,}{+^{(\mathrm{s})}_{k,k+1}}=
(-1)^{\delta^{\,}_{1,\mathrm{sgn}(l-k)}}\,
(-1)^{l},
\\
&
\mel*{+^{(\mathrm{a})}_{k,k+1}}{\,\widehat{X}^{\,}_{l}\,}{+^{(\mathrm{a})}_{k,k+1}}=
(-1)^{\delta^{\,}_{1,\mathrm{sgn}(l-k)}}\,
(-1)^{l}.
\end{align}
\end{subequations}

We consider the corner $(\pi/2,1)$ from Fig.~\ref{suppfig:phase diagram}.
We choose the ground state 
\begin{subequations}
\label{suppeq:NeelSPTx}
\begin{equation}
\ket*{\mathrm{\hbox{N\'eel}}^{\mathrm{SPT}}_{x}, -,\uparrow\downarrow}\coloneqq
|\
\uparrow\
\Leftarrow\Rightarrow\cdots
\Leftarrow\Rightarrow\
\Leftarrow\Rightarrow\
\Leftarrow\Rightarrow\cdots\
\Leftarrow\Rightarrow\
\downarrow\
\rangle
\label{suppeq:NeelSPTx a}
\end{equation}
when selecting OBC
in the basis
(\ref{suppeq:NeelXSPT states 00})
and
(\ref{suppeq:NeelXSPT states 11})
for sites $2,\cdots,2N-1$
and in the basis that diagonalizes
$\widehat{Z}^{\,}_{1}$ and $\widehat{Z}^{\,}_{2N}$
for sites 1 and $2N$.
We choose the ground state 
\begin{equation}
\ket*{\mathrm{\hbox{N\'eel}}^{\mathrm{SPT}}_{x}, -}\coloneqq
|\
\Rightarrow\
\Leftarrow\Rightarrow\cdots
\Leftarrow\Rightarrow\
\Leftarrow\Rightarrow\
\Leftarrow\Rightarrow\cdots\
\Leftarrow\Rightarrow\
\Leftarrow\
\rangle
\label{suppeq:NeelSPTx b}
\end{equation}
\end{subequations}
when selecting PBC
in the basis
(\ref{suppeq:NeelXSPT states 00})
and
(\ref{suppeq:NeelXSPT states 11})
for all sites.
Equipped with the ground state (\ref{suppeq:NeelSPTx a}),
we define for any $k=2,\cdots,2N-1$
the pair of orthonormal excited eigenstates
\begin{subequations}
\begin{align}
&
\ket*{\Psi^{\,}_{8}}^{\,}_{k}\coloneqq
|\,
\uparrow\,
\Leftarrow\Rightarrow\,
\cdots\,
\Leftarrow\red{\!\!\Rightarrow}
\red{\Rightarrow\!\!}\Leftarrow\,
\Rightarrow\Leftarrow\,
\cdots\,
\Rightarrow\Leftarrow\,
\downarrow\,
\rangle\equiv
\prod_{j=k}^{2N-1}
\widehat{Z}^{\,}_{j}\,
\ket*{\mathrm{\hbox{N\'eel}}^{\mathrm{SPT}}_{x}, -,\uparrow\downarrow},
\\
&
\ket*{\Psi^{\,}_{9}}^{\,}_{k+1}\coloneqq
|\,
\uparrow\,
\Leftarrow\Rightarrow\,
\cdots\,
\Leftarrow\Rightarrow
\red{\!\!\Leftarrow}
\red{\Leftarrow}
\Rightarrow\Leftarrow\,
\cdots\,
\Rightarrow\Leftarrow\,
\downarrow\,
\rangle\equiv
\prod_{j=k+1}^{2N-1}
\widehat{Z}^{\,}_{j}\,
\ket*{\mathrm{\hbox{N\'eel}}^{\mathrm{SPT}}_{x}, -,\uparrow\downarrow},
\end{align}
\end{subequations}
for OBC. These are two excited states
out of the $8(2N-3)$-dimensional eigenspace with eigenenergy $E^{\,}_{8}$
that support one domain wall
[compare with Eq.~(\ref{suppeq:degeneracy lowest excited states Ising OBC})].
Equipped with the ground state (\ref{suppeq:NeelSPTx b}),
we define for any $k=1,\cdots,2N-1$
the pair of orthonormal excited eigenstates
\begin{subequations}
\begin{align}
&
\ket*{\Psi^{\,}_{2}}^{\,}_{k}\coloneqq
|\,
\Rightarrow\,
\Leftarrow\Rightarrow\,
\cdots\,
\Leftarrow\red{\!\!\Rightarrow}
\red{\Rightarrow\!\!}\Leftarrow\,
\Rightarrow\Leftarrow\,
\cdots\,
\Rightarrow\blue{\!\Leftarrow}\,
\blue{\Leftarrow}\,
\rangle\equiv
\prod_{j=k}^{2N-1}
\widehat{Z}^{\,}_{j}\,
\ket*{\mathrm{\hbox{N\'eel}}^{\mathrm{SPT}}_{x}, -},
\\
&
\ket*{\Psi^{\,}_{3}}^{\,}_{k+1}\coloneqq
|\,\,
\blue{\Rightarrow}
\Leftarrow\Rightarrow\,
\cdots\,
\Leftarrow\Rightarrow
\red{\!\!\Leftarrow}
\red{\Leftarrow}
\Rightarrow\Leftarrow\,
\cdots\,
\Rightarrow\Leftarrow\,
\blue{\!\!\Rightarrow}\,
\rangle\equiv
\prod_{j=k+1}^{2N}
\widehat{Z}^{\,}_{j}\,
\ket*{\mathrm{\hbox{N\'eel}}^{\mathrm{SPT}}_{x}, -},
\end{align}
\end{subequations}
for PBC.
These are two excited states
out of the $2N(2N-1)$-dimensional eigenspace with eigenenergy $E^{\,}_{2}$
that support two domain walls
[compare with Eq.~(\ref{suppeq:degeneracy lowest excited states Ising PBC})].
We define the symmetric and antisymmetric linear combinations
\begin{subequations}
\begin{equation}
\ket*{+^{\,}_{\mathrm{OBC}}}^{\,}_{k,k+1}\coloneqq
\frac{1}{\sqrt{2}}
\left(
\ket*{\Psi^{\,}_{8}}^{\,}_{k}
+
\ket*{\Psi^{\,}_{9}}^{\,}_{k+1}
\right),
\qquad
\ket*{-^{\,}_{\mathrm{OBC}}}^{\,}_{k,k+1}\coloneqq
\frac{1}{\sqrt{2}}
\left(
\ket*{\Psi^{\,}_{8}}^{\,}_{k}
-
\ket*{\Psi^{\,}_{9}}^{\,}_{k+1}
\right),
\end{equation}
for OBC and
\begin{equation}
\ket*{+^{\,}_{\mathrm{PBC}}}^{\,}_{k,k+1}\coloneqq
\frac{1}{\sqrt{2}}
\left(
\ket*{\Psi^{\,}_{2}}^{\,}_{k}
+
\ket*{\Psi^{\,}_{3}}^{\,}_{k+1}
\right),
\qquad
\ket*{-^{\,}_{\mathrm{PBC}}}^{\,}_{k,k+1}\coloneqq
\frac{1}{\sqrt{2}}
\left(
\ket*{\Psi^{\,}_{2}}^{\,}_{k}
-
\ket*{\Psi^{\,}_{3}}^{\,}_{k+1}
\right),
\end{equation}
\end{subequations}
for PBC.
One verifies that
\begin{subequations}
\begin{align}
&
\widehat{Z}^{\,}_{k}\,
\ket*{\pm^{\,}_{\mathrm{OBC}}}^{\,}_{k,k+1}=
\pm\ket*{\pm^{\,}_{\mathrm{OBC}}}^{\,}_{k,k+1},
\\
&
\widehat{Z}^{\,}_{k}\,
\ket*{\pm^{\,}_{\mathrm{PBC}}}^{\,}_{k,k+1}=
\pm\ket*{\pm^{\,}_{\mathrm{PBC}}}^{\,}_{k,k+1},
\quad
\widehat{Z}^{\,}_{2N}\,
\ket*{\pm^{\,}_{\mathrm{PBC}}}^{\,}_{k,k+1}=
\pm\ket*{\pm^{\,}_{\mathrm{PBC}}}^{\,}_{k,k+1},
\end{align}
\end{subequations}
for $k=2,\cdots,2N-1$, from which it follows that
\begin{subequations}
\begin{align}
&
{}^{\,}_{k,k+1}\!\bra*{\pm^{\,}_{\mathrm{OBC}}}\,
\widehat{Z}^{\,}_{k}\,
\ket*{\pm^{\,}_{\mathrm{OBC}}}^{\,}_{k,k+1}=
\pm1,
\\
&
{}^{\,}_{k,k+1}\!\bra*{\pm^{\,}_{\mathrm{PBC}}}\,
\widehat{Z}^{\,}_{k}\,
\ket*{\pm^{\,}_{\mathrm{PBC}}}^{\,}_{k,k+1}=
\pm1,
\qquad
{}^{\,}_{k,k+1}\!\bra*{\pm^{\,}_{\mathrm{PBC}}}\,
\widehat{Z}^{\,}_{2N}\,
\ket*{\pm^{\,}_{\mathrm{PBC}}}^{\,}_{k,k+1}=
\pm1.
\end{align}
\end{subequations}
Moreover, because
$\widehat{Z}^{\,}_{k}$
and
$\widehat{Z}^{\,}_{k-1}\,\widehat{X}^{\,}_{k}\,\widehat{Z}^{\,}_{k+1}$
anticommute, the effect of
$\widehat{Z}^{\,}_{k-1}\,\widehat{X}^{\,}_{k}\,\widehat{Z}^{\,}_{k+1}$
on the eigenstates 
of $\widehat{Z}^{\,}_{k}$
with eigenvalues differing by a sign
is to interchange them. We then deduce that
\begin{subequations}
\begin{align}
&
{}^{\,}_{k,k+1}\!\bra*{\pm^{\,}_{\mathrm{OBC}}}\,
\widehat{Z}^{\,}_{k-1}\,\widehat{X}^{\,}_{k}\,\widehat{Z}^{\,}_{k+1}\,
\ket*{\pm^{\,}_{\mathrm{OBC}}}^{\,}_{k,k+1}=
0,
\\
&
{}^{\,}_{k,k+1}\!\bra*{\pm^{\,}_{\mathrm{PBC}}}\,
\widehat{Z}^{\,}_{k-1}\,\widehat{X}^{\,}_{k}\,\widehat{Z}^{\,}_{k+1}\,
\ket*{\pm^{\,}_{\mathrm{PBC}}}^{\,}_{k,k+1}=
0,
\qquad
{}^{\,}_{k,k+1}\!\bra*{\pm^{\,}_{\mathrm{PBC}}}\,
\widehat{Z}^{\,}_{2N-1}\,\widehat{X}^{\,}_{2N}\,\widehat{Z}^{\,}_{1}\,
\ket*{\pm^{\,}_{\mathrm{PBC}}}^{\,}_{k,k+1}=
0.
\end{align}
\end{subequations}
On the other hand, for any $l\neq k,2N$, we have
\begin{subequations}
\begin{align}
&
{}^{\,}_{k,k+1}\!\bra*{\pm^{\,}_{\mathrm{OBC}}}\,
\widehat{Z}^{\,}_{l}\,
\ket*{\pm^{\,}_{\mathrm{OBC}}}^{\,}_{k,k+1}=
0,
\\
&
{}^{\,}_{k,k+1}\!\bra*{\pm^{\,}_{\mathrm{PBC}}}\,
\widehat{Z}^{\,}_{l}\,
\ket*{\pm^{\,}_{\mathrm{PBC}}}^{\,}_{k,k+1}=
0,
\qquad
{}^{\,}_{k,k+1}\!\bra*{\pm^{\,}_{\mathrm{PBC}}}\,
\widehat{Z}^{\,}_{2N}\,
\ket*{\pm^{\,}_{\mathrm{PBC}}}^{\,}_{k,k+1}=
0,
\end{align}
\end{subequations}
and
\begin{subequations}
\begin{align}
&
{}^{\,}_{k,k+1}\!\bra*{\pm^{\,}_{\mathrm{OBC}}}\,
\widehat{Z}^{\,}_{l-1}\,\widehat{X}^{\,}_{l}\,\widehat{Z}^{\,}_{l+1}\,
\ket*{\pm^{\,}_{\mathrm{OBC}}}^{\,}_{k,k+1}=
(-1)^{\delta^{\,}_{1,\mathrm{sgn}(l-k)}}\,
(-1)^{l+1},
\\
&
{}^{\,}_{k,k+1}\!\bra*{\pm^{\,}_{\mathrm{PBC}}}\,
\widehat{Z}^{\,}_{l-1}\,\widehat{X}^{\,}_{l}\,\widehat{Z}^{\,}_{l+1}\,
\ket*{\pm^{\,}_{\mathrm{PBC}}}^{\,}_{k,k+1}=
(-1)^{\delta^{\,}_{1,\mathrm{sgn}(l-k)}}\,
(-1)^{l+1}.
\end{align}
\end{subequations}
These expectation values for OBC explain what is observed numerically in
Fig.~\ref{suppfig:DMRG OBC NeelXSPT spins b}.

At $(19\pi/20,9/10)$ from Fig.~\ref{suppfig:phase diagram},
expectation values in the ground state $|\Psi^{\,}_{0}\rangle$
have the following features.
The dependence on $j=2,\cdots,2N-1$
of the local cluster expectation value
$\langle
\widehat{Z}^{\,}_{j-1}\,
\widehat{X}^{\,}_{j}\,
\widehat{Z}^{\,}_{j+1}
\rangle^{\,}_{\Psi^{\,}_{0}}$
is $0.99\times(-1)^{j+1}$,
while
$\langle
\widehat{Z}^{\,}_{j-1}\,
\widehat{Y}^{\,}_{j}\,
\widehat{Z}^{\,}_{j+1}
\rangle^{\,}_{\Psi^{\,}_{0}}$,
$\langle
\widehat{Z}^{\,}_{j-1}\,
\widehat{Z}^{\,}_{j}\,
\widehat{Z}^{\,}_{j+1}
\rangle^{\,}_{\Psi^{\,}_{0}}$,
$\langle\widehat{X}^{\,}_{j}\rangle^{\,}_{\Psi^{\,}_{0}}$,
$\langle\widehat{Y}^{\,}_{j}\rangle^{\,}_{\Psi^{\,}_{0}}$,
and
$\langle\widehat{Z}^{\,}_{j}\rangle^{\,}_{\Psi^{\,}_{0}}$
all vanish
for both OBC $(b=0)$ in Fig.~\ref{suppfig:DMRG OBC NeelXSPT spins c}
and PBC $(b=1$) in Fig.~\ref{suppfig:DMRG PBC NeelXSPT spins c}.
For OBC,
$\langle\widehat{Z}^{\,}_{1}\rangle^{\,}_{\Psi^{\,}_{0}}$,
and
$\langle\widehat{Z}^{\,}_{32}\rangle^{\,}_{\Psi^{\,}_{0}}$
acquire non-vanishing expectation values at the end points
$j=1$ and $j=32$, unlike for the PBC case.

At $(19\pi/20,9/10)$ from Fig.~\ref{suppfig:phase diagram},
expectation values in the excited state $|\Psi^{\,}_{8}\rangle$
have the following features. For PBC,
$\langle
\widehat{Z}^{\,}_{j-1}\,
\widehat{X}^{\,}_{j}\,
\widehat{Z}^{\,}_{j+1}
\rangle^{\,}_{\Psi^{\,}_{2}}$
is $0.87\times(-1)^{j}$,
while
$\langle
\widehat{Z}^{\,}_{j-1}\,
\widehat{Y}^{\,}_{j}\,
\widehat{Z}^{\,}_{j+1}
\rangle^{\,}_{\Psi^{\,}_{2}}$,
$\langle
\widehat{Z}^{\,}_{j-1}\,
\widehat{Z}^{\,}_{j}\,
\widehat{Z}^{\,}_{j+1}
\rangle^{\,}_{\Psi^{\,}_{2}}$,
$\langle\widehat{X}^{\,}_{j}\rangle^{\,}_{\Psi^{\,}_{2}}$,
$\langle\widehat{Y}^{\,}_{j}\rangle^{\,}_{\Psi^{\,}_{2}}$,
and
$\langle\widehat{Z}^{\,}_{j}\rangle^{\,}_{\Psi^{\,}_{2}}$
all vanish according to Fig.~\ref{suppfig:DMRG PBC NeelXSPT spins d}.
For OBC,
the non-vanishing values of
$\langle 
\widehat{Z}^{\,}_{j-1}\,
\widehat{X}^{\,}_{j}\,
\widehat{Z}^{\,}_{j+1}
\rangle^{\,}_{\Psi^{\,}_{8}}$,
$\langle
\widehat{Z}^{\,}_{j-1}\,
\widehat{Z}^{\,}_{j}\,
\widehat{Z}^{\,}_{j+1}
\rangle^{\,}_{\Psi^{\,}_{8}}$,
and
$\langle\widehat{Z}^{\,}_{j}\rangle^{\,}_{\Psi^{\,}_{8}}$
$\langle\widehat{Z}^{\,}_{j}\rangle^{\,}_{\Psi^{\,}_{8}}$
decay exponentially fast in magnitude from the boundaries
at $j=1$ and $j=32$
according to Fig.~\ref{suppfig:DMRG OBC NeelXSPT spins d}.

\subsubsection{The $\mathrm{\hbox{N\'eel}}^{\mathrm{SPT}}_{y}$ phase}
\label{suppsubsubsec:NeelYSPT}

The $\mathrm{\hbox{N\'eel}}^{\mathrm{SPT}}_{y}$
phase is the only one of the four gapped phases
for which it is possible to find a ground state $|\Psi\rangle$ such that,
in the thermodynamic limit,
\begin{equation}
\overline{G^{2,x}}[\Psi]=0,
\qquad
\overline{G^{2,y}}[\Psi]=0,
\qquad
\overline{G^{4,x}}[\Psi]\neq0,
\qquad
\overline{G^{4,y}}[\Psi]=0,
\end{equation}
in Fig.~\ref{suppfig:DMRG OBC Gs}.

The unitary transformation $\widehat{U}^{\,}_{\mathrm{R}}$
defined in Eq.~\eqref{suppeq:two unitaries}
realizes the reflection symmetry of
the phase diagram in Fig.~\ref{suppfig:phase diagram}
with respect to $\theta=\pi/4$,
which holds for both OBC and PBC.
More specifically, the Hamiltonian \eqref{suppeq:def H}
transforms under conjugation by $\widehat{U}^{\,}_{\mathrm{R}}$ as
\begin{equation}
\widehat{U}^{\,}_{\mathrm{R}}\,
\widehat{H}^{\,}_{b}(\theta,\lambda)\,
\widehat{U}^{\dagger}_{\mathrm{R}}
=\widehat{H}^{\,}_{b}(\pi/2-\theta,\lambda),
\end{equation}
for $b=0,1$.
Consequently, the description of the $\mathrm{\hbox{N\'eel}}^{\mathrm{SPT}}_{x}$ phase
in Sec.~\ref{suppsubsubsec:NeelXSPT}
equally holds for
the $\mathrm{\hbox{N\'eel}}^{\mathrm{SPT}}_{y}$ phase if the following replacements
\begin{equation}
(\theta,\lambda) \to (\pi/2-\theta,\lambda),\qquad
\widehat{X}^{\,}_{j} \to \widehat{Y}^{\,}_{j}, \qquad
m^{x}_{\mathrm{sta,SPT}}\to m^{y}_{\mathrm{sta,SPT}},
\label{suppeq:changes NeelXSPT to NeelYSPT}
\end{equation}
are made. We have verified numerically the equivalence between
the properties of
the $\mathrm{\hbox{N\'eel}}^{\mathrm{SPT}}_{x}$
and the $\mathrm{\hbox{N\'eel}}^{\mathrm{SPT}}_{y}$ phases
once the identifications
\eqref{suppeq:changes NeelXSPT to NeelYSPT} are performed.

\begin{figure}[t!]
\centering
\includegraphics[width=1\columnwidth]{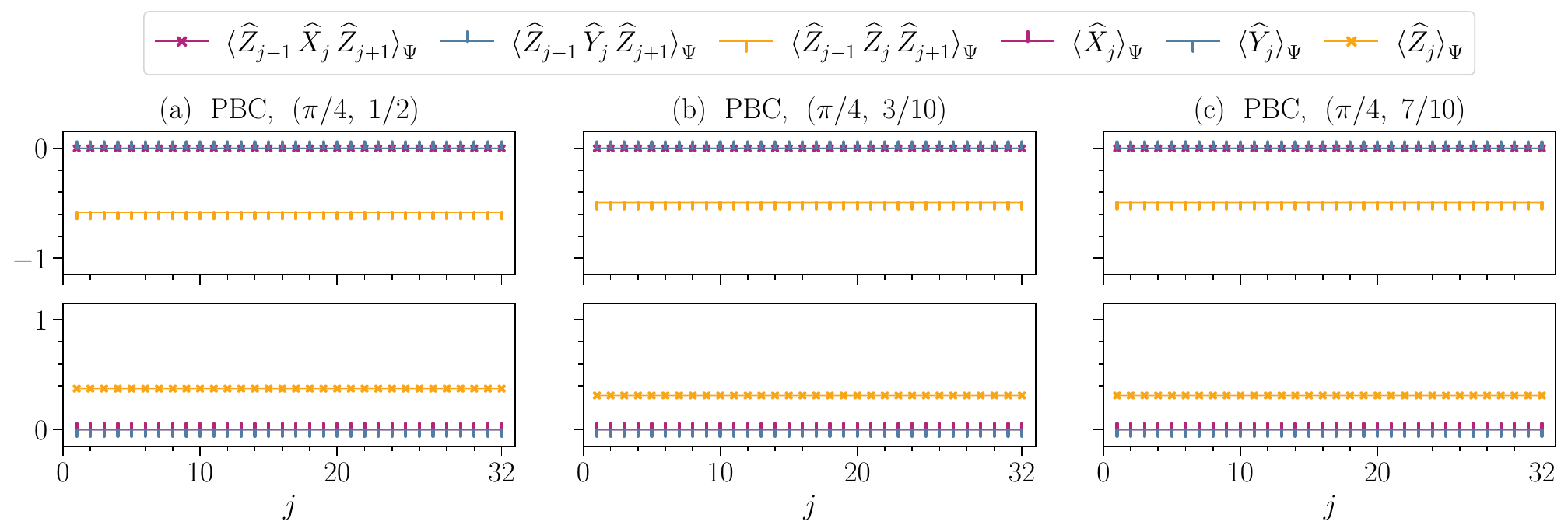}
\subfigure{\label{suppfig:DMRG PBC FMZ a} \relax}
\subfigure{\label{suppfig:DMRG PBC FMZ b} \relax}
\subfigure{\label{suppfig:DMRG PBC FMZ c} \relax}
\subfigure{\label{suppfig:DMRG PBC FMZ d} \relax}
\caption{(Color online)
The case of PBC.
Dependence on the site index $j=1,\cdots,2N$ of the local
cluster and spin expectation values obtained
using the DMRG approximation
($\chi=128$ and $2N=32$) to
the ground state $|\Psi^{\,}_{0}\rangle$ whose energy is
$E^{\,}_{0}$ at
(a) $(\pi/4,1/2)$,
(b) $(\pi/4,3/10)$, and
(c) $(\pi/4,7/10)$. 
}
\label{suppfig:DMRG PBC FMZ}
\end{figure}

\subsubsection{The $\mathrm{FM}^{\,}_{z}$ phase}
\label{suppsec:Numerics/Five/FMZ}

According to Fig.~\ref{suppfig:ed_10},
a fifth phase, centered around the point
$(\pi/4,1/2)$, is observed.
This phase has a diamond-like shape and
exhibits a twofold degeneracy for both PBC and OBC.
We label this phase the $\mathrm{FM}^{\,}_{z}$ phase.
We now explain why this interpretation is appropriate.
We consider first the case of PBC.

As is illustrated in Fig.~\ref{suppfig:DMRG PBC FMZ}
for PBC,
the local spin expectation value
$\ev*{\widehat{Z}^{\,}_{j}}^{\,}_{\Psi^{\,}_{0}}$
($\ev*{\widehat{X}^{\,}_{j}}^{\,}_{\Psi^{\,}_{0}}$,
$\ev*{\widehat{Y}^{\,}_{j}}^{\,}_{\Psi^{\,}_{0}}$)
along the $Z$-axis ($X$-, $Y$-axis)
in spin space
is non-vanishing (vanishing) and independent of $j=1,\cdots,2N$,
for any point in the $\mathrm{FM}^{\,}_{z}$ phase
from Fig.~\ref{suppfig:phase diagram}
with $\Psi^{\,}_{0}$ the ground state approximated by DMRG.
For example, we read from Fig.~\ref{suppfig:DMRG PBC FMZ} the values
\begin{align}  
m^{z}_{\mathrm{uni}}[\Psi^{\,}_{0}]=
\begin{cases}
0.3125,&\hbox{ at $(\pi/4,3/10)$,}\\
0.3750,&\hbox{ at $(\pi/4,1/2)$,}\\
0.3125,&\hbox{ at $(\pi/4,7/10)$.}\\
\end{cases}
\label{suppeq:mean values mzuni PPB}
\end{align}
The magnetization $m^{z}_{\mathrm{uni}}[\Psi^{\,}_{0}]$
defined in Eq.~\eqref{suppeq:muni}
is observed to reach its maximum $\approx0.4$
in absolute value at the center 
$(\pi/4,1/2)$ of Fig.\
\ref{suppfig:phase diagram},
whereby, the larger the system size $2N$,
the closer to $0.4$ the magnetization
$m^{z}_{\mathrm{uni}}[\Psi^{\,}_{0}]$.
The magnetization $m^{z}_{\mathrm{uni}}[\Psi^{\,}_{0}]$
converges to zero upon approaching the boundaries of the
$\mathrm{FM}^{\,}_{z}$ phase.
The quantum fluctuations are maximum at the
center $(\pi/4,1/2)$ of Fig.~\ref{suppfig:phase diagram}
due to the non-commuting nature
of the competing terms in
Hamiltonian (\ref{suppeq:def H})
and the enhanced symmetries of
Hamiltonian (\ref{suppeq:def H}) at this point.
We interpret the fact that the
magnetization $m^{z}_{\mathrm{uni}}[\Psi^{\,}_{0}]$
peaks in absolute value at $(\pi/4,1/2)$
as a manifestation of order by disorder.

As is also illustrated in 
Fig.~\ref{suppfig:DMRG PBC FMZ}
for PBC,  the cluster expectation value
$
\ev*{\widehat{Z}^{\,}_{j-1}\,
\widehat{Z}^{\,}_{j}\,
\widehat{Z}^{\,}_{j+1}}^{\,}_{\Psi^{\,}_{0}}
$
($\ev*{\widehat{Z}^{\,}_{j-1}\,
\widehat{X}^{\,}_{j}\,
\widehat{Z}^{\,}_{j+1}}^{\,}_{\Psi^{\,}_{0}}$,
$\ev*{\widehat{Z}^{\,}_{j-1}\,
\widehat{Y}^{\,}_{j}\,
\widehat{Z}^{\,}_{j+1}}^{\,}_{\Psi^{\,}_{0}}$)
is non-vanishing (vanishing) and independent of $j=1,\cdots,2N$,
for any point in the $\mathrm{FM}^{\,}_{z}$ phase
from Fig.~\ref{suppfig:phase diagram}
with $\Psi^{\,}_{0}$ the ground state approximated by DMRG.
The uniform SPT magnetization defined in
Eq.~\eqref{suppeq:muniSPT}
has the opposite sign from that of the uniform magnetization
at any point in the $\mathrm{FM}^{\,}_{z}$ phase. From
Fig.~\ref{suppfig:DMRG PBC FMZ}
we read the values
\begin{align}  
m^{z}_{\mathrm{uni,SPT}}[\Psi^{\,}_{0}]=
\begin{cases}
-0.493,&\hbox{ at $(\pi/4,3/10)$,}\\
-0.582,&\hbox{ at $(\pi/4,1/2)$,}\\
-0.493,&\hbox{ at $(\pi/4,7/10)$.}\\
\end{cases}
\label{suppeq:mean values mzuniSPT PPB}
\end{align}

In summary, the $\mathrm{FM}^{\,}_{z}$ phase under PBC
supports long-range order along the
$Z$-axis in spin space, with a uniform magnetization
no greater than $40\%$ of the uniform magnetization of 
the ferromagnetic product state
\begin{equation}
|\mathrm{Ferro}^{\,}_{z}\rangle\equiv
\ket*{\uparrow\ \uparrow\ \cdots\ \uparrow},
\qquad
\widehat{Z}^{\,}_{j}\,|\uparrow\rangle^{\,}_{j}=+
|\uparrow\rangle^{\,}_{j},
\qquad 
\widehat{Z}^{\,}_{j}\,|\downarrow\rangle^{\,}_{j}=-
|\downarrow\rangle^{\,}_{j}.
\end{equation}   
The $\mathrm{FM}^{\,}_{z}$ phase has
two orthonormal ground states
\begin{subequations}
\begin{equation}  
|\Psi^{\,}_{0}\rangle,
\qquad
|\Psi^{\,}_{1}\rangle=\widehat{T}\,|\Psi^{\,}_{0}\rangle,
\end{equation}
that support a uniform magnetization
of opposite signs and of magnitude
\begin{equation}
m^{z}_{\mathrm{uni}}[\Psi^{\,}_{0}]=
-m^{z}_{\mathrm{uni}}[\Psi^{\,}_{1}]\neq0.
\end{equation}
\end{subequations}

\begin{figure}[t!]
\centering
\includegraphics[width=0.5\columnwidth]{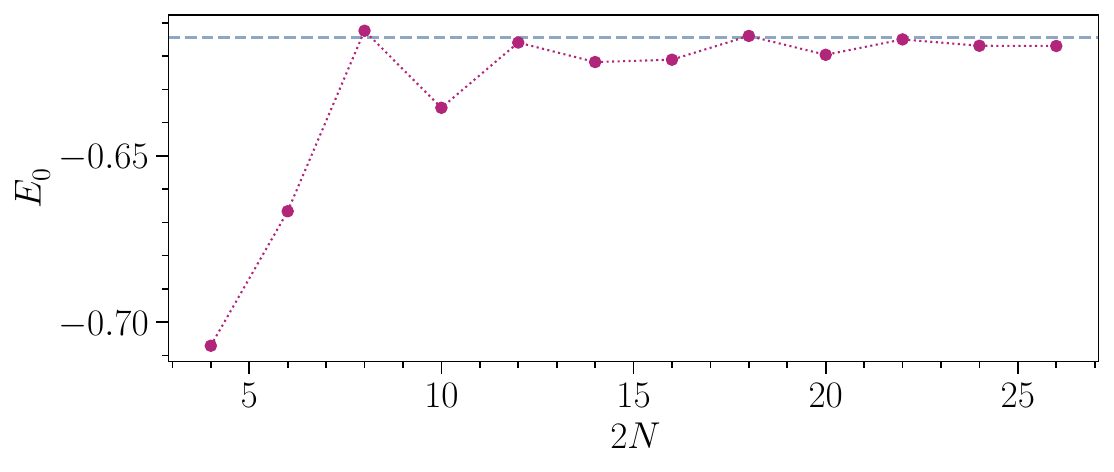}
\caption{(Color online)
The case of PBC.
Ground-state energy $E^{\,}_{0}$ per site 
as a function of the system size $2N$
of Hamiltonian \eqref{suppeq:def H}
at the point $(\pi/4,1/2)$.
This figure agrees with Fig. 5 from \cite{Zadnik21a},
provided $E^{\,}_{0}$ is substituted with
$\sqrt{2}\,E^{\,}_{0}$.
The dashed blue line is the value of
$\lim\limits_{2N\uparrow\infty}(E^{\,}_{0}/2N)$
computed from the thermodynamic Bethe Ansatz
in Ref.~\cite{Pozsgay21} (i.e., $-4\sqrt{2} \times 0.2172$).
}
\label{suppfig:ED energy center diamond}
\end{figure}

It has been shown in Refs.~\cite{Zadnik21a, Zadnik21b, Pozsgay21}
that Hamiltonian \eqref{suppeq:def H}
at $(\pi/4,1/2)$ is Bethe-Ansatz integrable,
realizes a conformal field theory with
central charge $\mathsf{c}=1$,
and supports ferromagnetic long range order.
The ground state energy per site that we have computed numerically
and plotted as a function of system size in
Fig.~\ref{suppfig:ED energy center diamond},
agrees with the prediction for it made with the Bethe Ansatz
in Refs.~\cite{Zadnik21a, Zadnik21b, Pozsgay21}.
It was also shown in Ref.~\cite{Yang20} that
Hamiltonian \eqref{suppeq:def H}
at $(\pi/4,1/2)$ exhibits Hilbert-space fragmentation.

The case of OBC displays strong boundary effects
which are tied to criticality and incommensuration effects, as we now discuss.

\subsection{Criticality and incommensuration of the $\mathrm{FM}^{\,}_{z}$ phase}
\label{suppsubsec:Criticality and incommensuration of the FMz phase}

Of all five phases in Fig.~\ref{suppfig:phase diagram},
it is only the $\mathrm{FM}^{\,}_{z}$ phase that has a distinctive coloring in
all panels from 
Fig.~\ref{suppfig:DMRG OBC Gs}.
The interpretation of
Figs.~\ref{suppfig:DMRG OBC Gs c},~\ref{suppfig:DMRG OBC Gs f},~\ref{suppfig:DMRG OBC Gs g},~\ref{suppfig:DMRG OBC Gs h},~\ref{suppfig:DMRG OBC Gs i},
and~\ref{suppfig:DMRG PBC FMZ}
is that the
$\mathrm{FM}^{\,}_{z}$ phase exhibits 
long-range ferromagnetic (FM) order along the $Z$-axis
in spin space.

The fact that $\overline{G^{r,z}}[\Psi^{\,}_{0}]$,
with $|\Psi^{\,}_{0}\rangle$ the 
DMRG approximation to the
ground state of Hamiltonian \eqref{suppeq:def H},
varies and changes sign as a function of
$r$ strongly suggests
the existence of oscillations in the spin-spin correlations along the
$Z$-axis in spin space,
with the period of oscillations
controlled by
$\lambda$.  This observation aligns with the 
empirical fact that $\ev*{\widehat{Z}^{\,}_{j}}^{\,}_{\Psi^{\,}_{0}}$
displays oscillations as a function of $j$ 
when OBC are selected, as we shall discuss in section
\ref{suppsec:LRO IC FMZ}. 

Figures~\ref{suppfig:DMRG OBC Gs a},~\ref{suppfig:DMRG OBC Gs b},~\ref{suppfig:DMRG OBC Gs d}, and~\ref{suppfig:DMRG OBC Gs e},
imply the non-vanishing values of
\begin{equation}
\overline{G^{2,x}}[\Psi^{\,}_{0}]\neq0, \qquad 
\overline{G^{2,y}}[\Psi^{\,}_{0}]\neq0, \qquad 
\overline{G^{4,x}}[\Psi^{\,}_{0}]\neq0, \qquad
\overline{G^{4,y}}[\Psi^{\,}_{0}]\neq0,
\end{equation}
in the $\mathrm{FM}^{\,}_{z}$ phase.
This empirical fact suggests
that the $\mathrm{FM}^{\,}_{z}$ phase
supports, in addition to
long-range FM order along the $Z$-axis in spin space,
quasi-long-range AFM
order for the local spins and local clusters
along the $X$- and $Y$-axes in spin space.
We are going to give in what follows supporting evidences for the claim that
long-range FM order coexists with AFM criticality
in the $\mathrm{FM}^{\,}_{z}$ phase.

\begin{figure}[t!]
	\centering
	\subfigure[]{\includegraphics[width=0.5\columnwidth]{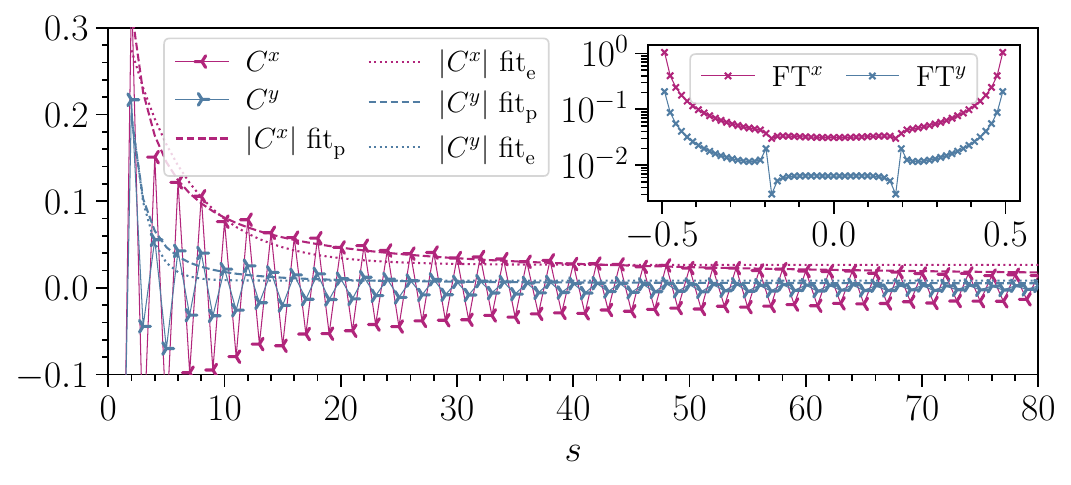} \label{suppfig:DMRG corr diamond xy a}}%
	\subfigure[]{\includegraphics[width=0.495\columnwidth]{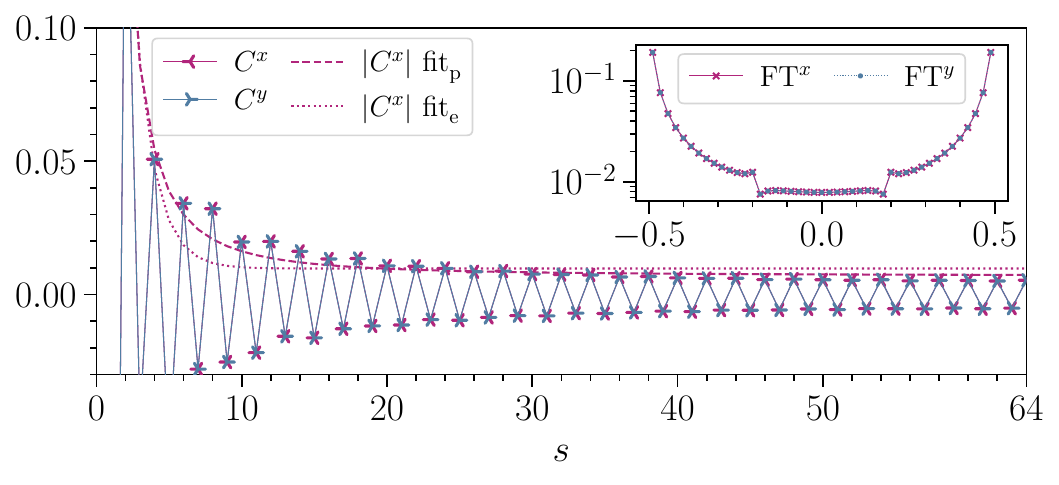} \label{suppfig:DMRG corr diamond xy b}} \\
	\subfigure[]{\includegraphics[width=0.5\columnwidth]{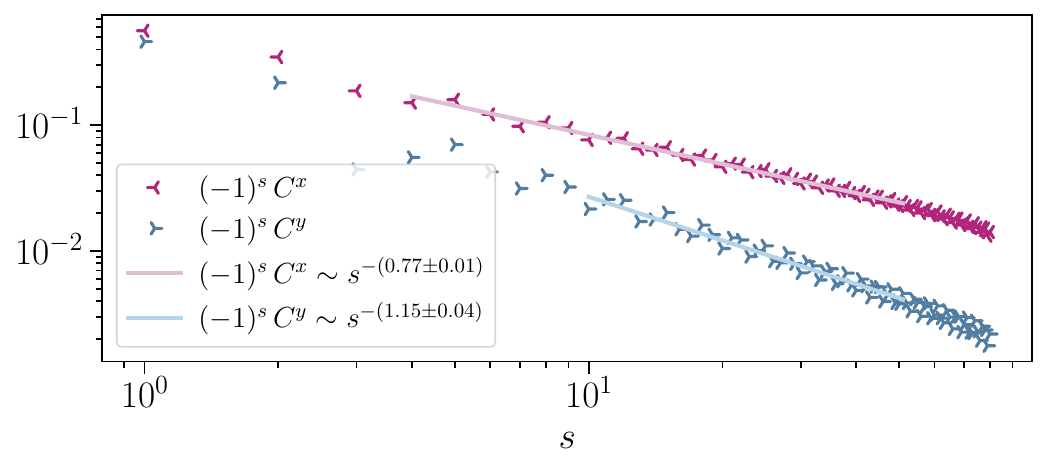} \label{suppfig:DMRG corr diamond xy c}}%
	\subfigure[]{\includegraphics[width=0.495\columnwidth]{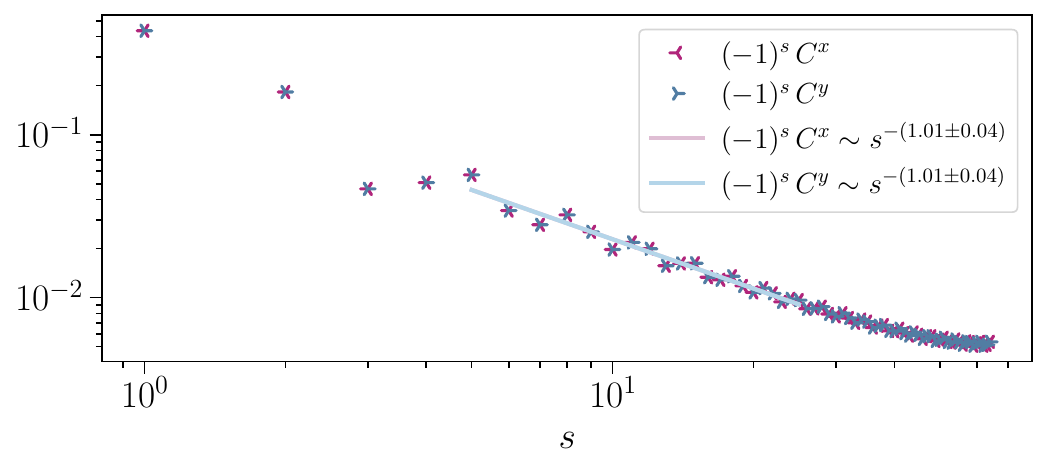} \label{suppfig:DMRG corr diamond xy d}}%
	\caption{(Color online)
		Dependence on $s$ given $j^{\,}_{0}$ for the
		spin-spin correlation functions
		$C^{x}_{j^{\,}_{0},j^{\,}_{0}+s}[\Psi^{\,}_{0}]$
		and
		$C^{y}_{j^{\,}_{0},j^{\,}_{0}+s}[\Psi^{\,}_{0}]$,  
		defined in Eq.~\eqref{suppeq:Calpha},
		computed using the DMRG approximation 
		to the ground state
		$|\Psi^{\,}_{0}\rangle$
		of Hamiltonian \eqref{suppeq:def H}
		(a)
		under OBC,
		at $(\theta,\lambda)=(\pi/4-0.1,0.4)$,
		with $j^{\,}_{0}=(2N-80)/2$,
		$\chi=128$ and $2N=128$ and
		(b)
		under PBC,
		at $(\theta,\lambda)=(\pi/4,1/2)$,
		with $j^{\,}_{0}=(2N-64)/2$,
		$\chi=256$ and $2N=128$.
		Fits used are labeled
		``p'' for power law,
		``e'' for exponential,
		Goodness of fit for ``p'' is always greater than that for ``e''.
		The best fits for $\eta^{x}$ and $\eta^{y}$
		are independent of the choice for $j^{\,}_{0}$.
		The insets show the Fourier transforms of the respective
		correlations, where the abscissa is the frequency
		$f\equiv q/(2\pi)$
		and the ordinate is the expansion coefficient $A(f)$.
		(c) Same as (a) as a log-log plot, and
		(d) same as (b) as a log-log plot.
	}
	\label{suppfig:DMRG corr diamond xy}
\end{figure}

\subsubsection{Criticality}
\label{suppeq:subsubsection Criticality} 

The criticality of the $\mathrm{FM}^{\,}_{z}$ phase 
in Fig.~\ref{suppfig:phase diagram} 
is corroborated by three observations.
\begin{itemize}
\item[$a$.]
First, the two transverse components
$C^{\alpha}_{j^{\,}_{0},j^{\,}_{0}+s}[\Psi^{\,}_{0}]$
of the spin-spin correlation functions 
in the DMRG approximation to the ground state
$|\Psi^{\,}_{0}\rangle$
of Hamiltonian \eqref{suppeq:def H}
defined in Eq.~\eqref{suppeq:Calpha} 
decay algebraically as
\begin{equation}
C^{\alpha}_{j^{\,}_{0},j^{\,}_{0}+s}[\Psi^{\,}_{0}]\propto
(-1)^{s}\,{s}^{-\eta^{\alpha}[\Psi^{\,}_{0}]},
\qquad
1\ll j^{\,}_{0},s,j^{\,}_{0}+s\ll2N,
\qquad
\alpha=x,y,
\label{suppeq:algebraic decay transverse spin-spin in FMZ phase}
\end{equation}  
with both PBC and OBC for any point 
in the $\mathrm{FM}^{\,}_{z}$ phase from Fig.~\ref{suppfig:phase diagram}.
\item[$b$.]
Second, any point in
the $\mathrm{FM}^{\,}_{z}$ phase from Fig.~\ref{suppfig:phase diagram}
realizes a CFT with the central charge $\mathsf{c}=1$.
\item[$c$.]
Finally, any energy spacing between non-degenerate
consecutive eigenenergies 
scales as $1/(2N)$ in the thermodynamic limit
for any point in the $\mathrm{FM}^{\,}_{z}$ phase from Fig.~\ref{suppfig:phase diagram}.
\end{itemize}

\paragraph{Quasi-long-range AFM order
in the $\mathrm{FM}^{\,}_{z}$ phase from Fig.~\ref{suppfig:phase diagram} ---}
Representative examples of the dependence on the separation $s$
of the transverse spin-spin correlation functions
\eqref{suppeq:Calpha}
in the $\mathrm{FM}^{\,}_{z}$ phase from Fig.~\ref{suppfig:phase diagram}
are shown in
Figs.~\ref{suppfig:DMRG corr diamond xy a} and
\ref{suppfig:DMRG corr diamond xy c}
for OBC,
and
Figs.~\ref{suppfig:DMRG corr diamond xy b} and
\ref{suppfig:DMRG corr diamond xy d}
for PBC.
The best fit selects the algebraic decay
(\ref{suppeq:algebraic decay transverse spin-spin in FMZ phase})
for the transverse components,
whereby
\begin{subequations}
\begin{equation}
\eta^{x}[\Psi^{\,}_{0}]=0.77\pm0.01, 
\qquad
\eta^{y}[\Psi^{\,}_{0}]=1.15\pm0.04,
\end{equation}
for $(\theta,\lambda)=(\pi/4-0.1,0.4)$ and under OBC,
while
\begin{equation}
\eta^{x}[\Psi^{\,}_{0}]=
\eta^{y}[\Psi^{\,}_{0}]=
1.01\pm0.04,
\end{equation}
\end{subequations}
for $(\theta,\lambda)=(\pi/4,1/2)$ and under PBC.

\begin{figure}[t!]
\centering
\subfigure[]{\includegraphics[width=0.5\columnwidth]{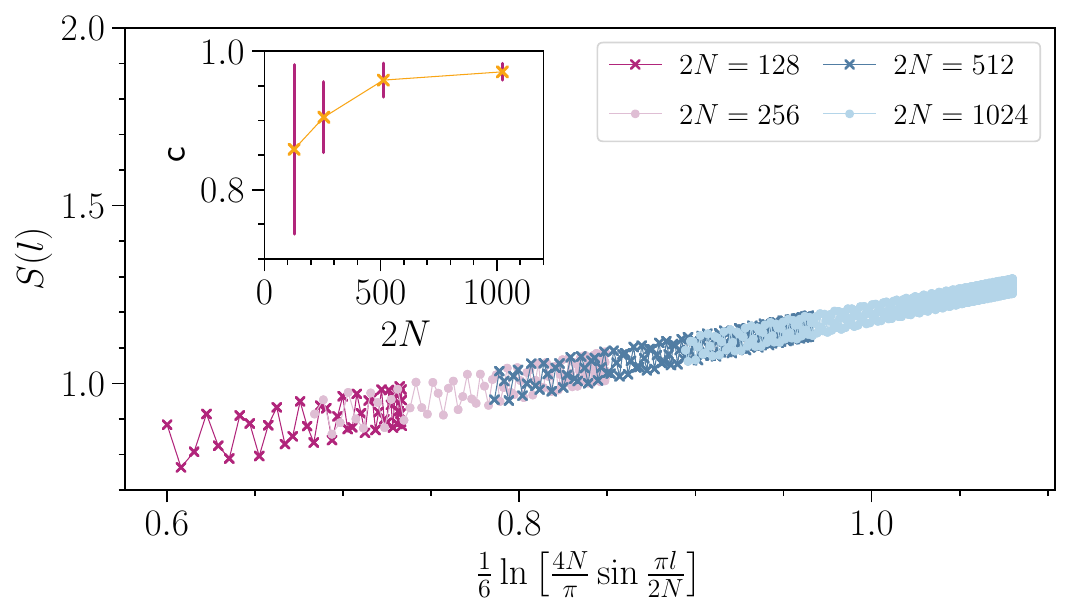} \label{suppfig:DMRG c diamond a}}%
\subfigure[]{\includegraphics[width=0.5\columnwidth]{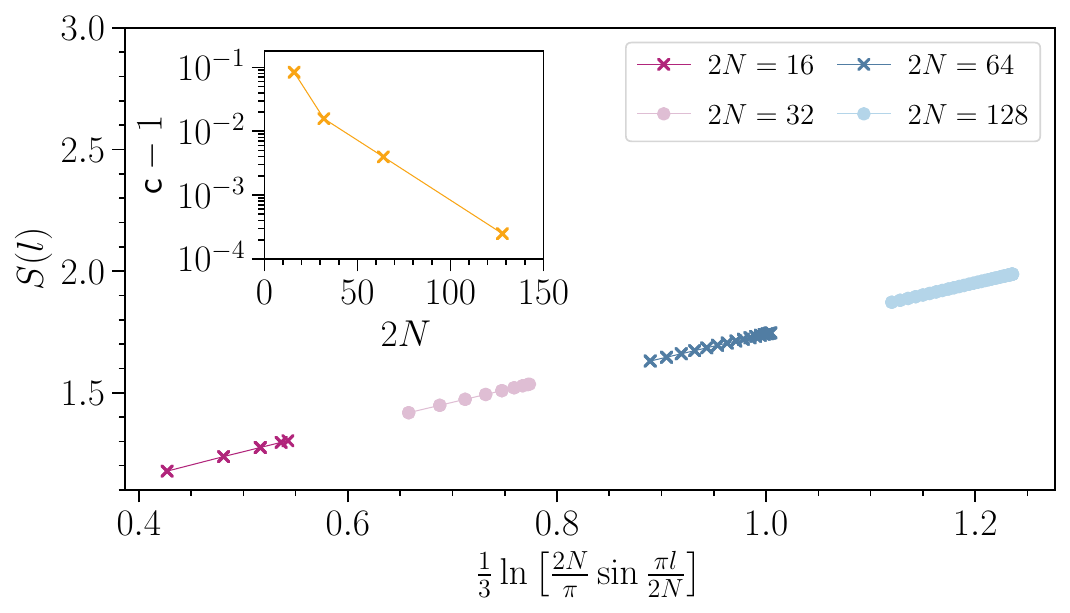} \label{suppfig:DMRG c diamond b}}
\caption{(Color online)
Dependence on the number $2N$ of sites of the
entanglement entropy $S^{\,}_{\mathrm{MPS},2N}({l})$
at $(\pi/4,1/2)$. 
The entanglement entropy $S^{\,}_{\mathrm{MPS},2N}({l})$ is defined in
Eq.~(\ref{suppeq:def S(ell) for MPS}) 
for a matrix product state
obtained using DMRG with
(a) OBC and $\chi=256$ and
(b) PBC and $\chi=512$.
The slope of the linear fit performed for each system
size $2N$ gives
the central charge $\mathsf{c}$ in the insets of
each panel. Error bars are given
in the inset of (a).
For all panels, the value of
$\delta{l}$ defined in Eq.~(\ref{suppeq:def delta{l}})
is $2N/2$.  
}
\label{suppfig:DMRG c diamond}
\end{figure}

\paragraph{CFT with central charge $\mathsf{c}=1$ ---}
The $\mathrm{FM}^{\,}_{z}$ phase exhibits quasi-long-range order along the $X$-
and $Y$-axes, suggesting that each point within this phase realizes a
CFT with central charge $\mathsf{c}$.
We have determined 
the value of the central charge
\begin{subequations}
\label{suppeq:CFT prediction for S2N(l)}
\begin{equation}
\mathsf{c}\coloneqq\lim_{2N\to\infty}\mathsf{c}^{\,}_{2N}
\label{suppeq:CFT prediction for S2N(l) a}
\end{equation}
in the $\mathrm{FM}^{\,}_{z}$ phase from Fig.~\ref{suppfig:phase diagram}
by comparing the CFT prediction
\cite{calabrese2004}
\begin{equation}
S^{\,}_{\mathrm{CFT},2N}({l})=
\begin{cases}
\frac{\mathsf{c}^{\,}_{2N}}{3}\,
\ln
\left(
\frac{2N}{\pi}\,
\sin\left(\frac{\pi{l}}{2N}\right)
\right)
+
\hbox{ const},
&
\hbox{ for PBC},
\\&\\
\frac{\mathsf{c}^{\,}_{2N}}{6}\,
\ln
\left(
\frac{4N}{\pi}\,
\sin\left(\frac{\pi{l}}{2N}\right)
\right)
+
\hbox{ const},
&
\hbox{ for OBC},
\end{cases}
\label{suppeq:CFT prediction for S2N(l) b}
\end{equation}
\end{subequations}
for the bipartite entanglement entropy $S^{\,}_{2N}({l})$
against its numerical evaluation $S^{\,}_{\mathrm{MPS},2N}({l})$.
Hereby, we define the entanglement
entropy $S^{\,}_{\mathrm{MPS},2N}({l})$
on the bond ${l}$ of the matrix product state that
approximates the ground state of Hamiltonian
\eqref{suppeq:def H} as follows. 
The chain of length $2N$ is partitioned into the
sites $j=1,\cdots,2N^{\,}_{A}$ that define subsystem $A$
with the Hilbert space
\begin{subequations}
\begin{equation}
\mathfrak{H}^{\,}_{A}\cong\mathbb{C}^{2^{4N^{\,}_{A}}}=
\mathrm{span}
\left\{
|\mathfrak{b}^{\,}_{A}\rangle\ \Big|\
\langle\mathfrak{b}^{\,}_{A}|\mathfrak{b}^{\prime}_{A}\rangle=
\delta^{\,}_{\mathfrak{b}^{\,}_{A},\mathfrak{b}^{\prime}_{A}},
\qquad
\mathfrak{b}^{\,}_{A},\mathfrak{b}^{\prime}_{A}=1,\cdots,2^{4N^{\,}_{A}}
\right\}
\end{equation}
and the sites $2N^{\,}_{A}+1,\cdots,2N$ that define subsystem $B$
with the Hilbert space
\begin{equation}
\mathfrak{H}^{\,}_{B}\cong\mathbb{C}^{2^{4N-4N^{\,}_{A}}}=
\mathrm{span}
\left\{
|\mathfrak{b}^{\,}_{B}\rangle\ \Big|\
\langle\mathfrak{b}^{\,}_{B}|\mathfrak{b}^{\prime}_{B}\rangle=
\delta^{\,}_{\mathfrak{b}^{\,}_{B},\mathfrak{b}^{\prime}_{B}},
\qquad
\mathfrak{b}^{\,}_{B},\mathfrak{b}^{\prime}_{B}=1,\cdots,2^{4N-4N^{\,}_{A}}
\right\}.
\end{equation}
Any state $|\Psi\rangle\in\mathbb{C}^{2^{4N}}$
can be expanded in the
tensor basis of $\mathfrak{H}^{\,}_{A}\otimes\mathfrak{H}^{\,}_{B}$,
\begin{equation}
|\Psi\rangle=
\sum_{\mathfrak{b}^{\,}_{A}=1}^{2^{4N^{\,}_{A}}}
\sum_{\mathfrak{b}^{\,}_{B}=1}^{2^{4N-4N^{\,}_{A}}}
M^{\,}_{\mathfrak{b}^{\,}_{A},\mathfrak{b}^{\,}_{B}}\,    
|\mathfrak{b}^{\,}_{A}\rangle
\otimes
|\mathfrak{b}^{\,}_{B}\rangle,
\qquad
M^{\,}_{\mathfrak{b}^{\,}_{A},\mathfrak{b}^{\,}_{B}}=
M^{\,}_{\mathfrak{b}^{\,}_{B},\mathfrak{b}^{\,}_{A}}\in\mathbb{C}.    
\end{equation}
If we denote with $s$ the singular values of
the matrix with the matrix elements
$M^{\,}_{\mathfrak{b}^{\,}_{A},\mathfrak{b}^{\,}_{B}}$,
then the bipartite von Neumann entropy of the state $|\Psi\rangle$
is defined by
\begin{equation}
S^{\,}_{N^{\,}_{A},N^{\,}_{B}}[\psi]\coloneqq
-\sum_{s}s^{2}\ln s^{2}.
\end{equation}
\end{subequations}      
For our matrix product state obtained from DMRG,
assuming that it is
in the so-called mixed canonical form on the bond
${l}$ with $1<{l}<2N$ and bond dimension $\chi$,
its entanglement entropy is
\begin{equation}
S^{\,}_{\mathrm{MPS},2N}({l})=
-
\sum_{i=1}^{\chi}
s^{2}_{i}({l})\,
\ln s^{2}_{i}({l}),
\label{suppeq:def S(ell) for MPS}
\end{equation}
where $s^{\,}_{i}({l})$ is the $i^{\mathrm{th}}$ singular value
associated to a bipartition of cardinality ${l}\,\equiv 2N^{\,}_{A}$ for subsystem $A$
and cardinality $2N-{l}$ for subsystem $B$.

We have used DMRG to evaluate
the bipartite entanglement entropy
$S^{\,}_{2N}(l)$
in the $\mathrm{FM}^{\,}_{z}$ phase of Fig.~\ref{suppfig:phase diagram}.
By plotting the dependence of
$S^{\,}_{\mathrm{MPS},2N}(l)$
on
$
\ln
\left(
\frac{4N}{\pi}\,
\sin\left(\frac{\pi{l}}{2N}\right)
\right)
$,
we were able to confirm the validity of
the CFT prediction
(\ref{suppeq:CFT prediction for S2N(l)})
with $\mathsf{c}=1$.
A representative example of such a procedure
at the center $(\pi/4,1/2)$
of the phase diagram
\ref{suppfig:phase diagram}
is found in Fig.~\ref{suppfig:DMRG c diamond}.
The range over which 
$S^{\,}_{\mathrm{MPS},2N}({l})$
defined in Eq.~(\ref{suppeq:def S(ell) for MPS})
is fitted by
$S^{\,}_{\mathrm{CFT},2N}({l})$
defined in Eq.~(\ref{suppeq:CFT prediction for S2N(l) b})
is
\begin{subequations}
\label{suppeq:def delta{l}}
\begin{equation}
\delta{l}\equiv
{l}^{\,}_{\mathrm{max}}
-
{l}^{\,}_{\mathrm{min}},
\qquad
1\leq{l}^{\,}_{\mathrm{min}}<{l}^{\,}_{\mathrm{max}}\leq2N.
\end{equation}
Here, we choose
\begin{equation}
\delta{l}\approx
2N/2,
\qquad
{l}^{\,}_{\mathrm{min}}\approx
2N/4,
\qquad
2N-{l}^{\,}_{\mathrm{max}}\approx
2N/4,
\end{equation}
\end{subequations}
so that the bonds ${l}^{\,}_{\mathrm{min}}$ and ${l}^{\,}_{\mathrm{max}}$ 
are each a distance of $2N/4$ away from
the first bond ${l}=1$ and the last bond ${l}=2N-1$
entering the matrix product state. The range $\delta{l}$ must be a
finite fraction of the number $2N$ of sites because the initial
matrix product state used for DMRG under both PBC and OBC
breaks explicitly the translation symmetry.

In Fig.\ \ref{suppfig:DMRG c diamond a}, the bipartite entanglement entropy
$S^{\,}_{\mathrm{MPS},2N}({l})$
shows oscillations as a function of ${l}$. This oscillatory behaviour is a the 
signature of the incommensuration in the $\mathrm{FM}^{\,}_{z}$ phase~\cite{Ye150205049}, which we shall explore in Sec.~\ref{suppsec:LRO IC FMZ}. 
It is well known that extracting the central charge in finite systems
is more accurate with PBC, as finite-size effects are strongly
suppressed compared to OBC.  
However, choosing PBC significantly
increases the computational cost of the DMRG algorithm \cite{schollwock2011}
\footnote{
Specifically, if DMRG converges to a ground state with variance $V$
using a bond dimension $\chi$ with OBC, achieving the same variance
$V$ with PBC requires a bond dimension of $\chi^{2}$. This increases
the computational cost of DMRG by $\mathcal{O}(\chi^{3})$ when using
PBC over OBC}. 
Imposing PBC, we find that $(\pi/4,1/2)$ realizes a CFT with
$\mathsf{c}=1$.  We computed $\mathsf{c}$ for other points inside
the $\mathrm{FM}^{\,}_{z}$ phase and consistently found $\mathsf{c}=1$ with the
same level of precision as in Fig.~\ref{suppfig:DMRG c diamond b}.

\begin{figure}[t!]
\centering
\subfigure{\relax \label{suppfig:DMRG gaps OBC 1over2,piover4 a}}
\subfigure{\includegraphics[width=0.395\columnwidth]{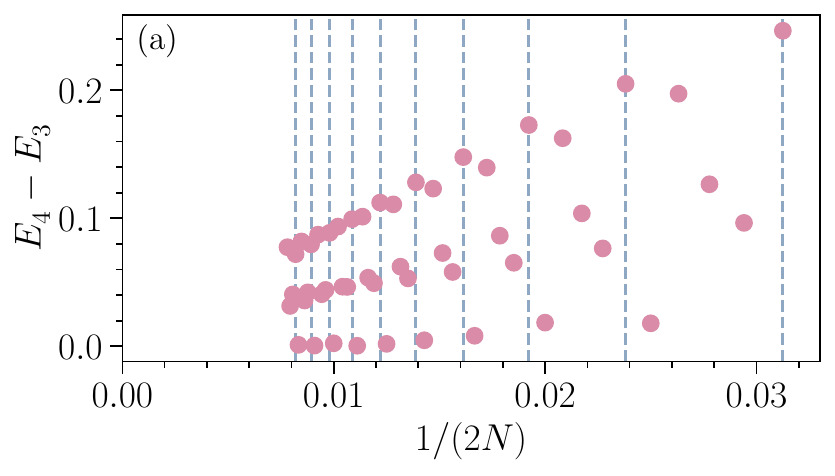} \label{suppfig:DMRG gaps OBC 1over2,piover4 b}}
\subfigure{\includegraphics[width=0.5\columnwidth]{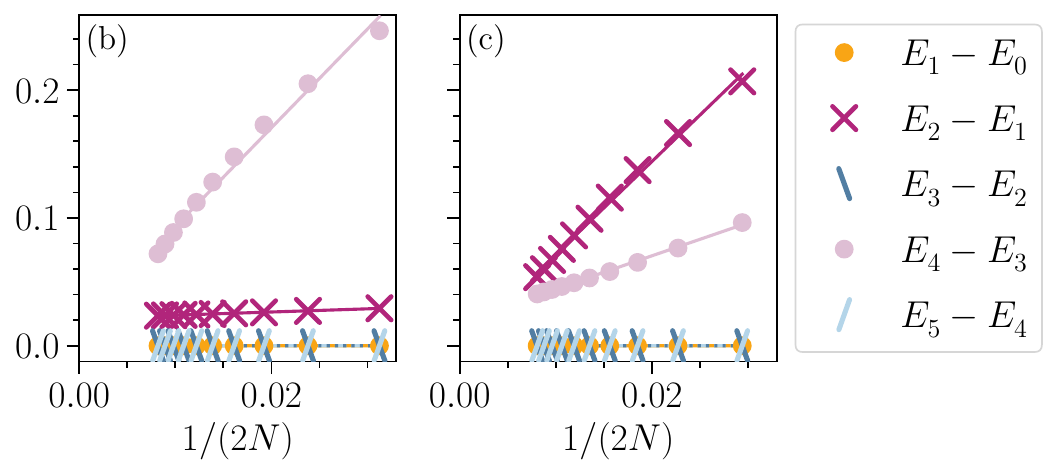} \label{suppfig:DMRG gaps OBC 1over2,piover4 c}}
\caption{(Color online)
The case of OBC.
Dependence on $1/(2N)$ of the
energy differences $E^{\,}_{n}-E^{\,}_{n-1}$
for $n=1,2,3,4,5$, at the center $(\pi/4,1/2)$
of the $\mathrm{FM}^{\,}_{z}$ phase from Fig.~\ref{suppfig:phase diagram}.
Energy eigenvalues are obtained using DMRG $(\chi=128)$
(a) Dependence on $1/(2N)$ of the energy difference $E^{\,}_{4}-E^{\,}_{3}$
is a non-monotonic decrease of the form
(\ref{suppeq:empirical dependence on 2N of delta n}).
(b)
Dependence on $1/(2N)$ with $32\leq 2N=32+10k\leq128$, $k=0,1,2,3,\cdots$
of the energy differences $E^{\,}_{n}-E^{\,}_{n-1}$
for $n=1,2,3,4,5$.
(c)
Dependence on $1/(2N)$ with $34\leq 2N=34+10k\leq128$, $k=0,1,2,3,\cdots$
of the energy differences $E^{\,}_{n}-E^{\,}_{n-1}$
for $n=1,2,3,4,5$.
Linear fits of $E^{\,}_{n}-E^{\,}_{n-1}$ for each $n$ in panels (b) and (c)
are represented by lines whose colors correspond
to the colors of the
data points in the legend.
}
\label{suppfig:DMRG gaps OBC 1over2,piover4}
\end{figure}

\paragraph{Gapless excitations ---}
An alternative strategy in support of the existence
of a CFT that encodes criticality
for any point in the $\mathrm{FM}^{\,}_{z}$ phase from Fig.~\ref{suppfig:phase diagram} 
involves analyzing the dependence on the system size $2N$
in the limit $2N\to\infty$
of the differences
\begin{equation}
\delta^{\,}_{n}(2N)\coloneqq
E^{\,}_{n}(2N)
-
E^{\,}_{n-1}(2N),
\qquad
n=1,2,\cdots,
\label{suppeq:def delta n}
\end{equation}
between consecutive low-lying energy eigenvalues. 
For these low-lying excited energy eigenvalues
\footnote{Only the low lying excitations 
associated to the primary fields
(as opposed to their descendants, say) 
are considered.  
},
CFT implies the dependence on the length $L\equiv2N\,\mathfrak{a}$
of the chain with the lattice spacing $\mathfrak{a}$ given by
\begin{equation}
E^{\,}_{n}(L)-E^{\,}_{0}(L)=
\sum_{m=1}^{n}\delta^{\,}_{m}(L)
\sim
\frac{2\pi}{L}\,
\hbar\,v\,
\Delta^{\,}_{n}
+
\mathcal{O}\left(\frac{1}{L\,\ln L}\right),
\label{suppeq:Cardy formula}
\end{equation}
where $E^{\,}_{n}(L)$ is the $n^{\mathrm{th}}$ 
excited energy above the ground-state energy 
$E^{\,}_{0}(L)$ when PBC are selected,
$v$ plays the role of the ``speed of light'' in the CFT,
and $\Delta^{\,}_{n}$ denotes the scaling dimension 
of the primary fields
$\phi^{\,}_{n}$
in the CFT
that is associated to the eigenenergy $E^{\,}_{n}$ \cite{cardy1988},
i.e.,
the  scaling dimension $\Delta^{\,}_{n}$ is the power that characterizes
the algebraic decay of the two-point function for
the primary field $\phi^{\,}_{n}$. Numerical evaluation of
the dependence on $1/L$ of Eq.~(\ref{suppeq:Cardy formula})
and of the dependence on the separation $s$ entering the two-point
function for the primary field $\phi^{\,}_{n}$ allows
to extract the ``speed of light'' $v$
and the scaling dimension $\Delta^{\,}_{n}$.
The central charge $\mathsf{c}$ of the CFT can then be 
determined by the numerical estimate of the dependence on $L$
of the ground-state energy 
\begin{equation}
E^{\,}_{0}(L)\sim
e^{\,}_{0}\times L
-
\frac{\pi\,\hbar\,v\,\mathsf{c}}{6\,L}
+
\mathcal{O}\left(\frac{1}{L\,\ln L}\right),
\end{equation}
where $e^{\,}_{0}$ 
is the ground-state energy per site 
in the thermodynamic limit $L\equiv2N\,\mathfrak{a}\to\infty$
\cite{cardy1988}. 
However, an accurate estimate
for the ``speed of light''
$v$ proved challenging.

Indeed, we were unable to measure the product
$v\,\Delta^{\,}_{n}$
on the right-hand side of Eq.~(\ref{suppeq:Cardy formula}),
because we found that the dependence on $2N$ of the energy difference
(\ref{suppeq:def delta n})
is of the form
\begin{equation}
\delta^{\,}_{n}(2N)\sim
\frac{\cos(Q^{\,}_{n}\times2N)}{2N},
\qquad
Q^{\,}_{n}\approx\frac{2\pi}{10},
\label{suppeq:empirical dependence on 2N of delta n} 
\end{equation}
in the $\mathrm{FM}^{\,}_{z}$ phase from Fig.~\ref{suppfig:phase diagram},
as is illustrated in Fig.~\ref{suppfig:DMRG gaps OBC 1over2,piover4}.
Figure~\ref{suppfig:DMRG gaps OBC 1over2,piover4} applies to
the center $(\pi/4,1/2)$ of Fig.~\ref{suppfig:phase diagram}.
Figure~\ref{suppfig:DMRG gaps OBC 1over2,piover4 a}
shows that the dependence on the number $2N$ of lattice points
of the energy difference $E^{\,}_{4}-E^{\,}_{3}$
takes the same form as the right-hand side of
Eq.~(\ref{suppeq:empirical dependence on 2N of delta n}).
A periodic oscillation is superimposed over the power law
decay $1/(2N)$.
Figures~\ref{suppfig:DMRG gaps OBC 1over2,piover4 b}
and~\ref{suppfig:DMRG gaps OBC 1over2,piover4 c}
for $E^{\,}_{2m+1}-E^{\,}_{2m}$ with $m=0,1,2$
show that $E^{\,}_{2m+1}-E^{\,}_{2m}$ is exponentially suppressed
as a function of $2N$, i.e., that
$E^{\,}_{2m+1}$ and $E^{\,}_{2m}$ become degenerate in the thermodynamic limit.
Figures~\ref{suppfig:DMRG gaps OBC 1over2,piover4 b}
and~\ref{suppfig:DMRG gaps OBC 1over2,piover4 c}
for $E^{\,}_{2m}-E^{\,}_{2m-1}$ with $m=1,2$,
show that $E^{\,}_{2m}-E^{\,}_{2m-1}$
is of the form (\ref{suppeq:empirical dependence on 2N of delta n})
with a characteristic periodicity of order $2\pi\,Q^{-1}_{n}\approx10$
that prevents access to sufficiently large systems sizes
for an accurate estimate of
$v\,\Delta^{\,}_{n}$.

\begin{figure}[t!]
\centering
\includegraphics[width=1\columnwidth]{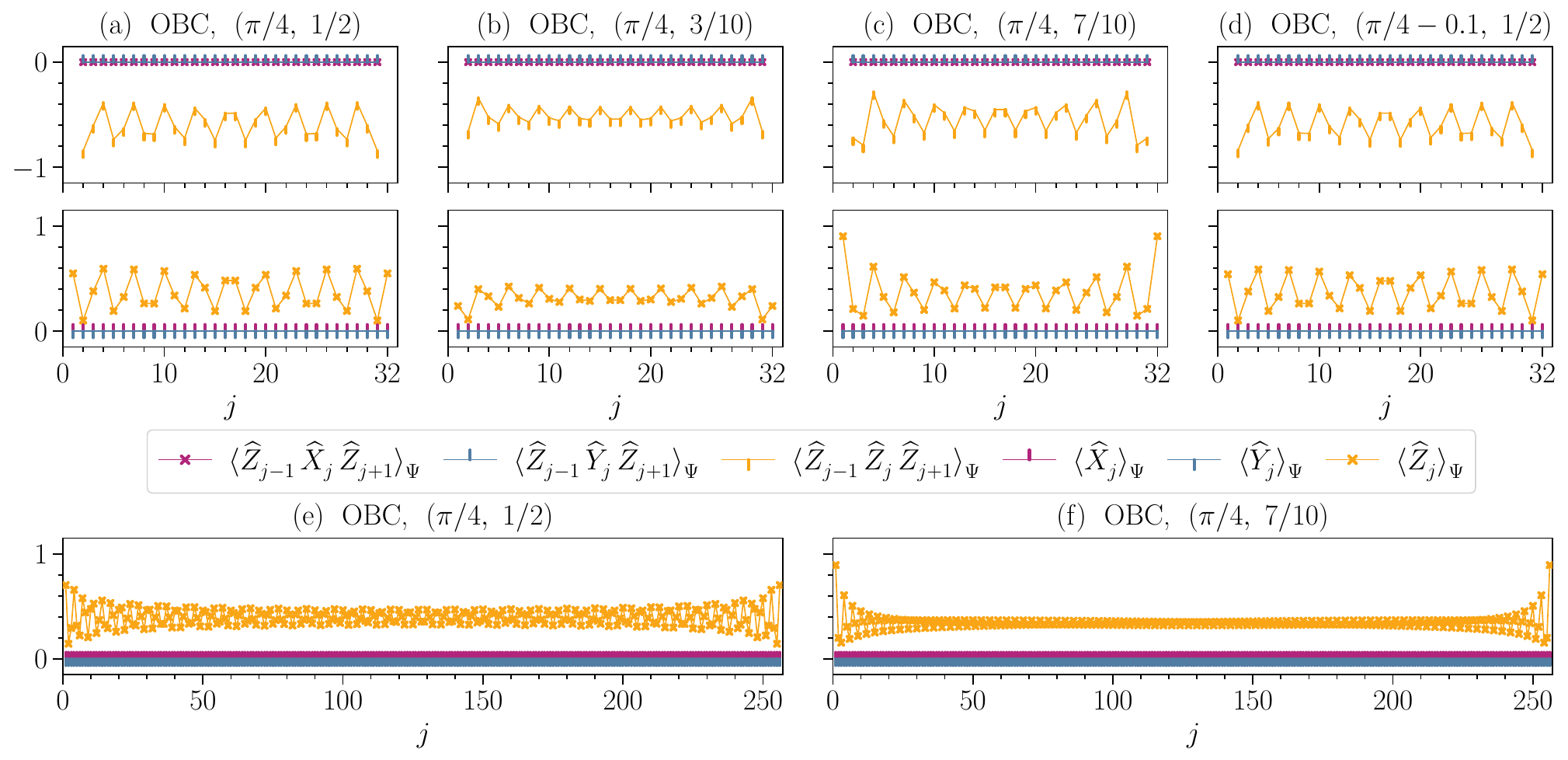}
\subfigure{\label{suppfig:DMRG OBC FMZ a} \relax}
\subfigure{\label{suppfig:DMRG OBC FMZ b} \relax}
\subfigure{\label{suppfig:DMRG OBC FMZ c} \relax}
\subfigure{\label{suppfig:DMRG OBC FMZ d} \relax}
\subfigure{\label{suppfig:DMRG OBC FMZ e} \relax}
\subfigure{\label{suppfig:DMRG OBC FMZ f} \relax}
\caption{(Color online)
The case of OBC.
Dependence on the site index $j=1,\cdots,2N$ of the local
cluster and spin expectation values obtained using the DMRG approximation
($\chi=128$ and $2N=32$) 
to the ground state $|\Psi^{\,}_{0}\rangle$ whose energy is $E^{\,}_{0}$ at
(a) $(\pi/4,1/2)$,
(b) $(\pi/4,3/10)$,
(c) $(\pi/4,7/10)$, and
(d) $(\pi/4-0.1,1/2)$
from Fig.~\ref{suppfig:phase diagram}.
Dependence on $j=1,\cdots,2N$ of the local
spin expectations values computed using the DMRG approximation
($\chi=128$ and $2N=256$) 
to the ground state $|\Psi^{\,}_{0}\rangle$
whose energy is $E^{\,}_{0}$ at
(e) $(\pi/4,1/2)$ and
(f) $(\pi/4,7/10)$
from Fig.~\ref{suppfig:phase diagram}.
}
\label{suppfig:DMRG OBC FMZ}
\end{figure}

\subsubsection{Incommensuration}
\label{suppsec:LRO IC FMZ}

Incommensuration effects become manifest upon using OBC as opposed
to PBC. For example, the non-vanishing local expectation values 
from Fig.~\ref{suppfig:DMRG PBC FMZ} are independent of the site index
when PBC are selected, while they become strongly site-dependent
when OBC are selected as is
illustrated in Fig.~\ref{suppfig:DMRG OBC FMZ}.

According to Fig.~\ref{suppfig:DMRG OBC FMZ},
the only non-vanishing local expectation values are
the ground-state expectation values of
$\widehat{Z}^{\,}_{j}$
and
$\widehat{Z}^{\,}_{j-1}\,\widehat{Z}^{\,}_{j}\,\widehat{Z}^{\,}_{j+1}$.
For a chain made of $2N=32$ sites under OBC,
they oscillate around the mean values
\begin{subequations}
\label{suppeq:mean values mzuni and mzuniSPT OPB}
\begin{align}
\label{suppeq:mean values mzuni and mzuniSPT OPB a}
m^{z}_{\mathrm{uni}}[\Psi^{\,}_{0}]=
\begin{cases}
0.373,&\hbox{ at $(\pi/4-0.1,1/2)$},
\\
0.375,&\hbox{ at $(\pi/4,7/10)$},
\\
0.375,&\hbox{ at $(\pi/4,1/2)$},
\\
0.312,&\hbox{ at $(\pi/4,3/10)$},
\\
0.373,&\hbox{ at $(\pi/4+0.1,1/2)$},
\end{cases}
\end{align}
defined in Eq.~(\ref{suppeq:muni})
and
\begin{align}
\label{suppeq:mean values mzuni and mzuniSPT OPB b}
m^{z}_{\mathrm{uni},\mathrm{SPT}}[\Psi^{\,}_{0}]=
\begin{cases}
-0.591,&\hbox{ at $(\pi/4-0.1,1/2)$},
\\
-0.543,&\hbox{ at $(\pi/4,7/10)$},
\\
-0.594,&\hbox{ at $(\pi/4,1/2)$},
\\
-0.507,&\hbox{ at $(\pi/4,3/10)$},
\\
-0.591,&\hbox{ at $(\pi/4+0.1,1/2)$},
\end{cases}
\end{align}
\end{subequations}
defined in Eq.~(\ref{suppeq:muniSPT}).
As was the case with
Eq.~(\ref{suppeq:mean values mzuni PPB}),
these mean values
reach their maxima in absolute value at the center
$(\pi/4,1/2)$ of the phase diagram \ref{suppfig:phase diagram}.
As was the case with
Eq.~(\ref{suppeq:mean values mzuni PPB}),
they are symmetric with respect to the mirror symmetry about the
$\mathrm{U}(1)$-symmetry axis $\theta=\pi/4$.
Unlike was the case with
Eq.~(\ref{suppeq:mean values mzuni PPB}),
they are not symmetric
about the axis $\lambda=1/2$. 
For a chain of length $2N=32$,
the ground-state expectation values of
$\widehat{Z}^{\,}_{j}$
and
$\widehat{Z}^{\,}_{j-1}\,\widehat{Z}^{\,}_{j}\,\widehat{Z}^{\,}_{j+1}$
can be fitted with the pair of Ansaetze
\begin{subequations}
\label{suppeq:oscillation ansatz in FMZ phase}
\begin{equation}
\langle\widehat{Z}^{\,}_{j}\rangle^{\,}_{\Psi^{\,}_{0}}=
m^{z}_{\mathrm{uni}}[\Psi^{\,}_{0}]
+
A^{z}_{\mathrm{loc}}[\Psi^{\,}_{0}]\,
\cos
\big(
q^{z}_{\mathrm{loc}}[\Psi^{\,}_{0}]\,j
+
\phi^{z}_{\mathrm{loc}}[\Psi^{\,}_{0}]
\big),
\qquad
1\ll j\ll2N,
\qquad\qquad\qquad\quad
\label{suppeq:oscillation ansatz in FMZ phase a}
\end{equation}
and
\begin{equation}
\langle
\widehat{Z}^{\,}_{j-1}\,\widehat{Z}^{\,}_{j}\,\widehat{Z}^{\,}_{j+1}
\rangle^{\,}_{\Psi^{\,}_{0}}=
m^{z}_{\mathrm{uni,SPT}}[\Psi^{\,}_{0}]
+
A^{z}_{\mathrm{loc,SPT}}[\Psi^{\,}_{0}]\,
\cos
\big(
q^{z}_{\mathrm{loc,SPT}}[\Psi^{\,}_{0}]\,j
+
\phi^{z}_{\mathrm{loc,SPT}}[\Psi^{\,}_{0}]
\big),
\qquad
1\ll j\ll2N,
\label{suppeq:oscillation ansatz in FMZ phase b}
\end{equation}
\end{subequations}
respectively,
whereby the wave numbers of the oscillations are
\begin{subequations}
\begin{align}
\frac{q^{z}_{\mathrm{loc}}[\Psi^{\,}_{0}]}{2\pi}=
\begin{cases}
0.313,&\hbox{ at $(\pi/4-0.1,7/10)$},
\\
0.324,&\hbox{ at $(\pi/4,7/10)$},
\\
0.313,&\hbox{ at $(\pi/4,1/2)$},
\\
0.323,&\hbox{ at $(\pi/4,3/10)$},
\\
0.313,&\hbox{ at $(\pi/4+0.1,7/10)$},
\end{cases}
\end{align}
for
$\langle\widehat{Z}^{\,}_{j}\rangle^{\,}_{\Psi^{\,}_{0}}$
and
\begin{align}
\frac{q^{z}_{\mathrm{loc,SPT}}[\Psi^{\,}_{0}]}{2\pi}=
\begin{cases}
0.313,&\hbox{ at $(\pi/4-0.1,7/10)$},
\\
0.324,&\hbox{ at $(\pi/4,7/10)$},
\\
0.313,&\hbox{ at $(\pi/4,1/2)$},
\\
0.325,&\hbox{ at $(\pi/4,3/10)$},
\\
0.313,&\hbox{ at $(\pi/4+0.1,7/10)$},
\end{cases}
\end{align}
\end{subequations}
for
$
\langle
\widehat{Z}^{\,}_{j-1}\,\widehat{Z}^{\,}_{j}\,\widehat{Z}^{\,}_{j+1}
\rangle^{\,}_{\Psi^{\,}_{0}}
$.
We empirically observe that
$q^{z}_{\mathrm{loc}}[\Psi^{\,}_{0}]$
and
$q^{z}_{\mathrm{loc,SPT}}[\Psi^{\,}_{0}]$
are a function of $|1/2-\lambda|$ only.

To obtain a more precise estimate of
$m^{z}_{\mathrm{uni}}[\Psi^{\,}_{0}]$
and $q^{z}_{\mathrm{loc}}[\Psi^{\,}_{0}]$,
we compute these quantities on a chain
of length $2N=256$ [see Figs.~\ref{suppfig:DMRG OBC FMZ e}
and~\ref{suppfig:DMRG OBC FMZ f}]. The local spin expectation
values $\ev*{\widehat{Z}^{\,}_{j}}^{\,}_{\Psi^{\,}_{0}}$
oscillates around the mean values
\begin{subequations}
\begin{align}
\label{suppeq:mean values mzuni 256}
m^{z}_{\mathrm{uni}}[\Psi^{\,}_{0}]=
\begin{cases}
0.340,&\hbox{ at $(\pi/4,7/10)$},
\\
0.398,&\hbox{ at $(\pi/4,1/2)$},
\\
0.337,&\hbox{ at $(\pi/4,3/10)$},
\end{cases}
\end{align}
with the wave number of oscillations
\begin{align}
\frac{q^{z}_{\mathrm{loc}}[\Psi^{\,}_{0}]}{2\pi}=
\begin{cases}
0.323,&\hbox{ at $(\pi/4,7/10)$},
\\
0.301,&\hbox{ at $(\pi/4,1/2)$},
\\
0.328,&\hbox{ at $(\pi/4,3/10)$}.
\end{cases}
\end{align}
\end{subequations}

We observe in 
Eq.~\eqref{suppeq:mean values mzuni 256}
that the values of the uniform magnetization
$m^{z}_{\mathrm{uni}}[\Psi^{\,}_{0}]$
for a chain of $2N=256$ sites
is approximately equal at the points
$(\pi/4,3/10)$ and $(\pi/4,7/10)$
as opposed to 
Eq.~\eqref{suppeq:mean values mzuni and mzuniSPT OPB a}
for a chain of $2N=32$ sites.
This observation is explained as follows.
(i) The unitary transformation 
$\widehat{U}^{\,}_{\mathrm{E}}$
defined in Eq.~\eqref{suppeq:two unitaries}
does not realize the reflection symmetry about
the line $\lambda=1/2$ under OBC.
(ii) The difference in $m^{z}_{\mathrm{uni}}[\Psi^{\,}_{0}]$
at the points
$(\pi/4,3/10)$ and $(\pi/4,7/10)$
is attributed to the effect of the
free spin-$1/2$ degrees of freedom
at both ends of the chain.
(iii) The difference in $m^{z}_{\mathrm{uni}}[\Psi^{\,}_{0}]$
at the points
$(\pi/4,3/10)$ and $(\pi/4,7/10)$
vanishes in the thermodynamic limit
$2N\to\infty$.

\begin{figure}[t!]
\centering
\subfigure[]{\includegraphics[width=0.5\columnwidth]{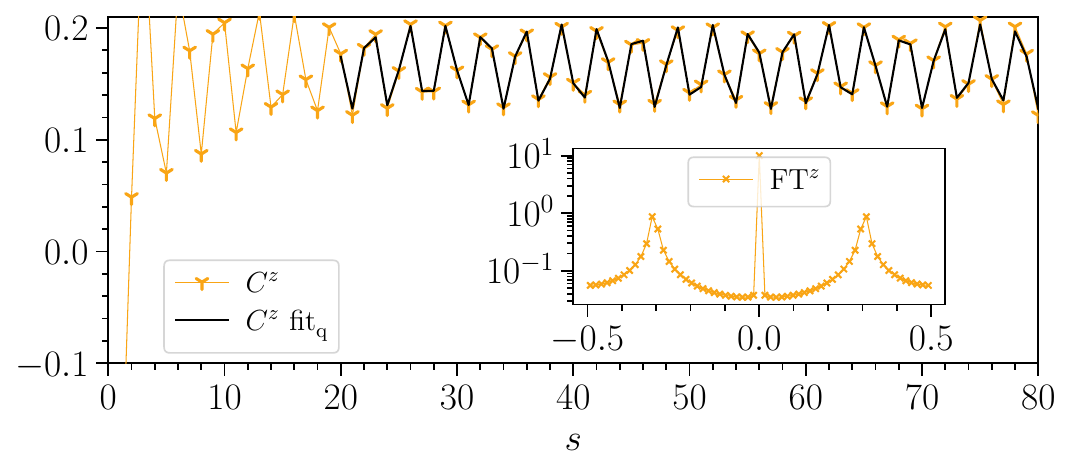} \label{suppfig:DMRG OBC corr diamond z a}}%
\subfigure[]{\includegraphics[width=0.5\columnwidth]{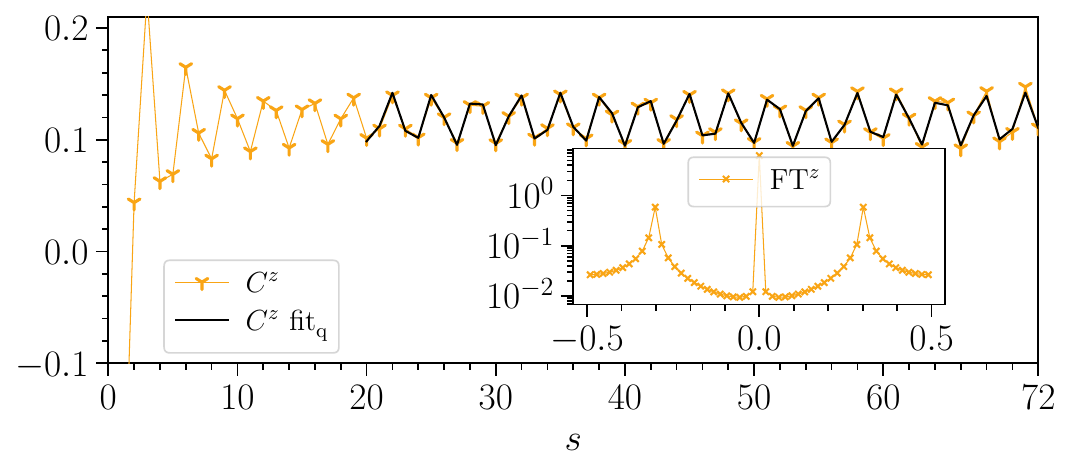} \label{suppfig:DMRG OBC corr diamond z b}}%
\caption{(Color online)
The case of OBC.
Dependence on $s$ given $j^{\,}_{0}$ for the
spin-spin correlation functions
$C^{\alpha}_{j^{\,}_{0},j^{\,}_{0}+s}[\Psi^{\,}_{0}]$
defined in Eq.~\eqref{suppeq:Calpha}
and computed using the DMRG approximation ($\chi=128$ and $2N=128$)
to the ground state
$|\Psi^{\,}_{0}\rangle$
of Hamiltonian \eqref{suppeq:def H}
at $(\theta,\lambda)=(\pi/4-0.1,0.4)$.
(a)
$C^{z}_{j^{\,}_{0},j^{\,}_{0}+s}[\Psi^{\,}_{0}]$
for
$j^{\,}_{0}=(2N-80)/2$.
(b)
$C^{z}_{j^{\,}_{0},j^{\,}_{0}+s}[\Psi^{\,}_{0}]$
for
$j^{\,}_{0}=(2N-72)/2$.
Fit used is ``q'' for cosine form.
In both cases,
$q^{z}\approx2\pi\,\frac{3}{10}$.
The best fit for $q^{z}$ 
is independent of the choice for $j^{\,}_{0}$.
The insets show the Fourier transforms of the respective
correlations, where the abscissa is the frequency
$f^{z}\equiv q^{z}/(2\pi)$
and the ordinate is the expansion coefficient $A(f^{z})$.
}
\label{suppfig:DMRG OBC corr diamond z}
\end{figure}

\begin{figure}[t!]
\centering
\subfigure{\relax \label{suppfig:DMRG OBC corr connected diamond a}}%
\subfigure{\relax \label{suppfig:DMRG OBC corr connected diamond b}}%
\subfigure{\includegraphics[width=0.5\columnwidth]{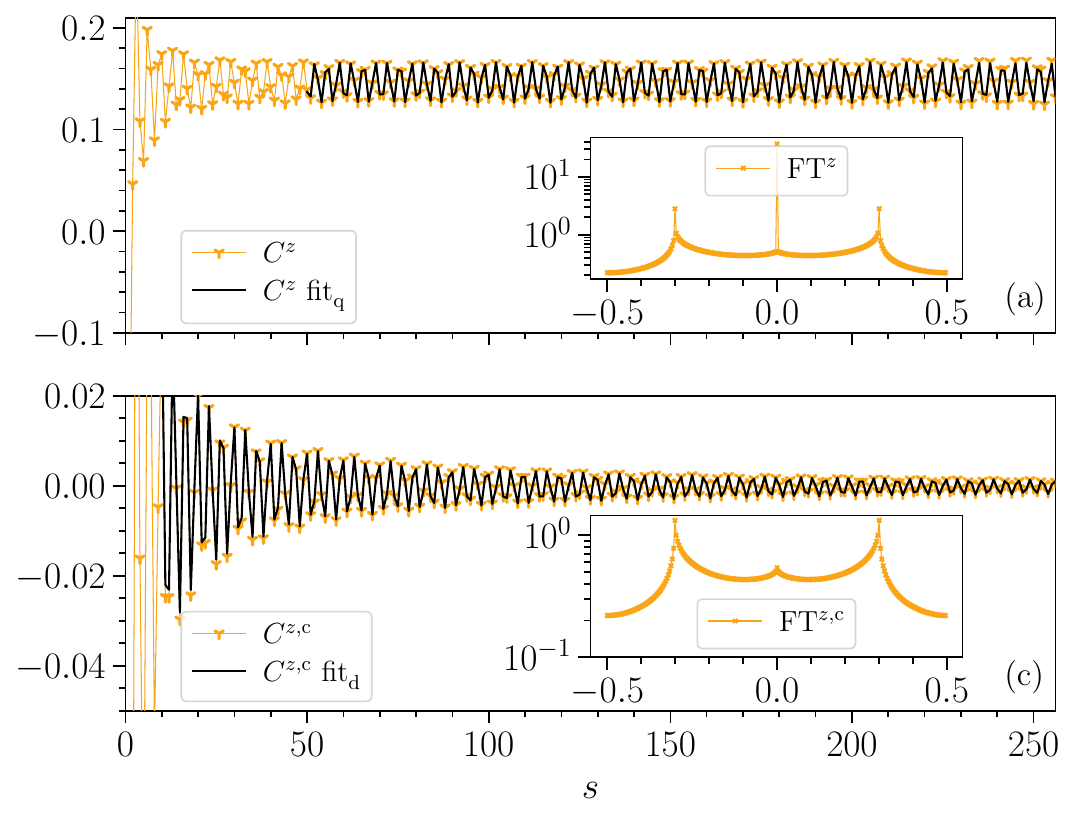} \label{suppfig:DMRG OBC corr connected diamond c}}%
\subfigure{\includegraphics[width=0.5\columnwidth]{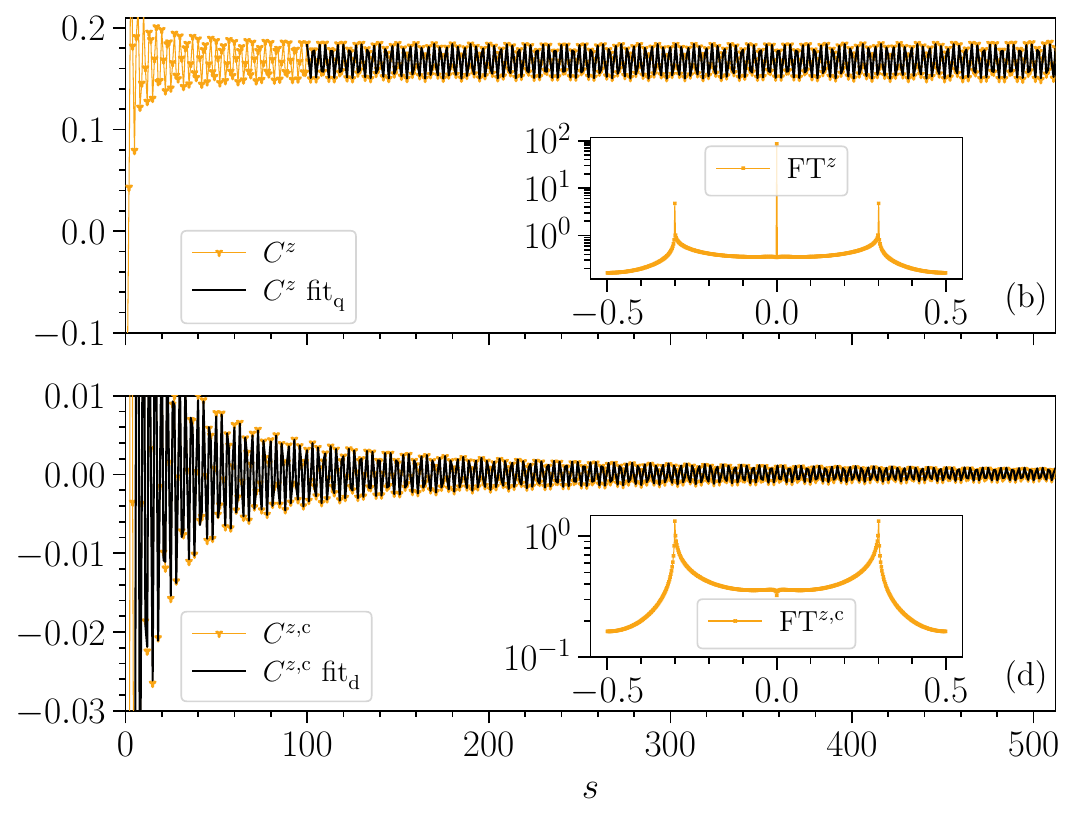} \label{suppfig:DMRG OBC corr connected diamond d}}
\caption{(Color online)
The case of OBC.
Dependence on the separation $s$
of the spin-spin correlation function
$C^{z}_{j^{\,}_{0},j^{\,}_{0}+s}[\Psi^{\,}_{0}]$
defined in Eq.~\eqref{suppeq:Calpha}
in the ground state $|\Psi^{\,}_{0}\rangle$
of Hamiltonian \eqref{suppeq:def H}
at the center $(\pi/4,1/2)$ 
of the $\mathrm{FM}^{\,}_{z}$ phase from Fig.~\ref{suppfig:phase diagram}.
The ground state is approximated by the DMRG ($\chi = 256$)
matrix product state, with
(a) $2N=512$ and
$j^{\,}_{0}=(2N-256)/2$ given and
(b) $2N=1024$ and
$j^{\,}_{0}=(2N-512)/2$ given.
Connected part
$C^{z,\mathrm{c}}_{j^{\,}_{0},j^{\,}_{0}+s}[\Psi^{\,}_{0}]$
defined in Eq.~\eqref{suppeq:CalphaConn} with
(a) $2N=512$ and
$j^{\,}_{0}=(2N-256)/2$ given and
(b) $2N=1024$ and
$j^{\,}_{0}=(2N-512)/2$ given.
Fits used are labeled ``q''
for cosine form and ``d'' for an algebraically damped cosine.
In all cases, $q^{z}\simeq2\pi\,\frac{3}{10}$.
The insets show the Fourier transforms of the respective correlations,
where the abscissa is the frequency $f^{z}\equiv q^{z}/(2\pi)$
and the ordinate is the expansion coefficient $A(f^{z})$.
}
\label{suppfig:DMRG OBC corr connected diamond}
\end{figure}

\begin{figure}[t!]
\subfigure[]{\includegraphics[width=0.5\columnwidth]{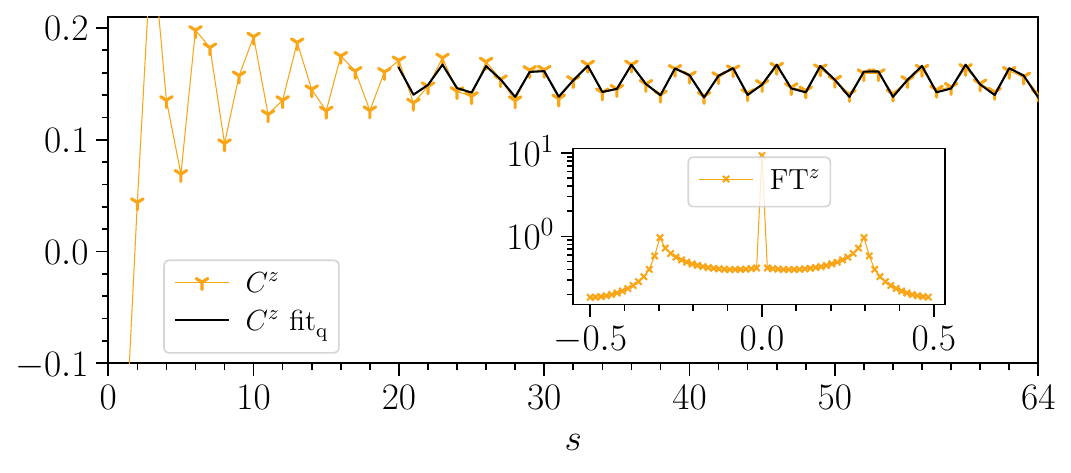} \label{suppfig:DMRG PBC corr diamond z a}}%
\subfigure[]{\includegraphics[width=0.5\columnwidth]{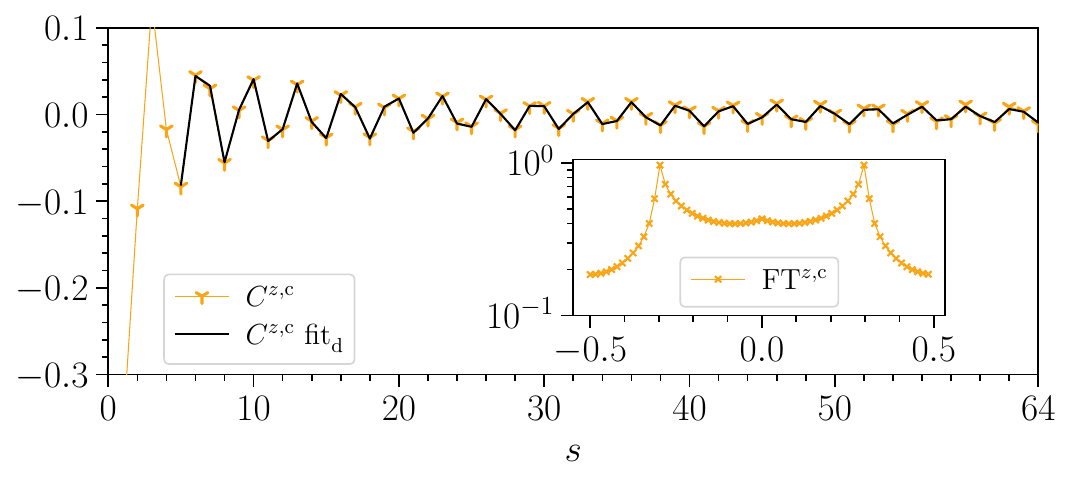} \label{suppfig:DMRG PBC corr diamond z b}}
\caption{
(Color online)
The case of PBC.
(a)
Dependence on the separation $s$
of the spin-spin correlation function
$C^{z}_{j^{\,}_{0},j^{\,}_{0}+s}[\Psi^{\,}_{0}]$
defined in Eq.~\eqref{suppeq:Calpha}
and computed using the DMRG ($\chi=512$ and $2N=128$) approximation 
to the ground state
$|\Psi^{\,}_{0}\rangle$
of Hamiltonian \eqref{suppeq:def H}
at $(\theta,\lambda)=(\pi/4,1/2)$.
(b)
Connected part 
$C^{z,\mathrm{c}}_{j^{\,}_{0},j^{\,}_{0}+s}[\Psi^{\,}_{0}]$
of
$C^{z}_{j^{\,}_{0},j^{\,}_{0}+s}[\Psi^{\,}_{0}]$
for a DMRG with $\chi=512$ and $2N=128$.
Fits used are labeled
``q'' for cosine form and 
``d'' for an algebraically damped cosine form. In both cases,
$q^{z}\simeq2\pi\,\frac{3}{10}$.
The insets show the Fourier transforms of the respective
correlations, where the abscissa is the frequency $f^{z}\equiv q^{z}/(2\pi)$
and the ordinate is the expansion coefficient $A(f^{z})$.
}
\label{suppfig:DMRG PBC corr diamond z}
\end{figure}

The oscillatory behavior observed in
Fig.~\ref{suppfig:DMRG OBC FMZ}
is a signature of both the incommensurability present in the
$\mathrm{FM}^{\,}_{z}$ phase in Fig.~\ref{suppfig:phase diagram},
and criticality as 
$m^{z}_{\mathrm{uni}}[\Psi^{\,}_{0}]$
and
$m^{z}_{\mathrm{uni,SPT}}[\Psi^{\,}_{0}]$
are observed to be sensitive to the choice between PBC and OBC
in the thermodynamic limit.
We now analyze the two-point correlation functions to establish
quantitatively the incommensuration of the $\mathrm{FM}^{\,}_{z}$ phase,
illustrated in Figs.~\ref{suppfig:DMRG OBC corr diamond z}
and~\ref{suppfig:DMRG OBC corr connected diamond} for OBC,
and in Fig.~\ref{suppfig:DMRG PBC corr diamond z} for PBC.

The longitudinal component
$C^{z}_{j^{\,}_{0},j^{\,}_{0}+s}[\Psi^{\,}_{0}]$
of the spin-spin correlation function
in the DMRG approximation to the ground state
$|\Psi^{\,}_{0}\rangle$
of Hamiltonian \eqref{suppeq:def H}
defined in Eq.~\eqref{suppeq:Calpha} 
can be fitted by the periodic Ansatz
\begin{equation}
C^{z}_{j^{\,}_{0},j^{\,}_{0}+s}[\Psi^{\,}_{0}]=
\Gamma^{z}_{j^{\,}_{0}}[\Psi^{\,}_{0}]
+
A^{z}_{j^{\,}_{0}}[\Psi^{\,}_{0}]\,
\cos
\big(
q^{z}[\Psi^{\,}_{0}]\,s
+
\phi^{z}_{j^{\,}_{0}}[\Psi^{\,}_{0}]
\big),
\qquad
1\ll j^{\,}_{0},s,j^{\,}_{0}+s\ll2N,
\label{suppeq:algebraic decay longitudinal spin-spin in FMZ phase}
\end{equation}
with the non-vanishing constant
$\Gamma^{z}_{j^{\,}_{0}}[\Psi^{\,}_{0}]$,
the amplitude $A^{z}_{j^{\,}_{0}}[\Psi^{\,}_{0}]$,
the wave number $q^{z}[\Psi^{\,}_{0}]$
and the phase $\phi^{z}_{j^{\,}_{0}}[\Psi^{\,}_{0}]$
real valued.
Under PBC, on the one hand, translation symmetry dictates that
$\Gamma^{z}_{j^{\,}_{0}}[\Psi^{\,}_{0}]$,
$A^{z}_{j^{\,}_{0}}[\Psi^{\,}_{0}]$,
$q^{z}[\Psi^{\,}_{0}]$
and $\phi^{z}_{j^{\,}_{0}}[\Psi^{\,}_{0}]$
are independent of $j^{\,}_{0}$
unless translation symmetry is spontaneously broken.
We observe that there is no dependence on $j^{\,}_{0}$
for any of the terms on the right-hand side of
Eq.~(\ref{suppeq:algebraic decay longitudinal spin-spin in FMZ phase})
under PBC. Under OBC, on the other hand, we have anticipated that
$q^{z}[\Psi^{\,}_{0}]$
is the only quantity 
on the right-hand side of
Eq.~(\ref{suppeq:algebraic decay longitudinal spin-spin in FMZ phase})
that remains independent of $j^{\,}_{0}$ in the thermodynamic limit.
This educated guess is confirmed by our numerics.

The best fit to the DMRG data selects the Ansatz
(\ref{suppeq:algebraic decay longitudinal spin-spin in FMZ phase})
for the longitudinal component of the spin-spin correlation functions
at $(\theta,\lambda)=(\pi/4-0.1,0.4)$,
whereby we read from Fig.~\ref{suppfig:DMRG OBC corr diamond z a}
\begin{subequations}
\begin{equation}
\Gamma^{z}_{j^{\,}_{0}}[\Psi^{\,}_{0}]=0.1653\pm0.0003,
\qquad
q^{z}[\Psi^{\,}_{0}]=
2\pi
\left(
\frac{3}{10}+0.005\pm9\times 10^{-5}
\right),
\end{equation}
for $j^{\,}_{0}=(128-80)/2$ ,
while we read from Fig.~\ref{suppfig:DMRG OBC corr diamond z b}
\begin{equation}
\Gamma^{z}_{j^{\,}_{0}}[\Psi^{\,}_{0}]=0.1183\pm0.0002,
\qquad
q^{z}[\Psi^{\,}_{0}]=
2\pi
\left(
\frac{3}{10}+0.006\pm14\times 10^{-5}
\right),
\end{equation}
\end{subequations}  
for $j^{\,}_{0}=(128-72)/2$.
We do not quote the value taken by $A^{z}_{j^{\,}_{0}}[\Psi^{\,}_{0}]$
in Eq.~\eqref{suppeq:algebraic decay longitudinal spin-spin in FMZ phase}
as it depends both on the choice of boundary
conditions and on the choice of $j^{\,}_{0}$.
We notice that the value taken by
$\Gamma^{z}_{j^{\,}_{0}}[\Psi^{\,}_{0}]$
for the choice we made of $j^{\,}_{0}$
does not equal the square of the uniform magnetization
$m^{z}_{\mathrm{uni}}[\Psi^{\,}_{0}]$
in Eq.~(\ref{suppeq:muni}), that we found
in Fig.~\ref{suppfig:DMRG PBC FMZ}
to be  approximately $0.4^{2}=0.16$.
This is a consequence of the observation that
the dependence on $j^{\,}_{0}$ of
$\Gamma^{z}_{j^{\,}_{0}}[\Psi^{\,}_{0}]$ 
also displays a periodicity [of approximately 10 lattice spacings
at $(\pi/4,1/2)$]. Upon averaging $\Gamma^{z}_{j^{\,}_{0}}[\Psi^{\,}_{0}]$
over $j^{\,}_{0}$ to account for this periodicity,
we recover the anticipated relation according to
which this average is the square of $m^{z}_{\mathrm{uni}}[\Psi^{\,}_{0}]$.

The insets in each of the panels in Fig.~\ref{suppfig:DMRG OBC corr diamond z}
represent the discrete Fourier transforms of the
correlations.  We establish with confidence that the oscillations in
$C^{z}_{j^{\,}_{0},j^{\,}_{0}+s}[\Psi^{\,}_{0}]$ are modulated by a
wave-vector $q^{z} \simeq 2\pi \times 3/10$.  However, the discrete
nature of the Fourier transforms limits the precision of $q^{z}$ due
to the discretization of the wave-vector interval $[- \pi, \pi]$ into
$n$ values, where $n$ is the size of the range of separations $s$ used
for the fit.  For example, $n = 60$ in Fig.~\ref{suppfig:DMRG OBC corr diamond z a}.
To mitigate this limitation, larger system sizes are
analyzed in Fig.~\ref{suppfig:DMRG OBC corr connected diamond}.

In Fig.~\ref{suppfig:DMRG OBC corr connected diamond},
we show the correlations
$C^{z}_{j^{\,}_{0},j^{\,}_{0}+s}[\Psi^{\,}_{0}]$
and
$C^{z,\mathrm{c}}_{j^{\,}_{0},j^{\,}_{0}+s}[\Psi^{\,}_{0}]$, defined
in \eqref{suppeq:Calpha} and \eqref{suppeq:CalphaConn}, respectively,
for larger system sizes $2N=512$ and $2N=1024$, at the center
$(\pi/4,1/2)$ of the $\mathrm{FM}^{\,}_{z}$ phase.
This allows us to obtain
\begin{align}
\frac{q^{z}[\Psi^{\,}_{0}]}{2\pi}=
\begin{cases}
0.30116 \pm 1\cdot10^{-5},&\hbox{ for $2N=512$},
\\
0.300804 \pm 3\cdot10^{-6},&\hbox{ for $2N=1024$},
\end{cases}
\label{suppeq:qz for Cz 2N=512,1024}
\end{align}
as shown in Figs.~\ref{suppfig:DMRG OBC corr connected diamond a} 
and~\ref{suppfig:DMRG OBC corr connected diamond b}, respectively, using
the Ansatz
\eqref{suppeq:algebraic decay longitudinal spin-spin in FMZ phase}
for the two-point correlation function
$C^{z}_{j^{\,}_{0},j^{\,}_{0}+s}[\Psi^{\,}_{0}]$.

To decide if the oscillations in
$C^{z}_{j^{\,}_{0},j^{\,}_{0}+s}[\Psi^{\,}_{0}]$ originate from the
oscillations in $\ev*{\widehat{Z}^{\,}_{j}}^{\,}_{\Psi^{\,}_{0}}$,
we compute the second cumulant of the operator $\widehat{Z}^{\,}_{j}$,
i.e., the connected correlations
$C^{z,\mathrm{c}}_{j^{\,}_{0},j^{\,}_{0}+s}[\Psi^{\,}_{0}]$ defined in
\eqref{suppeq:CalphaConn}.
Figures~\ref{suppfig:DMRG OBC corr connected diamond c}
and~\ref{suppfig:DMRG OBC corr connected diamond d} show
$C^{z,\mathrm{c}}_{j^{\,}_{0},j^{\,}_{0}+s}[\Psi^{\,}_{0}]$ for
$2N=512$ and $2N=1024$, respectively.

In addition to the oscillations that persist, we observe a decay as a
function of $s$. The fit$^{\,}_{\mathrm{d}}$ that we use is given by
\begin{equation}
C^{z,\mathrm{c}}_{j^{\,}_{0},j^{\,}_{0}+s}[\Psi^{\,}_{0}]=
\Gamma^{z,\mathrm{c}}[\Psi^{\,}_{0}]
+
A^{z,\mathrm{c}}_{j^{\,}_{0}}[\Psi^{\,}_{0}]\,
s^{-\eta^{z,\mathrm{c}}[\Psi^{\,}_{0}]}\,
\cos\big(q^{z,\mathrm{c}}[\Psi^{\,}_{0}]\,s
+\phi^{z,\mathrm{c}}_{j^{\,}_{0}}[\Psi^{\,}_{0}]\big),
\qquad
1\ll j^{\,}_{0},s,j^{\,}_{0}+s\ll2N,
\label{suppeq:fit q connected}
\end{equation}
where only $A^{z,\mathrm{c}}_{j^{\,}_{0}}[\Psi^{\,}_{0}]$ and
$\phi^{z,\mathrm{c}}_{j^{\,}_{0}}[\Psi^{\,}_{0}]$ depend on $j^{\,}_{0}$.
We find the exponent of the algebraic decay to take the values
\begin{equation}
\eta^{z,\mathrm{c}}[\Psi^{\,}_{0}]=
\begin{cases}
1.026 \pm 0.006,&\hbox{ for $2N = 512$},
\\
1.002 \pm 0.002,&\hbox{ for $2N = 1024$}, 
\end{cases}
\end{equation}
at the center $(\pi/4,1/2)$ of the $\mathrm{FM}^{\,}_{z}$ phase.
The wave-vector $q^{z,\mathrm{c}}[\Psi^{\,}_{0}]$ obtained for
$C^{z,\mathrm{c}}_{j^{\,}_{0},j^{\,}_{0}+s}[\Psi^{\,}_{0}]$
lies within the uncertainty interval of the wave-vector
$q^{z}[\Psi^{\,}_{0}]$ obtained for
$C^{z}_{j^{\,}_{0},j^{\,}_{0}+s}[\Psi^{\,}_{0}]$
in Eq.~\eqref{suppeq:qz for Cz 2N=512,1024}.
We thus claim that $C^{z,\mathrm{c}}_{j^{\,}_{0},j^{\,}_{0}+s}[\Psi^{\,}_{0}]$
is an algebraically decaying function of the form
$s^{-\eta^{z,\mathrm{c}}[\Psi^{\,}_{0}]}$
with exponent $\eta^{z,\mathrm{c}}[\Psi^{\,}_{0}]=1$,
modulated by oscillations as a function of $s$
with wave-vector $q^{z,\mathrm{c}}[\Psi^{\,}_{0}]=2\pi\times3/10$,
i.e.,
\begin{equation}
q^{z}[\Psi^{\,}_{0}]=q^{z,\mathrm{c}}[\Psi^{\,}_{0}].
\end{equation}

We now revisit the case of PBC. Within our numerical accuracy,
$\ev*{\widehat{Z}^{\,}_{j}}^{\,}_{\Psi^{\,}_{0}}$ is independent of $j$.
However, an oscillatory dependence on $s$ of the two-point function
$C^{z}_{j^{\,}_{0},j^{\,}_{0}+s}[\Psi^{\,}_{0}]$
along the $Z$-axis in spin space 
is present as is illustrated in Fig.~\ref{suppfig:DMRG PBC corr diamond z a}
at the center $(\pi/4,1/2)$ in Fig.~\ref{suppfig:phase diagram}.
The fit of the form
(\ref{suppeq:oscillation ansatz in FMZ phase a})
applied to $\ev*{\widehat{Z}^{\,}_{j}}^{\,}_{\Psi^{\,}_{0}}$ under PBC
confirms that the wave-vector of the oscillations
is approximately equal to that observed 
when OBC are chosen, i.e.,
$q^{z}_{\mathrm{loc}}[\Psi^{\,}_{0}]\simeq2\pi\times3/10$.
The insets showing the Fourier transforms are consistent 
with $q^{z}_{\mathrm{loc}}[\Psi^{\,}_{0}]\simeq2\pi\times3/10$,
together with a non-vanishing and non-decaying amplitude of oscillations.

We also analyze under PBC the connected correlations 
$C^{z,\mathrm{c}}_{j^{\,}_{0},j^{\,}_{0}+s}[\Psi^{\,}_{0}]$ 
defined in Eq.~\eqref{suppeq:CalphaConn}
by applying the fit of the form \eqref{suppeq:fit q connected}, 
obtaining $\eta^{z,\mathrm{c}}[\Psi^{\,}_{0}]=1.08\pm0.02$ 
[see Fig.~\ref{suppfig:DMRG PBC corr diamond z b}]
at the center $(\pi/4,1/2)$ in Fig.~\ref{suppfig:phase diagram}.
This result is consistent, albeit less accurate, 
compared to the one obtained using OBC.
These findings further confirm the incommensuration of the 
spin-spin correlations along the $Z$-axis in spin space of
the $\mathrm{FM}^{\,}_{z}$ phase.

\begin{figure}[t!]
\centering
\subfigure[]{\includegraphics[width=0.245\columnwidth]{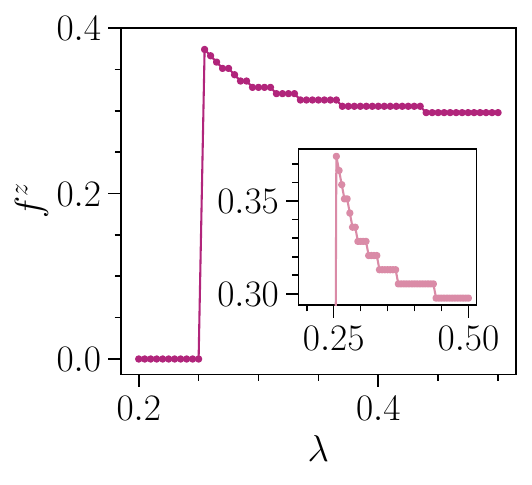} \label{suppfig:DMRG OBC corr ft diamond a}}%
\subfigure[]{\includegraphics[width=0.245\columnwidth]{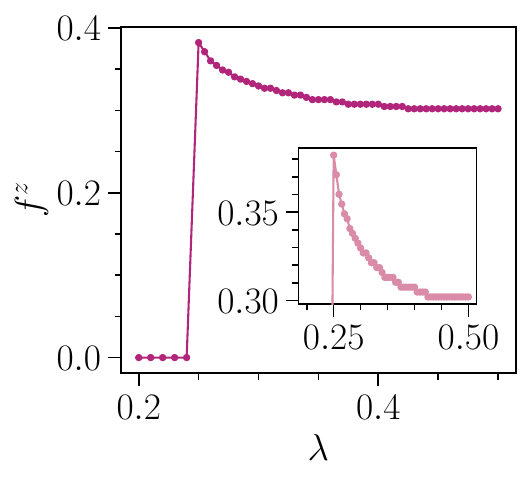} \label{suppfig:DMRG OBC corr ft diamond b}}%
\subfigure[]{\includegraphics[width=0.245\columnwidth]{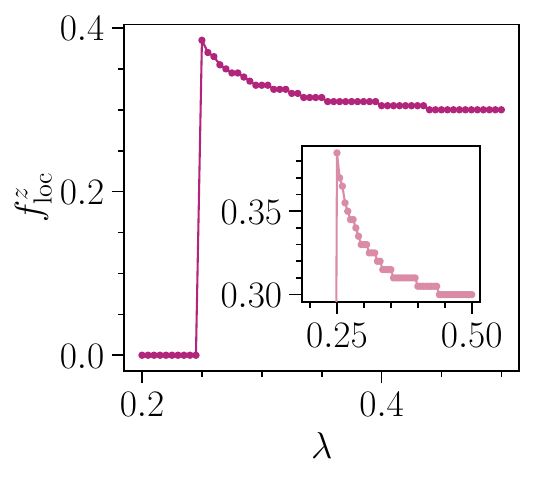} \label{suppfig:DMRG OBC corr ft diamond c}}%
\subfigure[]{\includegraphics[width=0.245\columnwidth]{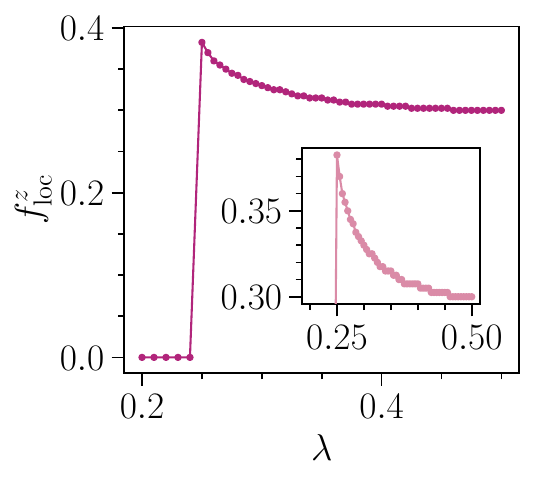} \label{suppfig:DMRG OBC corr ft diamond d}}
\caption{(Color online)
The case of OBC.
Dependence on $\lambda$ for the given value of $\theta=\pi/4$
of the frequency $f^{z}$ of the Fourier transform of
(a) the spin-spin correlations
$C^{z}_{j^{\,}_{0},j^{\,}_{0}+s}[\Psi^{\,}_{0}]$
for $2N=256$,
(b) same as for (a) except for $2N=512$,
(c) the local spin expectation values
$\bracket{\widehat{Z}^{\,}_{j}}^{\,}_{\Psi^{\,}_{0}}$
for $2N=256$,
and
(d) same as for (c) except for $2N=512$,
in the bulk obtained by approximating the ground state
$|\Psi^{\,}_{0}\rangle$ 
with the DMRG ($\chi=128$)
matrix product state.
The insets are zooms within the window
for which $f^{z}$ is non-vanishing.
}
\label{suppfig:DMRG OBC corr ft diamond}
\end{figure}

The insets in Figs.~\ref{suppfig:DMRG OBC corr diamond z} 
and~\ref{suppfig:DMRG OBC corr connected diamond}
display the Fourier transforms with respect to the $s$ dependence
of the spin-spin correlations
$C^{z}_{j^{\,}_{0},j^{\,}_{o}+s}[\Psi^{\,}_{0}]$
under OBC
defined in Eq.~\eqref{suppeq:Calpha}.
For each  $\lambda$ on the U(1)-symmetric line $\theta=\pi/4$,
we extract the non-vanishing frequency
\begin{equation}
f^{z}[\Psi^{\,}_{0}]\equiv
q^{z}[\Psi^{\,}_{0}]/(2\pi),
\end{equation}
whose Fourier amplitude is the largest among non-vanishing frequencies.
(The vanishing frequency corresponding to the long-range order
is always tied to a Fourier amplitude that is a local extremum.)
We report the values of
$f^{z}[\Psi^{\,}_{0}]$
as a function of $\lambda$
in Figs.~\ref{suppfig:DMRG OBC corr ft diamond a}
and~\ref{suppfig:DMRG OBC corr ft diamond b}.

We also compute the Fourier transforms
with respect to the $j$ dependence
of the spin expectation values
$\ev*{\widehat{Z}^{\,}_{j}}^{\,}_{\Psi^{\,}_{0}}$ under OBC.
For each  $\lambda$ on the U(1)-symmetric line $\theta=\pi/4$,
we extract the non-vanishing frequency
\begin{equation}
f^{z}_{\mathrm{loc}}[\Psi^{\,}_{0}]\equiv
q^{z}_{\mathrm{loc}}[\Psi^{\,}_{0}]/(2\pi),
\end{equation}
whose Fourier amplitude is the largest among non-vanishing frequencies.
(The vanishing frequency corresponding to the long-range order
is always tied to a Fourier amplitude that is a local extremum.)
We report the values of
$f^{z}_{\mathrm{loc}}[\Psi^{\,}_{0}]$
as a function of $\lambda$
in Figs.~\ref{suppfig:DMRG OBC corr ft diamond c}
and~\ref{suppfig:DMRG OBC corr ft diamond d}.

We observe from Figs.~\ref{suppfig:DMRG OBC corr ft diamond a}
and~\ref{suppfig:DMRG OBC corr ft diamond b}
that
$q^{z}[\Psi^{\,}_{0}]\equiv2\pi f^{z}[\Psi^{\,}_{0}]$
as a function of $\lambda$ is
(i) an increasing function of $\lambda$
upon approaching the boundary from the inside of the $\mathrm{FM}^{\,}_{z}$ phase, and
(ii) a vanishing function of $\lambda$
upon approaching the boundary from the outside of the $\mathrm{FM}^{\,}_{z}$ phase.
Due to the discrete nature of the Fourier transform, 
$q^{z}[\Psi^{\,}_{0}]$
does not visually appear to be a smooth function of $\lambda$ 
inside the $\mathrm{FM}^{\,}_{z}$ phase.
However, our numerics suggest that
$q^{z}[\Psi^{\,}_{0}]$ is inherently smooth 
as a function of $\lambda$ taking values
corresponding to the inside of the $\mathrm{FM}^{\,}_{z}$ phase, 
decreasing to the minimum
value $q^{z}[\Psi^{\,}_{0}]\simeq2\pi\times 3/10$
at the center $(\pi/4,1/2)$ in Fig.~\ref{suppfig:phase diagram}.
The continuous variation of $q^{z}[\Psi^{\,}_{0}]$
as a function of $\lambda$ taking values
corresponding to the inside of the $\mathrm{FM}^{\,}_{z}$ phase
confirms the existence of incommensuration 
of the spin-spin correlations along the $Z$-axis in spin space.
We also observe by comparing
Figs.~\ref{suppfig:DMRG OBC corr ft diamond a}
and~\ref{suppfig:DMRG OBC corr ft diamond b}
with
Figs.~\ref{suppfig:DMRG OBC corr ft diamond c}
and~\ref{suppfig:DMRG OBC corr ft diamond d}
that
\begin{equation}
q^{z}[\Psi^{\,}_{0}]=
q^{z}_{\mathrm{loc}}[\Psi^{\,}_{0}]
\label{suppeq:equality between fz and fzloc in FMz phase for theta=pi/4}
\end{equation}
as a function of $\lambda$ at $\theta=\pi/4$
(the equality holds up to the second digit with our DMRG
estimates of
$q^{z}[\Psi^{\,}_{0}]$
and
$q^{z}_{\mathrm{loc}}[\Psi^{\,}_{0}]$).

It is an empirical observation based on
the values quoted in Eq.\
(\ref{suppeq:oscillation ansatz in FMZ phase})
that Eq.\
(\ref{suppeq:equality between fz and fzloc in FMz phase for theta=pi/4})
is not limited to the vertical cut $\theta=\pi/4$ in the FM${}_{z}$ phase,
but applies throughout the FM${}_{z}$ phase, i.e.,
$q^{z}[\Psi^{\,}_{0}]$
and
$q^{z}_{\mathrm{loc}}[\Psi^{\,}_{0}]$
only depend on $\lambda$ (not on $\theta$).
The same is true of the absolute value of the uniform
magnetization $m^{z}_{\mathrm{uni}}[\Psi^{\,}_{0}]$ 
defined in Eq.~\eqref{suppeq:muni}.
It is a function of $\lambda$ only,
as is implied by the values quoted in Eq.\
(\ref{suppeq:mean values mzuni and mzuniSPT OPB}).

\begin{figure}[t!]
\subfigure[]{\includegraphics[width=0.31\columnwidth]{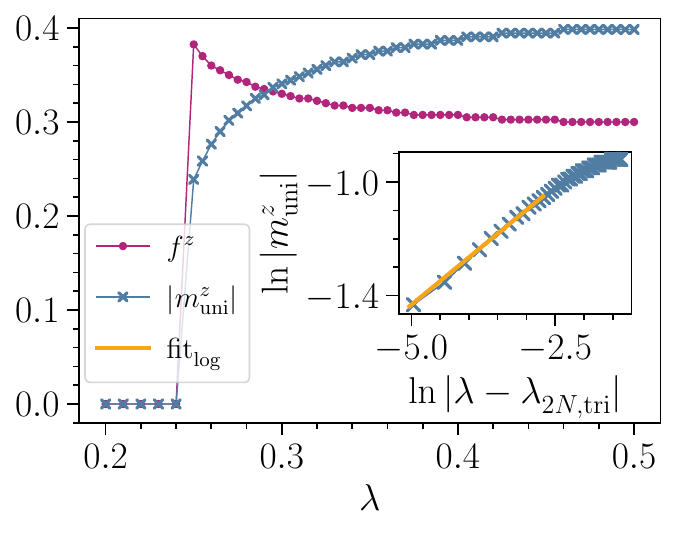} \label{suppfig:DMRG OBC incommensuration vs magnetization a}}%
\subfigure[]{\includegraphics[width=0.251\columnwidth]{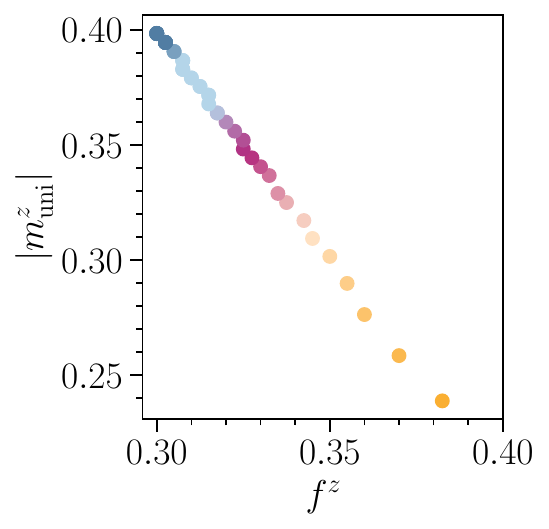} \label{suppfig:DMRG OBC incommensuration vs magnetization b}}%
\hspace{-5pt}\subfigure[]{\includegraphics[width=0.186\columnwidth]{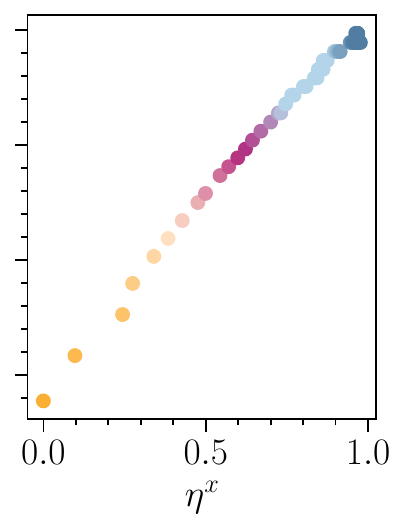} \label{suppfig:DMRG OBC incommensuration vs magnetization c}}%
\hspace{-5pt}\subfigure[]{\includegraphics[width=0.245\columnwidth]{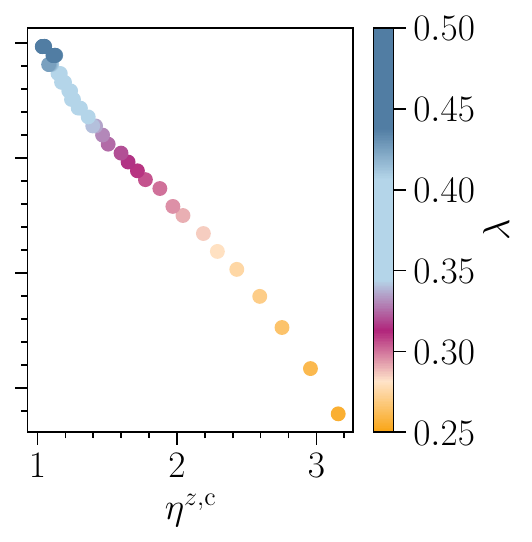} \label{suppfig:DMRG OBC incommensuration vs magnetization d}}%
\caption{(Color online)
The case of OBC.
(a) Dependence on $\lambda$ of the frequency $f^{z}$ of the Fourier transform 
of the spin-spin correlations $C^{z}_{j^{\,}_{0},j^{\,}_{0}+s}[\Psi^{\,}_{0}]$
defined in  Eq.~\eqref{suppeq:Calpha}
and of the absolute value of the uniform
magnetization $m^{z}_{\mathrm{uni}}[\Psi^{\,}_{0}]$ defined in
Eq.\ \eqref{suppeq:muni} in the bulk
of a chain with $2N=512$ sites obtained using DMRG ($\chi=128$) 
with $j^{\,}_{0}=(2N-512)/2$ at $\theta=\pi/4$. 
The inset shows the logarithm of the
nonvanishing values of $m^{z}_{\mathrm{uni}}[\Psi^{\,}_{0}]$
for $\lambda>\lambda^{\,}_{\mathrm{2N,tri}}$, where $\lambda^{\,}_{\mathrm{2N,tri}}$
is obtained through the $\mathrm{fit}^{\,}_{\mathrm{log}}$ Ansatz
\eqref{suppeq:fitlog}.
(b) The dependence of $m^{z}_{\mathrm{uni}}[\Psi^{\,}_{0}]$ on
the frequency $f^{z}[\Psi^{\,}_{0}]$ of the Fourier transform 
of the spin-spin correlations $C^{z}_{j^{\,}_{0},j^{\,}_{0}+s}[\Psi^{\,}_{0}]$,
whereby both quantities are computed with DMRG
[$2N=512$, $\chi=128$, $j^{\,}_{0}=(2N-512)/2$]
along the vertical line $\theta=\pi/4$ in the $\mathrm{FM}^{\,}_{z}$ phase.
(c) The dependence of $m^{z}_{\mathrm{uni}}[\Psi^{\,}_{0}]$ on
the exponent $\eta^{x}[\Psi^{\,}_{0}]$
defined in Eq.\
\eqref{suppeq:algebraic decay transverse spin-spin in FMZ phase}
of the algebraic decay of the 
spin-spin correlations $C^{x}_{j^{\,}_{0},j^{\,}_{0}+s}[\Psi^{\,}_{0}]$,
whereby both quantities are computed with DMRG
[$2N=512$, $\chi=128$, $j^{\,}_{0}=(2N-512)/2$]
along the vertical line $\theta=\pi/4$ in the $\mathrm{FM}^{\,}_{z}$ phase.
(d) The dependence of $m^{z}_{\mathrm{uni}}[\Psi^{\,}_{0}]$ on
the exponent $\eta^{z,\mathrm{c}}[\Psi^{\,}_{0}]$
defined in Eq.\ \eqref{suppeq:fit q connected}
of the algebraic decay of the 
spin-spin correlations $C^{z,\mathrm{c}}_{j^{\,}_{0},j^{\,}_{0}+s}[\Psi^{\,}_{0}]$,
whereby both quantities are computed with DMRG
[$2N=512$, $\chi=128$, $j^{\,}_{0}=(2N-512)/2$]
along the vertical line $\theta=\pi/4$ in the $\mathrm{FM}^{\,}_{z}$ phase.
The different values of $\lambda$
along the line $\theta=\pi/4$ in the $\mathrm{FM}^{\,}_{z}$ phase
are specified by different colors.
}
\label{suppfig:DMRG OBC incommensuration vs magnetization}
\end{figure}

Moreover, along the vertical line $\theta=\pi/4$ of the
$\mathrm{FM}^{\,}_{z}$ phase, we deduce from Fig.\
\ref{suppfig:DMRG OBC incommensuration vs magnetization}
that $|m^{z}_{\mathrm{uni}}[\Psi^{\,}_{0}]|$ is
approximately proportional
to two scaling exponents and three wave vectors, namely
\begin{equation}
|m^{z}_{\mathrm{uni}}[\Psi^{\,}_{0}]|\propto
\eta^{x}[\Psi^{\,}_{0}]\propto
\eta^{z,\mathrm{c}}[\Psi^{\,}_{0}]\propto
q^{z}[\Psi^{\,}_{0}]=
q^{z,\mathrm{c}}[\Psi^{\,}_{0}]=
q^{z}_{\mathrm{loc}}[\Psi^{\,}_{0}],
\qquad
0\leq\lambda\leq0.5.
\label{suppeq:propto mz}
\end{equation}
Relation (\ref{suppeq:propto mz})
is reminiscent of the critical phase called the
Tomonaga-Luttinger-Liquid phase with central charge one (TLL1) 
in the study done in Ref.\ \cite{Hikihara10}
of the quantum spin-1/2 chain
with nearest- and next-nearest-neighbor Heisenberg exchange couplings
in a uniform magnetic field.

The dependence on $\lambda$ of
$|m^{z}_{\mathrm{uni}}[\Psi^{\,}_{0}]|$
along the vertical cut $(\pi/4,\lambda)$
shown in Fig.~\ref{suppfig:DMRG OBC incommensuration vs magnetization a}
can be fitted with the $\mathrm{fit}^{\,}_{\mathrm{log}}$ Ansatz
\begin{subequations}
\begin{equation}
m^{z}_{\mathrm{fit}^{\,}_{\mathrm{log}}}(\lambda)\equiv
m^{z}_{2N,\mathrm{tri}}\,  
\Theta(\lambda-\lambda^{\,}_{2N,\mathrm{tri}})\,
(\lambda-\lambda^{\,}_{2N,\mathrm{tri}})^{\beta^{\,}_{\mathrm{tri}}},
\label{suppeq:fitlog}
\end{equation}
for $\lambda$ in the vicinity of $\lambda^{\,}_{2N,\mathrm{tri}}$,
where the fitting parameters are
$m^{z}_{2N,\mathrm{tri}}>0$,
$\lambda^{\,}_{2N,\mathrm{tri}}>0$,
and the scaling exponent
$\beta^{\,}_{\mathrm{tri}}>0$.
We have found the values
\begin{equation}
m^{z}_{2N,\mathrm{tri}}=0.550\pm 0.008,
\qquad
\lambda^{\,}_{2N,\mathrm{tri}}=0.2434\pm0.0006,
\qquad
\beta^{\,}_{\mathrm{tri}}=1/6\pm0.005.
\end{equation}
\end{subequations}

From Figs.\
\ref{suppfig:DMRG OBC incommensuration vs magnetization b},
\ref{suppfig:DMRG OBC incommensuration vs magnetization c},
and
\ref{suppfig:DMRG OBC incommensuration vs magnetization d},
we refine the relation \eqref{suppeq:propto mz}
to
\begin{equation}
|m^{z}_{\mathrm{uni}}[\Psi^{\,}_{0}]|=
\begin{cases}
-(1.98\pm0.02)\,f^{z}[\Psi^{\,}_{0}]
+(0.99\pm0.01),\\
\\
+(0.158\pm0.002)\,\eta^{x}[\Psi^{\,}_{0}]
+(0.247\pm0.001),\\
\\
-(0.104\pm0.005)\,\eta^{z,\mathrm{c}}[\Psi^{\,}_{0}]
+(0.505\pm0.006),
\end{cases}
\label{suppeq:equation mz}
\end{equation}
respectively.  Consequently, we conjecture the relation
\begin{equation}
q^{z}[\Psi^{\,}_{0}]\equiv
2\pi\,f^{z}[\Psi^{\,}_{0}]=
\pi\,\big(1-|m^{z}_{\mathrm{uni}}[\Psi^{\,}_{0}]|\big),
\label{suppeq:q against mz}
\end{equation}
which is the only one in Eq.\
\eqref{suppeq:equation mz} that
is not subject to systematic errors upon the choice
of the fitting range of the correlations
$C^{x}_{j^{\,}_{0},j^{\,}_{0}+s}[\Psi^{\,}_{0}]$
and
$C^{z,\mathrm{c}}_{j^{\,}_{0},j^{\,}_{0}+s}[\Psi^{\,}_{0}]$
in Figs.\
\ref{suppfig:DMRG OBC incommensuration vs magnetization c}
and
\ref{suppfig:DMRG OBC incommensuration vs magnetization d},
respectively. Hereto, we conjecture the following inverse relations
\begin{subequations}
\label{suppeq:etas against mz}
\begin{align}
&\eta^{x}[\Psi^{\,}_{0}]=
\frac{5}{3}\,\big(4\,|m^{z}_{\mathrm{uni}}[\Psi^{\,}_{0}]|-1\big),\\
&\eta^{z,\mathrm{c}}[\Psi^{\,}_{0}]=
5\,\big(1-2\,|m^{z}_{\mathrm{uni}}[\Psi^{\,}_{0}]|\big),
\end{align}
\end{subequations}
that lie within the uncertainty interval of the fit used
to obtain $\eta^{x}[\Psi^{\,}_{0}]$
and $\eta^{z,\mathrm{c}}[\Psi^{\,}_{0}]$ for different
choices of the fitting range of the correlations
$C^{x}_{j^{\,}_{0},j^{\,}_{0}+s}[\Psi^{\,}_{0}]$
and
$C^{z,\mathrm{c}}_{j^{\,}_{0},j^{\,}_{0}+s}[\Psi^{\,}_{0}]$,
respectively.  We observe that, at the point
$(\pi/4,1/2)$ of the phase diagram in Fig.\
\ref{suppfig:phase diagram}
where $|m^{z}_{\mathrm{uni}}[\Psi^{\,}_{0}]|=2/5$, relations
\eqref{suppeq:q against mz}
and
\eqref{suppeq:etas against mz} are consistent with the DMRG results
\begin{equation}
q^{z}[\Psi^{\,}_{0}]=2\pi\times3/10,
\qquad
\eta^{x}[\Psi^{\,}_{0}]=1,
\qquad
\eta^{z,\mathrm{c}}[\Psi^{\,}_{0}]=1,
\end{equation}
at the point$(\pi/4,1/2)$ of the phase diagram in Fig.\
\ref{suppfig:phase diagram}.

\subsection{Phase boundaries}

We have established numerically the nature of 
the extended $\mathrm{\hbox{N\'eel}}^{\,}_{x}$, $\mathrm{\hbox{N\'eel}}^{\,}_{y}$, 
$\mathrm{\hbox{N\'eel}}^{\mathrm{SPT}}_{x}$, $\mathrm{\hbox{N\'eel}}^{\mathrm{SPT}}_{y}$, 
and $\mathrm{FM}^{\,}_{z}$ phases.
We now turn our attention to the study 
of the transitions between any two of these phases 
that share a common boundary.

\subsubsection{Between $\mathrm{\hbox{N\'eel}}^{\,}_{x}$
and $\mathrm{\hbox{N\'eel}}^{\,}_{y}$}
\label{suppsec:trans neelx neely}

\paragraph{Location of the phase boundary ---}
The energy difference
\begin{equation}
E^{\,}_{1}-E^{\,}_{0}
\end{equation}
between the first excited state 
with energy $E^{\,}_{1}$ and the ground state
with energy $E^{\,}_{0}$ in either the $\mathrm{\hbox{N\'eel}}^{\,}_{x}$ 
or $\mathrm{\hbox{N\'eel}}^{\,}_{y}$ phase is a function 
of $2N$ that converges to zero
\begin{itemize}
\item[(i)] exponentially fast with the length $2N$ 
of the chain in the thermodynamic limit $N\to\infty$ 
deep in a gapped phase,
\item[(ii)] algebraically fast as $(2N)^{-z}$ 
with the length $2N$ of the chain in the 
thermodynamic limit $N\to\infty$ upon exiting 
a gapped phase through a gap-closing transition.
\end{itemize}
A gap closing with the dynamical scaling exponent $z=1$ 
between two gapped phases of matter can thus be identified 
through a cusp singularity when plotting the energy difference
\begin{equation}
2N(E^{\,}_{1}-E^{\,}_{0})
\label{suppeq: N times (E1-E0)} 
\end{equation}
as a function of the coupling driving a gap-closing 
phase transition between two gapped phases.
Accordingly, Figure~\ref{suppfig:ED evidence for DQCP 0,piover4} 
is consistent with the existence of a 
deconfined quantum critical point (DQCP) 
with $z=1$ located at $(\pi/4,0)$.

\begin{figure}[t!]
\centering
\includegraphics[width=1\columnwidth]{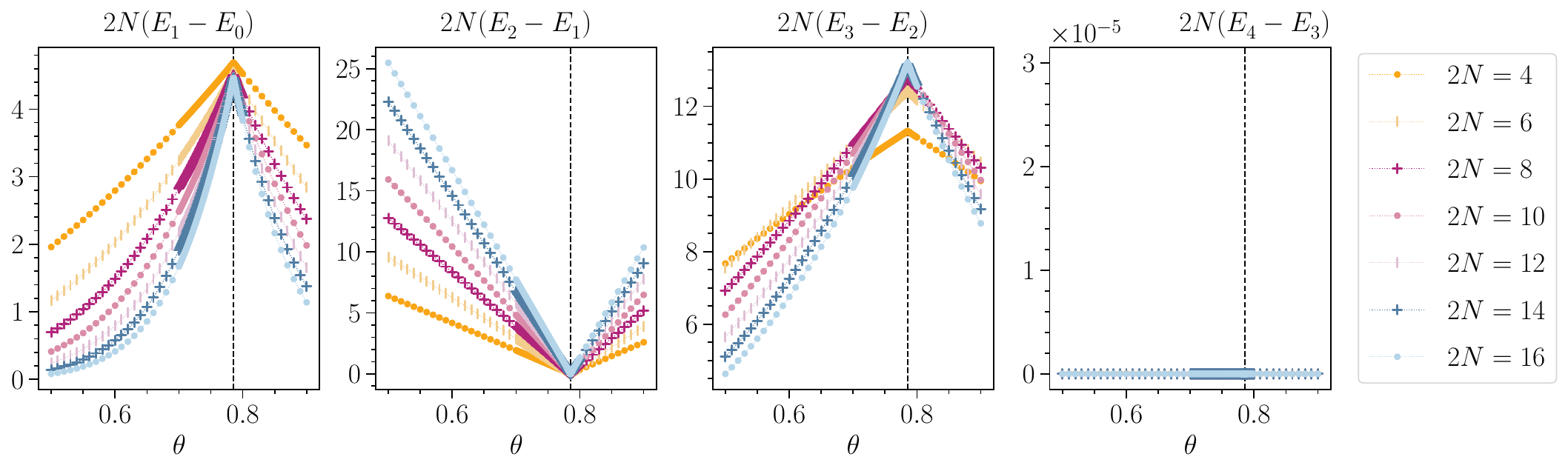}
\caption{
(Color online)
The case of PBC.
Dependence on $\theta$ of the energy spacings
$E^{\,}_{n}-E^{\,}_{n-1}$, $n=1,2,3,4$, above the ground
state energy $E^{\,}_{0}$ obtained with ED at $\lambda=0$ 
for different system sizes $2N$. The fact
that all lines reach an extremum at $\theta=\pi/4$ (dashed black line)
is consistent with a gap closing as $1/(2N)$ in the thermodynamic limit.
}
\label{suppfig:ED evidence for DQCP 0,piover4}
\end{figure}

We have located numerically
the transition between the $\mathrm{\hbox{N\'eel}}^{\,}_{x}$ 
and $\mathrm{\hbox{N\'eel}}^{\,}_{y}$ phases at $(\pi/4,0)$ by observing
the cusps in the energy gap $E^{\,}_{1}-E^{\,}_{0}$ 
shown in Fig.~\ref{suppfig:ED evidence for DQCP 0,piover4}.
We now present an alternative method to locate this phase transition, 
based on the computation of the order parameters.
This method will be particularly 
useful to identify the phase 
transition upon exiting the $\mathrm{FM}^{\,}_{z}$ phase.

It is known on analytical grounds \cite{Lieb61} that
the point $(\pi/4,0)$ with the $\mathrm{O}(2)$-symmetry
of Hamiltonian (\ref{suppeq:def H})
realizes a continuous phase transition
along the line $\lambda=0$
in the phase diagram from Fig.~\ref{suppfig:phase diagram}.
Verifying this prediction numerically faces two hurdles.
First, a (quantum)
phase transition can only take place in the thermodynamic limit
defined by taking the limit $L\to\infty$
for the length of the chain $L=2N\,\mathfrak{a}$,
while numerics demand that $2N$ is a finite number, i.e., that $L$ is finite.
Second, we approximate exact eigenstates of
Hamiltonian (\ref{suppeq:def H})
using DMRG.
We review the consequences of both limitations.

Suppose that the correlation length $\xi^{\,}_{\mathrm{phys}}(t)$
in the thermodynamic limit is known exactly as a function of $t$.
Here, $t\equiv\theta-\pi/4$
is the coupling with
the $\mathrm{\hbox{N\'eel}}^{\,}_{x}$ fixed point at $t=-\pi/4$,
the quantum critical point at $t=0$,
and the $\mathrm{\hbox{N\'eel}}^{\,}_{y}$ fixed point at $t=+\pi/4$.
The correlation length $\xi^{\,}_{\mathrm{phys}}(t)$
is an even positive function of $t$ that diverges at $t=0$,
has two absolute minima at $t=\pm\pi/4$ of order $\mathfrak{a}$,
and is a strictly monotonic function of $t$ on either side of $t=0$.
Suppose that it is possible to compute exactly the expectation value
$\langle\widehat{O}\rangle^{\,}_{\Psi^{\,}_{0,L}}$
of any operator $\widehat{O}$
in the ground state $|\Psi^{\,}_{0,L}\rangle$
of Hamiltonian (\ref{suppeq:def H})
for a given finite system size $L$. 
If $L\gg\xi^{\,}_{\mathrm{phys}}(t)$, then
$\langle\widehat{O}\rangle^{\,}_{\Psi^{\,}_{0,L}}$
is a good approximation
to $\lim\limits_{L\to\infty}\langle\widehat{O}\rangle^{\,}_{\Psi^{\,}_{0,L}}$,
as one has good control over the finite-size corrections when
$\xi^{\,}_{\mathrm{phys}}(t)/L\ll1$.
However, if $L\ll\xi^{\,}_{\mathrm{phys}}(t)$, then
$\langle\widehat{O}\rangle^{\,}_{\Psi^{\,}_{0,L}}$
will share the same behavior as the correlation function at criticality,
as one has no control over the finite-size corrections when
$\xi^{\,}_{\mathrm{phys}}(t)/L\gg1$.
The signature of the quantum phase transition
at $t=0$ when using exact diagonalization to compute
the $\mathrm{\hbox{N\'eel}}^{\,}_{x}$ and $\mathrm{\hbox{N\'eel}}^{\,}_{y}$ order parameters
is an interval of size $\delta t(L)$
centered around $t=0$ in which the order parameters
vanish, whereby $\delta t(L)$ is a positive function of $L$
that converges to the value 0 in the thermodynamic limit $L\to\infty$.

In practice, it is not possible for large system sizes $L\gg\mathfrak{a}$
to obtain an exact eigenstate of Hamiltonian (\ref{suppeq:def H}).
Approximations to an exact eigenstate of Hamiltonian (\ref{suppeq:def H})
such as DMRG are used instead. At criticality, here $t=0$,
DMRG tries to approximate a critical state
with a matrix product state characterized by a correlation length
$\xi^{\,}_{\mathrm{MPS}}(t=0,\chi)$. Here,
the correlation length $\xi^{\,}_{\mathrm{MPS}}(t,\chi)$
depends on the physical coupling $t$ entering the Hamiltonian
(\ref{suppeq:def H})
and an unphysical parameter, the bond dimension
$\chi$, that fixes the space of variational matrix product states
from which DMRG selects the approximation to the exact ground state
$|\Psi^{\,}_{0}\rangle$, say. This space of variational wave functions
is incomplete as a basis of the Hilbert space
for any given $\chi$.
The correlation length
$\xi^{\,}_{\mathrm{MPS}}(t,\chi)$
of the DMRG state is an increasing function of $\chi$
that is bounded from above by the physical correlation length
$\xi^{\,}_{\mathrm{phys}}(t)$. At criticality,
$\xi^{\,}_{\mathrm{phys}}(t=0)=\infty$
and all order parameters vanish.
On the one hand,
if
$\xi^{\,}_{\mathrm{MPS}}(t=0,\chi)\ll L$,
then DMRG predicts erroneously non-vanishing values for all order parameters.
On the other hand,
if
$\xi^{\,}_{\mathrm{MPS}}(t=0,\chi)\gg L$,
then DMRG predicts correctly vanishing values for all order parameters.

In summary,
there are three regimes for which finite-size effects  
are not under control.
There is the regime defined by
\begin{subequations}
\label{suppeq:three regimes close to criticality}
\begin{equation}
L\ll\xi^{\,}_{\mathrm{phys}}(t),
\label{suppeq:three regimes close to criticality a}
\end{equation}
for which exact diagonalization predicts that all order parameters vanish.
There is the regime defined by
\begin{equation}
L\ll\xi^{\,}_{\mathrm{MPS}}(t,\chi)\ll\xi^{\,}_{\mathrm{phys}}(t)
\label{suppeq:three regimes close to criticality b}
\end{equation}
for which DMRG agrees with exact diagonalization in that it also
predicts that all order parameters vanish.
There is the regime defined by
\begin{equation}
\xi^{\,}_{\mathrm{MPS}}(t,\chi)\ll L\ll\xi^{\,}_{\mathrm{phys}}(t),
\label{suppeq:three regimes close to criticality c}  
\end{equation}
\end{subequations}
for which DMRG disagrees with exact diagonalization in that it
predicts that some order parameters are non-vanishing.
It is the regime
(\ref{suppeq:three regimes close to criticality c})
that is problematic, for it should cross over to the regime
(\ref{suppeq:three regimes close to criticality b})
in the limit $\chi\to\infty$. An example of this crossover
is illustrated in Fig.~\ref{suppfig:DMRG OBC OP piover4,0 chi}
in which the value of the sum of the
absolute values of the
$\mathrm{\hbox{N\'eel}}^{\,}_{x}$
and
$\mathrm{\hbox{N\'eel}}^{\,}_{y}$
order parameters at criticality $t=0$
are plotted as a function of increasing
$\chi$ for fixed system sizes $2N$. For the three quoted values of $2N$,
a crossover from the regime
(\ref{suppeq:three regimes close to criticality c})
to the regime
(\ref{suppeq:three regimes close to criticality b})
is apparent. In the discussion of Fig.~\ref{suppfig:DMRG OBC OP 0,piover4 window} that follows,
$\chi$ is chosen sufficiently large for all system sizes to ensure that
regime (\ref{suppeq:three regimes close to criticality b})
holds when approaching the critical point at $t=0$.

The $\mathrm{\hbox{N\'eel}}^{\,}_{x}$ phase is characterized 
by a non-vanishing value of the order parameter 
$m^{x}_{\mathrm{sta}}[\Psi^{\,}_{0}]$, defined in Eq.~\eqref{suppeq:msta}, 
when computed with the approximation 
to the ground state 
$\ket*{\Psi^{\,}_{0}}$ of Hamiltonian \eqref{suppeq:def H}
obtained using DMRG.
The $\mathrm{\hbox{N\'eel}}^{\,}_{y}$ phase is characterized by 
a non-vanishing value of the order parameter 
$m^{y}_{\mathrm{sta}}[\Psi^{\,}_{0}]$, also defined in \eqref{suppeq:msta}, 
when computed with the approximation 
to the ground state 
$\ket*{\Psi^{\,}_{0}}$ of Hamiltonian \eqref{suppeq:def H}
obtained using DMRG.
Figure~\ref{suppfig:DMRG OBC OP 0,piover4 window}
shows the absolute values of 
$m^{x}_{\mathrm{sta}}[\Psi^{\,}_{0}]$ 
and 
$m^{y}_{\mathrm{sta}}[\Psi^{\,}_{0}]$
along the line 
$\lambda=0$ for different system sizes $2N$.  
We observe a window in parameter space 
around $\theta=\pi/4$ where both order parameters vanish.
The size of this window decreases as the system size increases,
as is expected for a quantum phase transition taking place at
$\theta=\pi/4$.

\begin{figure}[t!]
\centering
\includegraphics[width=0.44\columnwidth]{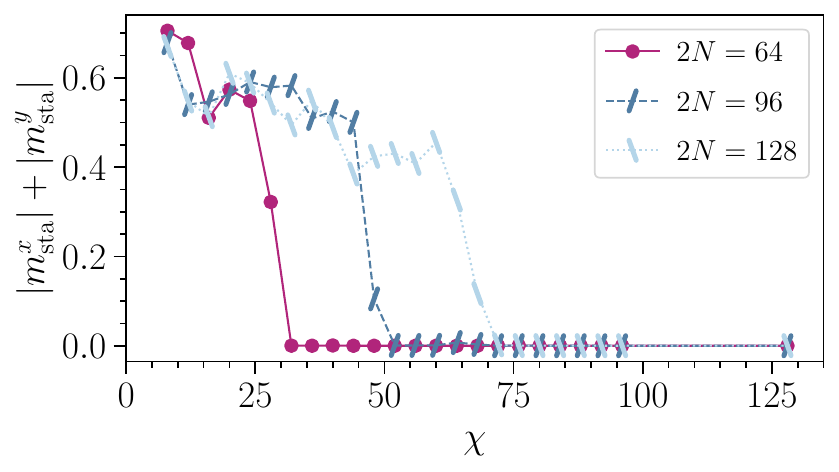}
\caption{
(Color online)
The case of PBC. DMRG estimate for the 
absolute value 
$|m^{x}_{\mathrm{sta}}[\Psi^{\,}_{0}]|
+
|m^{y}_{\mathrm{sta}}[\Psi^{\,}_{0}]|$ 
of the relevant order parameters at $(\pi/4,0)$ 
at fixed system sizes $2N$ as a function of the 
bond dimension $\chi$.
}
\label{suppfig:DMRG OBC OP piover4,0 chi}
\end{figure}

\begin{figure}[t!]
\centering
\includegraphics[width=0.5\columnwidth]{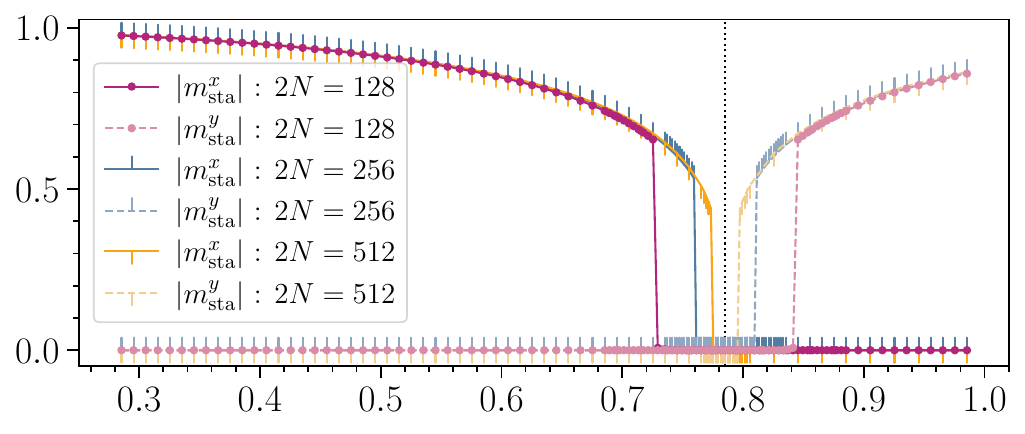}
\caption{
(Color online)
The case of OBC.
Relevant (non-vanishing) order parameters 
defined in \eqref{suppeq:def order parameters} obtained with the 
DMRG approximation ($\chi = 128$) across a cut around
$\theta=\pi/4$ at $\lambda=0$.
The dotted black line indicates $\theta=\pi/4$.
}
\label{suppfig:DMRG OBC OP 0,piover4 window}
\end{figure}

\begin{figure}[t!]
\centering
\subfigure[]{\includegraphics[width=1\columnwidth]{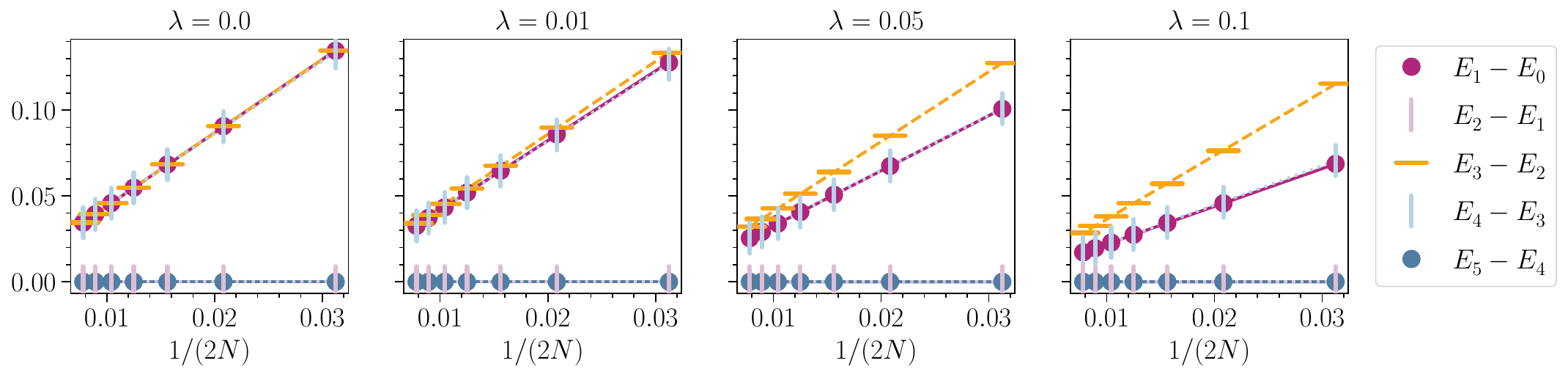}} \\
\subfigure[]{\includegraphics[width=1\columnwidth]{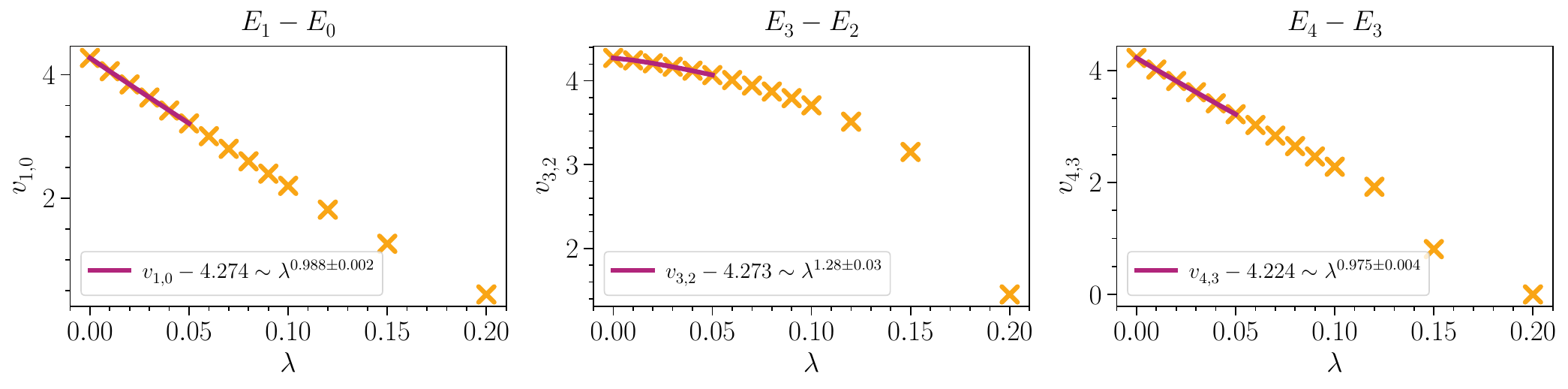}}
\caption{
(Color online)
The case of OBC.
(a) Dependence on 
$1/(2N)$ of the energy differences
$E^{\,}_{n}-E^{\,}_{n-1}$, $n=1,2,3,4,5$, above 
the ground state energy $E^{\,}_{0}$ obtained with DMRG 
($\chi = 128$) at $\theta=\pi/4$ for different $\lambda$. 
Linear fits for each $n$ are represented as lines whose 
color corresponds to the color of the data points in the legend.
(b) Dependence on $\lambda$ of the relevant nonvanishing 
velocities $v^{\,}_{n,n-1} \coloneqq \partial (E^{\,}_{n}-E^{\,}_{n-1}) 
/ \partial (2N)^{-1}$, $n=1,3,4$, computed with DMRG 
($\chi = 128$) at $\theta=\pi/4$ for systems sizes 
$2N \in [32, 128]$. Fits of the type $v^{\,}_{n,n-1} 
= \alpha \, \lambda^{\beta}+\gamma$ for each $n$ are computed.
}
\label{suppfig:DMRG DQCP OBC 0-0.2,piover4}
\end{figure} 

\paragraph{Degeneracies at the phase boundary ---}
We have already verified numerically that the ground-state energy
$E^{\,}_{0}$ is onefold degenerate with OBC at the
DQCP $(\pi/4,0)$
in that the energy difference between the first excited state
$E^{\,}_{1}$ and the ground state $E^{\,}_{0}$ decreases like
$1/(2N)$ as opposed to being exponentially small in $2N$.
We have verified that
this onefold degeneracy persists for 
all points on the critical line at $\theta=\pi/4$ emanating 
from $(\pi/4,0)$
[see Fig.~\ref{suppfig:DMRG DQCP OBC 0-0.2,piover4}]
until the $\mathrm{FM}^{\,}_{z}$ phase is reached.  

The rate at which two consecutive energy eigenvalues approach each other
as the system size is increased is a useful
characterization of excited states. To this end, we define
the ``velocities'' $v^{\,}_{n,n-1}$ by taking the derivative
with respect of the system size of two consecutive energy eigenvalues,
namely we define
\begin{subequations}
\label{suppeq:velocities}
\begin{equation}
v^{\,}_{n,n-1}\coloneqq
\pdv{\,(E^{\,}_{n}-E^{\,}_{n-1})}{\,(2N)^{-1}},
\label{suppeq:velocities a}
\end{equation}
assuming the partial ordering
\begin{equation}
E^{\,}_{n-1}\leq E^{\,}_{n},
\qquad
n=1,2,3,\cdots,
\label{suppeq:velocities b}
\end{equation}
\end{subequations}
of the exact energy eigenvalues given the finite-dimensional
Hilbert space $\mathbb{C}^{2^{2N}}$.
A ``velocity'' $v^{\,}_{n,n-1}$ vanishes if
$E^{\,}_{n}-E^{\,}_{n-1}$ is independent of $2N$ as occurs
if $E^{\,}_{n}$ and $E^{\,}_{n-1}$
are degenerate for any choice of $2N$, say.
A ``velocity'' $v^{\,}_{n,n-1}$ vanishes in the thermodynamic limit
$2N\to\infty$
if it decays faster than $(2N)^{-1}$.
The DMRG estimate of the ``velocities'' $v^{\,}_{n,n-1}$
under OBC for different values of $\lambda$ along the vertical boundary
emanating from $(\pi/4,0)$ are shown in
Fig.~\ref{suppfig:DMRG DQCP OBC 0-0.2,piover4}.
At $(\pi/4,0)$, it is observed that all the 
non-vanishing velocities are equal, i.e., 
\begin{equation}
v^{\,}_{1,0}= 
v^{\,}_{3,2}=
v^{\,}_{4,3}\simeq
4.274.
\end{equation}
Moving away from $(\pi/4,0)$, the velocities 
$v^{\,}_{1,0}$ and $v^{\,}_{4,3}$
remain equal up to numerical precision and 
decrease linearly as a function of $\lambda$.
However, these velocities differ from $v^{\,}_{3,2}$, 
which decreases non-linearly as a function of $\lambda$.  
The strength of the finite-size effects
close to the tricritical point at
$(\pi/4,\lambda^{\,}_{\mathrm{tri}})$, 
$\lambda^{\,}_{\mathrm{tri}}\approx0.25$,
prevent a reliable numerical estimate of these velocities
at and in the vicinity of the tricritical point.

A similar study using PBC as opposed to OBC 
is summarized in Fig.~\ref{suppfig:DMRG DQCP PBC 0-0.2,piover4}.
The ground-state degeneracy is confirmed to be onefold 
along the critical line at fixed $\theta=\pi/4$ emanating 
from $(\pi/4,0)$, consistent with the results 
obtained for OBC. This agreement in ground-state 
degeneracy between PBC and OBC is expected.
However, the excitation spectrum with PBC differs 
from that with OBC. For PBC, 
the non-vanishing velocities $v^{\,}_{1,0}$ and $v^{\,}_{3,2}$ 
both decrease linearly as a function of $\lambda$ for small $\lambda$.
Unlike with OBC in Fig.~\ref{suppfig:DMRG DQCP OBC 0-0.2,piover4}, 
these velocities are no longer 
identical for any value of $\lambda$.
The strength of the finite-size effects
close to the tricritical point at
$(\pi/4,\lambda^{\,}_{\mathrm{tri}})$, 
$\lambda^{\,}_{\mathrm{tri}}\approx0.25$,
prevent a reliable numerical estimate of these velocities
at and in the vicinity of the tricritical point.

\begin{figure}[t!]
\centering
\subfigure[]{\includegraphics[width=1\columnwidth]{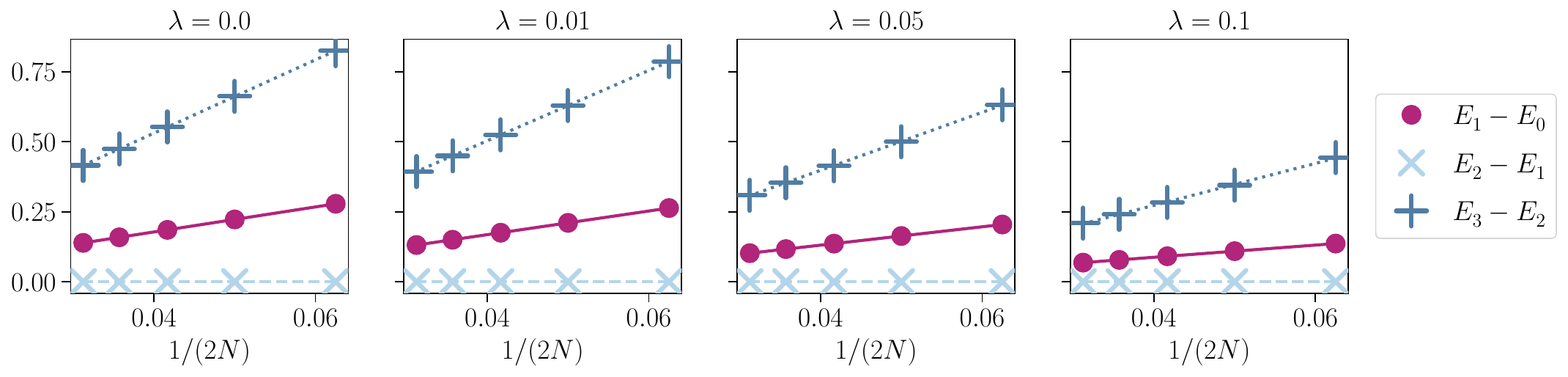}} \\
\subfigure[]{\includegraphics[width=1\columnwidth]{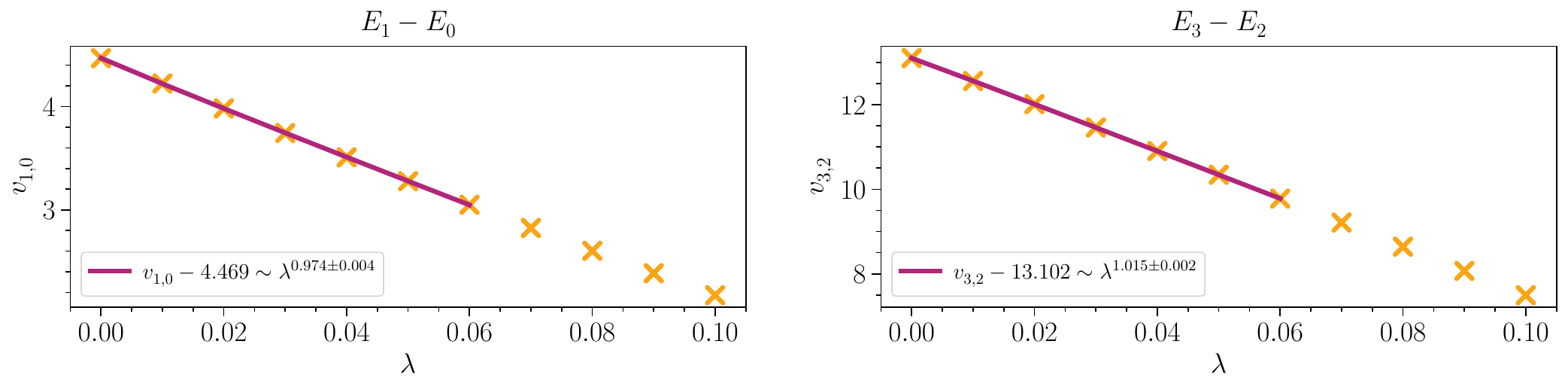}}
\caption{
(Color online)
The case of PBC.
(a) Dependence on
$1/(2N)$ of the energy differences
$E^{\,}_{n}-E^{\,}_{n-1}$, $n=1,2,3$, above the 
ground state energy $E^{\,}_{0}$ obtained with DMRG 
and PBC at $\theta=\pi/4$ for different $\lambda$. 
Linear fits for each $n$ are represented as lines whose 
color corresponds
to the color of the data points in the legend.
(b) Dependence on $\lambda$ of the relevant nonvanishing 
velocities
$v^{\,}_{n,n-1}\coloneqq
\partial(E^{\,}_{n}-E^{\,}_{n-1})/\partial(2N)^{-1}$,
$n=1,3$, computed with DMRG and PBC at $\theta=\pi/4$ 
for systems sizes $2N\in[16,32]$.
Fits of the type
$v^{\,}_{n,n-1}=\alpha\,\lambda^{\beta}+\gamma$
for each $n$ are computed.
}
\label{suppfig:DMRG DQCP PBC 0-0.2,piover4}
\end{figure}

\paragraph{Correlations at the phase boundary ---}
The point $(\pi/4,0)$
realizes the DQCP between the
$\mathrm{\hbox{N\'eel}}^{\,}_{x}$ 
and
$\mathrm{\hbox{N\'eel}}^{\,}_{y}$ phases.
Its low-energy theory is described by the free-fermion 
CFT with the central charge
$\mathsf{c}=1$ \cite{Lieb61}.
Numerical results confirm that the two-point correlation 
$C^{x}_{j^{\,}_{0},j^{\,}_{0}+s}[\Psi^{\,}_{0}]$ 
and 
$C^{y}_{j^{\,}_{0},j^{\,}_{0}+s}[\Psi^{\,}_{0}]$, 
defined in Eq.~\eqref{suppeq:Calpha} for OBC,
are equal due to the $\mathrm{O}(2)$-symmetry of Hamiltonian
\eqref{suppeq:def H}
when $\theta=\pi/4$.
These correlations follow the algebraic decay
\eqref{suppeq:algebraic decay transverse spin-spin in FMZ phase},
i.e., $(-1)^{s}\,s^{-\eta(\lambda=0)}$
with the exponent $\eta(\lambda=0)=1/2$
at $(\pi/4,0)$
[see Fig.~\ref{suppfig:DMRG OBC corr 0,piover4 a}].  

Figure~\ref{suppfig:DMRG OBC corr 0,piover4 b} 
shows the dependence on $\lambda$ of the scaling exponent
$\eta(\lambda)$ along the vertical boundary
$(\pi/4,\lambda)$ with $0\leq\lambda\leq0.22$.
Close to $\lambda=0$, 
the exponent $\eta(\lambda)$ decreases linearly with $\lambda$.
This linear relationship, however, breaks down upon 
approaching the tricritical point $(\pi/4,\lambda^{\,}_{\mathrm{tri}})$
with $\lambda^{\,}_{\mathrm{tri}}\approx0.25$
(see Sec.~\ref{suppsubsec:Tricritical points})
at which the
$\mathrm{\hbox{N\'eel}}^{\,}_{x}$,
$\mathrm{\hbox{N\'eel}}^{\,}_{y}$,
and $\mathrm{FM}^{\,}_{z}$
phases meet.
At around $\lambda=0.18$,
$\eta(\lambda)$ reaches a minimum. For larger values of $\lambda$,
the scaling exponent $\eta(\lambda)$ is an increasing function of
$\lambda$ that converges smoothly to the value
$\eta(\lambda=1/2)=1$ at the center $(\pi/4,1/2)$
of the phase diagram from Fig.~\ref{suppfig:phase diagram}.

\begin{figure}[t!]
\centering
\subfigure[]{\includegraphics[width=0.5\columnwidth]{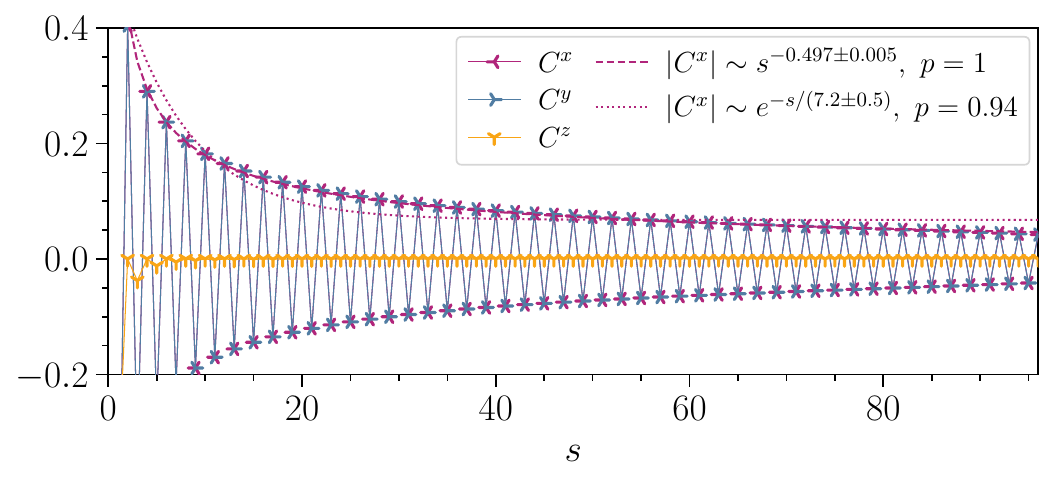} \label{suppfig:DMRG OBC corr 0,piover4 a}}%
\subfigure[]{\includegraphics[width=0.25\columnwidth]{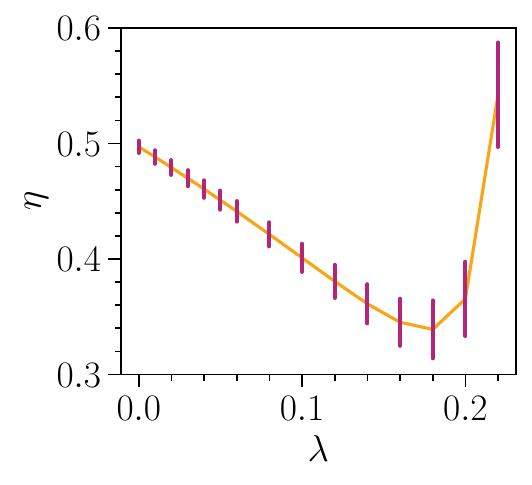} \label{suppfig:DMRG OBC corr 0,piover4 b}}
\caption{
(Color online)
The case of OBC.
(a)
Dependence on the separation $s$ in the
spin-spin correlation functions $C^{\alpha}_{j^{\,}_{0},j^{\,}_{0}+s}[\Psi^{\,}_{0}]$
with
$j^{\,}_{0}=(2N-96)/2$
and
$\alpha=x,y,z$
defined in Eq.~\eqref{suppeq:Calpha}
computed using DMRG 
($\chi=128$ and $2N=128$)
at $(\pi/4,0)$ in Fig.~\ref{suppfig:phase diagram}.
The fit for
$C^{x}_{j^{\,}_{0},j^{\,}_{0}+s}[\Psi^{\,}_{0}]=C^{y}_{j^{\,}_{0},j^{\,}_{0}+s}[\Psi^{\,}_{0}]$
is an algebraic decay.
The $p$-value
for each fit is given.
(b) Scaling exponents $\eta(\lambda)$
for the algebraic decay of 
$C^{x}_{j^{\,}_{0},j^{\,}_{0}+s}[\Psi^{\,}_{0}]=C^{y}_{j^{\,}_{0},j^{\,}_{0}+s}[\Psi^{\,}_{0}]\sim(-1)^{s} 
s^{-\eta(\lambda)}$
as a function of $\lambda$ holding $\theta=\pi/4$ fixed.
Uncertainties on $\eta(\lambda)$ are shown in magenta.
}
\label{suppfig:DMRG OBC corr 0,piover4}
\end{figure}

\paragraph{Central charge at the phase boundary ---}
The use of PBC with DMRG is best suited for the computation
of the central charge $\mathsf{c}$.
With OBC, determining $\mathsf{c}$ is heavily 
influenced by finite-size effects, often necessitating 
extrapolation to the thermodynamic limit using large values of
$2N$.
In contrast, with PBC, $\mathsf{c}$ can be extracted 
with reasonable accuracy even for relatively small system sizes, 
typically involving only a few tens of sites. 

Using PBC, we can fit
$S^{\,}_{\mathrm{MPS},2N}(l)$
defined in Eq.~(\ref{suppeq:def S(ell) for MPS}) 
with the CFT Ansatz
\eqref{suppeq:CFT prediction for S2N(l)}
to extract the DMRG estimate for the central charge
$\mathsf{c}^{\,}_{2N}$
and then perform the extrapolation
$\mathsf{c}\coloneqq\lim\limits_{2N\to\infty}\mathsf{c}^{\,}_{2N}$.
At $(\pi/4,0)$, we already find that
$2N=64$ delivers the value $\mathsf{c}=1\pm0.001$
according to Fig.~\ref{suppfig:DMRG PBC central charge 0,piover4 a}.
Finite-size corrections for the central charge
$\mathsf{c}^{\,}_{2N}$
increase along the vertical boundary
$(\pi/4,\lambda)$ with $0\leq\lambda\leq0.2$
according to Fig.~\ref{suppfig:DMRG PBC central charge 0,piover4 b}.
However, these finite-size corrections are consistent with the limiting value
$\mathsf{c}\coloneqq\lim\limits_{2N\to\infty}\mathsf{c}^{\,}_{2N}=1$
for $(\pi/4,\lambda)$ with $0\leq\lambda\leq0.2$.
Figure~\ref{suppfig:DMRG PBC central charge 0,piover4 c}
is consistent with the limiting value 
$\mathsf{c}\coloneqq\lim\limits_{2N\to\infty}\mathsf{c}^{\,}_{2N}=0$
for $(\theta,0)$ with $\theta\neq\pi/4$ in the closed interval $[0,\pi/2]$.
The value $\mathsf{c}=1$ for
$(\pi/4,\lambda)$ with $0\leq\lambda\leq0.2$
is the same as the value found in
Sec.~\ref{suppeq:subsubsection Criticality}
for the central charge inside the $\mathrm{FM}^{\,}_{z}$ phase.
The strength of the finite-size effects
close to the tricritical point at
$(\pi/4,\lambda^{\,}_{\mathrm{tri}})$, 
$\lambda^{\,}_{\mathrm{tri}}\approx0.25$,
prevent a reliable numerical estimate of the central charges
at and in the vicinity of the tricritical point.

In spite of the strong finite-size effects close to the tricritical point 
$(\pi/4,\lambda^{\,}_{\mathrm{tri}})$, $\lambda^{\,}_{\mathrm{tri}}\approx0.25$, 
the results obtained with DMRG on the chain
of $2N=64$ sites
support the conjecture that every point 
on the line $(\pi/4,\lambda)$, $0\leq\lambda\leq1$ realizes a CFT
with central charge $\mathsf{c}=1$.

\begin{figure}[t!]
\centering
\subfigure[]{\includegraphics[width=0.36\columnwidth]{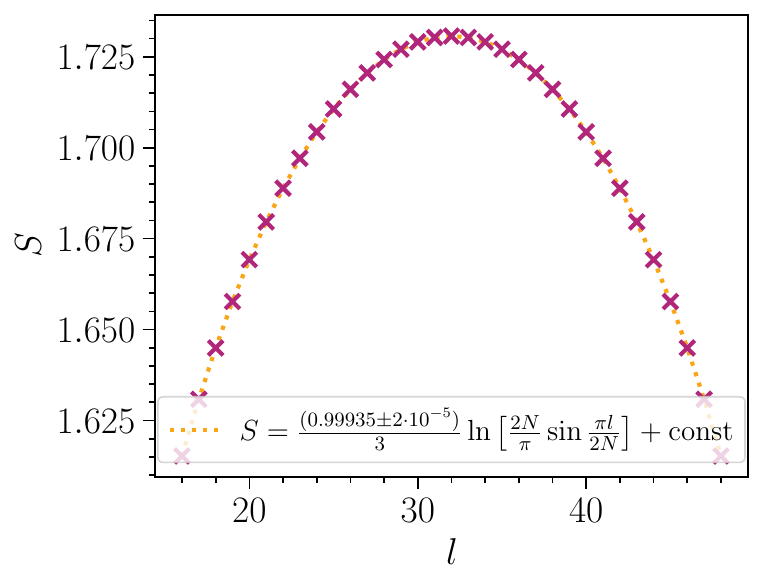} \label{suppfig:DMRG PBC central charge 0,piover4 a}}%
\subfigure[]{\includegraphics[width=0.3\columnwidth]{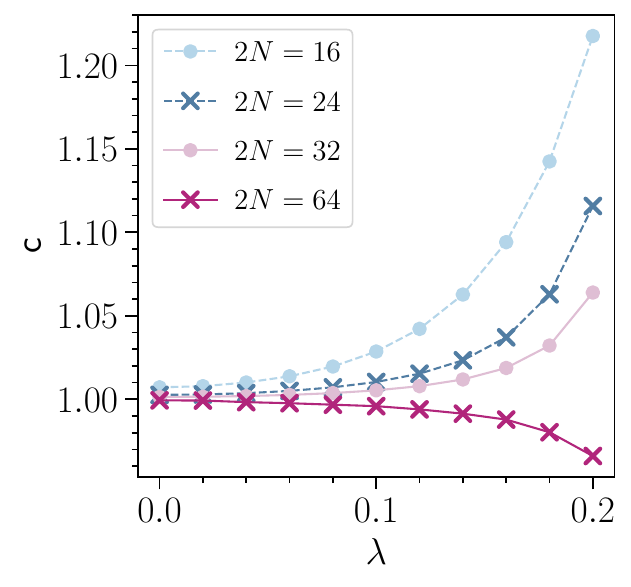} \label{suppfig:DMRG PBC central charge 0,piover4 b}}%
\subfigure[]{\includegraphics[width=0.293\columnwidth]{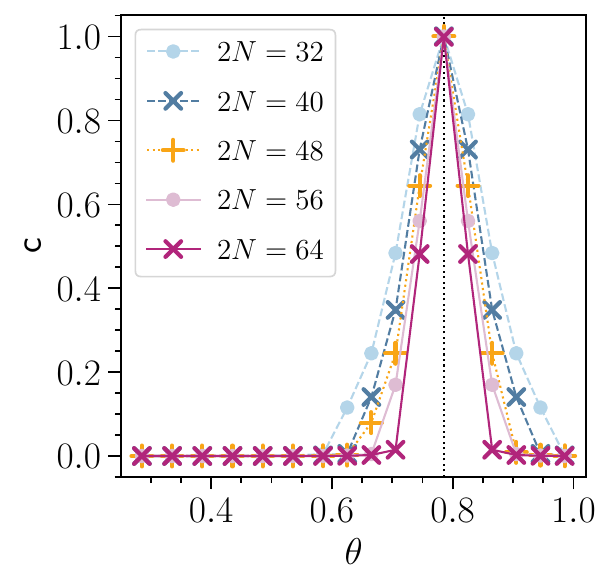} \label{suppfig:DMRG PBC central charge 0,piover4 c}}
\caption{
(Color online)
The case of PBC.
(a) Fit as a function of $l$ with the CFT Ansatz
\eqref{suppeq:CFT prediction for S2N(l)},
whereby the central charge is $\mathsf{c}^{\,}_{2N}=0.99935\pm2\cdot10^{-5}$,
to
$S^{\,}_{\mathrm{MPS},2N}(l)$
defined in Eq.~(\ref{suppeq:def S(ell) for MPS}) 
and obtained with DMRG ($\chi=256$ and $2N=64$) at $(\pi/4,0)$.
(b)
Central charge $\mathsf{c}^{\,}_{2N}$
as a function of $\lambda$ holding $\theta=\pi/4$ fixed
for different values of $2N$
computed by fitting the entanglement entropy
$S^{\,}_{\mathrm{MPS},2N}(l)$.
(c)
Central charge $\mathsf{c}^{\,}_{2N}$
as a function of $\theta$ holding $\lambda=0$ fixed
for different values of $2N$
computed by fitting the entanglement entropy
$S^{\,}_{\mathrm{MPS},2N}(l)$.
In panels (b) and (c),
$\delta{l}$ defined in Eq.~(\ref{suppeq:def delta{l}})
is $2N/2$.
}
\label{suppfig:DMRG PBC central charge 0,piover4}
\end{figure}

\subsubsection{Between $\mathrm{\hbox{N\'eel}}^{\mathrm{SPT}}_{x}$
and $\mathrm{\hbox{N\'eel}}^{\mathrm{SPT}}_{y}$}

\paragraph{Location of the phase boundary ---}
Under PBC, the phase boundary between the
$\mathrm{\hbox{N\'eel}}^{\mathrm{SPT}}_{x}$
and $\mathrm{\hbox{N\'eel}}^{\mathrm{SPT}}_{y}$
is the vertical interval
$(\pi/4,1-\lambda)$,
$0\leq\lambda\leq\lambda^{\,}_{\mathrm{tri}}$,
with $\lambda^{\,}_{\mathrm{tri}}\approx0.25$
in Fig.~\ref{suppfig:phase diagram}
(see Secs.~\ref{suppsec:Symmetries}
and~\ref{suppsec:trans neelx neely}).

\paragraph{Degeneracies at the phase boundary ---}
Under PBC, the DQCP at $(\pi/4,0)$ and $(\pi/4,1)$
are unitarily equivalent by the transformation
$\widehat{U}^{\,}_{\mathrm{E}}$ in Eq.~\eqref{suppeq:two unitaries b}.
The same is true for the vertical boundaries
emanating from $(\pi/4,0)$ and $(\pi/4,1)$.
Therefore, any property of the vertical boundary
emanating from $(\pi/4,0)$
carries over to
the vertical boundary emanating from $(\pi/4,1)$.
The former properties have been derived in 
Sec.~\ref{suppsec:trans neelx neely}.
Remarkably, the converse is not true.
Not all properties from the vertical boundary
emanating from $(\pi/4,1)$
carry over to the vertical boundary
emanating from $(\pi/4,0)$, owing to the sensitivity
of the vertical boundary
emanating from $(\pi/4,1)$ to the change of boundary conditions
from periodic to open ones.

When OBC are selected,
the fixed point Hamiltonians
(\ref{suppeq:Hamiltonians at lambda=1 a})
and
(\ref{suppeq:Hamiltonians at lambda=1 b})
each have eightfold degenerate ground states
for any finite length $L=2N\,\mathfrak{a}$ of the chain.
This degeneracy arises because 
(i)
Hamiltonians
(\ref{suppeq:Hamiltonians at lambda=1 a})
and
(\ref{suppeq:Hamiltonians at lambda=1 b})
each are the sum of $2N-3$ pairwise commuting operators,
(ii)
there is a twofold degeneracy
inherited from the case of PBC
and
(iii)
there is an additional fourfold degeneracy
that arises because of the existence of two
independent boundary spin-1/2 degrees of freedom that
decouple from the Hamiltonian
[recall Eq.~(\ref{suppeq:def left triplet spin operators})].
The twofold degeneracy shared 
by the fixed point Hamiltonians
(\ref{suppeq:Hamiltonians at lambda=1 a})
and
(\ref{suppeq:Hamiltonians at lambda=1 b})
under both PBC and OBC for any $L$ is tied to time-reversal symmetry.

Prior to taking the thermodynamic limit
$L\to\infty$
and for any
$0<\theta<\pi/2$
and
$\theta\neq\pi/4$
under PBC,
the twofold degeneracy of the ground states
of the fixed point Hamiltonians
(\ref{suppeq:Hamiltonians at lambda=1 a})
and
(\ref{suppeq:Hamiltonians at lambda=1 b})
is lifted
along the horizontal line $(\theta,1)$
from Fig.~\ref{suppfig:phase diagram}
by an amount exponentially small in the ratio
\begin{subequations}
\label{suppeq:xiPBC if lambda=1}
\begin{equation}
\frac{L}{\xi^{\,}_{\mathrm{PBC}}(\theta,1)},
\qquad
\xi^{\,}_{\mathrm{PBC}}(\theta,1)\coloneqq
\frac{\hbar\,v^{\,}_{\mathrm{PBC}}(\theta,1)}{\Delta(\theta,1)},
\label{suppeq:xiPBC if lambda=1 a}
\end{equation}  
where $v^{\,}_{\mathrm{PBC}}(\theta,1)$
is a characteristic non-vanishing speed proportional to the ``velocity''
\begin{equation}
v^{\,}_{3,2}(\theta,1)\coloneqq
\frac{
\partial\left[E^{\,}_{\mathrm{PBC},3}(\theta,1)-E^{\,}_{\mathrm{PBC},2}(\theta,1)\right]
}{\partial(2N)^{-1}}
\label{suppeq:xiPBC if lambda=1 b}
\end{equation}
and
$\Delta(\theta,1)$ is the thermodynamic energy gap between
the ground states and all the excited states in the partially ordered set
\begin{equation}
E^{\,}_{\mathrm{PBC},0}(\theta,1)<
E^{\,}_{\mathrm{PBC},1}(\theta,1)\leq
E^{\,}_{\mathrm{PBC},2}(\theta,1)\leq
E^{\,}_{\mathrm{PBC},3}(\theta,1)\leq
E^{\,}_{\mathrm{PBC},4}(\theta,1)\leq
\cdots.
\label{suppeq:xiPBC if lambda=1 bc}
\end{equation}
At the critical point
$(\pi/4,1)$
with PBC prior to taking the thermodynamic limit $L\to\infty$,
the energy difference between the energies
$E^{\,}_{\mathrm{PBC},0}(\pi/4)$
and 
$E^{\,}_{\mathrm{PBC},1}(\pi/4)$
closes as the power law $L^{-1}$
[see Fig.~\ref{suppfig:DMRG DQCP PBC 1-0.9,piover4 a}].
We may then substitute
$E^{\,}_{\mathrm{PBC},3}(\pi/4,1)-E^{\,}_{\mathrm{PBC},2}(\pi/4,1)$
by
$E^{\,}_{\mathrm{PBC},1}(\pi/4,1)-E^{\,}_{\mathrm{PBC},0}(\pi/4,1)$
in the numerator of Eq.~(\ref{suppeq:xiPBC if lambda=1 b})
[see Fig.~\ref{suppfig:DMRG DQCP PBC 1-0.9,piover4 b}]
with the limiting behavior
\begin{equation}
\lim\limits_{\theta\to\pi/4}\xi^{\,}_{\mathrm{PBC}}(\pi/4,1)=\infty,
\label{suppeq:xiPBC if lambda=1 d}
\end{equation}
\end{subequations}
due to the closing of the thermodynamic gap
[see Fig.~\ref{suppfig:DMRG DQCP PBC 1-0.9,piover4 a}].

\begin{figure}[t!]
\centering
\subfigure[]{\includegraphics[width=1\columnwidth]{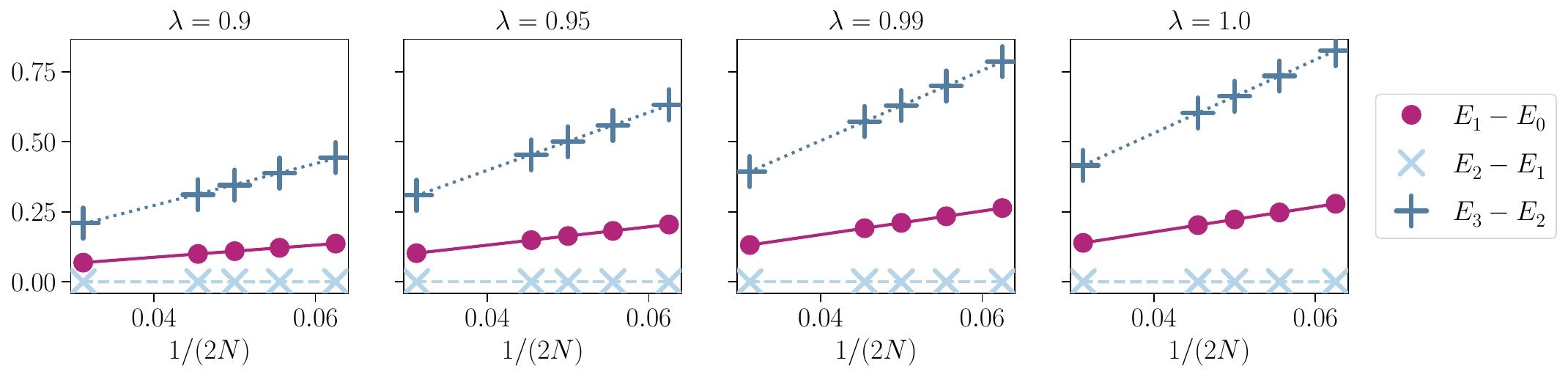} \label{suppfig:DMRG DQCP PBC 1-0.9,piover4 a}} \\
\subfigure[]{\includegraphics[width=1\columnwidth]{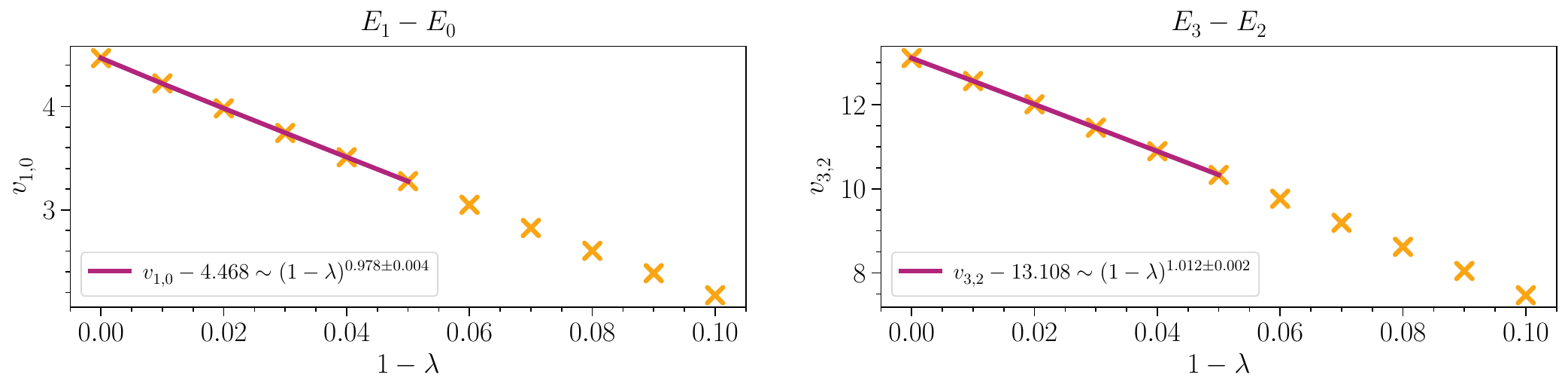} \label{suppfig:DMRG DQCP PBC 1-0.9,piover4 b}}
\caption{
(Color online)
The case of PBC.
(a) Dependence on 
$1/(2N)$ of the energy differences
$E^{\,}_{n}-E^{\,}_{n-1}$, 
$n=1,2,3$, above the ground state energy 
$E^{\,}_{0}$ obtained with DMRG 
at $\theta=\pi/4$ for different $\lambda$. 
Linear fits for each $n$ are represented as 
lines whose color corresponds to the color 
of the data points in the legend.
(b) 
Dependence on $1-\lambda\equiv\lambda^{\prime}$ of the relevant 
nonvanishing velocities $v^{\,}_{n,n-1} \coloneqq \partial 
(E^{\,}_{n}-E^{\,}_{n-1}) / \partial (2N)^{-1}$, 
$n=1,3$, computed with DMRG at $\theta=\pi/4$ 
for systems sizes $2N \in [16, 32]$. 
Fits of the form
$v^{\,}_{n,n-1}=
\alpha\,(1-\lambda)^{\beta}+\gamma$
for each $n$ are shown in red
for a restricted range of $1-\lambda\equiv\lambda^{\prime}$.
}
\label{suppfig:DMRG DQCP PBC 1-0.9,piover4}
\end{figure}

\begin{figure}[t!]
\centering
\subfigure[]{\includegraphics[width=1\columnwidth]{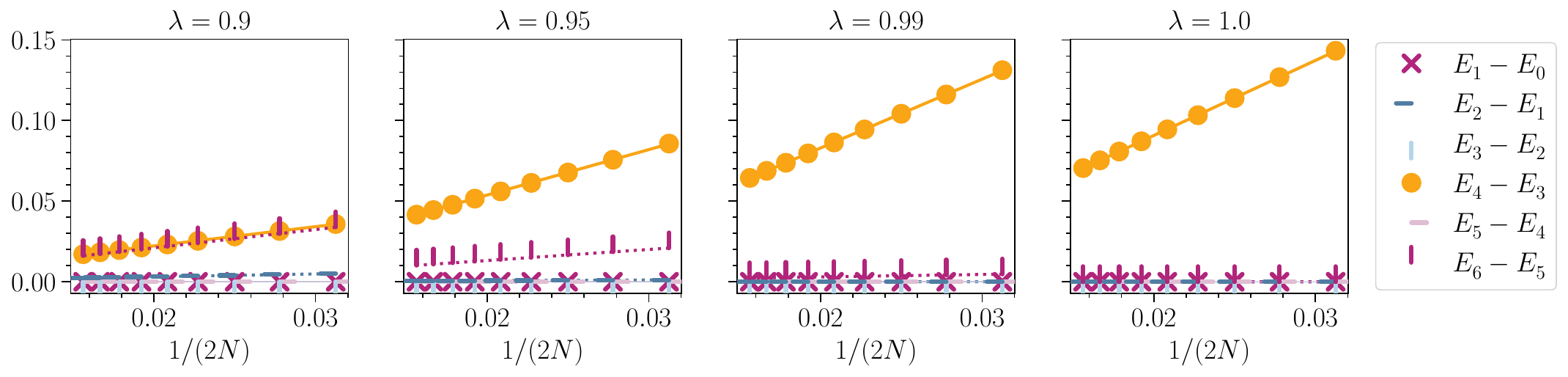} \label{suppfig:DMRG DQCP OBC 1-0.9,piover4 a}} \\
\subfigure[]{\includegraphics[width=1\columnwidth]{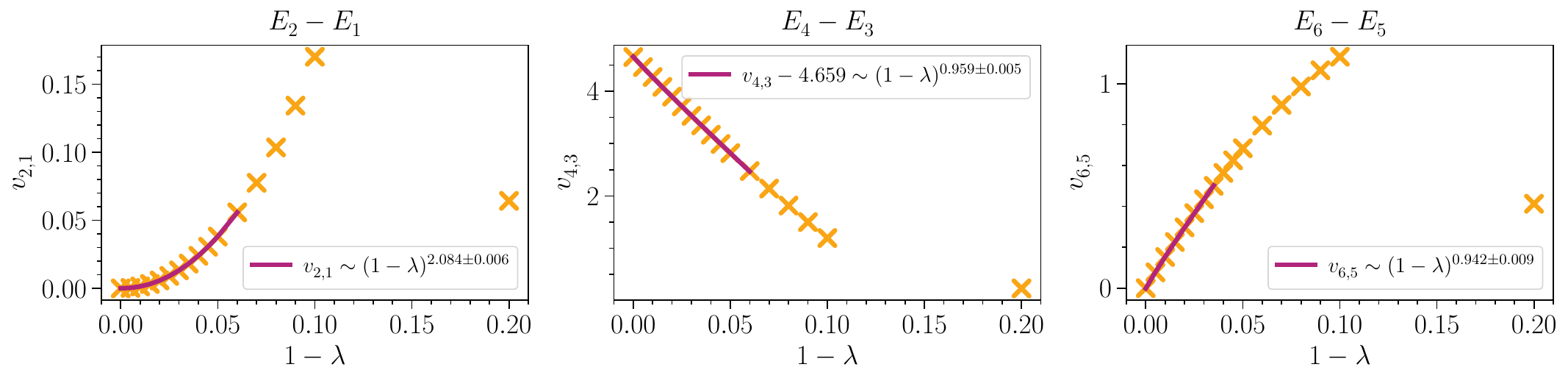} \label{suppfig:DMRG DQCP OBC 1-0.9,piover4 b}}
\caption{
(Color online)
The case of OBC.
(a) Dependence on 
$1/(2N)$ of the energy differences
$E^{\,}_{n}-E^{\,}_{n-1}$, 
$n=1,2,3,4,5,6$, above the ground state 
energy $E^{\,}_{0}$ obtained with DMRG 
($\chi = 128$) at $\theta=\pi/4$ for 
different $\lambda$. Linear fits for each $n$ 
are represented as lines whose color correspond
to the color of the data points in the legend.
(b) 
Dependence on $1-\lambda\equiv\lambda^{\prime}$ of the relevant nonvanishing
velocities
$v^{\,}_{n,n-1}\coloneqq
\partial(E^{\,}_{n}-E^{\,}_{n-1})/\partial(2N)^{-1}$, 
$n=2,4,6$, computed with DMRG ($\chi = 128$) 
at $\theta=\pi/4$ for systems sizes
$2N\in[32,64]$. 
Fits of the form
$v^{\,}_{n,n-1}=\alpha\,(1-\lambda)^{\beta}+\gamma$
for each $n$ are shown in red
for a restricted range of $1-\lambda\equiv\lambda^{\prime}$.
}
\label{suppfig:DMRG DQCP OBC 1-0.9,piover4}
\end{figure}

When selecting OBC prior to taking the thermodynamic limit $L\to\infty$,
we have seen in Secs.~\ref{suppsubsubsec:NeelXSPT}
and~\ref{suppsubsubsec:NeelYSPT}
that the lifting of the degeneracy between the eight degenerate ground states
of the fixed point Hamiltonians
(\ref{suppeq:Hamiltonians at lambda=1 a})
and
(\ref{suppeq:Hamiltonians at lambda=1 b})
incurred by choosing
$0<\theta<\pi/2$
and
$\theta\neq\pi/4$
along the horizontal line $(\theta,1)$
from Fig.~\ref{suppfig:phase diagram}
is also exponentially small in the ratio
\begin{subequations}
\label{suppeq:xiOBC if lambda=1}
\begin{equation}
\frac{L}{\xi^{\,}_{\mathrm{OBC}}(\theta,1)},
\qquad
\xi^{\,}_{\mathrm{OBC}}(\theta,1)\coloneqq
\frac{\hbar\,v^{\,}_{\mathrm{OBC}}(\theta,1)}{\Delta(\theta,1)},
\label{suppeq:xiOBC if lambda=1 a}
\end{equation}
where $v^{\,}_{\mathrm{OBC}}(\theta,1)$
is a characteristic non-vanishing speed proportional to the ``velocity''
\begin{equation}
v^{\,}_{9,8}=
\frac{
\partial\left[E^{\,}_{\mathrm{OBC},9}(\theta,1)-E^{\,}_{\mathrm{OBC},8}(\theta,1)\right]
}{\partial(2N)^{-1}}
\label{suppeq:xiOBC if lambda=1 b}
\end{equation}
and $\Delta(\theta,1)$ is the thermodynamic energy gap between
the ground states and all the excited states in the partially ordered set
\begin{equation}
E^{\,}_{\mathrm{OBC},0}(\theta,1)<
E^{\,}_{\mathrm{OBC},1}(\theta,1)\leq
E^{\,}_{\mathrm{OBC},2}(\theta,1)\leq
\cdots\leq
E^{\,}_{\mathrm{OBC},8}(\theta,1)\leq
E^{\,}_{\mathrm{OBC},9}(\theta,1)\leq
E^{\,}_{\mathrm{OBC},10}(\theta,1)\leq
\cdots.
\label{suppeq:xiOBC if lambda=1 c}
\end{equation}
Prior to taking the thermodynamic limit $L\to\infty$
at the critical point
$(\pi/4,1)$
under OBC, 
the energy difference between 
any one of the four energies
$E^{\,}_{\mathrm{OBC},0}(\pi/4,1)$,
$E^{\,}_{\mathrm{OBC},1}(\pi/4,1)$,
$E^{\,}_{\mathrm{OBC},2}(\pi/4,1)$
or
$E^{\,}_{\mathrm{OBC},3}(\pi/4,1)$
and the energy $E^{\,}_{\mathrm{OBC},4}(\pi/4,1)$
closes as the power law $L^{-1}$
[see Fig.~\ref{suppfig:DMRG DQCP OBC 1-0.9,piover4 a}]
so that we may substitute
$E^{\,}_{\mathrm{OBC},9}(\pi/4,1)-E^{\,}_{\mathrm{OBC},8}(\pi/4,1)$
by
$E^{\,}_{\mathrm{OBC},4}(\pi/4,1)-E^{\,}_{\mathrm{OBC},3}(\pi/4,1)$
in the numerator of Eq.~(\ref{suppeq:xiOBC if lambda=1 b})
[see Fig.~\ref{suppfig:DMRG DQCP OBC 1-0.9,piover4 b}]
with the limiting behavior
\begin{equation}
\lim\limits_{\theta\to\pi/4}\xi^{\,}_{\mathrm{OBC}}(\pi/4,1)=\infty,
\label{suppeq:xiOBC if lambda=1 d}
\end{equation}
\end{subequations}
due to the closing of the thermodynamic gap
[see Fig.~\ref{suppfig:DMRG DQCP OBC 1-0.9,piover4 a}].

This is why, when selecting OBC, we observe numerically
for any finite chain length $L=2N\,\mathfrak{a}$
that Hamiltonian 
\eqref{suppeq:DQCP at 0 pi/4}
under OBC ($b=1$)
supports a non-degenerate 
ground state, whereas Hamiltonian \eqref{suppeq:DQCP at 1 pi/4}
under OBC ($b=1$)
supports fourfold-degenerate ground states due to the presence 
of boundary zero modes, whose energy differences cannot be resolved
in Fig.~\ref{suppfig:DMRG DQCP OBC 1-0.9,piover4 a}.

Prior to taking the thermodynamic limit $L\to\infty$
at the critical point
$(\pi/4,1-\lambda^{\prime})$
with $0<\lambda^{\prime}<0.1$ 
under OBC, 
the energy difference between 
any one of the two energies
$E^{\,}_{\mathrm{OBC},0}(\pi/4,1-\lambda^{\prime})$
or
$E^{\,}_{\mathrm{OBC},1}(\pi/4,1-\lambda^{\prime})$
and the energy $E^{\,}_{\mathrm{OBC},2}(\pi/4,1-\lambda^{\prime})$
closes as the power law $L^{-1}$
[see Fig.~\ref{suppfig:DMRG DQCP OBC 1-0.9,piover4 b}]
so that we may substitute
$E^{\,}_{\mathrm{OBC},9}(\pi/4,1-\lambda^{\prime})
-E^{\,}_{\mathrm{OBC},8}(\pi/4,1-\lambda^{\prime})$
by
$E^{\,}_{\mathrm{OBC},2}(\pi/4,1-\lambda^{\prime})
-E^{\,}_{\mathrm{OBC},1}(\pi/4,1-\lambda^{\prime})$
in the numerator of Eq.~(\ref{suppeq:xiOBC if lambda=1 b})
with the limiting behavior
\begin{equation}
\lim\limits_{\theta\to\pi/4}\xi^{\,}_{\mathrm{OBC}}(\pi/4,1-\lambda^{\prime})=
\infty,
\label{suppeq:xiOBC if lambda<1}
\end{equation}
due to the closing of the thermodynamic gap
[see Fig.~\ref{suppfig:DMRG DQCP OBC 1-0.9,piover4 a}].
Accordingly,
upon tuning $1-\lambda^{\prime}$ to values just below $1$,
the small perturbation 
\eqref{suppeq:DQCP at 0 pi/4} on the unperturbed Hamiltonian 
\eqref{suppeq:DQCP at 1 pi/4} lowers the ground-state 
degeneracy from four to two when OBC are selected
and after the thermodynamic limit $L\to\infty$ has been taken.

We chose to plot the dependence on $1-\lambda\equiv\lambda^{\prime}$ close
to $\lambda^{\prime}=0$ along the vertical boundary
$(\pi/4,1-\lambda^{\prime})$
from Fig.~\ref{suppfig:phase diagram}
of the ``velocities''
$v^{\,}_{n,n-1}$ defined in Eq.~(\ref{suppeq:velocities})
that are vanishing when $\lambda^{\prime}=0$ and non-vanishing for
any small deviation $0<\lambda^{\prime}\leq0.1$
in
Figs.~\ref{suppfig:DMRG DQCP PBC 1-0.9,piover4 b}
and~\ref{suppfig:DMRG DQCP OBC 1-0.9,piover4 b}.
The change from a vanishing to a non-vanishing ``velocity''
as a function of $\lambda^{\prime}$ is the signature of a change
in the degeneracy of two consecutive eigenvalues in the thermodynamic limit.
In Fig.~\ref{suppfig:DMRG DQCP OBC 1-0.9,piover4 b},
we chose the range $0\leq\lambda^{\prime}\leq0.1$ over which
these ``velocities'' are monotonic functions of $\lambda^{\prime}$.
We also report the values of these ``velocities''
at $\lambda^{\prime}=0.2$ to emphasize that this monotonic dependence
does not extend beyond the range $0\leq\lambda^{\prime}\leq0.1$
and that the dependence on $\lambda^{\prime}$ is non-linear.
The numerical estimate of these ``velocities''
becomes unreliable due to strong finite-size effects
along the vertical boundary
$(\pi/4,1-\lambda^{\prime})$
with $0.2<\lambda^{\prime}<\lambda^{\,}_{\mathrm{tri}}$ where
the value $1-\lambda^{\,}_{\mathrm{tri}}\approx0.75$
is the approximate location of the tricritical
point at which the
$\mathrm{\hbox{N\'eel}}^{\mathrm{SPT}}_{x}$,
$\mathrm{\hbox{N\'eel}}^{\mathrm{SPT}}_{y}$,
and $\mathrm{FM}^{\,}_{z}$
phases meet.

\begin{figure}[t!]
\centering
\subfigure[]{\includegraphics[width=0.39\columnwidth]{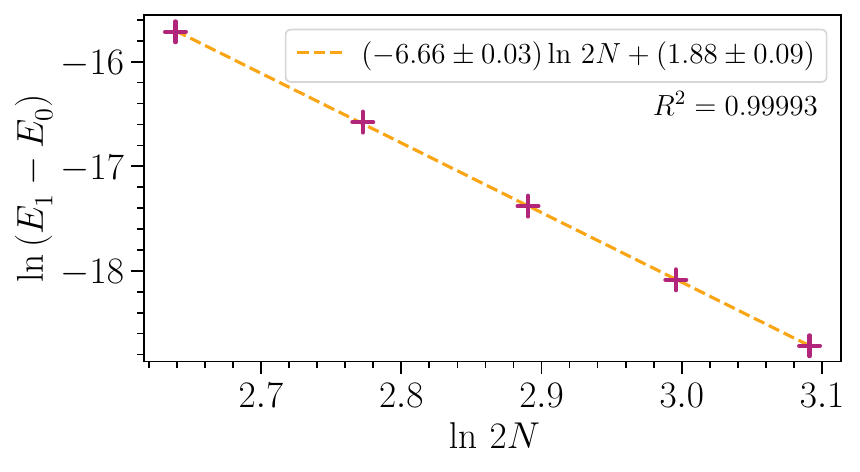} \label{suppfig:E1-E0 as a fct 2N using ED or Lanczos a}}\quad
\subfigure[]{\includegraphics[width=0.39\columnwidth]{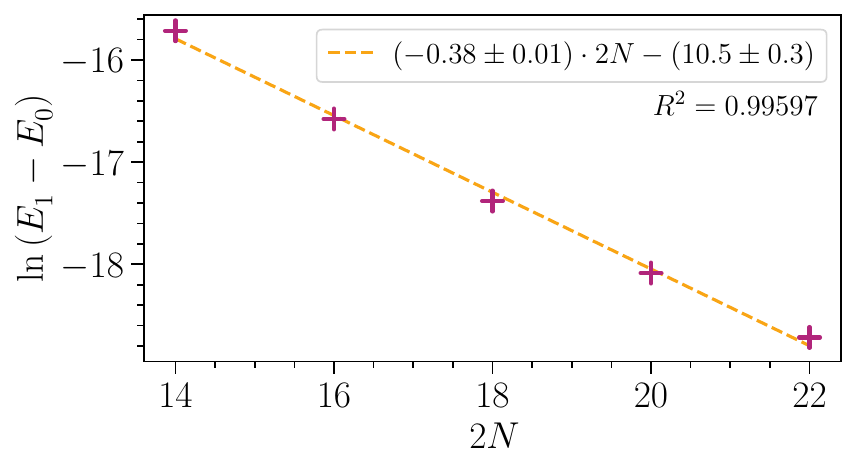} \label{suppfig:E1-E0 as a fct 2N using ED or Lanczos b}}%
\caption{
(Color online)
The case of OBC.
(a) Dependence on $2N$ of $\ln\,(E^{\,}_{1}-E^{\,}_{0})$
at $(\pi/4,0.95)$ for $2N=14,16,18,20,22$
obtained using the Lanczos algorithm.
(b) Dependence on $\ln\,(2N)$ of $\ln\,(E^{\,}_{1}-E^{\,}_{0})$
at $(\pi/4,0.95)$ for $2N=14,16,18,20,22$
obtained using the Lanczos algorithm.
        }
\label{suppfig:E1-E0 as a fct 2N using ED or Lanczos}
\end{figure}

The DMRG estimate for both
$E^{\,}_{1}-E^{\,}_{0}$
and
$E^{\,}_{3}-E^{\,}_{2}$
along the interval
$(\pi/4,1-\lambda^{\prime})$ with $0<\lambda^{\prime}\leq0.2$
is zero for $2N=6,8,\cdots$
within the numerical accuracy of $10^{-9}$
set by condition
(\ref{suppeq:three conditions for DMRG to have converged b}).
We have computed the lowest four energy eigenvalues
$E^{\,}_{0}\leq E^{\,}_{1}\leq E^{\,}_{2}\leq E^{\,}_{3}$
along the interval $0<\lambda^{\prime}\leq0.2$
for the system sizes $6,8,10,12,14$ with exact diagonalization
and for the system sizes $6,8,10,12,14,16,18,10,22$ with
the Lanczos algorithm. For a given value of $2N$
and $\lambda^{\prime}$,
the largest difference between the energy eigenvalue obtained
with exact diagonalization and the Lanczos algorithm
is of order $10^{-14}$.
We report in
Fig.~\ref{suppfig:E1-E0 as a fct 2N using ED or Lanczos}
the values of $E^{\,}_{1}-E^{\,}_{0}$ as a function of $2N$
at the point $(\pi/4,0.95)$ obtained using the Lanczos algorithm.
We find that the fit with the power law Ansatz
\begin{equation}
E^{\,}_{1}(2N)-E^{\,}_{0}(2N)=
A\,(2N)^{-\kappa},
\quad
A=6.6\pm0.6,
\quad
\kappa=6.66\pm0.03,
\label{suppeq:power law fit lambda'=0.05}
\end{equation}
for $2N=14,16,18,20,22$
is better than the fit with the exponential Ansatz
\begin{equation}
E^{\,}_{1}(2N)-E^{\,}_{0}(2N)=
A\,e^{-\frac{2N}{\xi}},
\qquad
A=(2.8\pm0.8)\cdot10^{-5},
\qquad
\label{suppeq:exp law fit lambda'=0.05}
\xi=2.63\pm0.07,
\end{equation}
We have verified that the dependence on $\lambda'$ of the scaling exponent
$\kappa$ in Eq.\ (\ref{suppeq:power law fit lambda'=0.05})
is weak, contrary to that of $A$.
Exact diagonalization,
the Lanczos algorithm,
and DMRG all measure
the decay
\begin{equation}
E^{\,}_{2}(2N)-E^{\,}_{1}(2N)\propto
E^{\,}_{4}(2N)-E^{\,}_{3}(2N)\propto
(2N)^{-1}
\end{equation}
along the interval $(\pi/4,1-\lambda^{\prime})$
with $0<\lambda^{\prime}\leq0.2$.
As two energy eigenvalues 
in a gapless spectrum
are said to be degenerate
provided their energy difference 
decays faster with the system size than the typical level spacing does,
we conclude that the interval
$(\pi/4,1-\lambda^{\prime})$ with $0<\lambda^{\prime}\leq0.2$
has a twofold degenerate gapless ground state.
Finally,
the energy spacing $E^{\,}_{3}(2N)-E^{\,}_{2}(2N)$ calculated
using the Lanczos algorithm for $2N=14,16,18,20,22$
is zero within the numerical accuracy.

\begin{figure}[t!]
\centering
\subfigure[]{\includegraphics[width=0.5\columnwidth]{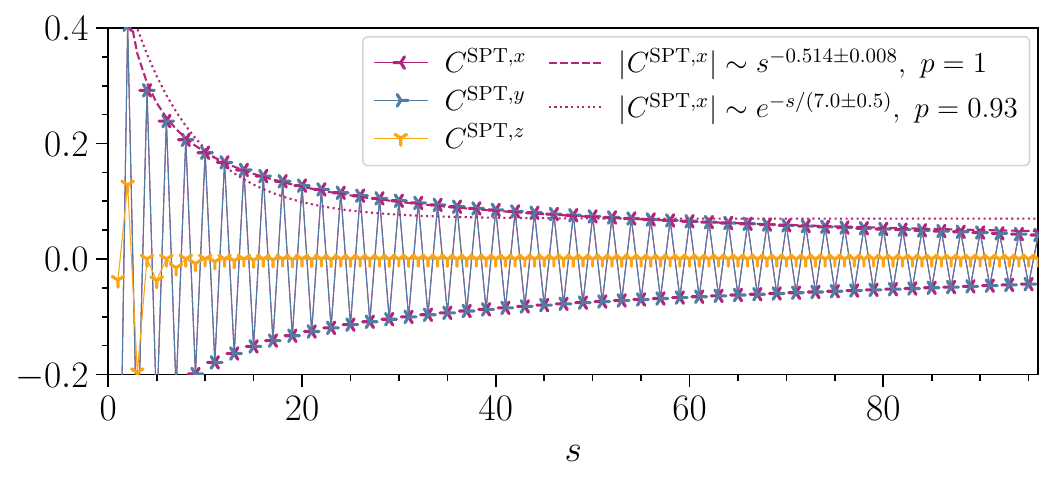} \label{suppfig:DMRG OBC corr 1,piover4 a}}%
\subfigure[]{\includegraphics[width=0.244\columnwidth]{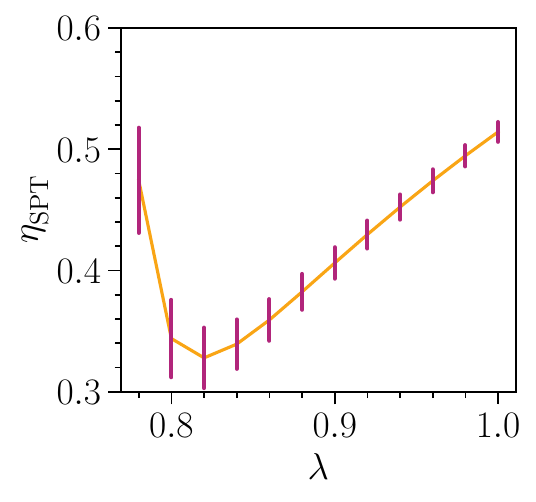} \label{suppfig:DMRG OBC corr 1,piover4 b}}
\caption{
(Color online)
The case of OBC.
(a)
Dependence on the separation $s$ in the
spin-spin correlation functions
$C^{\mathrm{SPT},\alpha}_{j^{\,}_{0},j^{\,}_{0}+s}[\Psi^{\,}_{0}]$
with $\alpha=x,y,z$
defined in Eq.~\eqref{suppeq:CalphaSPT} 
for $j^{\,}_{0}=(2N-96)/2$
computed using DMRG 
($\chi=128$ and $2N=128$)
at $(\pi/4,1)$ in Fig.~\ref{suppfig:phase diagram}.
The fit for
$C^{\mathrm{SPT},x}_{j^{\,}_{0},j^{\,}_{0}+s}[\Psi^{\,}_{0}]=
C^{\mathrm{SPT},y}_{j^{\,}_{0},j^{\,}_{0}+s}[\Psi^{\,}_{0}]$
is an algebraic decay.
The $p$-value
for each fit is given. 
(b)
Scaling
exponents $\eta^{\,}_{\mathrm{SPT}}(\lambda)$
for the algebraic decay of 
$C^{\mathrm{SPT},x}_{j^{\,}_{0},j^{\,}_{0}+s}
[\Psi^{\,}_{0}]=C^{\mathrm{SPT},y}_{j^{\,}_{0},j^{\,}_{0}+s}[\Psi^{\,}_{0}]\sim(-1)^{s} 
s^{-\eta^{\,}_{\mathrm{SPT}}(\lambda)}$
as a function of $\lambda$ holding $\theta=\pi/4$ fixed.
Uncertainties on $\eta^{\,}_{\mathrm{SPT}}(\lambda)$ are shown in magenta.
}
\label{suppfig:DMRG OBC corr 1,piover4}
\end{figure}

\paragraph{Correlations at the phase boundary ---}
The point $(\pi/4,1)$ realizes the DQCP between the
$\mathrm{\hbox{N\'eel}}^{\mathrm{SPT}}_{x}$
and
$\mathrm{\hbox{N\'eel}}^{\mathrm{SPT}}_{y}$ phases.
Its low-energy theory is described by the free-fermion CFT with
$\mathsf{c}=1$ when PBC are chosen
[recall Eqs.~\eqref{suppeq:two unitaries b} 
and~\eqref{suppeq:two symmetries of phase diagram}].

Numerical results confirm that the two-point correlation 
$C^{\mathrm{SPT},x}_{j^{\,}_{0},j^{\,}_{0}+s}[\Psi^{\,}_{0}]$ 
and 
$C^{\mathrm{SPT},y}_{j^{\,}_{0},j^{\,}_{0}+s}[\Psi^{\,}_{0}]$, 
defined in Eq.~\eqref{suppeq:CalphaSPT} for OBC,
are equal due to the $\mathrm{O}(2)$-symmetry of Hamiltonian
\eqref{suppeq:def H}
at the point $(\pi/4,1)$ from Fig.~\ref{suppfig:phase diagram}.
These correlations
follow the algebraic decay
\eqref{suppeq:algebraic decay transverse spin-spin in FMZ phase},
i.e.,
$(-1)^{s}s^{-\eta^{\,}_{\mathrm{SPT}}(\lambda=1)}$, 
with the exponent $\eta^{\,}_{\mathrm{SPT}}(\lambda=1)=1/2$
at $(\pi/4,1)$
[see Fig.~\ref{suppfig:DMRG OBC corr 1,piover4 a}].  

Figure~\ref{suppfig:DMRG OBC corr 1,piover4 b} 
shows the dependence on $\lambda$ of the scaling exponent
$\eta^{\,}_{\mathrm{SPT}}(\lambda)$
along the vertical boundary $(\pi/4,\lambda)$
from Fig.~\ref{suppfig:phase diagram}.
Close to $\lambda=1$, 
the exponent
$\eta^{\,}_{\mathrm{SPT}}(\lambda)=
\eta^{\,}_{\mathrm{SPT}}\big(1-(1-\lambda)\big)$
decreases linearly with $0\leq1-\lambda\ll1$
away from the value $1/2$.
This linear relationship, however, breaks down upon 
approaching the tricritical point
$\lambda^{\,}_{\mathrm{tri}}\approx0.75$
at which the
$\mathrm{\hbox{N\'eel}}^{\mathrm{SPT}}_{x}$,
$\mathrm{\hbox{N\'eel}}^{\mathrm{SPT}}_{y}$,
and $\mathrm{FM}^{\,}_{z}$
phases meet.
At around $\lambda=0.82$,
$\eta^{\,}_{\mathrm{SPT}}(\lambda)$
reaches a minimum. For smaller values of $\lambda$,
the scaling exponent $\eta^{\,}_{\mathrm{SPT}}(\lambda)$
is an increasing function of
$\lambda$ that converges smoothly to the value
$\eta^{\,}_{\mathrm{SPT}}(\lambda=1/2)=1$
at the center
$(\pi/4,1/2)$
of the phase diagram
\ref{suppfig:phase diagram}.

\begin{figure}[t!]
\centering
\subfigure[]{\includegraphics[width=0.34\columnwidth]{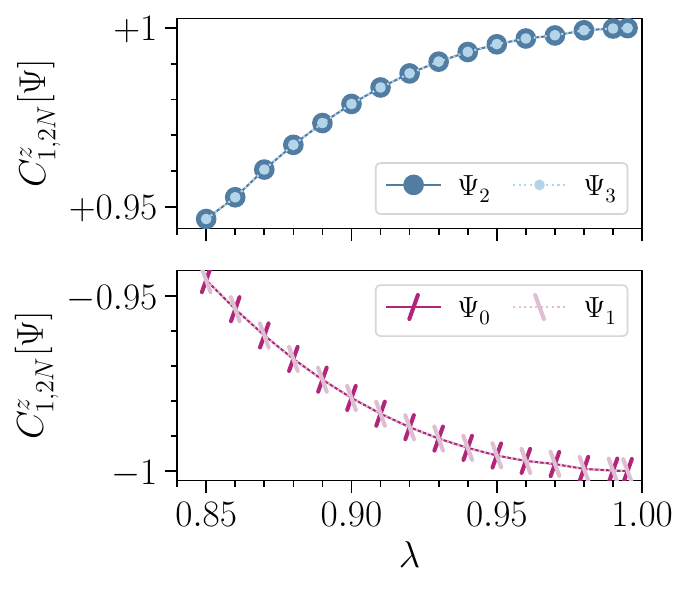} \label{suppfig:boundary spin Ising correlations a}}\quad
\subfigure[]{\includegraphics[width=0.431\columnwidth]{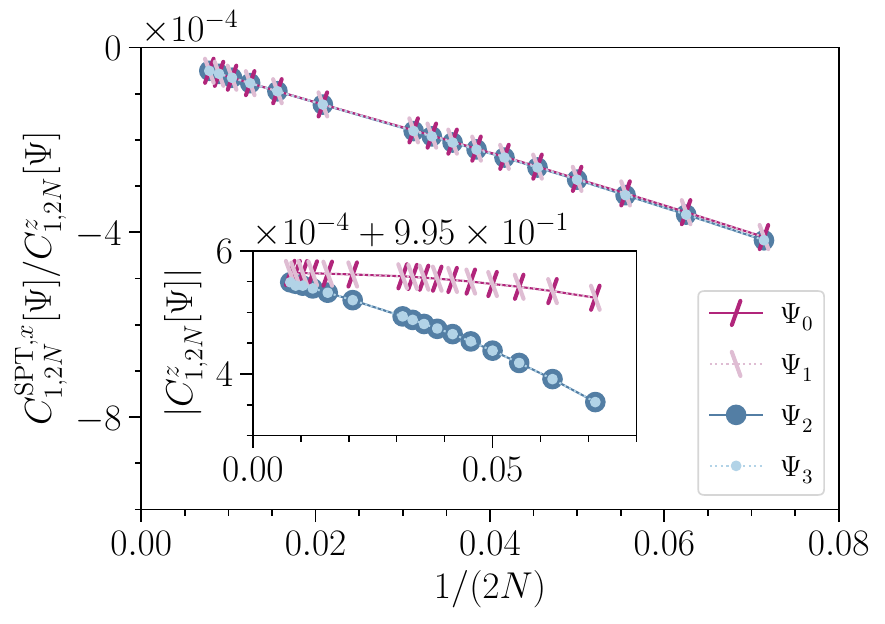} \label{suppfig:boundary spin Ising correlations b}}
\caption{
(Color online)
The case of OBC.
The symbol $\Psi$ labels any one of the first four energy eigenstates
$\Psi^{\,}_{0}$, $\Psi^{\,}_{1}$, $\Psi^{\,}_{2}$, and $\Psi^{\,}_{3}$
approximated by DMRG with $2N\in[14,128]$ and $\chi=128$. The value 
$\theta=\pi/4$ is fixed.
(a)
Two-point correlation function
$C^{z}_{1,2N}[\Psi]\equiv
\langle\Psi|\,\widehat{Z}^{\,}_{1}\,\widehat{Z}^{\,}_{2N}\,|\Psi\rangle$
as a function of $\lambda$ for $2N=64$.
(b)
Anisotropy ratio 
$C^{\mathrm{SPT},x}_{1,2N}[\Psi]/C^{z}_{1,2N}[\Psi]$
as a function of $1/(2N)$ at $(\pi/4,0.95)$
with $2N=14,16,18,20,22,24,26,28,30,32,48,64,80,96,112,128$,
where 
$C^{\mathrm{SPT},x}_{1,2N}[\Psi]\equiv
\langle\Psi|
(\widehat{X}^{\,}_{1}\,\widehat{Z}^{\,}_{2})\,
(\widehat{Z}^{\,}_{2N-1}\,\widehat{X}^{\,}_{2N})
|\Psi\rangle$.
A linear extrapolation
performed on each of the eigenstates $\Psi$
has the goodness of fit $R^{2}=0.9997$.
The inset shows the dependence of
$\big|C^{z}_{1,2N}[\Psi]\big|$ on $1/(2N)$ at $(\pi/4,0.95)$.
        } 
\label{suppfig:boundary spin Ising correlations}
\end{figure}

Whereas Fig.~\ref{suppfig:DMRG OBC corr 1,piover4}
presents the bulk SPT correlation functions,
Fig.~\ref{suppfig:boundary spin Ising correlations}
presents the correlations between
the two quantum spin-1/2 degrees of freedom 
\begin{equation}
\widehat{\bm{S}}^{\,}_{\mathrm{L}}\coloneqq
\begin{pmatrix}
\widehat{X}^{\,}_{1}\,\widehat{Z}^{\,}_{2}
\\
\widehat{Y}^{\,}_{1}\,\widehat{Z}^{\,}_{2}
\\
\widehat{Z}^{\,}_{1}
\end{pmatrix}
\equiv
\begin{pmatrix}
\widehat{S}^{X}_{\mathrm{L}}
\\
\widehat{S}^{Y}_{\mathrm{L}}
\\
\widehat{S}^{Z}_{\mathrm{L}}
\end{pmatrix},
\qquad
\widehat{\bm{S}}_{\mathrm{R}}\coloneqq
\begin{pmatrix}
\widehat{Z}^{\,}_{2N-1}\,
\widehat{X}^{\,}_{2N}
\\
\widehat{Z}^{\,}_{2N-1}\,
\widehat{Y}^{\,}_{2N}
\\
\widehat{Z}^{\,}_{2N}
\end{pmatrix}
\equiv
\begin{pmatrix}
\widehat{S}^{X}_{\mathrm{R}}
\\
\widehat{S}^{Y}_{\mathrm{R}}
\\
\widehat{S}^{Z}_{\mathrm{R}}
\end{pmatrix},
\label{suppeq:free spins}
\end{equation}
along the interval $(\pi/4,\lambda)$ with $0.85\leq\lambda<1$.
The quantum spin-1/2 operators
$\widehat{\bm{S}}^{\,}_{\mathrm{L}}$
and
$\widehat{\bm{S}}^{\,}_{\mathrm{R}}$
commute with the Hamiltonian
(\ref{suppeq:def H a})
at the point $(\pi/4,1)$
for any system size.
Any eigenvalue of Hamiltonian (\ref{suppeq:def H a}) must be at least
four-fold degenerate at the point $(\pi/4,1)$
for any system size.
Correspondingly, the spectrum is gapless with a minimal four-fold degeneracy
for all energy eigenvalues at the point $(\pi/4,1)$ in the thermodynamic limit.
Moreover, the two boundary operators
(\ref{suppeq:free spins}) are uncorrelated at the point $(\pi/4,1)$
for any system size. The quantum spin-1/2 operators
$\widehat{\bm{S}}^{\,}_{\mathrm{L}}$
and
$\widehat{\bm{S}}^{\,}_{\mathrm{R}}$
do not commute with the Hamiltonian
(\ref{suppeq:def H a})
along the interval $(\pi/4,\lambda)$ with $0\leq\lambda<1$.
Remarkably,
$\widehat{S}^{Z}_{\mathrm{L}}$ and $\widehat{S}^{Z}_{\mathrm{R}}$
are correlated until the tricritical point is reached in any one of the four
lowest lying energy eigenstates as is inferred from
Fig.~\ref{suppfig:boundary spin Ising correlations a}
when $2N=64$.
Indeed, according to
Fig.~\ref{suppfig:boundary spin Ising correlations a},
it is an empirical observation that
(nonvanishing antiferromagnetic Ising correlation)
\begin{subequations}
\begin{equation}
\langle\,\Psi^{\,}_{0}\,|\,
\widehat{S}^{Z}_{\mathrm{L}}\,
|\,\Psi^{\,}_{0}\,\rangle=
-
\langle\,\Psi^{\,}_{0}\,|\,
\widehat{S}^{Z}_{\mathrm{R}}\,
|\,\Psi^{\,}_{0}\,\rangle\neq0,
\qquad
\langle\,\Psi^{\,}_{1}\,|\,
\widehat{S}^{Z}_{\mathrm{L}}\,
|\,\Psi^{\,}_{1}\,\rangle=
-
\langle\,\Psi^{\,}_{1}\,|\,
\widehat{S}^{Z}_{\mathrm{R}}\,
|\,\Psi^{\,}_{1}\,\rangle\neq0,
\label{suppeq:AF ZZ bd correlation}
\end{equation}
while (nonvanishing ferromagnetic Ising correlation)
\begin{equation}
\langle\,\Psi^{\,}_{2}\,|\,
\widehat{S}^{Z}_{\mathrm{L}}\,
|\,\Psi^{\,}_{2}\,\rangle=
+
\langle\,\Psi^{\,}_{2}\,|\,
\widehat{S}^{Z}_{\mathrm{R}}\,
|\,\Psi^{\,}_{2}\,\rangle\neq0,
\qquad
\langle\,\Psi^{\,}_{3}\,|\,
\widehat{S}^{Z}_{\mathrm{L}}\,
|\,\Psi^{\,}_{3}\,\rangle=
+
\langle\,\Psi^{\,}_{3}\,|\,
\widehat{S}^{Z}_{\mathrm{R}}\,
|\,\Psi^{\,}_{3}\,\rangle\neq0,
\label{suppeq:F ZZ bd correlation}
\end{equation}
along the interval
$(\pi/4,\lambda)$ with $0.85\leq\lambda<1$ when $2N=64$.
Moreover, it is observed in the inset of
Fig.~\ref{suppfig:boundary spin Ising correlations b}
that
\begin{equation}
\big|\langle\,\Psi^{\,}_{0}\,|\,
\widehat{S}^{Z}_{\mathrm{L}}\,
|\,\Psi^{\,}_{0}\,\rangle\big|\approx
\big|\langle\,\Psi^{\,}_{1}\,|\,
\widehat{S}^{Z}_{\mathrm{L}}\,
|\,\Psi^{\,}_{1}\,\rangle\big|\approx
\big|\langle\,\Psi^{\,}_{2}\,|\,
\widehat{S}^{Z}_{\mathrm{L}}\,
|\,\Psi^{\,}_{2}\,\rangle\big|\approx
\big|\langle\,\Psi^{\,}_{3}\,|\,
\widehat{S}^{Z}_{\mathrm{L}}\,
|\,\Psi^{\,}_{3}\,\rangle\big|
\end{equation}
and the bounds
\begin{equation}
\sqrt{0.94}<
\big|
\langle\,\Psi^{\,}_{0}\,|\,
\widehat{S}^{Z}_{\mathrm{L}}\,
|\,\Psi^{\,}_{0}\,\rangle
\big|<1
\end{equation}
\end{subequations}
become independent of $2N$ for large $2N$
on the interval $(\pi/4,\lambda)$ with $0.85\leq\lambda<1$.
In contrast,
Fig.~\ref{suppfig:boundary spin Ising correlations b}
implies that 
\begin{equation}
\big|\langle\,\Psi^{\,}_{0}\,|\,
\widehat{S}^{X(Y)}_{\mathrm{L}}\,
|\,\Psi^{\,}_{0}\,\rangle\big|\approx
\big|\langle\,\Psi^{\,}_{1}\,|\,
\widehat{S}^{X(Y)}_{\mathrm{L}}\,
|\,\Psi^{\,}_{1}\,\rangle\big|\approx
\big|\langle\,\Psi^{\,}_{2}\,|\,
\widehat{S}^{X(Y)}_{\mathrm{L}}\,
|\,\Psi^{\,}_{2}\,\rangle\big|\approx
\big|\langle\,\Psi^{\,}_{3}\,|\,
\widehat{S}^{X(Y)}_{\mathrm{L}}\,
|\,\Psi^{\,}_{3}\,\rangle\big|,
\end{equation}
although non-vanishing for finite values of $2N$,
decay like $1/(2N)$ for $2N>12$.

Whereas the ``effective boundary Hamiltonian''
\begin{subequations}
\begin{equation}
\widehat{H}^{\,}_{\mathrm{eff}}(2N,\lambda)=
J^{XY}_{\mathrm{eff}}(2N,\lambda)\,
\left(
\widehat{S}^{X}_{\mathrm{L}}\,\widehat{S}^{X}_{\mathrm{R}}
+
\widehat{S}^{Y}_{\mathrm{L}}\,\widehat{S}^{Y}_{\mathrm{R}}
\right)
+
J^{Z}_{\mathrm{eff}}(2N,\lambda)\,  
\widehat{S}^{Z}_{\mathrm{L}}\,\widehat{S}^{Z}_{\mathrm{R}}
\end{equation}
with
\begin{equation}
J^{XY}_{\mathrm{eff}}(2N,\lambda)=
{A(\lambda)\over(2N)^{\kappa}},
\qquad
J^{Z}_{\mathrm{eff}}(2N,\lambda)=
{B(\lambda)\over2N},
\qquad
A(\lambda)\in\mathbb{R},
\qquad
B(\lambda)>0
\end{equation}
\end{subequations}
explains the dependence on $2N$ of the energy levels
$E^{\,}_{0}\leq E^{\,}_{1}\leq E^{\,}_{2}\leq E^{\,}_{3}$
along the interval $0<\lambda^{\prime}\leq0.2$ presented in
Fig.~\ref{suppfig:E1-E0 as a fct 2N using ED or Lanczos}, 
it fails to reproduce the dependence on $2N$ presented in
Fig.~\ref{suppfig:boundary spin Ising correlations}
of the correlation functions
$\langle\Psi|\,\widehat{Z}^{\,}_{1}\,\widehat{Z}^{\,}_{2N}\,|\Psi\rangle$
and
$\langle\Psi|
(\widehat{X}^{\,}_{1}\,\widehat{Z}^{\,}_{2})\,
(\widehat{Z}^{\,}_{2N-1}\,\widehat{X}^{\,}_{2N})
|\Psi\rangle$
with $\Psi$ labeling any one of the first four energy eigenstates labeled by
$\Psi^{\,}_{0}$, $\Psi^{\,}_{1}$, $\Psi^{\,}_{2}$, and $\Psi^{\,}_{3}$.

\begin{figure}[t!]
\centering
\subfigure[]{\includegraphics[width=0.36\columnwidth]{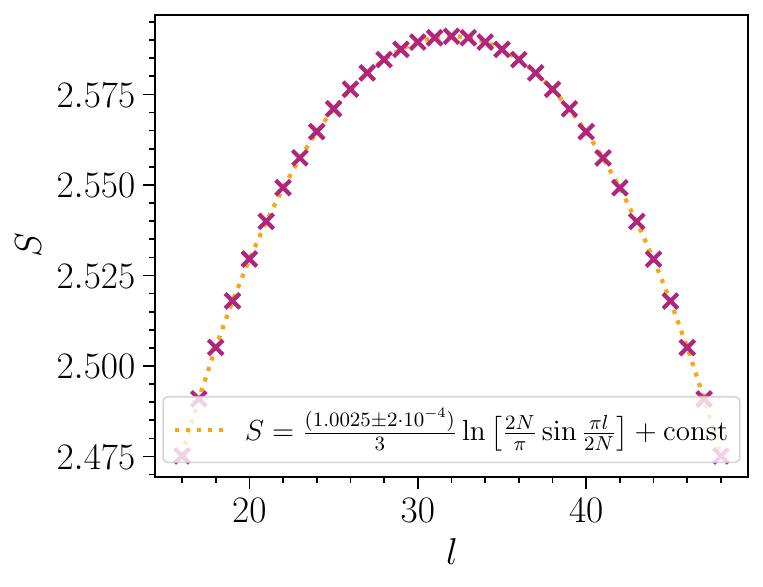} \label{suppfig:DMRG PBC central charge 1,piover4 a}}%
\subfigure[]{\includegraphics[width=0.3\columnwidth]{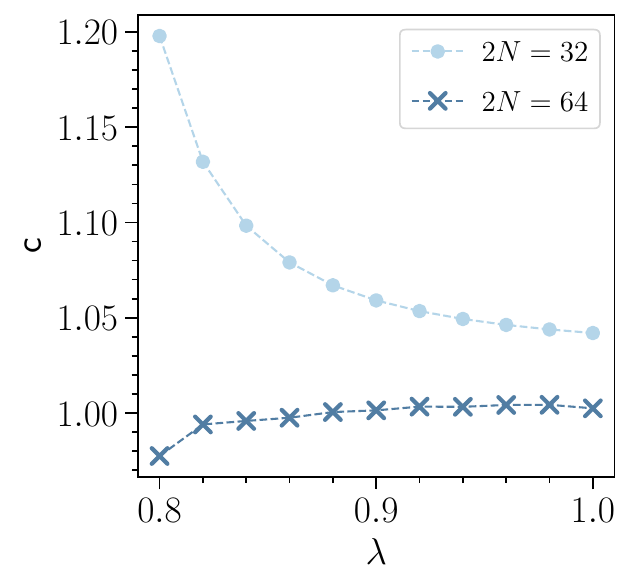} \label{suppfig:DMRG PBC central charge 1,piover4 b}}%
\subfigure[]{\includegraphics[width=0.293\columnwidth]{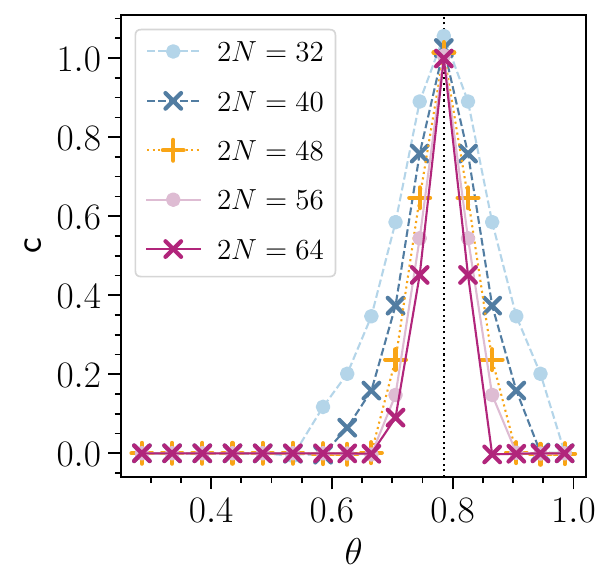} \label{suppfig:DMRG PBC central charge 1,piover4 c}}
\caption{
(Color online)
The case of PBC.
(a) Fit as a function of $l$ with the CFT Ansatz
\eqref{suppeq:CFT prediction for S2N(l)},
whereby the central charge is $\mathsf{c}^{\,}_{2N}=1.0025\pm2\cdot10^{-4}$,
to
$S^{\,}_{\mathrm{MPS},2N}(l)$
defined in Eq.~(\ref{suppeq:def S(ell) for MPS}) 
and obtained with DMRG ($\chi=256$ and $2N=64$) at $(\pi/4,1)$.
(b)
Central charge $\mathsf{c}^{\,}_{2N}$
as a function of $\lambda$ holding $\theta=\pi/4$ fixed
for different values of $2N$
computed by fitting the entanglement entropy $S^{\,}_{\mathrm{MPS},2N}(l)$.
(c)
Central charge $\mathsf{c}^{\,}_{2N}$
as a function of $\theta$ holding $\lambda=1$ fixed
for different values of $2N$
computed by fitting the entanglement entropy $S^{\,}_{\mathrm{MPS},2N}(l)$.
In panels (b) and (c),
$\delta{l}$ defined in Eq.~(\ref{suppeq:def delta{l}})
is $2N/2$.
}
\label{suppfig:DMRG PBC central charge 1,piover4}
\end{figure}

\paragraph{Central charge at the phase boundary ---}
The fixed point $(\pi/4,1)$
on the O(2)-symmetric vertical line $\theta=\pi/4$
from Fig.~\ref{suppfig:phase diagram} under PBC
realizes a $\mathsf{c}=1$ CFT
[recall Eqs.~\eqref{suppeq:two unitaries b} 
and~\eqref{suppeq:two symmetries of phase diagram}]. 
We have confirmed numerically this analytical prediction in
Fig.~\ref{suppfig:DMRG PBC central charge 1,piover4}
by computing the DMRG estimate for the central charge
$\mathsf{c}^{\,}_{2N}$ and then performing
the extrapolation
$\mathsf{c}\coloneqq\lim\limits_{2N\to\infty}\mathsf{c}^{\,}_{2N}$.
Figure~\ref{suppfig:DMRG PBC central charge 1,piover4 c}
is consistent with the limiting value
$\mathsf{c}(\theta,1)\coloneqq
\lim\limits_{2N\to\infty}\mathsf{c}^{\,}_{2N}(\theta,1)=0$
with $\theta\neq\pi/4$ in the closed interval $[0,\pi/2]$.
The DMRG estimates of $\mathsf{c}^{\,}_{2N}(\pi/4,\lambda)$
along the vertical boundary with $0.8\leq\lambda\leq1$
are shown in Fig.~\ref{suppfig:DMRG PBC central charge 1,piover4 b}.
They are consistent with the constant value
$\mathsf{c}(\pi/4,\lambda)=1$
for $0.8\leq\lambda\leq1$
in the thermodynamic limit $2N\to\infty$,
i.e., the same value found in
Sec.~\ref{suppeq:subsubsection Criticality}
for the central charge inside the $\mathrm{FM}^{\,}_{z}$ phase.
Finite-size effects do not allow a reliable estimate of
$\mathsf{c}(\pi/4,\lambda)$
in the window
$0.74<\lambda<0.8$
centered about the tricritical point
$(\pi/4,\lambda^{\,}_{\mathrm{tri}})$
with $\lambda^{\,}_{\mathrm{tri}}\approx0.75$.

\subsubsection{Between $\mathrm{\hbox{N\'eel}}^{\,}_{x}$ and
$\mathrm{\hbox{N\'eel}}^{\mathrm{SPT}}_{y}$}

The continuous phase transitions at
$(\pi/4,0)$ and $(\pi/4,1)$
in the phase diagram \ref{suppfig:phase diagram}
are both exactly soluble under PBC,
for (i) the Jordan-Wigner transformation
maps the quantum-spin-1/2 Hamiltonian
\eqref{suppeq:DQCP at 0 pi/4}
to a non-interacting chain of fermions hopping through a nearest-neighbor
hopping \cite{Lieb61}
and (ii) the lines $(\theta,0)$ and $(\theta,1)$
under PBC are unitarily equivalent.
The Jordan-Wigner transformation
applied to Hamiltonian
(\ref{suppeq:def H})
at $(0,1/2)$ delivers an interacting fermionic lattice Hamiltonian
that we could not solve exactly. We claim on the basis of
numerical arguments that the
$\mathrm{\hbox{N\'eel}}^{\,}_{x}$
and
$\mathrm{\hbox{N\'eel}}^{\mathrm{SPT}}_{y}$
phases meet at a single quantum critical point
that is located at $(0,1/2)$
along the line $(0,\lambda)$ with $0\leq\lambda\leq1$
in Fig.~\ref{suppfig:phase diagram}.
We were unable to compute the central charge at the putative
quantum critical point
at $(0,1/2)$
due to strong finite-size effects reminiscent of the strong
finite-size effects that prevented us from measuring the
central charges at the pair of tricritical points
$(\pi/4,\lambda^{\,}_{\mathrm{tri}})$
with
$\lambda^{\,}_{\mathrm{tri}}\approx0.25$
and
$\lambda^{\,}_{\mathrm{tri}}\approx0.75$.

\begin{figure}[t!]
\includegraphics[width=1\columnwidth]{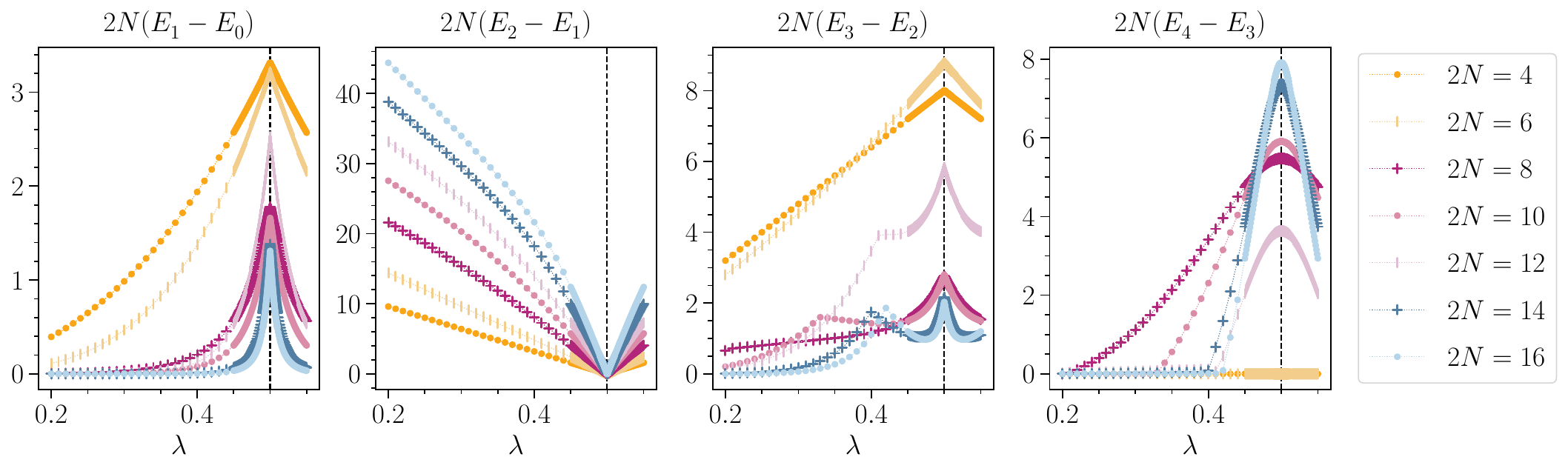}
\caption{
(Color online)
The case of PBC.
Dependence on $\lambda$ of the energy spacings
$E^{\,}_{n}-E^{\,}_{n-1}$, $n=1,2,3,4$, 
above the ground
state energy $E^{\,}_{0}$ obtained 
with ED at $\theta=0$. The fact
that all lines reach an extremum 
at $\lambda=1/2$ (dashed black line)
is consistent with a gap closing as $1/(2N)$ 
in the thermodynamic limit. However, 
oscillations as a function of $2N$ with the approximate
period $10$ prevent a quantitative measurement of the
``velocities'' 
$\partial(E^{\,}_{n}-E^{\,}_{n-1})/\partial(2N)^{-1}$.
}
\label{suppfig:ED evidence for DQCP 1over2,0}
\end{figure}

\paragraph{Location of the phase boundary ---}
When PBC are chosen, each corner 
of the phase diagram in Fig.~\ref{suppfig:phase diagram}
has twofold degenerate ground states
prior to taking the thermodynamic limit. This twofold degeneracy persists
after the thermodynamic limit has been taken
in all the gapped phases controlled by the four corners of the
the phase diagram \ref{suppfig:phase diagram}.
Hence, the energy difference
\begin{equation}
E^{\,}_{1}-E^{\,}_{0}
\end{equation}
between the first excited energy eigenstate with 
energy $E^{\,}_{1}$ and the ground state 
with energy $E^{\,}_{0}$ in either the 
$\mathrm{\hbox{N\'eel}}^{\,}_{x}$ or the $\mathrm{\hbox{N\'eel}}^{\mathrm{SPT}}_{y}$ phase 
is a function of the length $L=2N\,\mathfrak{a}$ of the chain
that converges to the value zero
\begin{enumerate}
\item[(i)]
exponentially fast with the length $L$ 
of the chain in the thermodynamic limit $L\to\infty$ deep in a gapped phase,
\item[(ii)]
algebraically fast as $L^{-z}$ in the thermodynamic limit $L\to\infty$ 
upon exiting the gapped phase through a gap-closing transition.
\end{enumerate}
A gap-closing transition between two 
gapped phases of matter can thus be identified 
through a cusp singularity when 
plotting the rescaled energy difference
\begin{equation}
(2N)^{z}\,(E^{\,}_{1}-E^{\,}_{0}),
\end{equation}
for some value of the dynamical exponent $z$
as a function of the coupling driving the gap-closing phase transition. 

In Fig.~\ref{suppfig:ED evidence for DQCP 1over2,0}, 
we present the dependence on $\lambda$ of 
\begin{equation}
(2N)\,
\left[
E^{\,}_{1}(0,\lambda)
-
E^{\,}_{0}(0,\lambda)
\right]
\end{equation}
under PBC at fixed values of $\theta=0$ and $2N$. 
The presence of a cusp at $\lambda=1/2$
is consistent with the existence of a quantum phase transition
with the dynamical exponent $z=1$
at $(0,1/2)$ along the line $(0,\lambda)$ with $0\leq\lambda\leq1$  
separating two gapped phases, each one being twofold degenerate.
The cusp values taken by
\begin{equation}
(2N)\,
\left[
E^{\,}_{1}(0,1/2)
-
E^{\,}_{0}(0,1/2)
\right]
\label{suppeq:cusp values at (0,1/2)}
\end{equation}
oscillate as a function of $2N$
with an approximate period of 10.
This is a sharp difference with the dependence on $2N$ of
\begin{equation}
(2N)\,
\left[
E^{\,}_{1}(\pi/4,0)
-
E^{\,}_{0}(\pi/4,0)
\right]
\label{suppeq:cusp values at (pi/4,0)}
\end{equation}
from Fig.~\ref{suppfig:ED evidence for DQCP 0,piover4},
which is strictly monotonic and rapidly convergent
as a function of $2N$.
We attribute the existence of a periodicity of approximately 10
as a function of $2N$
in the values (\ref{suppeq:cusp values at (0,1/2)})
as a signature of incommensuration effects
that are related to the incommensuration effects
observed in the $\mathrm{FM}^{\,}_{z}$ phase, for they share the same periodicity 
(see Sec.~\ref{suppsec:LRO IC FMZ}).
These oscillations impede the accurate estimate of the
``velocities''
\begin{equation}
v^{\,}_{n,n-1}(0,1/2)=
\frac{
\partial
\left[E^{\,}_{n}(0,1/2)-E^{\,}_{n-1}(0,1/2)\right]
}
{
\partial\,
(2N)^{-1}
},
\qquad
n=1,2,3,\cdots.
\end{equation}

\begin{figure}[t!]
\centering
\includegraphics[width=0.44\columnwidth]{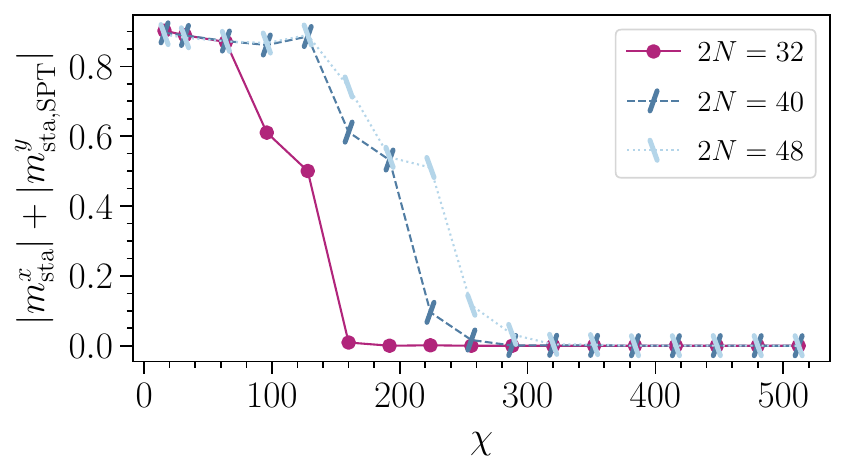}
\caption{
(Color online)
The case of PBC. DMRG estimate for the
absolute value $|m^{x}_{\mathrm{sta}}[\Psi^{\,}_{0}]|
+
|m^{y}_{\mathrm{sta,SPT}}[\Psi^{\,}_{0}]|$ 
of the relevant order parameters at $(0,1/2)$ at 
fixed system sizes $2N$ as a function of the bond
dimension $\chi$.
}
\label{suppfig:DMRG OBC OP 0,1over2 chi}
\end{figure}

The severity of the finite-size effects at
$(0,1/2)$ relative to those at $(\pi/4,0)$
is captured by comparing
Fig.~\ref{suppfig:DMRG OBC OP 0,1over2 chi}
against
Fig.~\ref{suppfig:DMRG OBC OP piover4,0 chi}.
The bond dimension $\chi$ needed to make sure that
it is the regime (\ref{suppeq:three regimes close to criticality b})  
and not the regime (\ref{suppeq:three regimes close to criticality c})
that holds when estimating the value of the order parameter at
criticality is larger than 30 for $2N=64$ in
Fig.~\ref{suppfig:DMRG OBC OP piover4,0 chi},
whereas it is larger than 300 for $2N=48$ in
Fig.~\ref{suppfig:DMRG OBC OP 0,1over2 chi}.
This why in Fig.~\ref{suppfig:DMRG OP 0,1over2 a}
when OBC are selected,
for lack of sufficiently large bond dimension $\chi$,
there is a window for the values of $\lambda$ centered about $\lambda=1/2$
for which the competing order parameters
$\mathrm{\hbox{N\'eel}}^{\,}_{x}$ and $\mathrm{\hbox{N\'eel}}^{\mathrm{SPT}}_{y}$
are simultaneously non-vanishing.
This problem is less severe in
Fig.~\ref{suppfig:DMRG OP 0,1over2 b}
when PBC are selected.
The strong finite size effects at $(0,1/2)$ are thus tied to
the existence of boundary states in the
$\mathrm{\hbox{N\'eel}}^{\mathrm{SPT}}_{y}$
phase.

Another important finite-size effect present in
Fig.~\ref{suppfig:DMRG OP 0,1over2 a}
but absent in
Fig.~\ref{suppfig:DMRG OP 0,1over2 b}
is that each window for which the
$\mathrm{\hbox{N\'eel}}^{\,}_{x}$ and $\mathrm{\hbox{N\'eel}}^{\mathrm{SPT}}_{y}$
order parameters coexist for a fixed system size in
Fig.~\ref{suppfig:DMRG OP 0,1over2 a}
is centered at a value $\lambda^{\,}_{2N,\star}$ larger than
$\lambda^{\,}_{\mathrm{c}}=1/2$, whereby the extrapolation
$\lim\limits_{2N\to\infty}\lambda^{\,}_{2N,\star}$ is consistent with the
limiting value $\lambda^{\,}_{\mathrm{c}}=1/2$.
In Fig.~\ref{suppfig:DMRG OP 0,1over2 b},
each window for which the
$\mathrm{\hbox{N\'eel}}^{\,}_{x}$ and $\mathrm{\hbox{N\'eel}}^{\mathrm{SPT}}_{y}$
order parameters simultaneously vanish for a fixed system size
is centered at $\lambda^{\,}_{\mathrm{c}}=1/2$.
This difference is not an artifact of DMRG.
It is already present with ED when comparing
Fig.~\ref{suppfig:ed_10 a}
and
Fig.~\ref{suppfig:ed_10 b}.
In Fig.~\ref{suppfig:ed_10 a},
the area of the
$\mathrm{\hbox{N\'eel}}^{\,}_{x}$
and
$\mathrm{\hbox{N\'eel}}^{\mathrm{SPT}}_{y}$
``phases'' are approximately the same,
with their ``phase boundary'' approximately
centered at $\lambda^{\,}_{\mathrm{c}}=1/2$.
In Fig.~\ref{suppfig:ed_10 b},
the area of the
$\mathrm{\hbox{N\'eel}}^{\,}_{x}$ ``phase''
is larger than that of the
$\mathrm{\hbox{N\'eel}}^{\mathrm{SPT}}_{y}$ ``phase.''
Their ``phase boundary'' is located 
at the value $\lambda^{\,}_{2N,\star}>\lambda^{\,}_{\mathrm{c}}=1/2$,
with $\lambda^{\,}_{2N,\star}$ a decreasing function of $2N$ for the values
of $2N$ for which ED is feasible. Hereto, we anticipate
$\lim\limits_{2N\to\infty}\lambda^{\,}_{2N,\star}=\lambda^{\,}_{\mathrm{c}}=1/2$,
it is only in the thermodynamic limit $2N\to\infty$
that the difference in the number of
terms present in
\begin{subequations}
\begin{equation}
\sum_{j=1}^{2N-b}
\left(
\cos\theta\,
\widehat{X}^{\,}_{j}\,
\widehat{X}^{\,}_{j+1}
+
\sin\theta\,
\widehat{Y}^{\,}_{j}\,
\widehat{Y}^{\,}_{j+1}
\right)
\end{equation}
and in
\begin{equation}
\sum_{j=1}^{2N-3b}
\widehat{Z}^{\,}_{j}\,
\left(
\cos\theta\,
\widehat{X}^{\,}_{j+1}\,
\widehat{X}^{\,}_{j+2}
+
\sin\theta\,
\widehat{Y}^{\,}_{j+1}\,
\widehat{Y}^{\,}_{j+2}
\right)
\widehat{Z}^{\,}_{j+3},
\end{equation}
\end{subequations}
becomes immaterial when OBC are selected through the choice $b=1$
in Hamiltonian (\ref{suppeq:def H a}).

By the symmetry arguments of
Sec.~\ref{suppsec:Symmetries},
we expect the phase boundary between the
$\mathrm{\hbox{N\'eel}}^{\,}_{x}$
and
$\mathrm{\hbox{N\'eel}}^{\mathrm{SPT}}_{y}$
phases to be the horizontal interval
$(\theta,1/2)$, $0\leq\theta\leq\theta^{\,}_{\mathrm{tri}}$,
with $\theta^{\,}_{\mathrm{tri}}\approx0.3$
in Fig.~\ref{suppfig:phase diagram}.
We have verified this expectation by computing
$E^{\,}_{1}-E^{\,}_{0}$ under PBC and the
order parameters under OBC.

\paragraph{Degeneracies at the phase boundary ---}
Finite-size effects under OBC
were too strong to study the ground-state degeneracy
for any point along the phase boundary
$(\theta,1/2)$, $0\leq\theta\leq\theta^{\,}_{\mathrm{tri}}$.
The ground states are non-degenerate under PBC
for any point along the phase boundary
$(\theta,1/2)$, $0\leq\theta\leq\theta^{\,}_{\mathrm{tri}}$,
by the symmetry arguments of
Sec.~\ref{suppsec:Symmetries}.

\paragraph{Correlations at the phase boundary ---}
Finite-size effects under OBC and PBC
were too strong to study the
correlation functions
for any point along the phase boundary
$(\theta,1/2)$, $0\leq\theta\leq\theta^{\,}_{\mathrm{tri}}$.

\begin{figure}[t!]
\subfigure[]{\includegraphics[width=0.5\columnwidth]{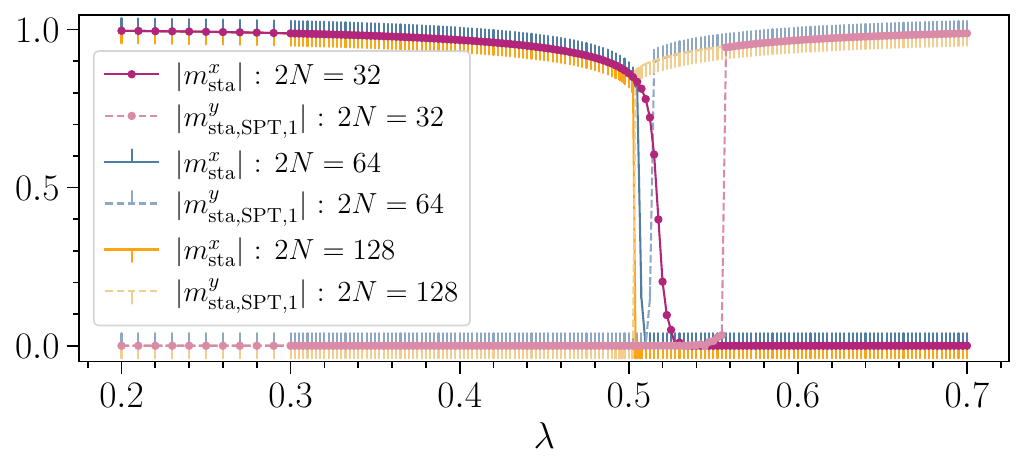} \label{suppfig:DMRG OP 0,1over2 a}}%
\subfigure[]{\includegraphics[width=0.5\columnwidth]{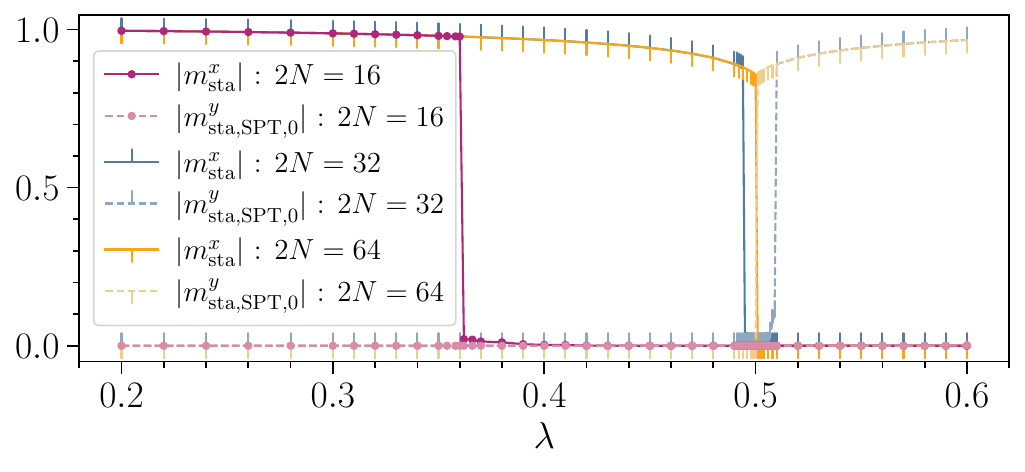} \label{suppfig:DMRG OP 0,1over2 b}}%
\caption{
(Color online)
Relevant (non-vanishing) order parameters 
defined in Eq.~\eqref{suppeq:def order parameters} 
obtained with DMRG across a cut around $\lambda=1/2$ at $\theta=0$ 
(a) 
with OBC ($\chi = 128$) and
(b)
with PBC ($\chi=256$).
}
\label{suppfig:DMRG OP 0,1over2}
\end{figure}

\begin{figure}[t!]
\subfigure[]{\includegraphics[width=0.27\columnwidth]{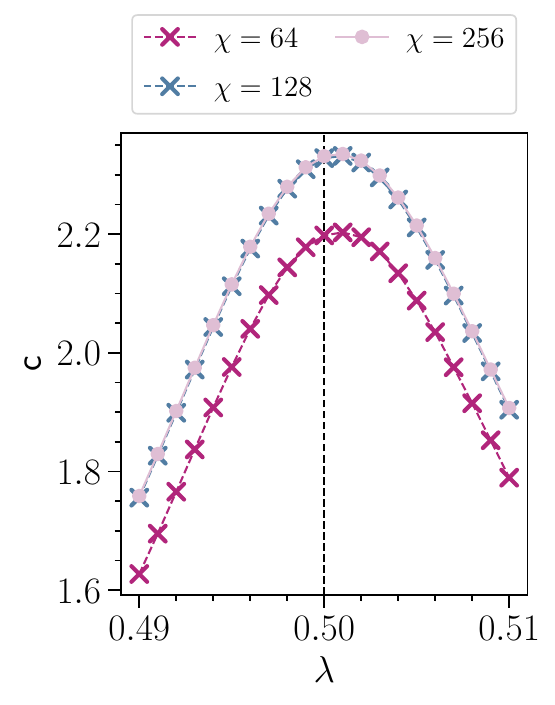} \label{suppfig:DMRG PBC central charge 1over2,0 a}}%
\hspace{-6pt}%
\subfigure[]{\includegraphics[width=0.27\columnwidth]{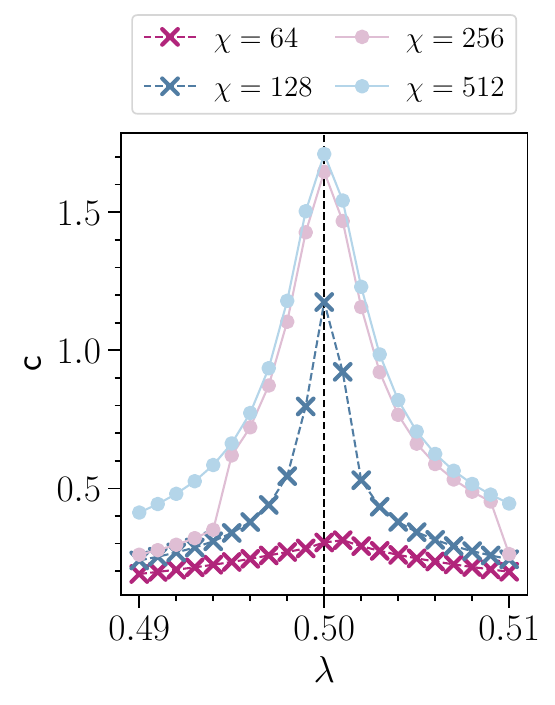} \label{suppfig:DMRG PBC central charge 1over2,0 b}}%
\hspace{-6pt}%
\subfigure[]{\includegraphics[width=0.27\columnwidth]{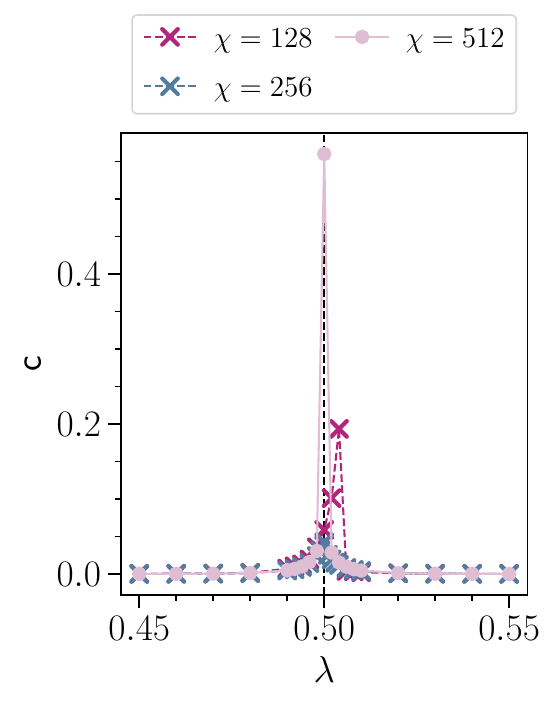} \label{suppfig:DMRG PBC central charge 1over2,0 c}}\\
\subfigure[]{\includegraphics[width=0.55\columnwidth]{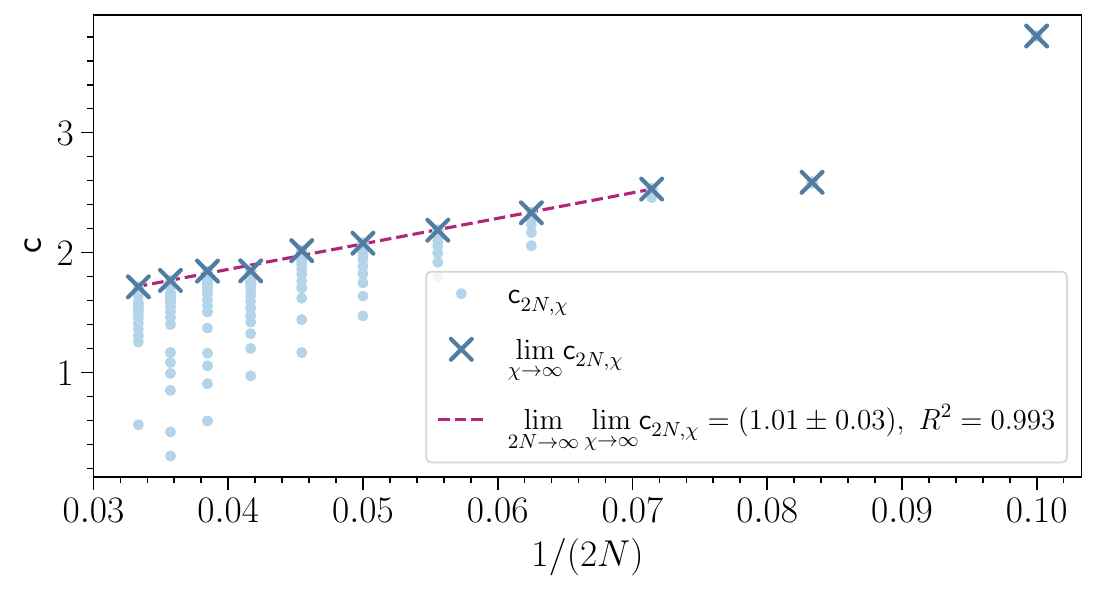} \label{suppfig:DMRG PBC central charge 1over2,0 d}}%
\caption{
(Color online)
The case of PBC.
Central charge $\mathsf{c}$ obtained with 
DMRG and PBC at $\theta=0$ for different $\lambda$
and bond dimensions $\chi$, and with 
(a) $2N=16$, 
(b) $2N=32$ and
(c) $2N=64$.
The dashed black line indicates the location of the
boundary between the $\mathrm{\hbox{N\'eel}}^{\,}_{x}$ and
the $\mathrm{\hbox{N\'eel}}^{\mathrm{SPT}}_{y}$ phases.
(d) Central charge $\mathsf{c}$ obtained with 
DMRG and PBC at $(0,1/2)$ in
Fig.~\ref{suppfig:phase diagram}
for different 
bond dimensions
$\chi\in[50,1200]$
and system sizes
$2N=10,12,\cdots,28,30$.
An extrapolation in the limit $2N\to\infty$ 
of the central charges $\lim^{\,}_{\chi\to\infty} 
\mathsf{c}^{\,}_{2N,\chi}$ 
obtained for each size $2N$ is performed.
For all panels, the value of
$\delta{l}$ defined in Eq.~(\ref{suppeq:def delta{l}})
is $2N/2$.
}
\label{suppfig:DMRG PBC central charge 1over2,0}
\end{figure}

\paragraph{Central charge at the phase boundary ---}
The strong finite-size effects
present in Figs.~\ref{suppfig:ED evidence for DQCP 1over2,0},~\ref{suppfig:DMRG OBC OP 0,1over2 chi},
and~\ref{suppfig:DMRG OP 0,1over2}
are also present in
Fig.~\ref{suppfig:DMRG PBC central charge 1over2,0}.
Figure~\ref{suppfig:DMRG PBC central charge 1over2,0}
is an estimate of the central charge
$\mathsf{c}(0,1/2)$
at the quantum critical
point $(0,1/2)$ along the line $(0,\lambda)$ with $0\leq\lambda\leq1$.
The computation of the central charge using DMRG involves taking the double
limit
\begin{equation}
\mathsf{c}(\theta,\lambda)=
\lim_{2N\to\infty}
\lim_{\chi\to\chi^{\,}_{2N}}
\mathsf{c}^{\,}_{2N,\chi}(\theta,\lambda),
\label{suppeq:c double limit}
\end{equation}
where, for given $2N$, $\chi^{\,}_{2N}$
is the bond dimension for which the regime
(\ref{suppeq:three regimes close to criticality b})  
holds. According to
Fig.~\ref{suppfig:DMRG PBC central charge 1over2,0}:
\begin{itemize}
\item[1.]
For $2N=16$ and $2N=32$, the central charge
$\mathsf{c}^{\,}_{2N,\chi}(0,\lambda)$
increases with increasing $\chi$ 
to some limiting value
for each $\lambda$.
\item[2.]
For $2N=64$,
the value
$\mathsf{c}^{\,}_{2N,\chi}(0,\lambda)$  
for a given $\lambda$ as a function of $\chi$
is non-monotonic in $\chi$.
\item[3.]
The extrapolation (\ref{suppeq:c double limit})
based on the system sizes $2N=10,12,\cdots,28,30$
with $\chi\in[50,1200]$ gives the value
\begin{equation}
\mathsf{c}(0,1/2)=1.01\pm0.03.
\label{suppeq:DMRG estimate mathsf{c}(0,1/2)}  
\end{equation}
\end{itemize}
Our computational resources do not allow a reliable extrapolation
of $\mathsf{c}^{\,}_{2N,\chi}(0,1/2)$
to the thermodynamic limit
$2N\to\infty$ close to the quantum critical point $(0,1/2)$
for the following reasons.
First, the choice for the range $\delta{l}$
of the values of the bonds $l$ in
Eq.~(\ref{suppeq:def S(ell) for MPS})
over which the CFT fit
(\ref{suppeq:CFT prediction for S2N(l) b})
is used [recall Eq.~(\ref{suppeq:def delta{l}})]
can be the cause of a systematic error that can be as large as 30$\%$ 
[as opposed to 1$\%$ at the point $(\pi/4,0)$
and 3$\%$ at the point $(\pi/4,1)$]
of the value
(\ref{suppeq:DMRG estimate mathsf{c}(0,1/2)}).
This systematic error dominates when $\delta{l}$ is
greater than $6N/4$, while the uncertainty on the CFT fit
(\ref{suppeq:CFT prediction for S2N(l) b}) dominates when
$\delta{l}$ is smaller than $6N/4$.
Second, we could not verify that the extrapolation
(\ref{suppeq:c double limit})
based on the system sizes $2N=32,34,\cdots,62,64$
agrees with that done in 
Fig.~\ref{suppfig:DMRG PBC central charge 1over2,0 d}.

\subsubsection{Between $\mathrm{\hbox{N\'eel}}^{\,}_{x}$ and $\mathrm{FM}^{\,}_{z}$}

We have argued for the existence of a 
fifth diamond-shaped phase, labeled $\mathrm{FM}^{\,}_{z}$, 
which supports simultaneously long-range ferromagnetic order 
along the $Z$-axis in spin space
and algebraic AF order along the $X$- and $Y$-axis in spin space.
With PBC,
the phase diagram is symmetric with respect to a 
$\pi$-rotation about the center of the 
$\mathrm{FM}^{\,}_{z}$ phase $(\pi/4,1/2)$ 
(as shown in Sec.~\ref{suppsec:Symmetries}).
Without loss of generality,
we are going to locate and characterize the transition
between the $\mathrm{\hbox{N\'eel}}^{\,}_{x}$ and $\mathrm{FM}^{\,}_{z}$ phases.

\paragraph{Location of the phase boundary ---}
When PBC are chosen, each corner 
of the phase diagram in Fig.~\ref{suppfig:phase diagram}
has twofold degenerate ground states
prior to taking the thermodynamic limit. This twofold degeneracy persists
after the thermodynamic limit has been taken
in all the gapped phases controlled by the four corners of the
the phase diagram \ref{suppfig:phase diagram}.
Moreover, after taking the thermodynamic limit,
the $\mathrm{FM}^{\,}_{z}$ phase
in Fig.~\ref{suppfig:phase diagram}
has a two-fold degenerate ground state for both PBC and OBC.
Hence, the energy difference
\begin{equation}
E^{\,}_{1}-E^{\,}_{0}
\end{equation}
between the first excited energy eigenstate with 
energy $E^{\,}_{1}$ and the ground state 
with energy $E^{\,}_{0}$ in either the 
$\mathrm{\hbox{N\'eel}}^{\,}_{x}$ or the $\mathrm{FM}^{\,}_{z}$ phase 
is a function of the length $L=2N\,\mathfrak{a}$ of the chain
that converges to the value zero
\begin{enumerate}
\item[(i)]
exponentially fast with the length $L$ 
of the chain in the thermodynamic limit $L\to\infty$ 
in either of the $\mathrm{\hbox{N\'eel}}^{\,}_{x}$ or the $\mathrm{FM}^{\,}_{z}$ phase,
\item[(ii)]
algebraically fast as $L^{-z}$ in the thermodynamic limit $L\to\infty$ 
upon exiting the gapped phase through a gap-closing transition.
\end{enumerate}
A gap-closing transition between two 
phases of matter can thus be identified 
through a cusp singularity when 
plotting the rescaled energy difference
\begin{equation}
(2N)^{z}\,(E^{\,}_{1}-E^{\,}_{0}),
\end{equation}
for some value of the dynamical exponent $z$
as a function of the coupling driving the gap-closing phase transition. 

\begin{figure}[t!]
\centering
\includegraphics[width=1\columnwidth]{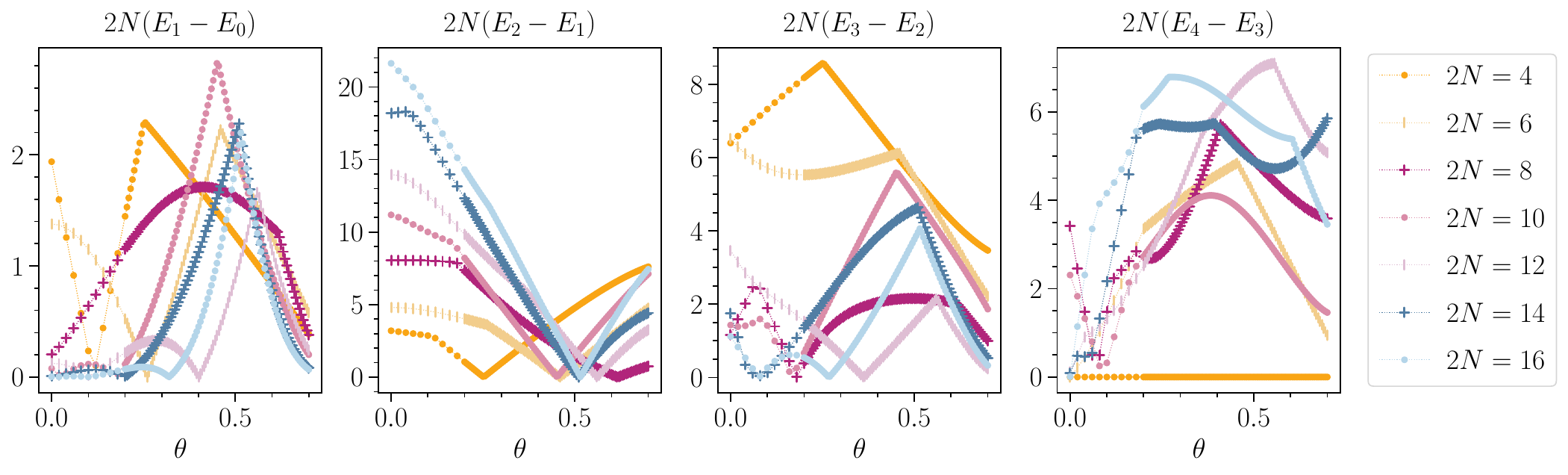}
\caption{
(Color online)
The case of PBC. Dependence on $\theta$ of the energy spacings
$E^{\,}_{n}-E^{\,}_{n-1}$, $n=1,2,3,4$, 
above the ground state energy $E^{\,}_{0}$ obtained with ED
at $\lambda=0.4$. All lines reach an extremum at 
different values of $\theta$, unlike for
Figs.~\ref{suppfig:ED evidence for DQCP 0,piover4}
and~\ref{suppfig:ED evidence for DQCP 1over2,0}.
}
\label{suppfig:ED evidence for DQCP 0.4}
\end{figure}

In Fig.~\ref{suppfig:ED evidence for DQCP 0.4}, 
we present the dependence on $\theta$ of 
\begin{equation}
(2N)\,
\left[
E^{\,}_{1}(\theta,0.4)
-
E^{\,}_{0}(\theta,0.4)
\right]
\end{equation}
under PBC
at fixed values of $\lambda=0.4$ and $2N$.
We observe that:
\begin{itemize}
\item[(i)]
The cusps in $2N(E^{\,}_{1}-E^{\,}_{0})$ vary 
in height with $2N$. This is
similar to the transition between the $\mathrm{\hbox{N\'eel}}^{\,}_{x}$ and
$\mathrm{\hbox{N\'eel}}^{\mathrm{SPT}}_{y}$ phases in Fig.~\ref{suppfig:ED evidence for DQCP 1over2,0}.
\item[(ii)]
The positions of the cusps in $2N(E^{\,}_{1}-E^{\,}_{0})$ 
occur at different values of $\theta$
for different values of $2N$.
This is unlike in Figs.~\ref{suppfig:ED evidence for DQCP 0,piover4}
and~\ref{suppfig:ED evidence for DQCP 1over2,0}.
\end{itemize}
The same behavior is observed for the energy spacings
$2N(E^{\,}_{2}-E^{\,}_{1})$, 
$2N(E^{\,}_{3}-E^{\,}_{2})$, and
$2N(E^{\,}_{4}-E^{\,}_{3})$
in Fig.~\ref{suppfig:ED evidence for DQCP 0.4}.
We interpret features (i) and (ii) as the signature of
incommensuration accompanied by strong finite-size effects.
These finite-size effects combined with our finite computing resources
prevent us from measuring precisely the position of the
putative continuous transition between the
$\mathrm{\hbox{N\'eel}}^{\,}_{x}$
and
$\mathrm{\hbox{N\'eel}}^{\mathrm{SPT}}_{y}$ phases using ED.

\begin{figure}[t!]
\centering
\includegraphics[width=0.5\columnwidth]{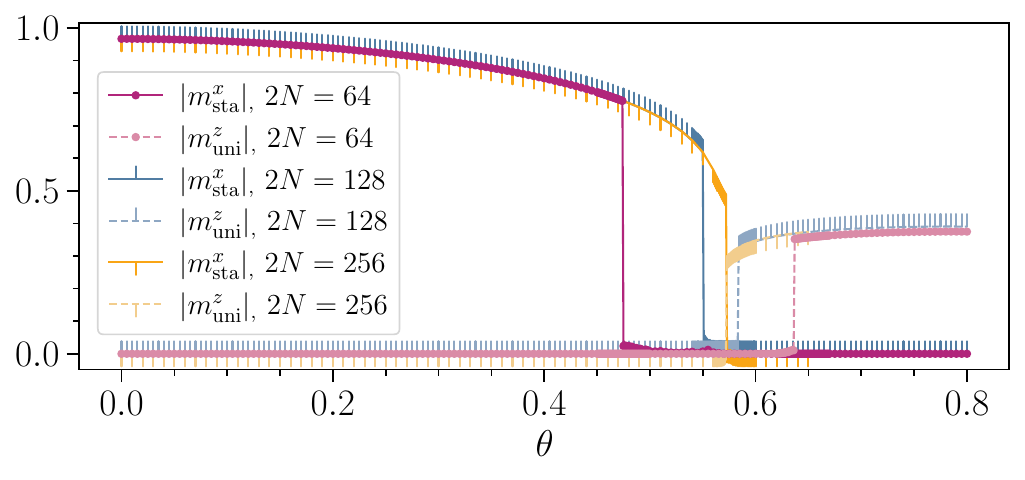}
\caption{
(Color online)
The case of OBC.
Magnitudes of the staggered magnetization per site along the
$X$-axis \eqref{suppeq:msta} and the uniform magnetization per site along
the $Z$-axis \eqref{suppeq:muni}, obtained along the line $\lambda=0.4$
using DMRG ($\chi=128$) for $N=64,128,256$ sites.
The sharp suppression
of the order parameters on both sides of the quantum phase transition
around $\theta=0.57$ indicates that the correlation lengths
in the gapped phases are
larger than the length of the chain
sufficiently close to the phase boundary.
}
\label{suppfig:DMRG OBC OP 0.4}
\end{figure}

The phase transition from $\mathrm{\hbox{N\'eel}}^{\,}_{x}$ to $\mathrm{FM}^{\,}_{z}$ 
phases can, however, be located by measuring
the order parameters 
$m^{x}_{\mathrm{sta}}[\Psi^{\,}_{0}]$
and 
$m^{z}_{\mathrm{uni}}[\Psi^{\,}_{0}]$
defined in
Eqs.~(\ref{suppeq:msta})
and~(\ref{suppeq:muni}),
respectively.
Figure~\ref{suppfig:DMRG OBC OP 0.4}
presents the dependence on $\theta$ of these order parameters
along the horizontal cut $(\theta,0.4)$, $0\leq\theta\leq0.8$,
in the phase diagram \ref{suppfig:phase diagram}
for different values of the number $2N$ of sites 
when OBC are selected.
The DMRG results obtained for the numbers
$2N=64,128$ of sites suggest that
(i)
there is a window 
\begin{equation}
\theta\in
\left[
\theta^{\,}_{\mathrm{c}}-(\delta\theta^{\,}_{\mathrm{c}})^{\,}_{2N},
\theta^{\,}_{\mathrm{c}}+(\delta\theta^{\,}_{\mathrm{c}})^{\,}_{2N}
\right]
\end{equation}
of size $2(\delta\theta^{\,}_{\mathrm{c}})^{\,}_{2N}$
along the horizontal cut
in which the
order parameters
$m^{x}_{\mathrm{sta}}[\Psi^{\,}_{0}]$
and 
$m^{z}_{\mathrm{uni}}[\Psi^{\,}_{0}]$
vanish
[recall the regime \eqref{suppeq:three regimes close to criticality b}]
and
(ii)
the size $2(\delta\theta^{\,}_{\mathrm{c}})^{\,}_{2N}$ of this window
decreases monotonically with
increasing $2N$ with the extrapolated limiting value
\begin{equation}
\lim_{2N\to\infty} 2(\delta\theta^{\,}_{\mathrm{c}})^{\,}_{2N}=0.
\end{equation}
Hereto, there are strong finite-size effect.
For example,
the DMRG results obtained for the number 
$2N=256$ of sites
follow the regime \eqref{suppeq:three regimes close to criticality c},
i.e., both order parameters
$m^{x}_{\mathrm{sta}}[\Psi^{\,}_{0}]$
and 
$m^{z}_{\mathrm{uni}}[\Psi^{\,}_{0}]$
are non-vanishing at the point
$(\theta^{\,}_{\mathrm{c}},0.4)$,
where $\theta^{\,}_{\mathrm{c}}\approx0.57$
is the approximate location of the boundary
between the $\mathrm{\hbox{N\'eel}}^{\,}_{x}$ and the $\mathrm{FM}^{\,}_{z}$ 
phases along the horizontal cut
$\lambda=0.4$
in the phase diagram \ref{suppfig:phase diagram}.
The finite-size effects are strong as is evidenced
by computing the crossover between the unphysical regime
\eqref{suppeq:three regimes close to criticality c}
and the physical regime
\eqref{suppeq:three regimes close to criticality b}
as a function of the bond dimension $\chi$ used
to perform the DMRG estimate of
$|m^{x}_{\mathrm{sta}}[\Psi^{\,}_{0}]|
+
|m^{z}_{\mathrm{uni}}[\Psi^{\,}_{0}]|$ 
at $(0.57,0.4)$.
According to Fig.~\ref{suppfig:DMRG OP 0.4,0.57 chi},
when $2N=192$, the bond dimension must be larger than
$\chi^{\,}_{192}=230$
for the matrix product state selected by DMRG to have a correlation
length larger than $192\,\mathfrak{a}$.

\begin{figure}[t!]
\centering
\includegraphics[width=0.44\columnwidth]{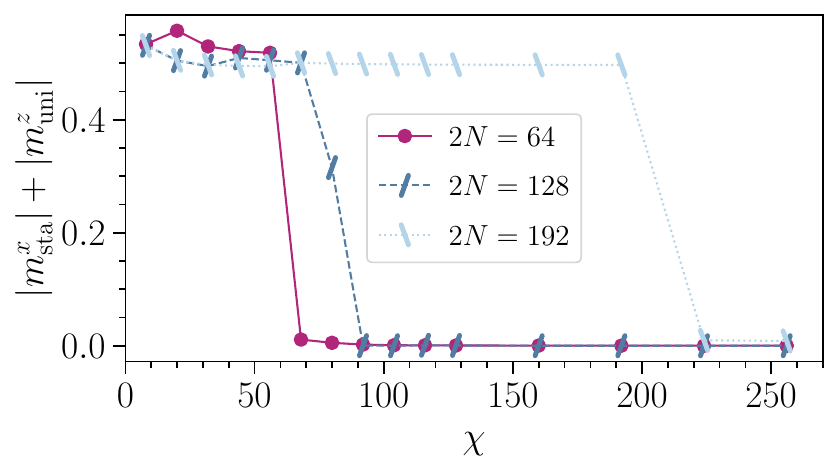}
\caption{
(Color online)
The case of OBC. DMRG estimate for the
absolute value 
$|m^{x}_{\mathrm{sta}}[\Psi^{\,}_{0}]|
+
|m^{z}_{\mathrm{uni}}[\Psi^{\,}_{0}]|$ 
of the relevant order parameters at $(0.57,0.4)$ 
at fixed system sizes $2N$ as a function of the 
bond dimension $\chi$.
}
\label{suppfig:DMRG OP 0.4,0.57 chi}
\end{figure}

\begin{figure}[t!]
\centering
\subfigure[]{\includegraphics[width=0.487\columnwidth]{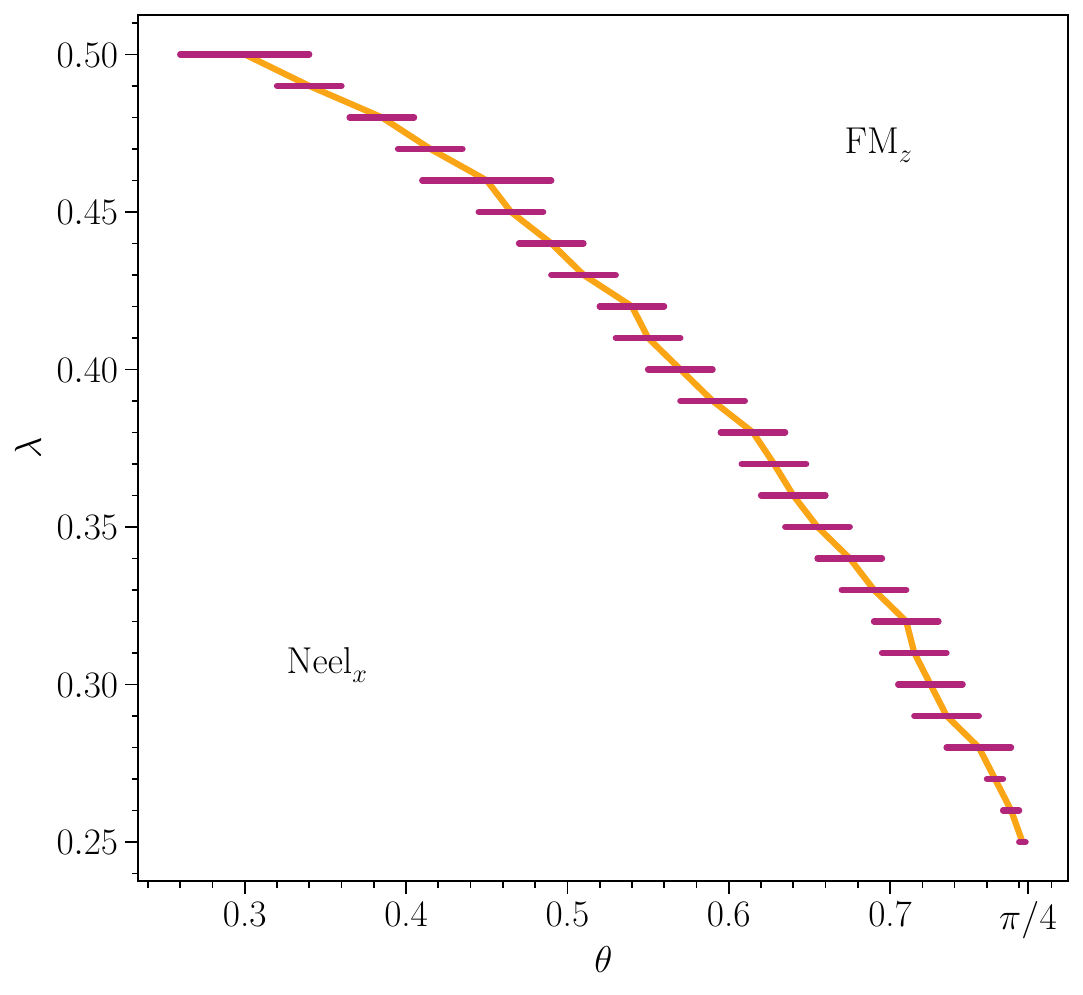} \label{suppfig:DMRG OBC OP transition from diamond a}}%
\subfigure[]{\includegraphics[width=0.5\columnwidth]{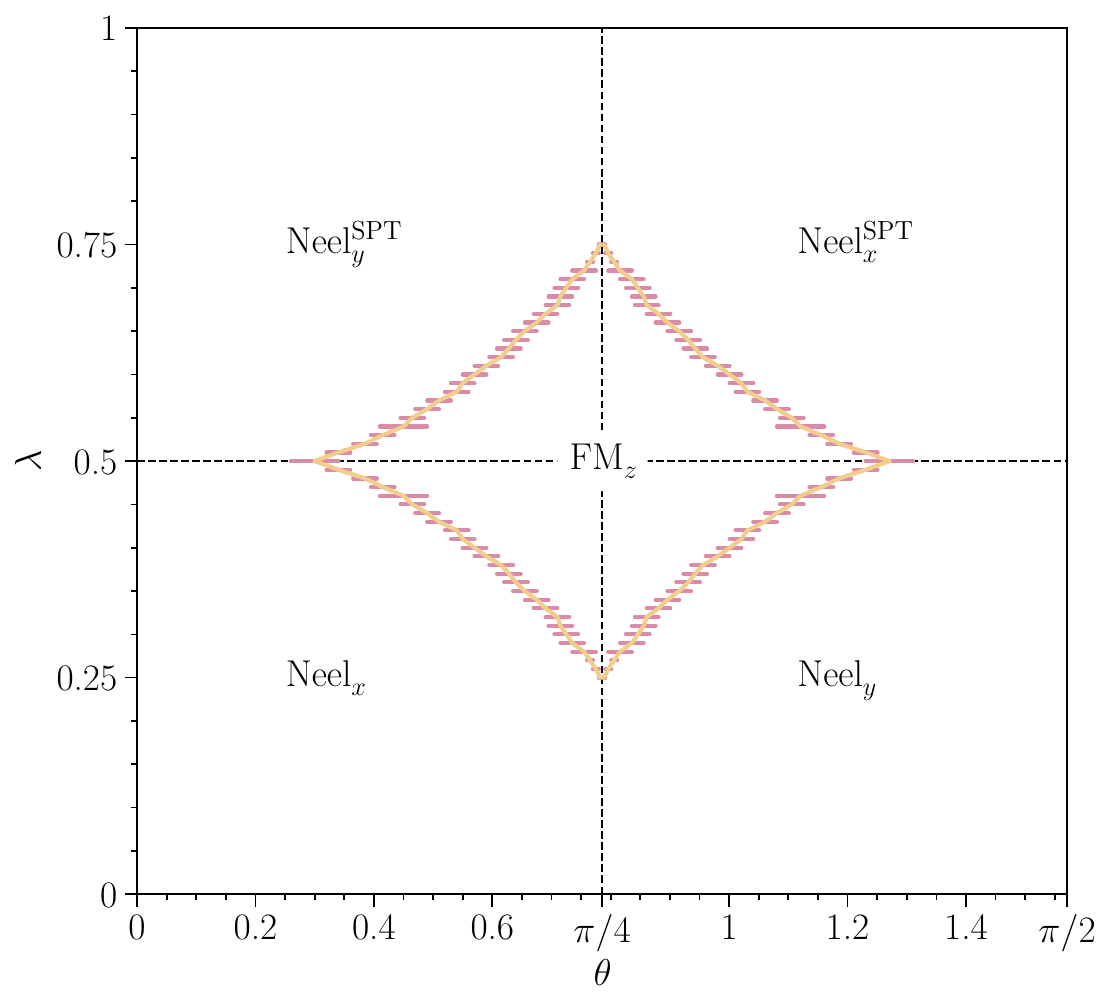} \label{suppfig:DMRG OBC OP transition from diamond b}}
\caption{
(Color online)
The case of OBC.
(a)
Phase boundary between the $\mathrm{\hbox{N\'eel}}^{\,}_{x}$ and $\mathrm{FM}^{\,}_{z}$ phases
in the lower left quadrant of Fig.~\ref{suppfig:phase diagram}.
This phase boundary is obtained using DMRG
($\chi=128$ and $2N=128$) by
locating the points in the phase diagram \ref{suppfig:phase diagram}
for which the $\mathrm{\hbox{N\'eel}}^{\,}_{x}$ and $\mathrm{FM}^{\,}_{z}$
order parameters vanish simultaneously,
as was done in Fig.~\ref{suppfig:DMRG OBC OP 0.4}. 
Panel (b) is obtained by mirroring the phase boundary in (a) 
using the operations represented by $\widehat{U}^{\,}_{\mathrm{R}}$ 
and $\widehat{U}^{\,}_{\mathrm{E}}$. The uncertainties
(horizontal bars colored in magenta)
are the sizes of the windows in $\theta$ for which
the two competing order parameters vanish as a function of $\theta$ 
holding $\lambda$ fixed
[recall Eq.~\eqref{suppeq:three regimes close to criticality b}].
}
\label{suppfig:DMRG OBC OP transition from diamond}
\end{figure}

We have measured
$|m^{x}_{\mathrm{sta}}[\Psi^{\,}_{0}]|$,
$|m^{z}_{\mathrm{uni}}[\Psi^{\,}_{0}]|$,
and
$|m^{x}_{\mathrm{sta}}[\Psi^{\,}_{0}]|
+
|m^{z}_{\mathrm{uni}}[\Psi^{\,}_{0}]|$ 
along several horizontal cuts
in the lower left quadrant of the phase diagram
\ref{suppfig:phase diagram}
to determine the location of the phase boundary
$(\theta^{\,}_{\mathrm{c}},\lambda_{\mathrm{c}})$
separating the $\mathrm{\hbox{N\'eel}}^{\,}_{x}$ from the $\mathrm{FM}^{\,}_{z}$ phase,
under the assumption that the phase transition is continuous.
The error on the value
$(\theta^{\,}_{\mathrm{c}},\lambda_{\mathrm{c}})$
is the size $2(\delta\theta^{\,}_{\mathrm{c}})^{\,}_{2N}$
of the window over which
$|m^{x}_{\mathrm{sta}}[\Psi^{\,}_{0}]|
+
|m^{z}_{\mathrm{uni}}[\Psi^{\,}_{0}]|$ 
vanishes when $2N=128$.
This phase boundary is presented in
Fig.~\ref{suppfig:DMRG OBC OP transition from diamond a}.
We have then used the symmetries
from Sec.~\ref{suppsec:Symmetries}
to deduce
in
Fig.~\ref{suppfig:DMRG OBC OP transition from diamond b}
the phase boundaries in the remaining three quadrants
of the phase diagram
\ref{suppfig:phase diagram}.

\paragraph{Degeneracies at the phase boundary ---}
The $\mathrm{FM}^{\,}_{z}$ phase shares boundaries 
with the four gapped phases in Fig.~\ref{suppfig:phase diagram}.
The phase transition from the $\mathrm{FM}^{\,}_{z}$ 
phase to any of the four gapped phases 
realizes a continuous phase transition
in the thermodynamic limit such that,
prior to taking the thermodynamic limit,
the non-degenerate ground state is
separated from the first excited state by
an energy difference that scales like $1/(2N)$
for a sufficiently large number $2N$ of sites.
This scaling is presented
in Fig.~\ref{suppfig:DMRG DQCP OBC 04,057}
at the point $(0.57,0.4)$ from the phase diagram
\ref{suppfig:phase diagram}.
As was the case with the dependence on $2N$
for the separation of two consecutive energies
deep in the FM${}_{z}$ phase presented in
Fig.~\ref{suppfig:DMRG gaps OBC 1over2,piover4},
the power law scaling $1/(2N)$
in Fig.~\ref{suppfig:DMRG DQCP OBC 04,057}
is non-monotonic.
There are oscillations in the separation of two consecutive energies
as a function of $2N$ that are made manifest in
Fig.~\ref{suppfig:DMRG DQCP OBC 04,057 a}.
These oscillations occur with the same frequency
$f^{z}[\Psi^{\,}_{0}]$ as
the oscillation frequency of the spin-spin
correlations
$C^{z}_{j^{\,}_{0},j^{\,}_{0}+s}[\Psi^{\,}_{0}]$
as a function of the separation $s$, and the
spin expectation values $\ev*{\sigma^{z}_{j}}_{\Psi^{\,}_{0}}$ as a function of
the site index $j$.
The frequency $f^{z}[\Psi^{\,}_{0}]$ is approximately $3/10$.

Since the system size $2N$ must be an integer, unmasking
these oscillations
necessitates considering the scaling with $2N$ of
$E^{\,}_{1}-E^{\,}_{0}$
for values of $2N$ that are integer multiple of 10
to account for three full periods of oscillations.
This adjustment yields an approximately linear
relationship between the non-vanishing energy spacings
and $1/(2N)$.
However, because $f^{z}\simeq 3/10$ is only an approximation,
the linearity is itself
approximate.
This is a partial explanation for the deviations from linear scaling
evident in Fig.~\ref{suppfig:DMRG DQCP OBC 04,057 b}.
A complementary explanation for the deviations from linearity
is the error committed in identifying the
critical value $\theta^{\,}_{\mathrm{c}}\approx0.57$.

Finally, finite-size effects
were too strong to study the degeneracies
in the thermodynamic limit under OBC
for any point along the phase boundary
between the $\mathrm{FM}^{\,}_{z}$ phase
and any of the four gapped  phases
in Fig.~\ref{suppfig:phase diagram},
in particular the degeneracies between
the $\mathrm{FM}^{\,}_{z}$ and the eightfold-degenerate
$\mathrm{\hbox{N\'eel}}^{\mathrm{SPT}}_{y}$
and
$\mathrm{\hbox{N\'eel}}^{\mathrm{SPT}}_{x}$
phases.

\begin{figure}[t!]
\centering
\subfigure[]{\includegraphics[width=0.38\columnwidth]{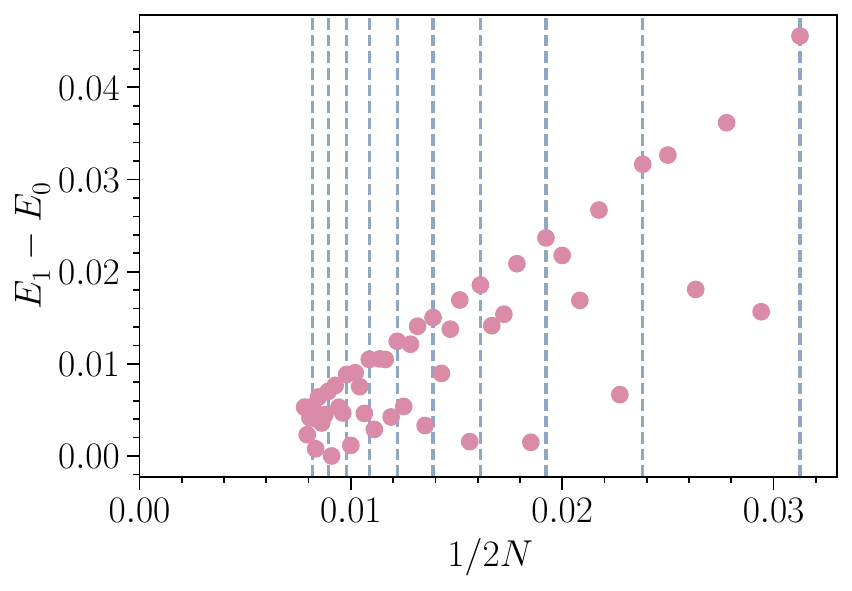} \label{suppfig:DMRG DQCP OBC 04,057 a}}%
\subfigure[]{\includegraphics[width=0.375\columnwidth]{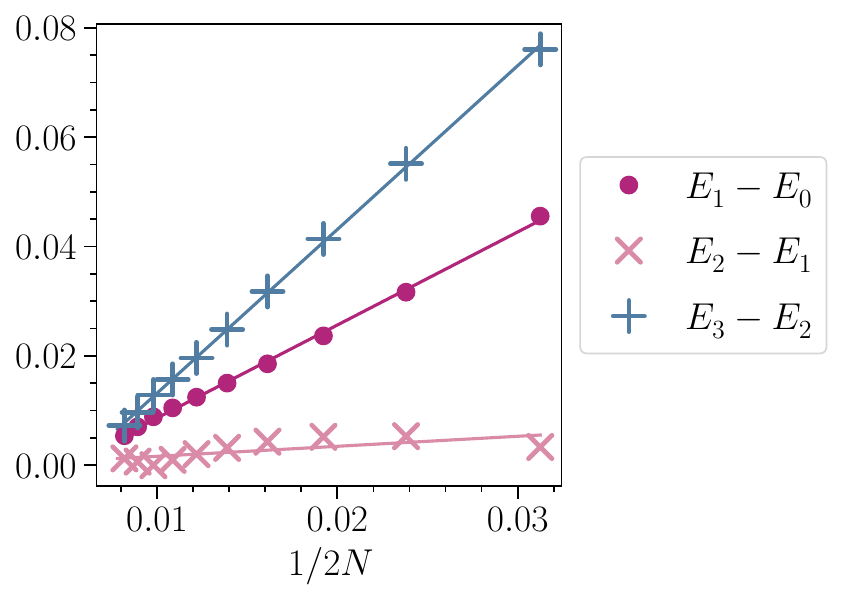} \label{suppfig:DMRG DQCP OBC 04,057 b}}
\caption{
(Color online)
The case of OBC.
(a)
Dependence on
$1/(2N)$ of the energy spacing
$E^{\,}_{1}-E^{\,}_{0}$ above the ground state energy $E^{\,}_{0}$
obtained with DMRG ($\chi=128$)
at the point
$(0.57,0.4)$ 
in the phase diagram \ref{suppfig:phase diagram}.
A non-monotonic dependence on $1/(2N)$ is visible.
(b) 
To make the oscillatory behavior with a period of $10$ in (a) explicit,
the values of $E^{\,}_{n}-E^{\,}_{n-1}$, $n=1,2,3$,
as a function of $1/(2N)$ are plotted
for $2N=32+10\,k$ with $k\in\mathbb{Z}$
in the range $[32,128]$.
Linear fits for each $n$ are represented as lines whose color
corresponds to the color of the data points in the legend.  
}
\label{suppfig:DMRG DQCP OBC 04,057}
\end{figure}

\paragraph{Correlation at the phase boundary ---}
Finite-size effects under OBC and PBC
were too strong to study the
correlation functions
for any point along the phase boundary
between the $\mathrm{FM}^{\,}_{z}$ phase 
and any of the four gapped phases in
Fig.~\ref{suppfig:phase diagram}.

\paragraph{Central charge at the phase boundary ---}
We have argued in Sec.~\ref{suppsubsec:Criticality and incommensuration of the FMz phase}
that each point inside the $\mathrm{FM}^{\,}_{z}$ phase is a 
gapless incommensurate phase that simultaneously
supports long-range 
order along the $Z$-axis in spin space
and realizes a $\mathsf{c}=1$ conformal field theory (CFT).
Finite size effects inside the $\mathrm{FM}^{\,}_{z}$ phase
did not hamper the extrapolation from the measured
$\mathsf{c}\approx 1$  to the claim that $\mathsf{c}=1$,
see Fig.~\ref{suppfig:DMRG c diamond b}.

\begin{figure}[t!]
\centering
\subfigure[]{\includegraphics[width=0.29\columnwidth]{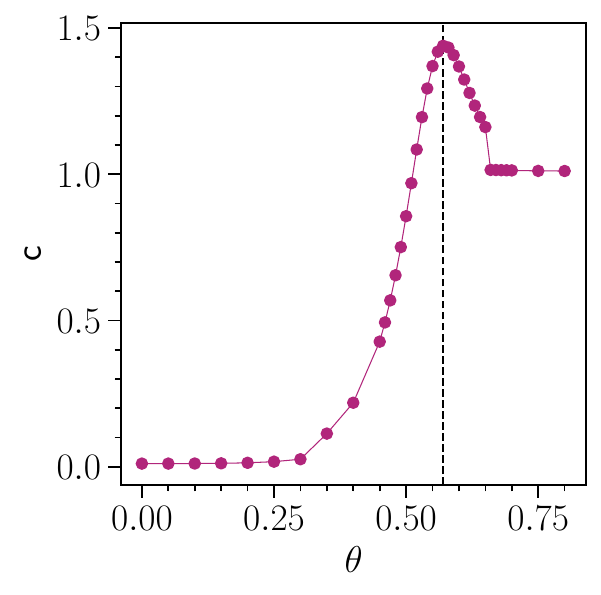} \label{suppfig:DMRG PBC central charge 0.4 a}}%
\subfigure[]{\includegraphics[width=0.29\columnwidth]{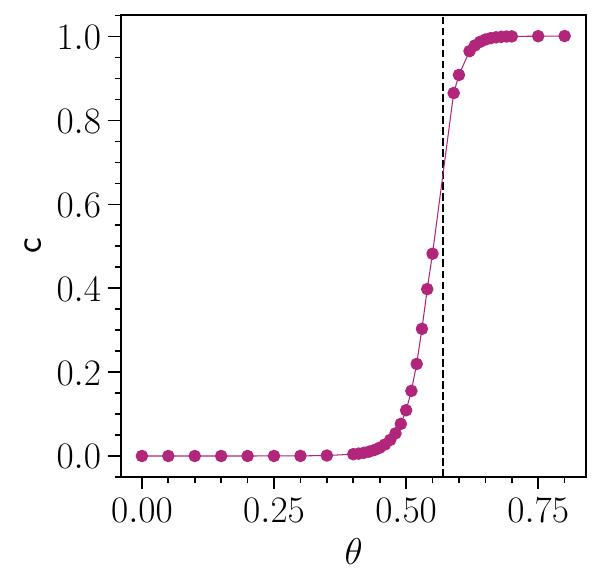} \label{suppfig:DMRG PBC central charge 0.4 b}}\\
\subfigure[]{\includegraphics[width=0.55\columnwidth]{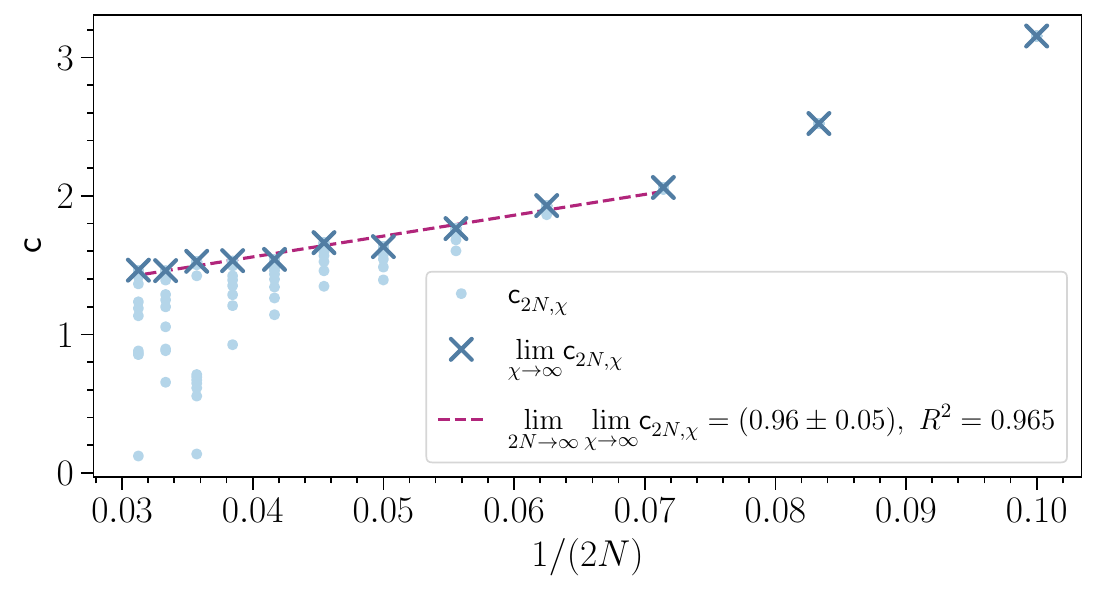} \label{suppfig:DMRG PBC central charge 0.4 c}}%
\caption{
(Color online)
The case of PBC. Dependence on $\theta$ of the DMRG ($\chi=256$)
estimate for the central charge $\mathsf{c}$
holding $\lambda=0.4$ fixed  with the number
(a) $2N=32$
and
(b) $2N=64$
of sites, respectively.
The dashed black line indicates the location of the
boundary between the $\mathrm{\hbox{N\'eel}}^{\,}_{x}$ and
the $\mathrm{FM}^{\,}_{z}$ phases, $\theta^{\,}_{\mathrm{c}}\approx0.57$.
(c) Central charge $\mathsf{c}$ obtained with 
DMRG and PBC at $(0.57,0.4)$ in
Fig.~\ref{suppfig:phase diagram}
for different 
bond dimensions $\chi\in[20,800]$
and system sizes $2N=10,12,14,\cdots,30,32$. 
An extrapolation in the limit $2N\to\infty$ 
of the central charges $\lim^{\,}_{\chi\to\infty} 
\mathsf{c}^{\,}_{2N,\chi}$ 
obtained for each size $2N$ is performed.
For all panels, the value of
$\delta{l}$ defined in Eq.~(\ref{suppeq:def delta{l}})
is $2N/2$.
}
\label{suppfig:DMRG PBC central charge 0.4}
\end{figure}

Finite-size effects using DMRG ($\chi=256$ and $2N=128$)
under PBC are much stronger at the point
$(0.57\pm0.02,0.4)$ in Fig.~\ref{suppfig:phase diagram}
on the boundary between the
$\mathrm{\hbox{N\'eel}}^{\,}_{x}$ and $\mathrm{FM}^{\,}_{z}$
phases compared to the finite effects
plaguing the estimate of the central charge using DMRG
deep inside the FM${}^{\,}_{z}$ as shown in Fig.~\ref{suppfig:DMRG c diamond b}.
If we estimate the dependence of the central charge $\mathsf{c}$
as a function of $\theta$ along the horizontal cut
$0\leq\theta\leq0.8$ with $\lambda=0.4$
in the phase diagram \ref{suppfig:phase diagram},
we find that:
\begin{enumerate}
\item[1.]
The DMRG ($\chi=256$ and $2N=32$)
estimate $\mathsf{c}^{\,}_{2N}(\theta,0.4)$ is a non-monotonic
function of $\theta$ that
overshoots by 50$\%$ the value 1 as is shown in Fig.~\ref{suppfig:DMRG PBC central charge 0.4 a}.
\item[2.]
The DMRG ($\chi=256$ and $2N=64$)
estimate $\mathsf{c}^{\,}_{2N}(\theta,0.4)$
as a function of $\theta$
is well fitted by a smooth and monotonic regularization of the
Heaviside step function
$\Theta(\theta-\theta^{\,}_{\mathrm{c}})$
with
$\theta^{\,}_{\mathrm{c}}=0.57\pm0.02$, as is shown
in Fig.~\ref{suppfig:DMRG PBC central charge 0.4 b}.
\end{enumerate}
Observe that the number of values for the central charge
$\mathsf{c}^{\,}_{2N}(\theta,0.4)$ with
$\theta$ in the vicinity of
the putative critical value $\theta^{\,}_{\mathrm{c}}\approx0.57$
is much less in Fig.~\ref{suppfig:DMRG PBC central charge 0.4 b}
than it is in Fig.~\ref{suppfig:DMRG PBC central charge 0.4 a}
due to the time needed for DMRG to converge when $2N=64$
instead of $2N=32$.
Moreover, as was the case with the analysis made with
Eq.~(\ref{suppeq:c double limit})
to interpret Fig.~\ref{suppfig:DMRG PBC central charge 1over2,0 d},
we find a strong dependence on the bond dimension $\chi$
for the values of $\mathsf{c}^{\,}_{2N,\chi}(\theta,0.4)$
when $2N$ approaches $32$ as is reported in Fig.~\ref{suppfig:DMRG PBC central charge 0.4 c}.
The extrapolation (\ref{suppeq:c double limit})
based on the system sizes $2N=10,12,\cdots,30,32$
with $\chi\in[20,800]$ gives the value
\begin{equation}
\mathsf{c}(0.57,0.4)=0.96\pm0.05.
\label{suppeq:DMRG estimate mathsf{c}(0.57,0.4)}  
\end{equation}
As was the case with the continuous
phase transition at $(0,1/2)$
in the phase diagram \ref{suppfig:phase diagram},
our computational resources do not allow a reliable extrapolation
of $\mathsf{c}^{\,}_{2N,\chi}(\theta^{\,}_{\mathrm{c}},\lambda)$
to the thermodynamic limit $2N\to\infty$
for the following reasons.
First, the choice for the range $\delta{l}$
of the values of the bonds $l$ in
Eq.~(\ref{suppeq:def S(ell) for MPS})
over which the CFT fit
(\ref{suppeq:CFT prediction for S2N(l) b})
is used [recall Eq.~(\ref{suppeq:def delta{l}})]
can be the cause of a systematic error that can be as large as 30$\%$
of the value [as opposed to 1$\%$ at the point $(\pi/4,0)$
and 3$\%$ at the point $(\pi/4,1)$]
of the value
(\ref{suppeq:DMRG estimate mathsf{c}(0.57,0.4)}).
This systematic error dominates when $\delta{l}$ is
greater than $6N/4$, while the uncertainty on the CFT fit
(\ref{suppeq:CFT prediction for S2N(l) b}) dominates when
$\delta{l}$ is smaller than $6N/4$.
Second, we could not verify that the extrapolation
(\ref{suppeq:c double limit})
based on the system sizes $2N=34,36,\cdots,62,64$
agrees with that done in 
Fig.~\ref{suppfig:DMRG PBC central charge 0.4 c}.

\subsection{Tricritical points}
\label{suppsubsec:Tricritical points}

There are four tricritical points in the phase diagram
phase diagram presented in Fig.~\ref{suppfig:phase diagram}.
Without loss of generality, in view of the symmetries
discussed in Sec.~\ref{suppsec:Symmetries},
it is sufficient to locate the pair of
tricritical points in the lower left quadrant of
Fig.~\ref{suppfig:phase diagram}.
This was done in Fig.~\ref{suppfig:DMRG OBC OP transition from diamond}
by computing the competing order parameters
that distinguish the
$\mathrm{\hbox{N\'eel}}^{\,}_{x}$
from the
$\mathrm{FM}^{\,}_{z}$ phases.
We find the estimate
\begin{equation}
(\theta^{\,}_{\mathrm{tri}},\lambda^{\,}_{\mathrm{tri}})=
(\pi/4,0.25\pm0.01)
\label{suppeq:estimate tri along theta=pi/4}  
\end{equation}
for the tricritical point along the vertical cut $\theta=\pi/4$
and the estimate
\begin{equation}
(\theta^{\,}_{\mathrm{tri}},\lambda^{\,}_{\mathrm{tri}})=
(0.3\pm0.04,1/2)
\label{suppeq:estimate tri along lambda=1/2}  
\end{equation}
for the tricritical point along the horizontal cut $\lambda=1/2$.

\begin{figure}[t!]
\centering
\subfigure[]{\includegraphics[width=0.5\columnwidth]{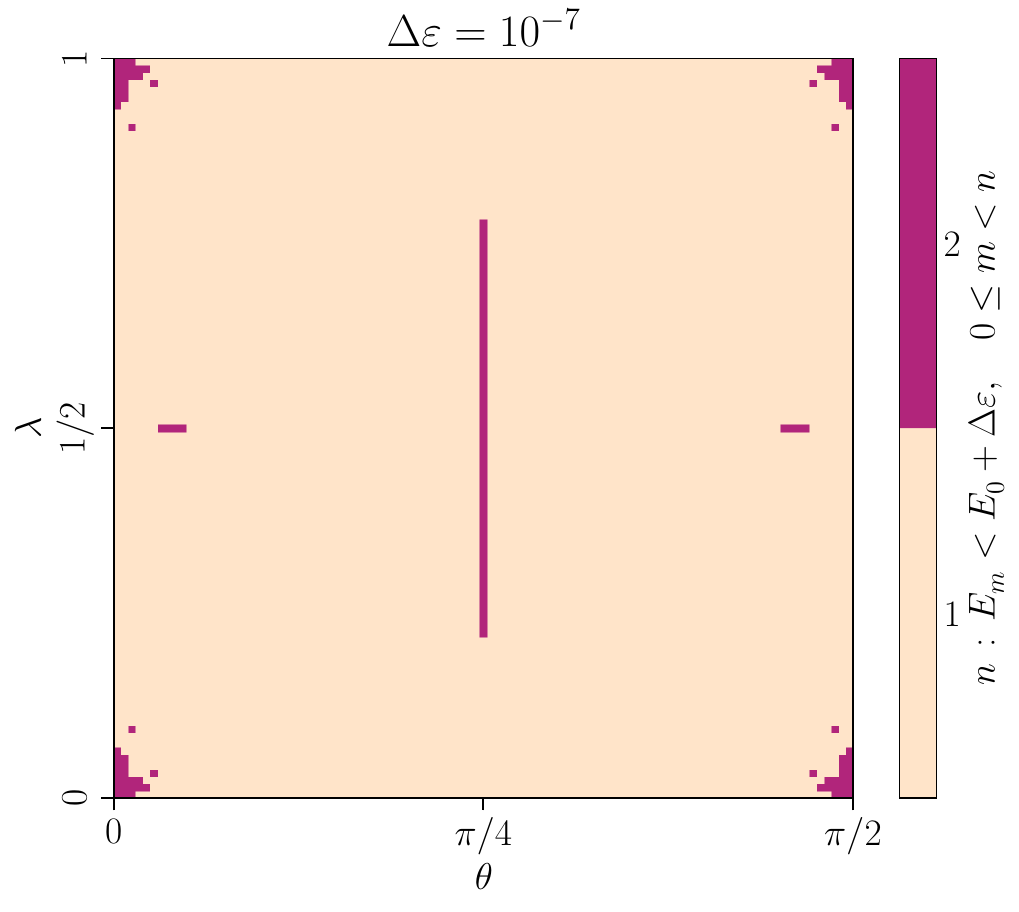} \label{suppfig:ED PBC transition a}}%
\subfigure[]{\includegraphics[width=0.32\columnwidth]{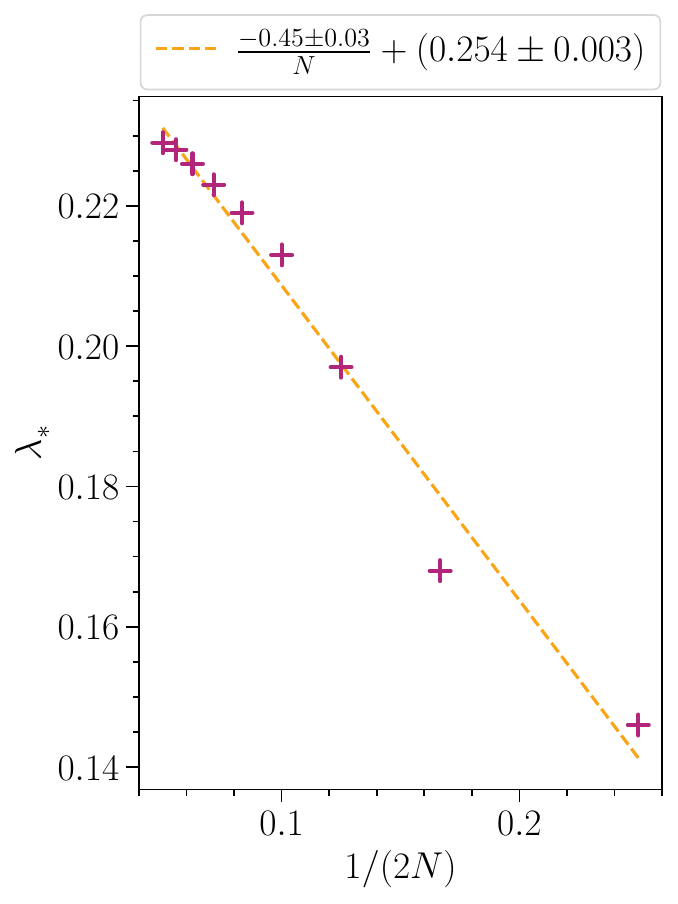} \label{suppfig:ED PBC transition b}}
\caption{
(Color online)
(a) Number $n$ of eigenenergies $E^{\,}_{m}$ 
with an energy smaller than
$E^{\,}_{0}+\Delta\varepsilon$, where $E^{\,}_{0}$ is the ground
state energy calculated using ED with PBC for $2N=10$ sites,
$\Delta\varepsilon=10^{-7}$,
and the integer $0\leq m < n$.
(b) Dependence on $1/(2N)$
with $2N$ taking the values $2N=4,\cdots,20$
of $\lambda^{\,}_{\star,2N}$ obtained with ED and PBC at 
$\theta=\pi/4$. We define $\lambda^{\,}_{\star,2N}$ for each 
$2N$ as the value of $\lambda$ at $\theta=\pi/4$
for which
the  energy gap $E^{\,}_{1}-E^{\,}_{0}$ changes from 
$E^{\,}_{1}-E^{\,}_{0}\sim10^{-1}$ to $E^{\,}_{1}-E^{\,}_{0}<10^{-7}$. 
}
\label{suppfig:ED PBC transition}
\end{figure}

The uncertainty for the tricritical point along the vertical cut
$\theta=\pi/4$ can be improved with the help of ED
as shown in Fig.~\ref{suppfig:ED PBC transition}.
For each values of $2N$ and for any point
along the the vertical cut $\theta=\pi/4$,
we obtain the partially ordered energy eigenvalues
\begin{equation}
E^{\,}_{0}<E^{\,}_{1}\leq\cdots\leq E^{\,}_{2^{2N}-1}<E^{\,}_{2^{2N}} 
\end{equation}
of Hamiltonian
(\ref{suppeq:def H}).
We define $\lambda^{\,}_{\star,2N}$ for each given number
$2N$ of sites as the value of $\lambda$ at $\theta=\pi/4$
for which the  energy gap $E^{\,}_{1}-E^{\,}_{0}$ changes from 
$E^{\,}_{1}-E^{\,}_{0}\sim10^{-1}$ to $E^{\,}_{1}-E^{\,}_{0}<10^{-7}$.
We then extrapolate the values of $\lambda^{\,}_{\star,2N}$
to its value in the thermodynamic limit $2N\to\infty$.
This procedure delivers the improved estimate
\begin{equation}
(\theta^{\,}_{\mathrm{tri}},\lambda^{\,}_{\mathrm{tri}})=
(\pi/4,0.254\pm0.003)    
\end{equation}
for the tricritical point along the vertical cut $\theta=\pi/4$.
This procedure does not improve the estimate
(\ref{suppeq:estimate tri along lambda=1/2}).

\end{document}